%%%%%%%%%%%%%%%%%%%%%%%%%%%%%%%%%%%%%%%%%%%%%%%%%%%%%%%%%%%%%%%%%%%%%%%%%%
%                                                                        %
%                                                                        % 
%                by P. Di Francesco and  E. Guitter                      %
%                TEX file, using harvmac.tex macros                      %
%                                                                        %
%                                                                        %
%%%%%%%%%%%%%%%%%%%%%%%%%%%%%%%%%%%%%%%%%%%%%%%%%%%%%%%%%%%%%%%%%%%%%%%%%%
\input harvmac 
\input epsf.tex
\input mssymb.tex

\overfullrule=0mm
\def\tvi{\vrule height 12pt depth 6pt width 0pt}
\def\tv{\tvi\vrule}

\def\file#1{#1}
\def\figbox#1#2{\epsfxsize=#1\vcenter{
\epsfbox{\file{#2}}}} 
\newcount\figno
\figno=0
\def\fig#1#2#3{
\par\begingroup\parindent=0pt\leftskip=1cm\rightskip=1cm\parindent=0pt
\baselineskip=11pt
\global\advance\figno by 1
\midinsert
\epsfxsize=#3
\centerline{\epsfbox{#2}}
\vskip 12pt
{\bf Fig.\the\figno:} #1\par
\endinsert\endgroup\par
}
\def\figlabel#1{\xdef#1{\the\figno}}
\def\encadremath#1{\vbox{\hrule\hbox{\vrule\kern8pt\vbox{\kern8pt
\hbox{$\displaystyle #1$}\kern8pt}
\kern8pt\vrule}\hrule}}

%Macros 
%%%%%%%%%%%%%%%%%%%%%%%%%%%%%%%%%%%%%%%%%%%%%%%%%%%%%%%%%%%%%%%%%

\def\IR{\relax{\rm I\kern-.18em R}}
\font\cmss=cmss10 \font\cmsss=cmss10 at 7pt

\font\cmss=cmss10 \font\cmsss=cmss10 at 7pt
\def\IZ{\relax\ifmmode\mathchoice
{\hbox{\cmss Z\kern-.4em Z}}{\hbox{\cmss Z\kern-.4em Z}}
{\lower.9pt\hbox{\cmsss Z\kern-.4em Z}}
{\lower1.2pt\hbox{\cmsss Z\kern-.4em Z}}\else{\cmss Z\kern-.4em Z}\fi}
\def\IN{\relax{\rm I\kern-.18em N}}

%%%%%%%%%%%%%%%%%%%%%%%%%%%%%%%%%%%%%%%%%%%%%%%%%%%%%%%%%%%%%%%%%

\Title{\vbox{\hsize=3.truecm \hbox{SPhT-T05/069}}}
{{\vbox {
%\centerline{}
\bigskip
\centerline{Geometrically constrained statistical systems}
\medskip
\centerline{on regular and random lattices:}
\medskip
\centerline{From folding to meanders}
}}}
\bigskip
\centerline{ 
P. Di Francesco\foot{philippe@spht.saclay.cea.fr},
E. Guitter\foot{guitter@spht.saclay.cea.fr}}
\medskip
\centerline{ \it Service de Physique Th\'eorique, CEA/DSM/SPhT}
\centerline{ \it Unit\'e de recherche associ\'ee au CNRS}
\centerline{ \it CEA/Saclay}
\centerline{ \it 91191 Gif sur Yvette Cedex, France}
\bigskip
%\vskip .5in
%abstract
We review a number a recent advances in the study of two-dimensional
statistical models with strong geometrical constraints. These include
folding problems of regular and random lattices as well as the famous
meander problem of enumerating the topologically inequivalent configurations
of a meandering road crossing a straight river through a given number of 
bridges.  All these problems turn out to have reformulations in terms of
fully packed loop models allowing for a unified Coulomb gas description
of their statistical properties. A number of exact results and physically
motivated conjectures are presented in detail, including the
remarkable meander configuration exponent 
$\alpha=(29+\sqrt{145})/12$.
%\draft
\Date{05/05}
\noindent {\bf INTRODUCTION} \leaderfill{3} \par
\noindent {\bf PART A: FOLDING OF REGULAR LATTICES} \par
\noindent {1.} {Phantom folding problems} \leaderfill{5} \par 
\noindent \quad{1.1.} {Folding: some trivial examples} \leaderfill{6} \par 
\noindent \quad{1.2.} {Compactly foldable 2D lattices} \leaderfill{12} \par 
\noindent {2.} {Folding of the triangular lattice} \leaderfill{13} \par 
\noindent \quad{2.1.} {Generalities} \leaderfill{13} \par 
\noindent \quad{2.2.} {Folding of the triangular lattice in $d=2$: an 11-vertex model} \leaderfill{15} \par 
\noindent \quad{2.3.} {Discrete folding of the triangular lattice in $d=3$: a 96-vertex model} \leaderfill{20} \par 
\noindent {\bf PART B: LOOP MODELS ON REGULAR LATTICES} \par
\noindent {3.} {Loop gas and height model reformulations of the triangular lattice folding} \leaderfill{28} \par 
\noindent \quad{3.1.} {Fully packed loop gas formulations of the 2d folding} \leaderfill{28} \par 
\noindent \quad{3.2.} {Fully packed loop gas formulations of the 3d folding} \leaderfill{30} \par 
\noindent \quad{3.3.} {Height models} \leaderfill{31} \par 
\noindent {4.} {Exact solutions via Bethe Ansatz: the example of the $FPL(1)$ model} \leaderfill{33} \par 
\noindent \quad{4.1.} {$FPL(1)$ and rhombus tiling of the plane} \leaderfill{34} \par 
\noindent \quad{4.2.} {Transfer matrix and eigenvalue equations} \leaderfill{35} \par 
\noindent \quad{4.3.} {Bethe Ansatz} \leaderfill{37} \par 
\noindent \quad{4.4.} {Continuum limit: largest eigenvalue} \leaderfill{38} \par 
\noindent \quad{4.5.} {Thermodynamic entropy and central charge} \leaderfill{39} \par 
\noindent {5.} {Effective field theory description via Coulomb gas} \leaderfill{40} \par 
\noindent \quad{5.1.} {Two-component Coulomb gas for $FPL(2)$} \leaderfill{40} \par 
\noindent \quad{5.2.} {General $FPL(n)$ and $O(n)$ cases} \leaderfill{44} \par 
\noindent {\bf PART C: LOOP MODELS ON RANDOM LATTICES} \par
\noindent {6.} {Folding of random lattices} \leaderfill{47} \par 
\noindent \quad{6.1.} {Foldability of triangulations} \leaderfill{47} \par 
\noindent \quad{6.2.} {Enumeration of foldable triangulations} \leaderfill{49} \par 
\noindent {7.} {Statistical models coupled to 2D Quantum Gravity} \leaderfill{55} \par 
\noindent {8.} {One-flavor fully packed loops} \leaderfill{57} \par 
\noindent \quad{8.1.} {Fully packed loops on random trivalent graphs} \leaderfill{57} \par 
\noindent \quad{8.2.} {Fully packed loops on random trivalent bipartite graphs} \leaderfill{60} \par 
\noindent {\bf PART D: MEANDERS} \par
\noindent {9.} {1D self-avoiding folding: Meanders} \leaderfill{66} \par 
\noindent \quad{9.1.} {The meander problem} \leaderfill{66} \par 
\noindent \quad{9.2.} {Meanders as a 1D self-avoiding folding problem} \leaderfill{67} \par 
\noindent \quad{9.3.} {A brief history of meanders} \leaderfill{68} \par 
\noindent {10.} {Solvable cases} \leaderfill{69} \par 
\noindent \quad{10.1.} {Generalization: multi-circuit meanders} \leaderfill{69} \par 
\noindent \quad{10.2.} {Combinatorial solutions at $q=\infty $, $1$ and $-1$: exact enumeration via arch statistics}\par \noindent \quad{} \leaderfill{71} \par 
\noindent \quad{10.3.} {Multi-river, multi-circuit meanders: asymptotic enumeration via matrix models} \par \noindent \quad{} \leaderfill{73} \par 
\noindent \quad{10.4.} {Meander determinant} \leaderfill{75} \par 
\noindent {11.} {Two-flavor fully packed loops: exponents of the meander problem} \leaderfill{78} \par 
\noindent \quad{11.1.} {Generalized meanders as random lattice loop models} \leaderfill{78} \par 
\noindent \quad{11.2.} {The $FPL^2$ model on the square lattice} \leaderfill{79} \par 
\noindent \quad{11.3.} {Coupling $FPL^2(n_1,n_2)$ to gravity: Effective field theory of meanders} \leaderfill{83} \par 
\noindent \quad{11.4.} {More meander exponents} \leaderfill{85} \par 
\noindent \quad{11.5.} {Multi-circuit meander exponents} \leaderfill{87} \par 
\noindent {12.} {Numerical checks} \leaderfill{88} \par 
\noindent \quad{12.1.} {Arch growth algorithm} \leaderfill{88} \par 
\noindent \quad{12.2.} {Transfer matrix algorithm} \leaderfill{91} \par 
\noindent \quad{12.3.} {Numerical results} \leaderfill{92} \par 
\noindent {13.} {Meanders as folding of random quadrangulations} \leaderfill{96} \par 
\noindent {\bf CONCLUSION} \leaderfill{103} \par

%\writetoc

%references
\nref\PF{{\it Protein Folding Handbook}, J. Buchner and T. Kiefhaber Eds.,
J. Wiley \& Sons (2005).}
\nref\SMMS{{\it Statistical mechanics of membranes and surfaces}
(2nd edition), D. Nelson, T. Piran and S. Weinberg Eds., Word Scientific 
Publishing, June 2004. }
\nref\TW{A. Lobkovsky, S. Gentges, H. Li, D. Morse and T. Witten,
{\it Scaling properties of stretching ridges in a crumpled elastic
sheet}, Science {\bf 270} (1995) 1482-1485.} 
%%%%%%%%%%%%%%%%%%%%%NOS REFERENCES PART A%%%%%%%%%%%%%%%%%%%%
\nref\GDF{P. Di Francesco and E. Guitter, {\it Entropy of folding of the 
triangular lattice}, Europhys. Lett. {\bf 26} (1994) 455-460, 
arXiv:cond-mat/9402058.}
\nref\Foltran{P. Di Francesco and E. Guitter, {\it Folding transition of 
the triangular lattice}, Phys. Rev. {\bf E 50} (1994) 4418-4426,
arXiv:cond-mat/9406041.}
\nref\BFGG{M. Bowick, P. Di Francesco, O. Golinelli and E. Guitter, 
{\it Three-dimensional folding of the triangular lattice},
Nucl. Phys. {\bf B 450} [FS] (1995) 463-494,
arXiv:cond-mat/9502063.}
\nref\QBR{P. Di Francesco, E. Guitter and S. Mori, 
{\it Folding of the triangular lattice with quenched random bending rigidity},
Phys. Rev. {\bf E 55} (1997) 237-251,
arXiv:cond-mat/9607077.}
\nref\BGGM{M. Bowick, O. Golinelli, E. Guitter and S. Mori, 
{\it Geometrical folding transitions of the triangular lattice
in the face-centered cubic lattice}, 
Nucl. Phys. {\bf B 495} [FS] (1997) 583-607,
arXiv:cond-mat/9611105.}
%%%%%%%%%%%%%%%%%%%%%%NOS REFERENCES PART C%%%%%%%%%%%%%%%%%%%%%%%%%
\nref\CRT{P. Di Francesco, B. Eynard and E. Guitter, {\it Coloring 
random triangulations}, Nucl. Phys. {\bf B 516} [FS] (1998) 543-587,
arXiv:cond-mat/9711050.}
\nref\EGK{B. Eynard, E. Guitter, C. Kristjansen, {\it Hamiltonian cycles 
on a random three-coordinate lattice}, Nucl.Phys. {\bf B 528} (1998) 523-532,
arXiv:cond-mat/9801281.}
\nref\GKN{E. Guitter, C. Kristjansen, and J. Nielsen, {\it Hamiltonian cycles 
on random Eulerian triangulations}, Nucl.Phys. {\bf B 546} (1999) 731-750,
arXiv:cond-mat/9811289.}
\nref\DGK{P. Di Francesco, E. Guitter and C. Kristjansen, {\it Fully 
packed O(n=1) model on random Eulerian triangulations}, 
Nucl.Phys. {\bf B 549} (1999) 657-667,
arXiv:cond-mat/9902082.}
%%%%%%%%%%%%%%%%%%%%%%NOS REFERENCES PART D%%%%%%%%%%%%%%%%%%%%%%%%%
\nref\MFAS{P. Di Francesco, O. Golinelli and E. Guitter, {\it Meander,
folding and arch statistics}, Mathl. Comput. Modelling {\bf 26} (1997) 97-147,
arXiv:hep-th/9506030.}
\nref\Detmean{P. Di Francesco, O. Golinelli and E. Guitter,
{\it Meanders and the Temperley-Lieb algebra},
Commun. Math. Phys. {\bf 186} (1997), 1-59,
arXiv:hep-th/9602025.}
\nref\NOUS{P. Di Francesco, O. Golinelli and E. Guitter, {\it Meanders:
a direct enumeration approach}, Nucl. Phys. {\bf B 482} [FS] (1996) 497-535,
arXiv:hep-th/9607039.}
\nref\ASY{P. Di Francesco, O. Golinelli and E. Guitter,
{\it Meanders: exact asymptotics}, Nucl.Phys. {\bf B 570} (2000) 699-712,
arXiv:cond-mat/9910453.}
\nref\DGJ{P. Di Francesco, E. Guitter and J. Jacobsen, {\it Exact meander 
asymptotics: a numerical check}, Nucl.Phys. {\bf B 580} (2000) 757-795,
arXiv:cond-mat/0003008.}
%%%%%%%%%%%%%%%%%%%%%%AUTRES REVUES%%%%%%%%%%%%%%%%%%%%%%%%%%%%%%%%%%%%%
\nref\FOLCO{P. Di Francesco, {\it Folding and coloring problems in 
mathematics and physics}, Bulletin of the AMS, Vol. {\bf 37}, 
No. {\bf 3} (2000) 251-307.}
\nref\MSRI{P. Di Francesco, {\it Matrix model combinatorics: applications
to folding and coloring}, in {\it Random matrices and their
applications}, P. Bleher, A. Its eds. (Cambridge University Press),
MSRI Publications {\bf 40} (2001), 111-170, arXiv:math-ph/9911002.}
\nref\HARM{P. Di Francesco, {\it Geometrically constrained statistical
models on fixed and random lattices: from hard squares to meanders},
arXiv:cond-mat/0211591.}
%%%%%%%%%%%%%%%%%%%%%REFS TEXTE%%%%%%%%%%%%%%%%%%%%%%%%%%%%%%%%%%%%%%%%
\nref\NP{D.R. Nelson and L. Peliti, {\it Fluctuations in membranes
with crystalline and hexatic order}, J. Physique {\bf 48} (1987) 1085-1092.}
\nref\KN{Y. Kantor and D.R. Nelson, {\it Crumpling transition
in polymerized membranes}, Phys. Rev. Lett. {\bf 58} (1987) 2774-2777
and {\it Phase transitions in flexible polymeric surfaces},
Phys. Rev.  {\bf A 36} (1987) 4020-4032.}
\nref\PKM{M. Paczuski, M. Kardar and D.R. Nelson, {\it Landau
theory of the crumpling transition}, Phys. Rev. Lett.
{\bf 60} (1988) 2638-2640.}
\nref\DG{F. David and E. Guitter, {\it Crumpling transition
in elastic membranes: renormalization group treatment},
Europhys. Lett. {\bf 5} (1988) 709-713.}
\nref\otherfold{P. Di Francesco, {\it Folding the square-diagonal lattice},
Nucl. Phys. {\bf B 525} (1998) 507-548, arXiv:cond-mat/9803051; 
{\it Folding transitions of 
the square-diagonal lattice}, Nucl Phys. {\bf B528} (1998) 609-634,
arXiv:cond-mat/9804073.}
\nref\KJ{Y. Kantor and M.V. Jari\'c, {\it Triangular lattice foldings: a
transfer matrix study}, Europhys. Lett. {\bf 11} (1990) 157-161.}
\nref\Baxt{R. J. Baxter, {\it Colorings of a hexagonal lattice}, J. Math. Phys. 
{\bf 11} (1970) 784-789.}
\nref\CGP{E. N. M. Cirillo, G. Gonnella and A. Pelizzola, 
{\it Folding transitions of the triangular lattice with defects},
Phys. Rev. E {\bf 53} (1996) 1479-1484, arXiv:hep-th/9507161.}
\nref\CIGOPE{E. N. M. Cirillo, G. Gonnella and A. Pelizzola, {\it Folding
transition of the triangular lattice in a discrete three-dimensional space},
Phys. Rev. E {\bf 53} (1996) 3253-3256, arXiv:hep-th/9512069.}
\nref\BN{ H.W.J.\ Bl\"{o}te and B.\ Nienhuis, {\it Fully packed loop
model on the honeycomb lattice}, Phys.\ Rev.
Lett. {\bf 72} (1994) 1372-1375.}
\nref\Baxbook{R. J. Baxter, {\it Exactly solved models in statistical 
mechanics}, Academic Press, London (1984).}
\nref\Nien{For an introduction to the Coulomb gas formalism, see
B.\ Nienhuis, {\it Phase transitions and critical phenomena},
Vol.\ 11, eds.\ C.\ Domb and J.L.\ Lebowitz, Academic Press (1987).}
\nref\origami{E. F. Shender, V. B. Cherepanov, P. C. W. Holdsworth, and 
A. J. Berlinsky, {\it Kagome antiferromagnet with defects: Satisfaction, 
frustration, and spin folding in a random spin system}, Phys. Rev. Lett.
{\bf 70} (1993), 3812-3815, arXiv:cond-mat/9303031.}
\nref\KGN{J. Kondev, J. de Gier and B. Nienhuis, {\it Operator spectrum
and exact exponents of the fully packed loop model}, 
J. Phys. {\bf A 29}: Math. Gen. (1996) 6489-6504, arXiv:cond-mat/9603170.}
\nref\CFT{for a general introduction to conformal field theory, see
P. Di Francesco, P. Mathieu and D. S\'en\'echal, {\it Conformal field theory},
Graduate Texts in Contemporary Physics,
Springer (1996) 1-890 (1st ed.) and Springer (1999) 1-890 (2nd ed.).}
\nref\MacM{P.A. Mac Mahon, {\it Combinatory analysis}, Vol. 2,
Cambridge University Press (1916), reprinted by Chelsea, New York (1960).}
\nref\BSY{M.T. Batchelor, J. Suzuki and C.M. Yung, 
{\it Exact results for Hamiltonian walks from the solution of the 
fully packed loop model on the honeycomb lattice}, Phys. Rev. Lett. {\bf 73}
(1994) 2646-2649, arXiv:cond-mat/9308083.}
\nref\QGRA{V. Kazakov, {\it Bilocal regularization of models of random
surfaces}, Phys. Lett. {\bf B150} (1985) 282-284; F. David, {\it Planar
diagrams, two-dimensional lattice gravity and surface models},
Nucl. Phys. {\bf B257} (1985) 45-58; J. Ambjorn, B. Durhuus and J. Fr\"ohlich, 
{\it Diseases of triangulated random surface models and possible cures}, 
Nucl. Phys. {\bf B257}(1985) 433-449; V. Kazakov, I. Kostov and A. Migdal
{\it Critical properties of randomly triangulated planar random surfaces},
Phys. Lett. {\bf B157} (1985) 295-300.} 
\nref\BIPZ{E. Br\'ezin, C. Itzykson, G. Parisi and J.-B. Zuber, {\it Planar
diagrams}, Comm. Math. Phys. {\bf 59} (1978) 35-51.}
\nref\DGZ{P. Di Francesco, P. Ginsparg
and J. Zinn--Justin, {\it 2D gravity and random matrices},
Physics Reports {\bf 254} (1995) 1-131, arXiv:hep-th/9306153.}
\nref\EY{B. Eynard, {\it Random matrices}, Saclay Lecture Notes (2000),
\hfill\break 
http://www-spht.cea.fr/lectures\_notes.shtml .}
\nref\KPZ{V.G. Knizhnik, A.M. Polyakov and A.B. Zamolodchikov, 
{\it Fractal structure of 2D quantum gravity}, Mod. Phys. Lett.
{\bf A3} (1988) 819-826; F. David, {\it Conformal field theories coupled 
to 2D gravity in the conformal gauge}, Mod. Phys. Lett. {\bf A3} (1988) 
1651-1656; J. Distler and H. Kawai, {\it Conformal field theory and 2D 
quantum gravity}, Nucl. Phys. {\bf B321} (1989) 509-527.}
\nref\Intel{This problem is discussed in the mathematical entertainments
section, edited by A. Shen, of the Mathematical Intelligencer Volume {\bf 19} 
number 4 (1997) 48.}
\nref\Tutt{W. Tutte,
%{\it A census of planar triangulations}
%Canad. Jour. of Math. 14 (1962) 21-38;
%{\it A census of Hamiltonian polygons}
%Canad. Jour. of Math. 14 (1962) 402-417;
%{\it A census of slicings}
%Canad. Jour. of Math. 14 (1962) 708-722;
{\it A census of planar maps},
Canad. Jour. of Math. 15 (1963) 249-271.}
\nref\RECT{P. Di Francesco, {\it Rectangular matrix models and 
combinatorics of colored graphs}, Nucl. Phys. {\bf B 648} (2003) 461-496,
arXiv:cond-mat/0208037.}
\nref\SCH{G. Schaeffer, {\it Bijective census and random
generation of Eulerian planar maps}, Electronic Journal of Combinatorics, 
vol. {\bf 4} (1997) R20; {\it Conjugaison d'arbres et cartes combinatoires 
al\'eatoires}, PhD Thesis, Universit\'e Bordeaux I (1998).}
\nref\CENSUS{J. Bouttier, P. Di Francesco and E. Guitter, 
{\it Census of planar maps: From the one-matrix model
solution  to a combinatorial proof}, Nucl.Phys. {\bf B645} (2002) 477-499,
arXiv:cond-mat/0207682.}
\nref\COL{J. Bouttier, P. Di Francesco and E. Guitter, 
{\it Counting colored random triangulations}, Nucl. Phys. {\bf B641[FS]} 
(2002) 519-532, arXiv:cond-mat/0206452.}
\nref\CONST{M. Bousquet-M\'elou and G. Schaeffer,
{\it Enumeration of planar constellations}, Adv. in Applied Math.,
{\bf 24} (2000) 337-368.}
\nref\EK{B. Eynard and C. Kristjansen, {\it An iterative solution of
the three-color problem on a random lattice}, Nucl. Phys. {\bf B 516} 
(1998) 529-542, arXiv:cond-mat/9710199.}
\nref\IK{I. Kostov, {\it Exact solution of the three-color problem
on a random lattice}, Phys. Lett. {\bf B 549} (2002) 245-252,
arXiv:hep-th/0005190.}
\nref\ON{I. Kostov, {\it $O(n)$ vector model on a planar random lattice:
spectrum of anomalous dimensions}, Mod. Phys. Lett. {\bf A 4} (1989) 217-226;
B. Eynard and C. Kristjansen, {\it More on the exact solution
of the $O(n)$ model on a random lattice and an investigation of the case
$\vert n \vert >2$}, Nucl.Phys. {\bf B 466} (1996) 463-487,
arXiv:hep-th/9512052.}
\nref\DK{B. Duplantier and I. Kostov, {\it Conformal spectra of polymers
on a random surface}, Phys. Rev. Lett. {\bf 61} (1988) 1433-1437.}
\nref\CDV{for a combinatorial interpretation of this formula, see
R. Cori, S. Dulucq and G. Viennot, {\it Shuffle of parenthesis systems
and Baxter permutations}, J. Combin. Theory {\bf A 43} (1986)
1-22.}
\nref\KZJ{V. Kazakov and P. Zinn-Justin, {\it Two-matrix model 
with ABAB interaction}, Nucl.Phys. {\bf B546} (1999) 647-668,
arXiv:hep-th/9808043.}
\nref\Lucas{E. Lucas {\it Th\'eorie des nombres I}, Gauthier-Villars, Paris
(1891); reedited by Blanchard, Paris, (1961).}
\nref\Saintelag{A. Sainte Lagu\"e {\it Avec des nombres et des lignes:
r\'ecr\'eations math\'ematiques}, p 147, Vuibert, Paris (1937).}
\nref\Touc{J. Touchard, {\it Contributions \`a l'\'etude du probl\`eme
des timbres poste}, Canad. J. Math. {\bf 2} (1950) 385-398.}
\nref\Koeh{J. Koehler, {\it Folding a strip of stamps}, J. Combin. 
Theory {\bf 5} (1968) 135-152.}
\nref\Lunn{W. Lunnon, {\it A map--folding problem},
Math. of Computation {\bf 22} (1968) 193-199.}
\nref\Arno{V. Arnold, {\it The branched covering of $CP_2 \to S_4$,
hyperbolicity and projective topology},
Siberian Math. Jour. {\bf 29} (1988) 717-726.}
\nref\LZ{S. Lando and A. Zvonkin, {\it Plane and projective meanders},
Theor. Comp.  Science {\bf 117} (1993) 227-241, and {\it Meanders},
Selecta Math. Sov. {\bf 11} (1992) 117-144.}
\nref\KOSMO{K.H. Ko, L. Smolinsky, {\it A combinatorial matrix in
$3$-manifold theory}, Pacific. J. Math {\bf 149} (1991) 319-336.}
\nref\HMRT{K. Hoffman, K. Mehlhorn, P. Rosenstiehl and R. Tarjan, {\it
Sorting Jordan sequences in linear time using level-linked search
trees}, Information and Control {\bf 68} (1986) 170-184.}
\nref\Phil{ A. Phillips, {\it The visual mind: Art and mathematics}, p 65,
MIT Press (1993); see also http://www.math.sunysb.edu/$\sim$tony/mazes/.}
\nref\CK{L. Chekhov and C. Kristjansen, {\it Hermitian matrix model with
plaquette interaction}, Nucl.Phys. {\bf B 479} [FS] (1996) 683-696,
arXiv:hep-th/9605013.}
\nref\MAK{Y. Makeenko, {\it Strings, matrix models, and meanders},
Nucl.Phys.Proc.Suppl. {\bf 49} (1996) 226-237, arXiv:hep-th/9512211;
G. Semenoff and R. Szabo, {\it Fermionic matrix models}, Int. J. Mod. Phys.
{\bf A12} (1997) 2135-2292, arXiv:hep-th/9605140.}
\nref\TL{H. Temperley and E. Lieb, {\it Relations between the percolation
and coloring problems and other graph-theoretical problems associated with
regular planar lattices: Some exact results for the percolation
problem}, Proc. Roy. Soc. {\bf A322} (1971) 251-280; see also the book
by P. Martin, {\it Potts models and related problems in statistical
mechanics}, World Scientific, Singapore (1991) for a review.}
\nref\MEDET{P. Di Francesco, {\it Meander determinants},
Commun. Math. Phys. {\bf 191} (1998) 543-583, arXiv:hep-th/9612026.}
\nref\SUM{P. Di Francesco, {\it SU(N) meander determinants},
Jour. Math. Phys. {\bf 38} (1997) 5905-5943, arXiv:hep-th/9702181.}
\nref\JACO{J. Jacobsen and J. Kondev, {\it Field theory of compact polymers
on the square lattice}, Nucl. Phys. {\bf B 532} [FS], (1998) 635-688,
arXiv:cond-mat/9804048.}
\nref\DCN{D. Dei Cont and B. Nienhuis, {\it The packing of two species
of polygons on the square lattice}, J. Phys. A: Math. Gen. {\bf 37}
(2004) 3085-3100, arXiv:cond-mat/0311244;
{\it Critical exponents for the $FPL^2$ model}, arXiv:cond-mat/0412018.}
\nref\JZ{J.L. Jacobsen and P. Zinn-Justin, {\it Algebraic Bethe Ansatz for
the $FPL^2$ model}, J. Phys. A: Math. Gen. {\bf 37}
(2004) 7213-7225, arXiv:math-ph/0402008.}
\nref\JKdense{J.L. Jacobsen and J. Kondev, 
{\it Transition from the compact to the dense phase of two-dimensional 
polymers}, J. Stat. Phys. {\bf 96}, (1999) 21-48, arXiv:cond-mat/9811085.}
\nref\JEN{I. Jensen, {\it Enumerations of plane meanders}, 
arXiv:cond-mat/9910313 (1999).}
\nref\POLEM{I. Jensen and A. Guttmann, {\it Critical exponents of plane 
meanders}, J. Phys. {\bf A 33} (2000) L187-L192, arXiv:cond-mat/0004321; 
I. Jensen, 
{\it A transfer matrix approach to the enumeration of
plane meanders}, J. Phys. {\bf A 33} (2000) 5953-5963, arXiv:cond-mat/0008178.}
\nref\Goli{O. Golinelli, {\it A Monte-Carlo study of meanders},
Eur. Phys. J. {\bf B14} (2000) 145-155, arXiv:cond-mat/9906329.}
\nref\PS{see for instance: D. Poulalhon and G. Schaeffer, {\it A note on
bipartite Eulerian planar maps}, http://www.loria.fr/$\sim$schaeffe/}
%%%%%%%%%%%%%%%%%%%%REFERENCES CONCLUSION%%%%%%%%%%%%%%%%%%%%%%%%%%%%%%%%
\nref\GF{D. Gaunt and M. Fisher, {\it Hard-Sphere lattice gases. I. 
Plane-square lattice}, J. Chem. Phys. {\bf 43} (1965) 2840-2863.}
\nref\RUN{L. Runnels, L. Combs and J. Salvant, {\it Exact finite 
methods of lattice statistics. II. Honeycomb-lattice gas of hard molecules}, 
J. Chem. Phys. {\bf 47} (1967) 4015-4020.}
\nref\BAXHS{R. J. Baxter, I. G. Enting and S.K. Tsang, {\it Hard square 
lattice gas}, J. Stat. Phys. {\bf 22} (1980) 465-489.}
\nref\BAXHH{R. J. Baxter, {\it Hard hexagons: Exact solution}, J. Phys. 
{\bf A 13} (1980) L61-L70; R. J. Baxter and S.K. Tsang, {\it Entropy of 
hard hexagons}, J. Phys. {\bf A 13} (1980) 1023-1030; see also
\Baxbook.}
\nref\BAX{R. J. Baxter, {\it Planar lattice gases with nearest-neighbour
exclusion}, Annals of Combin. No. {\bf 3} (1999) 191-203, 
arXiv:cond-mat/9811264.}
\nref\HARD{J. Bouttier, P. Di Francesco and E. Guitter, {\it Critical
and tricritical hard objects on bicolourable random lattices: exact solutions},
J. Phys. {\bf A 35}: Math. Gen. (2002) 3821-3854, arXiv:cond-mat/0201213.}
\nref\SZ{G. Schaeffer and P. Zinn--Justin, {\it On the asymptotic number of
plane curves and alternating knots}, Experimental Mathematics {\bf 13(4)}
(2005), 483-494, arXiv:math-ph/0304034.}
\nref\BJ{D. Bisch and V. Jones, {\it Algebras associated to
intermediate subfactors}, Inv. Math. {\bf 128} (1997) 89-157.}
\nref\LUN{W. Lunnon, {\it Multi-dimensional strip folding},
Computer J. {\bf 14} (1971) 75-79.}
\nref\CS{P. Chassaing and G. Schaeffer, {\it Random planar lattices and
integrated superBrownian excursion}, Probability Theory and Related Fields
{\bf 128(2)} (2004) 161-212, arXiv:math.CO/0205226.}
\nref\DUP{B. Duplantier, {\it Random walks and quantum gravity in two
dimensions}, Phys. Rev. Lett. {\bf 81} (1998) 5489-5492;
{\it Harmonic measure exponents for two-dimensional percolation},
Phys. Rev. Lett. {\bf 82} (1999) 3940-3943, arXiv:cond-mat/9901008.}
\nref\LSW{G. Lawler, O. Schramm and W. Werner, {\it Values of Brownian
intersection exponents I: half-plane exponents, II: Plane exponents,
III: two-sided exponents}, Acta Math {\bf 187} (2001) 237-273, 275-308,
Ann. Inst. Henri Poincar\'e PR {\bf 38} (2002) 109-123, 
arXiv:math.PR/9911084, 0003156, 0005294.}
\nref\MOB{J. Bouttier, P. Di Francesco and E. Guitter, {\it Planar maps
as labeled mobiles}, Elec. Jour. of Combinatorics {\bf 11(1)} (2004) R69,
arXiv:math.CO/0405099.}
%%%%%%%%%%%%%%%%%%%%%FIN DES REFERENCES%%%%%%%%%%%%%%%%%%%%%%%%%%%%%%%%%%

%text
\vfill\eject
\leftline{\bf INTRODUCTION}
\bigskip
The physics of folding encompasses a wide variety of topics
ranging from biology to pure mathematics. Folding problems arise for 
instance in the study of protein conformations \PF\ and in the physical 
modelling of biological or artificial membranes \SMMS. 
In a different context,
the physics of paper folding also displays interesting phenomena, in 
connection with the mechanical and statistical properties of thin elastic 
plates or membranes \TW. At a more abstract level, the very description of the 
folding degrees of freedom of a surface gives rise to very interesting 
statistical models and quite involved combinatorial problems.

In this wide subject, we choose to concentrate on this last type
of questions by focusing on statistical models of {\it discrete lattice
folding} and emphasizing the crucial role played by geometrical constraints
on the statistics of large folded objects.
More precisely, we will consider discrete models of ``solid" or ``fluid"
membranes in the form of folding problems for respectively {\it fixed}
and {\it random} lattices. An example of fixed lattice is a piece of
the two-dimensional triangular lattice, viewed as a regular collection
of rigid equilateral triangles whose edges may serve as hinges 
between neighboring triangles. An example of random lattice is a 
triangulation made of the same triangles arranged into a possibly
irregular and fluctuating surface tessellation, allowing for
curvature defects (more or less than 6 triangles around a node)
or topological defects (tessellations of a surface with handles).  
 
Our first task will be to characterize the folding degrees of freedom
in these models. This happens to be an extremely difficult problem
in the presence of self-avoidance. For this reason, we shall first
consider the much simpler ``phantom" folding problem in which 
the membrane is allowed to interpenetrate itself. A first 
description uses local normal vectors as
effective spin variables subject to various geometrical constraints. 
Although local, these constraints induce a non-local behavior with
propagating creases, resulting in a rich phase diagram even for 
regular lattices. 

A second and more powerful description uses loop configurations
on the lattice and allows to view folding as a particular instance 
of the more general problem of the ``fully packed" loop gas, that 
is a set of non-intersecting loops visiting all nodes of the lattice.
This approach, combined with field-theoretical techniques, allows for 
a number of predictions on the statistics of large folded lattices.
For regular lattices, these predictions rely either on exact
solutions via Bethe Ansatz (integrable case) or on effective
field-theoretical descriptions via Coulomb gas. 

In the case of random lattices, the loop gas interpretation leads 
in particular to precise predictions for various configurational 
exponents characterizing the asymptotics of the number of allowed 
folding or loop configurations. 
These predictions make extensive use of interpretations of the 
folding configurations in terms of discrete two-dimensional quantum gravity,
namely statistical models defined on dynamical tessellations of
surfaces. At large scales, these models are described by 
conformal field theories coupled to two-dimensional quantum gravity, 
and the configuration
exponents are obtained from a proper identification of the central
charge of the conformal field theory at hand. Alternatively, 
the discrete models of
folding may be represented via diagrammatic expansions of 
{\it matrix integrals}, allowing for a number of direct computations.

Returning to the crucial question of self-avoidance, we will
concentrate on the one-dimensional self-avoiding folding problem,
also known as the ``stamp folding problem". This problem
happens to be equivalent to a famous combinatorial problem,
the {\it meander problem} which may be stated as follows:
``enumerate all topologically inequivalent configurations
of a non-intersecting circuit crossing a straight river through a 
given number of bridges". 
Remarkably, in its meander formulation, the one-dimensional self-avoiding 
folding problem is described by a
two-dimensional gas of fully packed loops on a random lattice,
hence belongs to the class of two-dimensional phantom foldings
of random lattices. 
A remarkable outcome of our study is the predicted value for the meander 
configuration exponent
\eqn\predicme{\alpha=(29+\sqrt{145})/12}
which governs the large $n$ asymptotics of the number of meanders with
$2n$ bridges $\propto R^{2n}/n^\alpha$.

This review is based mainly on a series of articles by the authors 
[\xref\GDF-\xref\DGJ]. Other reviews may be found in 
[\xref\FOLCO-\xref\HARM]. 

\vfill\eject
\leftline{\bf PART A: FOLDING OF REGULAR LATTICES}
\bigskip
This part is devoted to the study of regular lattice folding
in the absence of self-avoidance (phantom folding).
Such foldings were considered for instance as discrete models for
polymers (1D) or for ``tethered" membranes (2D), equipped with an 
underlying regular 
elastic skeleton, such as crystalline or polymerized membranes
[\xref\NP-\xref\DG].
Our task is to first identify the folding degrees of freedom
and to finally derive explicit statistical properties of large
folded objects in the presence of bending rigidity.

\newsec{Phantom folding problems}

In all generality, we may define the {\it phantom folding} 
of a $D$-dimensional lattice in a $d$-dimensional target
space with $d\geq D$ as follows. Viewing the lattice 
as a simplicial complex with nodes, links, faces, ... 
$D$-cells, a folding is a map from the lattice to 
$\IR^d$ such that its restriction to each $D$-cell
is an isometry, i.e.\ preserves the distance between
any pair of points within the cell. The shapes of the
$D$-cells are therefore preserved in a folding, but
the relative position of two adjacent $D$-cells of the lattice
may change as long as they remain adjacent to ensure a unique
position of their common $(D-1)$-cell. In other words, the 
$(D-1)$-cell serves as a ``hinge" between them.  
Such a folding is phantom in the sense that we allow cells 
to interpenetrate one-another, as well as distinct points of the
lattice to occupy the same target position. A phantom folding 
is therefore defined by the corresponding folded configuration, 
without reference to an actual folding process which would
continuously deform the original lattice.  
All the foldings considered in parts A-C below will be phantom. Moreover,
we will restrict ourselves to the cases $D=1$, $d$ arbitrary or 
$D=2$ and $d=2$ or $3$. We are mainly interested in the statistical
properties of folded configurations of large portions of
lattices for various ensembles corresponding to possibly
restricted target spaces within $\IR^d$. For instance, we
may consider {\it continuous foldings} in which folding ``angles"
between adjacent $D$-cells vary continuously with specific
distributions, or {\it discrete foldings} in which these angles
only take finitely many values. For a good choice of these
discrete angles, this simply corresponds to discretizing
the target space into a $d$-dimensional lattice.

\subsec{Folding: some trivial examples}

In this Section, we review two particularly simple folding
problems, namely that of a 1D lattice in any dimension, and
that of the 2D square lattice in $d=2$.
\bigskip
\noindent $\diamond$ \underbar{\sl 1D folding}
\par\nobreak
\fig{Folding of a segment of 1D lattice in either continuous or
discrete target space. The associated bending energy takes the form
$-K\cos\theta$ for each pair of consecutive links with relative angle
$\theta$. An external force $F$ may also be exerted at the 
ends.}{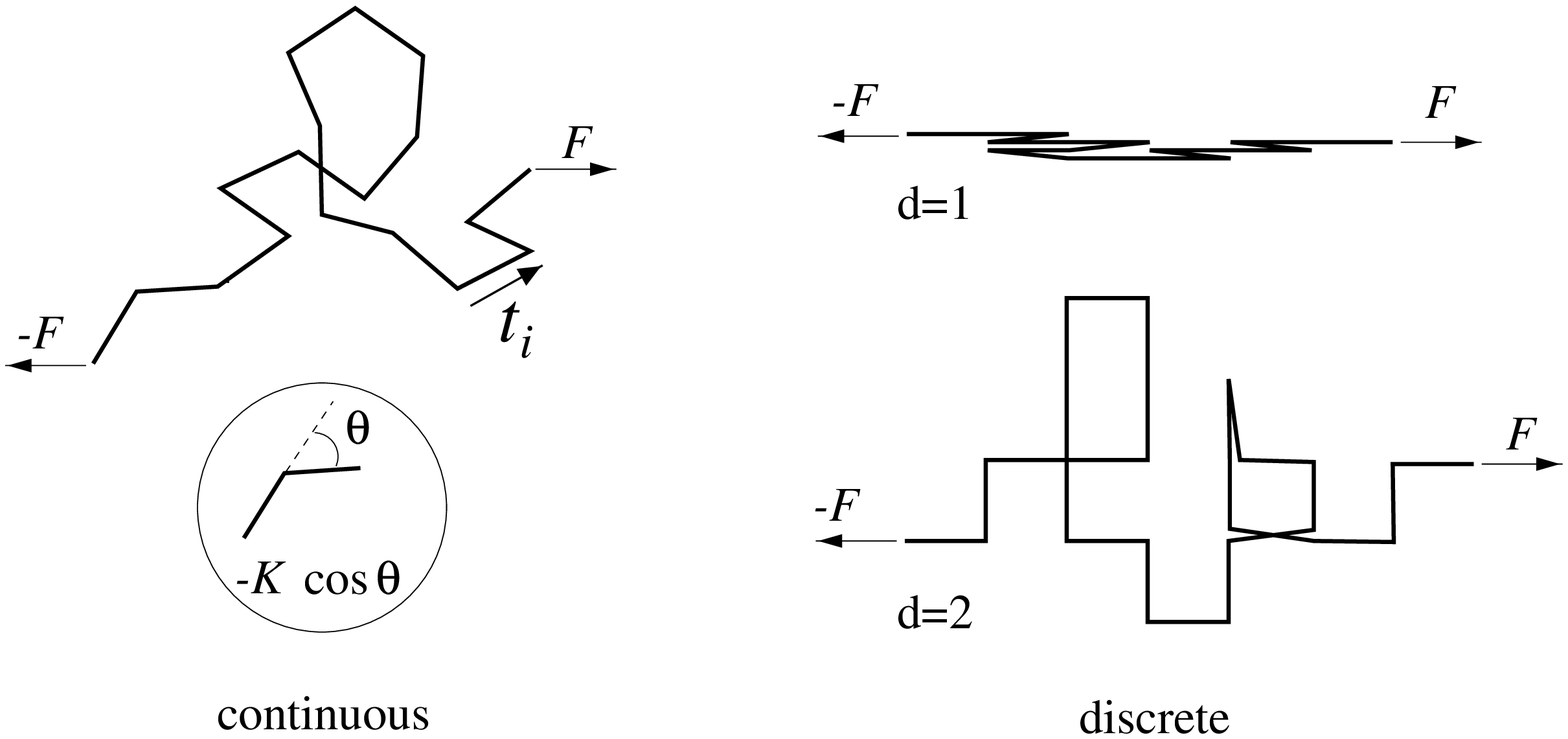}{12.cm}
\figlabel\oneDfolding
As a preliminary exercise, let us study the case $D=1$ of 
the phantom folding of a segment of length, say $L$ of the
lattice of relative integers $\IZ$. Viewing this segment of
lattice as made of a chain of $L+1$ nodes connected by
$L$ oriented links of unit length, a folding in $\IR^d$ is entirely
characterized, up to a global translation, by the sequence 
$\{t_i\}_{1\leq i\leq L}$ of 
``tangent" vectors, all of unit length, which are the images 
of the links after folding, themselves connected into a possibly
overlapping chain.  
We decide to attach to each folded configuration a weight
\eqn\energ{w(\{t_i\})\equiv \exp (-E(\{t_i\}))}
where the energy of the configuration reads
\eqn\enercon{E(\{t_i\})=- K \sum_{i=1}^{L-1} t_{i}\cdot t_{i+1} - 
F\cdot \sum_{i=1}^{L} t_i}
corresponding to a ``bending" energy with bending rigidity $K$ and an
external force $F$ between the extremities of the chain.
We may consider several statistical ensembles for the $t_i$'s.
Natural choices are (see Fig.\oneDfolding):
\item{(a)} A continuum equidistributed set of unit vectors $t_i$ in $\IR^d$.
\item{(b)} A discrete equidistributed set with
$t_i\in\{\pm e_1,\pm e_2,\cdots,\pm e_d\}$
where $\{e_i\}_{1\leq i\leq d}$ is the canonical basis of $\IR^d$. 
This choice alternatively
amounts to requiring that the nodes of the chain have their folded
images on that of a $d$-dimensional hypercubic lattice.
\par
Let us start with $d=1$, in which case only the discrete ensemble (b)
above makes sense. The model with tangent vectors $t_i=\pm 1$
and weights \energ\-\enercon\ is nothing but the 1D Ising model with
spin coupling $K$ and magnetic field $F$. The partition function
on a segment of length $L$ and with free boundary conditions reads:
\eqn\partIsing{\eqalign{Z_L(K,F) & = 
\left(\cosh{F}+{e^{-K}+e^{K}\sinh^2(F)\over \Delta}\right)
(e^{K}\cosh(F)+\Delta)^{L-1} \cr &\ \ \ +
\left(\cosh{F}-{e^{-K}+e^{K}\sinh^2(F)
\over \Delta} \right) (e^{K}\cosh(F)-\Delta)^{L-1}\cr}}
with $\Delta=\sqrt{e^{-2K}+e^{2K}\sinh^2(F)}$. We may define the
order parameter ${\cal M}$ as the average orientation of the links through 
${\cal M}=(1/L)\sum_i t_i$, which is nothing but the magnetization
in the Ising language, with expectation value $M\equiv \langle {\cal M}\rangle
=(1/L)\partial_F {\rm Log}\, Z_L(K,F)$. For large $L$ and small $F$, 
we easily get
\eqn\magnet{ M \sim e^{2K} F}
In particular, we recover the well known (1D Ising) result that there
is no spontaneous orientation, i.e.\ $M=0$
for $F\to 0$ for all finite values of $K$, expressing that the
chain is always ``crumpled" in the absence of external force. 
Still a ``flattening" transition occurs at $K=+\infty$ at which 
$M ={\rm sign}(F)=\pm 1$ for $F\to 0$, corresponding 
to a fully extended chain. 

This result is qualitatively the same for all target dimensions $d$.
For instance, if the target space is the $d$-dimensional hypercubic
lattice (case (b) above), we get an average orientation ${\cal M}$ in the
direction of the force $F$ with expectation value:
\eqn\magnetd{M \sim {d-1+e^{K}\over d( d-1+e^{-K})} F}
for small $F$, which displays the same flattening transition at 
$K=+\infty$.

Similarly, in the continuous case (a) above, we get
\eqn\magnetcont{M \sim {1+g(K)\over d(
1-g(K))} F}
where $g(K)=\partial_K {\rm Log}(\int d^dt \exp(K t\cdot e))$ 
for any fixed unit vector $e$. For instance, when $d=3$, we have
$g(K)=\coth(K)-1/K$. As $g(K)$ tends to $1$ only at $K=+\infty$,
we find again a flattening transition at $K=+\infty$.

\bigskip
\noindent $\diamond$ \underbar{\sl Folding of the square lattice 
in $d=2$}
\par\nobreak
\fig{The 2d phantom folding of a piece of square lattice factorizes
into the two 1D foldings of its south and west 
boundaries.}{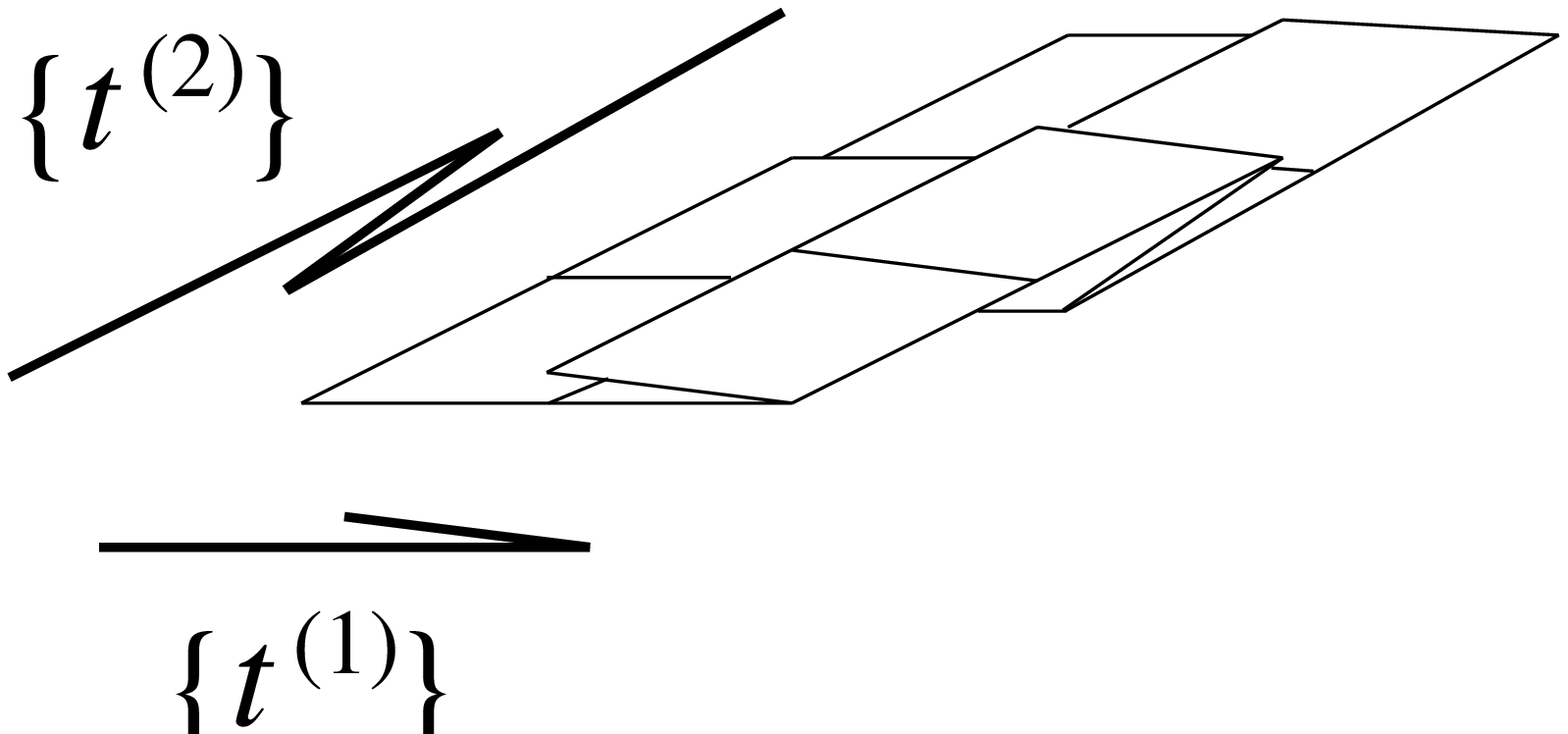}{8.cm}
\figlabel\pliagesq
Another particularly simple example is that of phantom folding
of a rectangular piece, say of size $L_1\times L_2$ of the 2D 
square lattice into a plane target. The image of each square cell 
is a unit square in the plane, entirely specified by the data of the two 
unit and orthogonal 
``tangent" vectors $t^{(1)}$ and $t^{(2)}$, images of the south and west 
boundary links of the cell, pointing from their common origin (SW node). 
It is easily seen 
that preserving the shape and connectivity of the square cells implies
that, up to a global rotation, $t^{(1)}$ may take only the
two values, say $\pm e_1$ while $t^{(2)}=\pm e_2$, and that moreover
the value of $t^{(1)}$ (resp. $t^{(2)}$) is {\it the same} for all
south (horizontal) oriented links in a column (resp. all 
west - vertical - oriented links in a row).
In other words, the folding constraints impose that the creases
must propagate along straight lines all the way across the lattice.
Each folded state is therefore entirely determined by the
south (resp. west) 1D tangent vectors along the southern
(resp. western) border of the $L_1\times L_2$ rectangle, namely
by two independent sets of Ising spins $\{t^{(1)}_i\}_{1\leq i\leq L_1}$
and $\{t^{(2)}_j\}_{1\leq j\leq L_2}$ (see Fig.\pliagesq).
In particular, the total number of folded states is $2^{L_1+L_2}$,
leading to a thermodynamic entropy per square of 
$s=\lim_{L_1,L_2\to \infty}1/(L_1L_2){\rm Log}(2^{L_1+L_2})=0$.
The above factorization also allows to write the bending energy of the
entire rectangle in terms of the spins as
\eqn\enertwoD{E_b=-K L_2 \sum_{i=1}^{L_1-1}t^{(1)}_i t^{(1)}_{i+1}
-K L_1 \sum_{j=1}^{L_2-1}t^{(2)}_j t^{(2)}_{j+1}}
As in the 1D case, we may also introduce a symmetry breaking
field $F$ coupled to the ``orientation" of the folded state. 
For all 2D folding problems and for $d=2$ or $3$, it is convenient 
to couple $F$ to the {\it projected area} on a given plane of
the target space\foot{Note that for a 1D chain, 
the coupling to the ``projected length" on a given line is equivalent
to a force exerted at the ends of the chain in the direction of
this line.}. For $d=2$, this amounts to adding an energy
\eqn\enered{E_p= - F \sum_{i=1}^{L_1}\sum_{j=1}^{L_2} t^{(1)}_i t^{(2)}_j
= -F S_1 S_2}
where $S_a=\sum_{l=1}^{L_a} t^{(a)}_l$ is the total (algebraic)
projected length in the direction $a$. 
The partition function may be written as
\eqn\partsq{Z_{L_1,L_2}(K,F)=
\sum_{S_1=-L_1\atop S_1=L_1\,{\rm mod}\,2}^{L_1}
\sum_{S_2=-L_2\atop S_2=L_2\,{\rm mod}\,2}^{L_2}
e^{ F S_1 S_2} {\hat Z}_{L_1}(K L_2, S_1) {\hat Z}_{L_2}(K L_1, S_2)}
where ${\hat Z}_{L}(K,S)$ is the coefficient of $e^{F\,S}$ in
$Z_L(K,F)$ as given by Eq.\partIsing, corresponding to the contribution
of total projected length $S$ to the 1D partition function.
In the thermodynamic limit, it is convenient to introduce
the free energy per square {\it at fixed projected lengths} 
$S_a=\sigma_a L_a$ 
\eqn\freeenerg{f(\sigma_1,\sigma_2)=\lim_{L_1,L_2\to \infty}
-{1\over L_1 L_2} {\rm Log}[e^{ F L_1 L_2 \sigma_1 \sigma_2} 
{\hat Z}_{L_1}(K L_2, \sigma_1 L_1) {\hat Z}_{L_2}(K L_1, \sigma_2 L_2)]}
as well as the total free energy per square 
\eqn\frenerg{f=\lim_{L_1,L_2\to \infty}
-{1\over L_1 L_2} {\rm Log}[Z_{L_1,L_2}(K,F)]=\min_{\sigma_1,\sigma_2\in
[-1,1]} f(\sigma_1,\sigma_2)}
\fig{A typical configuration (a) minimizing or (b) maximizing 
the number of folds at fixed projected length $S_1$. In case (a),
the number of folds is at most one while in case (b),
it is given by $L_1-|S_1|-\delta_{S_1,0}$.}{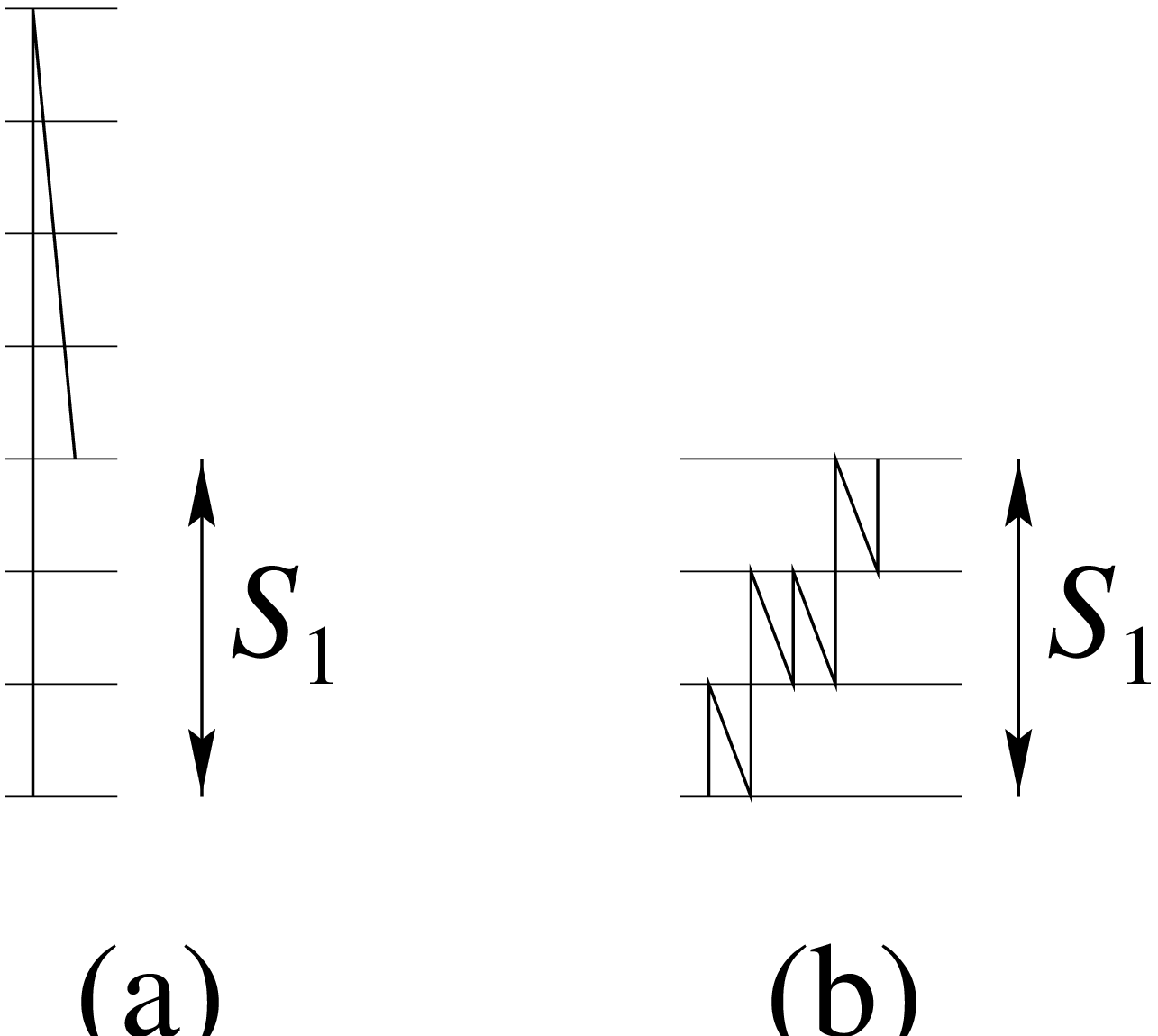}{7.cm}
\figlabel\confmax
\fig{Phase diagram of the phantom folding of the square lattice in 
2d, in the plane $(K,F)$ of bending rigidity $K$ and symmetry breaking
field $F$. First order transition lines (thick lines) separate
three regions with constant average projected area per square $M=0$, $+1$, or
$-1$. In each of these phases, the lattice is frozen in a unique
state, either completely flat ($M=\pm1$) or completely folded 
($M=0$).}{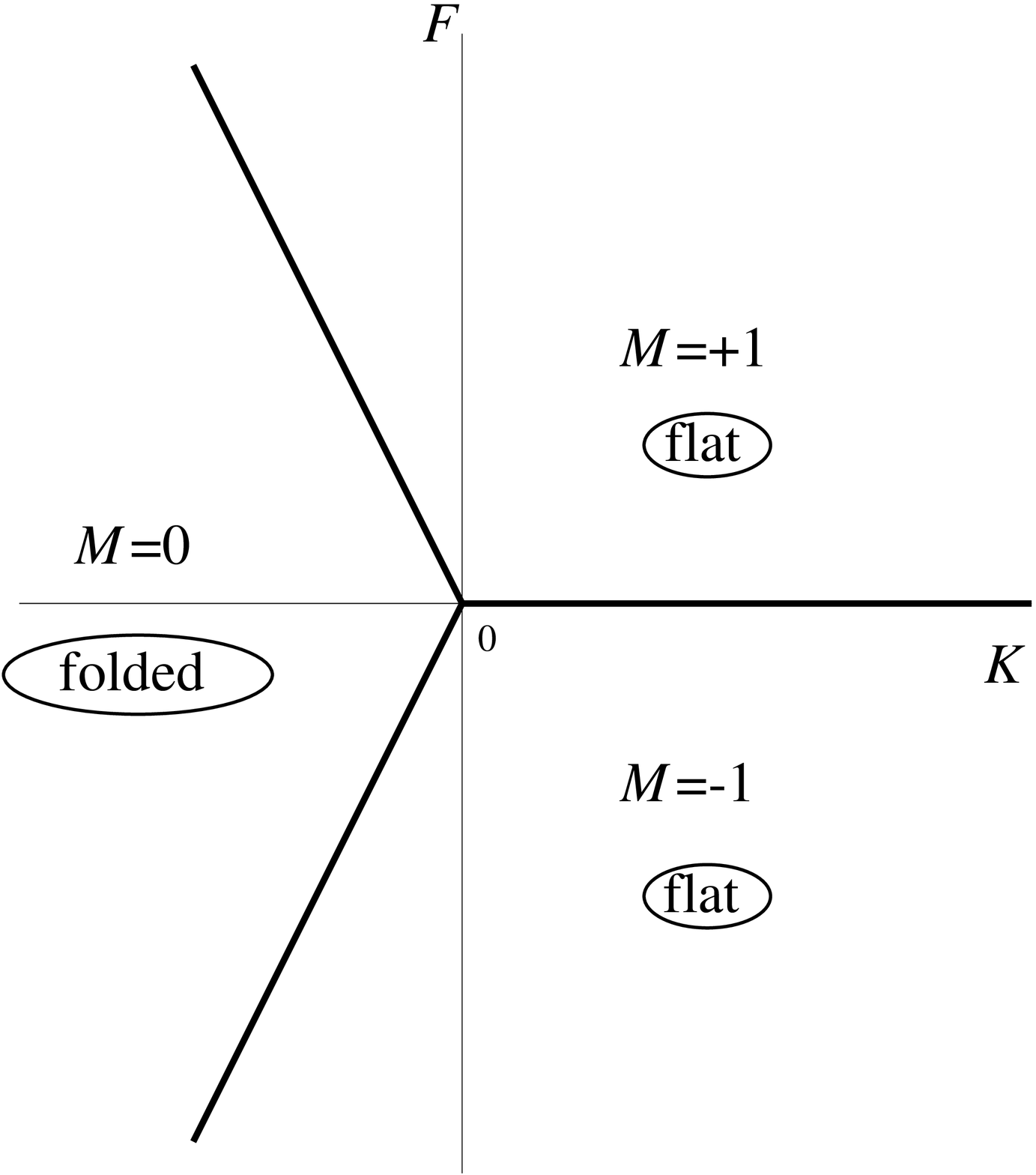}{8.cm}
\figlabel\phasediagsq
\noindent For large $L_1$ and $L_2$, the effective couplings $KL_1$ and $KL_2$
in \partsq\ tend to infinity.
For positive $K$, ${\hat Z}_{L_1}(KL_2,S_1=\sigma_1 L_1)$ 
is then dominated by two (symmetric) configurations with at most one fold in the
direction $1$ (see Fig.\confmax), hence, in the thermodynamic limit, 
we have
\eqn\fplus{f(\sigma_1,\sigma_2)=-2K -F\sigma_1\sigma_2}
The minimum of this free energy is attained at $\sigma_1= {\rm sign}(F)
\sigma_2=\pm 1$ expressing the complete flatness of the lattice in the 
thermodynamic limit for any $F$. The corresponding total free
energy reads
\eqn\ftotplus{f=-2K -|F|}

For negative $K$, ${\hat Z}_{L_1}(KL_2,S_1)$ is dominated by
the configurations which maximize the number of folds in 
the direction $1$ (see Fig.\confmax), hence with exactly 
$L_1-|S_1|-\delta_{S_1,0}$ such folds,
contributing by an energy $K(1-2|\sigma_1|)$ and by an 
entropy $0$ in the thermodynamic limit (the number of 
these configurations is clearly bounded by $2^{L_1}$).
This leads to a free energy 
\eqn\fmoins{f(\sigma_1,\sigma_2)=2K(1-|\sigma_1|-|\sigma_2|) 
-F\sigma_1\sigma_2}
The minimum of this free energy is now attained at $\sigma_1=\sigma_2=0$ for
$|F|<-4 K$, and at $\sigma_1={\rm sign}(F) \sigma_2= \pm 1$ otherwise,
leading to 
\eqn\ftotmoins{f=\min(-2K -|F|, 2K)}
For $K<0$ and in the thermodynamic limit, the lattice is 
therefore totally flat for $|F|>-4 K$, and folded for 
$|F|<-4 K$. More precisely, the folded phase is dominated by a
single configuration with maximal number of folds in both directions,
i.e.\ corresponds to the pure state of the completely folded lattice.
The transition across the lines $|F|=-4K$ is
first order, as well as that across $F=0$ for $K>0$ (see Fig.\phasediagsq). 
Note that, as opposed to the 1D case, the flattening transition now occurs
at $K=0$.

Although it may seem at first rather
pathological, the square lattice planar folding captures a number
of essential features of regular lattice folding: 
\item{-} the non-local nature of the creases which propagate throughout
the lattice;
\item{-} the consequent absence of local excitations of the pure
flat state;
\item{-} the existence of a first order transition separating a
completely flat phase made of this single flat state from a 
crumpled phase.
\par

\subsec{Compactly foldable 2D lattices}
\fig{The classification of two-dimensional compactly foldable
lattices.}{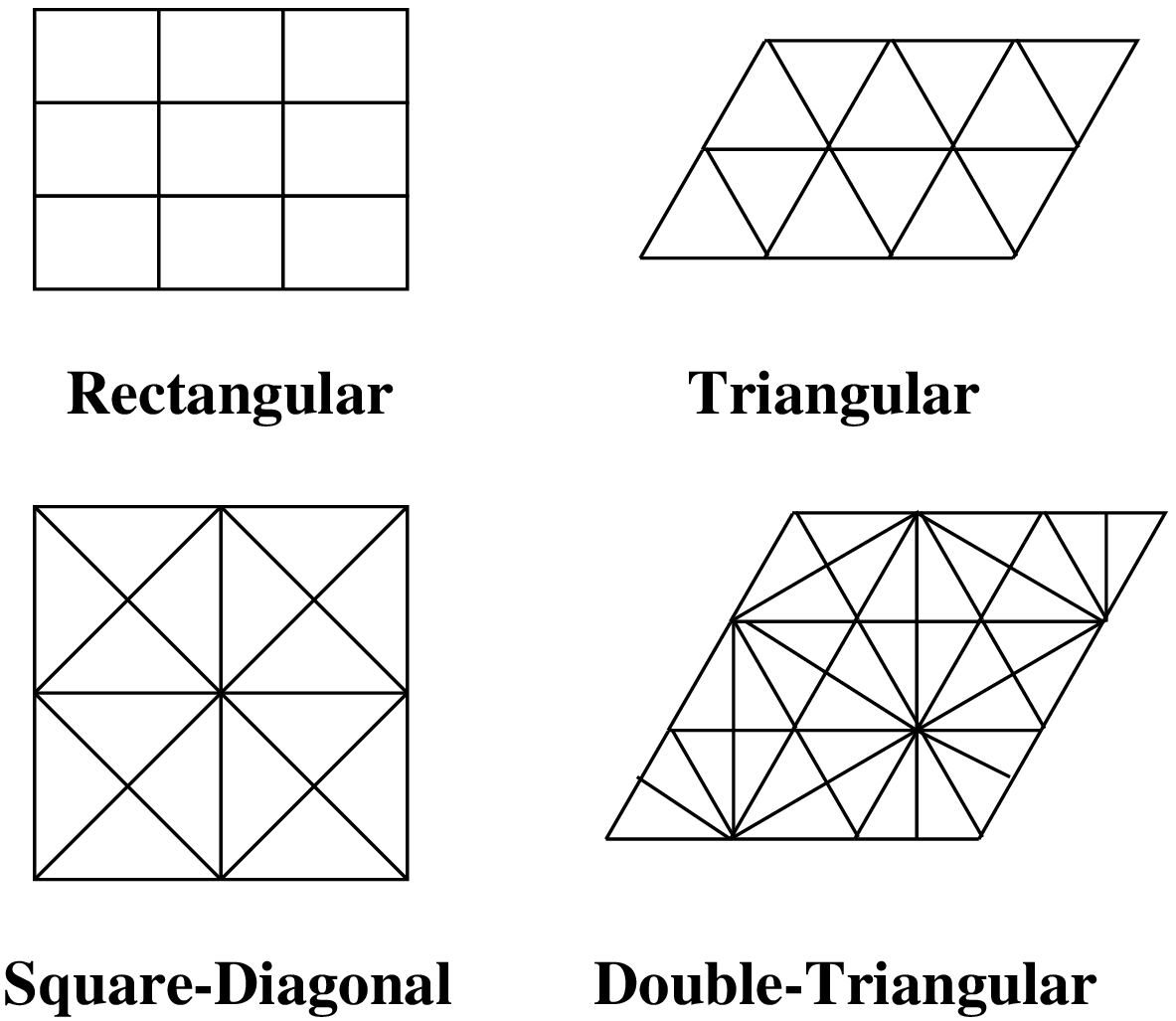}{9.cm}
\figlabel\classi

The details of folding are strongly lattice dependent. A good
criterion for choosing the lattices to be folded is that they
give rise to sufficiently many folded configurations. This is 
usually a prerequisite for allowing interesting (geometrical) 
phase transitions. For 2D lattices, a basic requirement is to demand that the 
lattice might be completely folded onto one of its faces. 
Such a lattice will be called {\it compactly foldable}. This restricts very 
strongly the possible lattices: indeed, the corresponding
lattice must have only one type of face, together with its finitely 
many possible rotations and reflections.
This gives rise to a classification of the two-dimensional 
compactly foldable lattices: they fall into the four cases 
depicted in Fig.\classi, namely {\it rectangular}, {\it triangular}, 
{\it square-diagonal} and {\it double-triangular}.

The proof of this classification goes as follows \FOLCO. Let us concentrate 
on a node of the lattice. Each adjacent edge may serve as a hinge 
in the folding of its two adjacent faces, hence bisects the wedge made by its two 
neighboring edges around the node. Moreover, as the normal vectors alternate
between neighboring faces in the completely folded state, the
number of faces adjacent to every node is even, as is the
number of edges adjacent to any node.
Each node $v$ is therefore the center of a regular star of 
say $2m_v$ edges forming angles of $\pi/m_v$, $m_v\geq 2$,
hence those angles are either right or acute. The faces are therefore polygons
with at most four edges, and they can have four only if they are rectangles.
This is the case where all $m_v=2$, the rectangular lattice.
Otherwise, all faces must be triangular, with right or acute angles.
Such a face has angles $\pi/m_1,\pi/m_2,\pi/m_3$, with $m_i\geq 2$, and
$\sum 1/m_i=1$. There are only three solutions up to permutation
for $(m_1,m_2,m_3)$ namely
\eqn\solo{ \eqalign{ (3,3,3)& \to \ {\rm Triangular} \cr
(2,4,4)& \to {\rm Square-Diagonal} \cr
(2,3,6)& \to {\rm Double-Triangular} \cr}}
This completes the proof.

It is clear that the folding of the rectangular lattice is equivalent
to that of the square lattice, in particular its 2d folding is trivial
as discussed in the previous Section, with a vanishing 
thermodynamic entropy of folding. As it turns out,
the three other cases share
the property of having a {\it non-zero} thermodynamic entropy of folding
per face [\xref\GDF,\xref\otherfold].  
In the following Sections, we will concentrate on the case of the
triangular lattice, which we will fold both in $d=2$ and $3$.

\newsec{Folding of the triangular lattice}

\subsec{Generalities}

In this Section, we address the question of the folding of the triangular 
lattice. This is the simplest non trivial example of regular
lattice folding and it captures all the generic properties of 
compactly foldable lattices. Another advantage of this model is
that it bears links to some exactly solvable integrable model.
Very generally, a folded configuration of the triangle lattice 
will consist in keeping the triangles equilateral of unit side, while
the edges serve as {\it hinges} between neighboring
triangles which may a priori form arbitrary angles.

In the following, we will study two particular examples of folding:

\item{(a)} The planar folding, i.e the case $d=2$.

\item{(b)} The folding on the Face Centered Cubic (FCC) lattice
corresponding to $d=3$ and suitably defined folding angles.

The foldings of the triangular lattice may be characterized
equivalently by link variables, corresponding to
tangent vectors, or face variables, corresponding to normal 
vectors.
\bigskip
\noindent $\diamond$ \underbar{\sl Link variables}
\par\nobreak
\fig{Orientation of the links of the triangular lattice. The
link ($t_\ell$) and face ($n_f$) variables.}{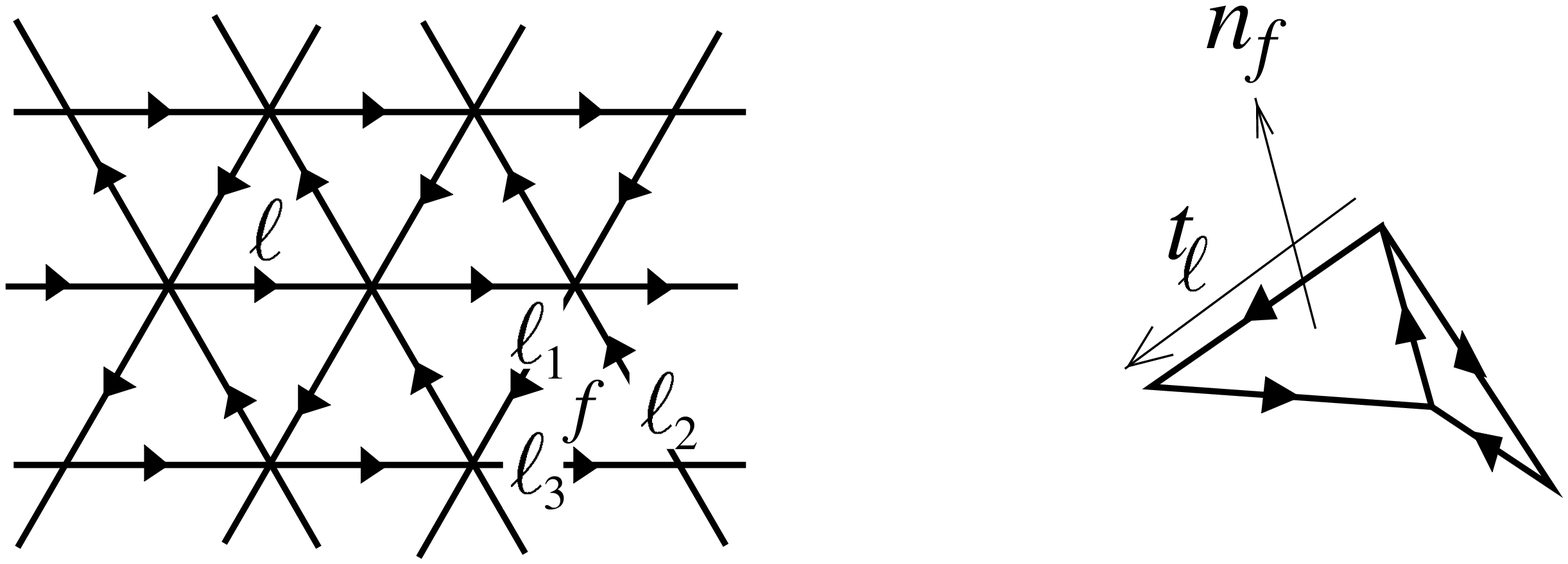}
{10.cm}
\figlabel\latticesem
We first start by {\it orienting} all the links of the triangular
lattice in a coherent way so that all triangles pointing up are
oriented counterclockwise and all triangles pointing down are
oriented clockwise (see Fig.\latticesem).
To each oriented link $\ell$, we associate its image 
$t_\ell$ in the folded configuration. To ensure that
any elementary triangle has for its image an
equilateral triangle of side $1$ in $\IR^d$,
it is sufficient to require that all $t_\ell$'s are unit vectors,
and that moreover
\eqn\constlien{t_{\ell_1}+t_{\ell_2}+t_{\ell_3}=0}
for the three oriented links $\ell_1$, $\ell_2$ and $\ell_3$ around each
triangle.
\bigskip
\noindent $\diamond$ \underbar{\sl Face variables}
\par\nobreak
We can also describe a folding by face variables by considering
the normal vector to any triangle $f$. For $d\leq 3$, it reads
\eqn\normal{n_f\equiv \epsilon{2\over \sqrt{3}}\  t_{\ell_1}\wedge t_{\ell_2}}
where $\ell_1$ and $\ell_2$ are two consecutive oriented links around
$f$ and with $\epsilon=+1$ (resp. $-1$) for triangles
pointing up (resp. down) on the lattice.

It is natural to think of this normal vector as a spin variable.
The introduction of a bending energy amounts to a spin-like coupling $K$ 
between normals of neighboring faces. As we shall see in the following,
such energy is responsible for various conformational transitions.

As for link variables, the face variables are not independent
on each triangle. The six normal vectors for the
six triangles around a node are correlated via a vertex-constraint.
A more precise formulation of this constraint  will be given
in the case (a) of the planar folding, where it leads to an
``$11$-vertex model", and in the case (b) of the folding on the FCC lattice,
where it leads to a ``$96$-vertex model".

\subsec{Folding of the triangular lattice in $d=2$: an 11-vertex model}

\fig{The $11$ possible folding environments of an elementary hexagon,
corresponding (from top to bottom) to : no fold, $1$ fold (3 configurations),
$2$ folds (6 configurations) or $3$ folds. The pictorial representation
on the left is only intended as a guide for the reader and only the
final (phantom) folded state is relevant. The folds are indicated
on the right by thick lines.}{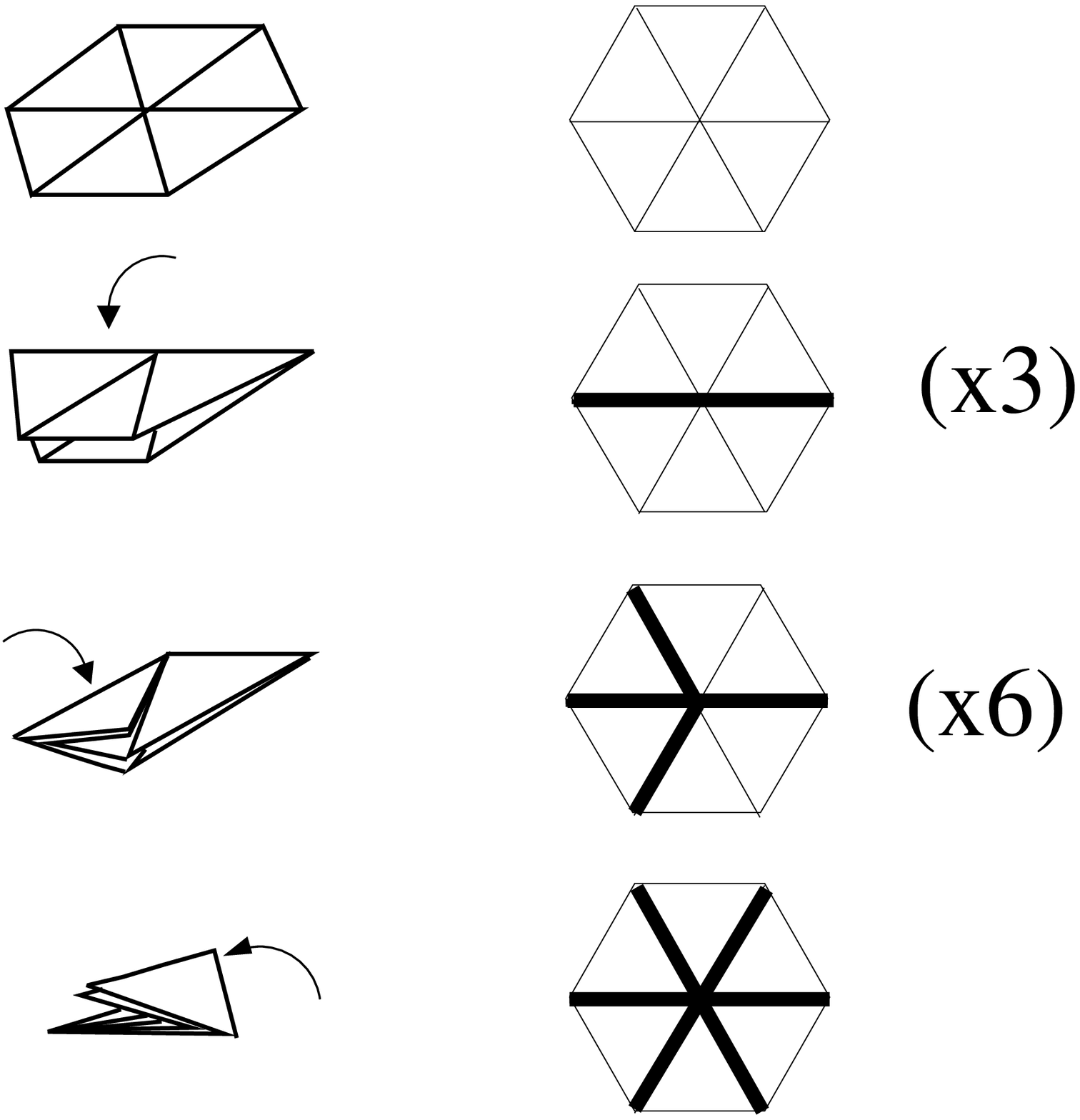}{8.cm}
\figlabel\orig

The simplest case of folding, first introduced by Kantor 
and Jari\'c \KJ, corresponds to the planar folding, 
i.e.\ the folding in the plane $\IR^2$.
In this case, the angle between two neighboring triangles\foot{We will
take here as definition of the angle between two triangles that between
their normal vectors.} is either $0^\circ$
(no fold), or $180^\circ$ (complete fold). Folding an elementary hexagon leads 
to the $11$ possible environments represented in Fig.\orig.
We recall that our definition of phantom folding is limited to the
{\it image} of the lattice and does not distinguish
between the different ways of reaching the same
configuration. One can easily convince oneself that all folding
constraints reduce to those of Fig.\orig\ around each
elementary hexagon. This will become clear below in the language 
of tangent vectors.
The problem of planar folding is therefore entirely characterized by the 
local rules
of Fig.\orig\ on elementary hexagons.
As such, planar folding is an $11$-vertex model.
\bigskip
\noindent$\diamond$ \underbar{\sl Face variables: constrained Ising spins}
\par\nobreak
\fig{The 11 possible spin environments (up to a global reversing of the spins)
around a node. In all cases, the number of $+1$ (resp. $-1$) spins is a multiple
of $3$.}{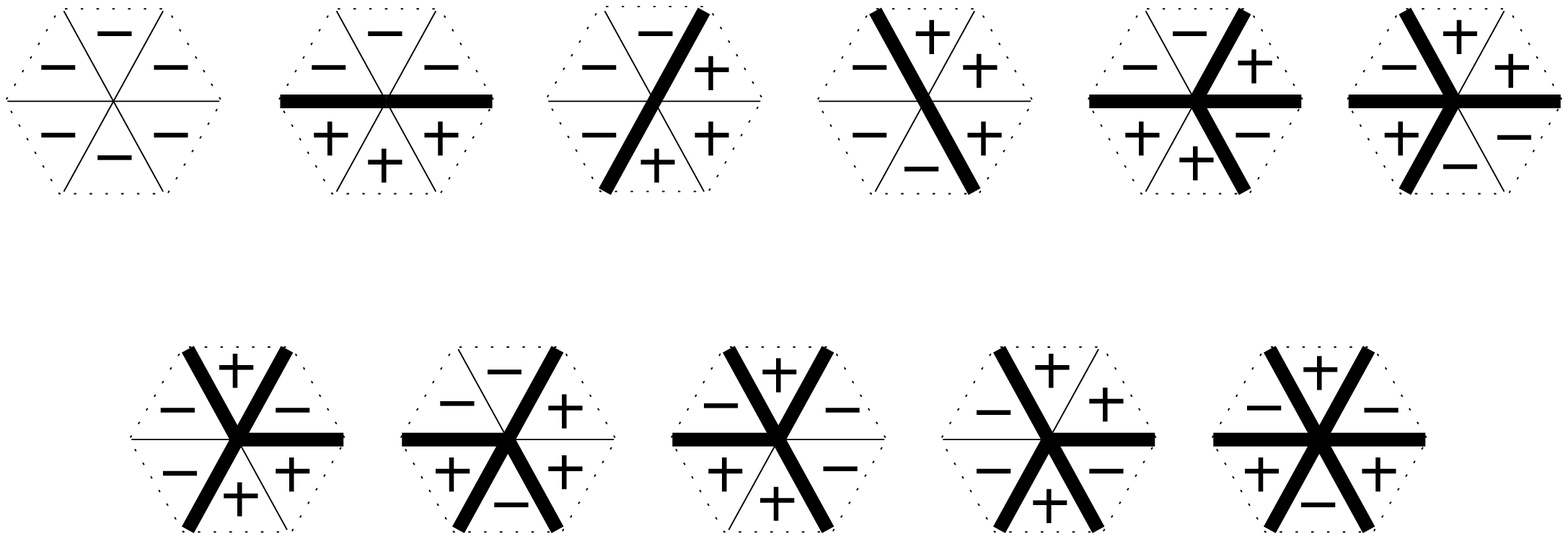}{12.cm}
\figlabel\spins
The normal vector to each triangle may be described by an Ising spin
which may take only two values $\sigma=+1$ or $\sigma=-1$ according
to whether the triangle has been flipped or not in the folding.
The folds are nothing but the domain walls for the spin variable.
The $11$ environments of Fig.\orig\ correspond to the $11$ possible
environments in terms of spins (up to a global reversal of the spins)
indicated in Fig.\spins. A simple
characterization of the allowed spin configurations \GDF\ is
that the number of spins $+1$ must be equal to $0$, $3$ or $6$,
i.e.\ must be a multiple of $3$, or equivalently:
\eqn\spinrule{\sum_{i=1}^6 \sigma_i= 0\quad  {\rm mod}\quad 3}
for the six spins around each node.
One easily checks that the number of such configurations is
${6 \choose 0}+{6\choose 3}+{6\choose 6}=22=2\times 11$ as it should.
\bigskip
\noindent$\diamond$ \underbar{\sl Link-variables: the three-color model}
\par\nobreak
A natural question concerns the computation of the folding {\it entropy}
which characterizes the exponential growth of the number $Z_N$ of
possible foldings as a function of the number $N$ of triangles (for
a finite sub-lattice of the triangular lattice) . One  defines the
folding entropy $s$ by\foot{In this definition, one expects that the
actual shape of the sub-lattice should not matter as long as the two dimensions
scale in the same way and provided we take free boundary conditions.}:
\eqn\entrop{s\equiv\lim_{N \to +\infty} {1 \over N} {\rm Log}\ Z_N  \equiv
{\rm Log} \ q \ .}
\fig{Starting from the completely folded configuration, we may
reverse the $6$ spins of some elementary hexagons
(here in grey) and still get an allowed configuration
provided the hexagons do not overlap (hard hexagons).}{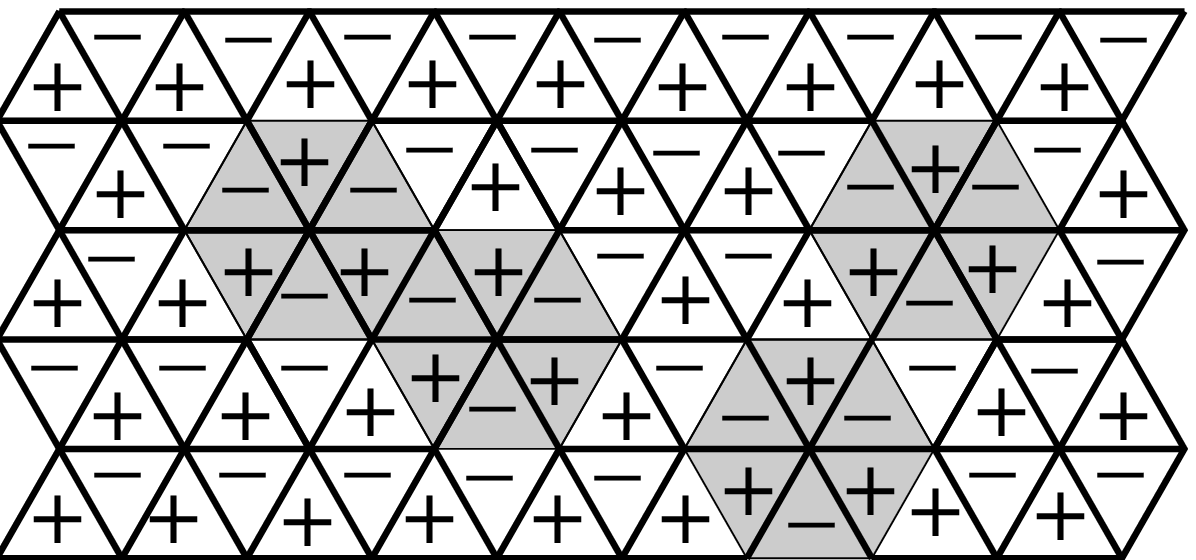}{8.cm}
\figlabel\hexagonbis
One can interpret $q$ as the number of degrees of freedom
per triangle, with clearly, for constrained Ising spins,
 $1\leq q \leq 2$. The existence of a non-vanishing entropy $s$ (i.e.\ $q>1$)
is not a priori obvious as, like in the square lattice case, 
there exists no configuration for which the folds are localized.
A simple way of seeing this is to cut each
configuration of Fig.\spins\ vertically in the middle and to notice
that whenever a fold is present on the left half of the hexagon,
then a fold is also present on the right side.
The folds therefore propagate all the way from and to infinity from
the left to the right.
In other words, despite the locality of the rules of Fig.\orig, 
the flat configuration (without fold) has no local excitation.
Numerically, Kantor and Jari\'c have shown \KJ\ that $q\sim 1.21$
and that there is therefore a non-vanishing folding entropy, as
opposed to the square lattice case. This result
may be understood by starting instead from the completely folded state,
with a perfect anti-ferromagnetic order for the spins. We may then reverse
globally all the spins of one single elementary hexagon and the configuration
remains admissible (see Fig.\hexagonbis). This operation may be
repeated for arbitrarily many hexagons as long as they do not overlap. 
There exist therefore
local excitations of the completely folded configuration, leading 
to a non-vanishing entropy.

\fig{The three allowed values (colors) for $t_\ell$
are $A$, $B={\cal R} A$ and $C={\cal R}^{-1} A$.
The relation between link variables $t_i$ and the face variables 
$\sigma_i=\pm 1$ around a node reads
 $t_{i+1}={\cal R}^{\sigma_i} t_i$.
The link variable is well defined if $t_7=t_1$, that is
if  \spinrule\ is satisfied.}
{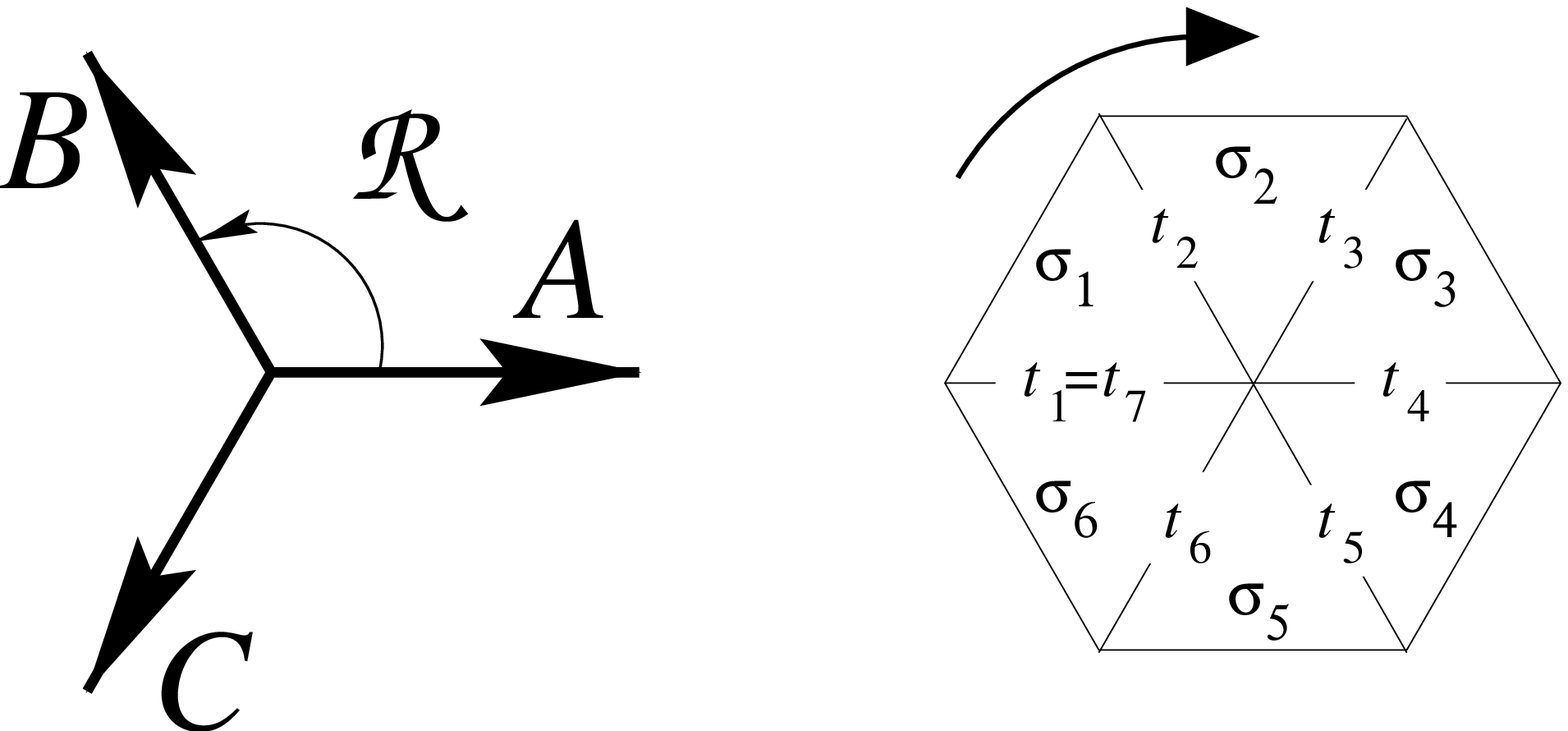}{10.cm}
\figlabel\spincol

The entropy may be computed exactly in the 
link-variable picture \GDF. Starting from a given triangle,
the vectors $t_{\ell_i}$, $i=1,2,3$ around this triangle are three unit
vectors in $\IR^2$ with vanishing sum, therefore of the form
$t_{\ell_1}=A$, $t_{\ell_2}=B\equiv {\cal R}A$ and $t_{\ell_3}=
C\equiv {\cal R}^2 A={\cal R}^{-1} A$ where $A$ is a unit vector and 
${\cal R}$ the rotation of angle $120^\circ$ (or $-120^\circ$).
One can then easily convince oneself that the vectors
$t_\ell$ for {\it all} the links of the lattice may only take
these same three values $A$, $B$ or $C$, with moreover the constraint that
the three ``colors" A, B and C must be present once and only once
around each triangle \GDF. A planar folding of the triangular lattice
is therefore equivalent to a {\it tricoloring} of the links of the lattice
with three colors which must be distinct around each triangle.
The color simply represents the orientation of the link in the
folded configuration, among three allowed values.

In this language, the spin of a face tells us about the cyclic order
of the colors around this face with $\sigma=+1$ (resp. $\sigma=-1$)
if the colors are in the order A,B,C (resp. A,C,B) in counterclockwise 
direction.
In particular (see Fig.\spincol), the variables $t_{i}$, $i=1,\cdots,6$
of the links around a node are related to the spins $\sigma_i$
by $t_{i+1}={\cal R}^{\sigma_i}t_{i}$.
As ${\cal R}^3=1$, the link variable is well defined after one turn around
the node if and only if the relation \spinrule\ is satisfied.
In this language, all folding constraints ensuring that
the link variable is well defined along any closed loop clearly reduce
to Eq.\spinrule\ around each elementary hexagon.

In the dual language, a planar folding therefore corresponds to the coloring
of the links of the {\it hexagonal lattice} with three colors, such that no two
adjacent links be of the same color.
The entropy of this three-color problem was obtained by
Baxter \Baxt, and can therefore be reinterpreted as the
planar folding entropy per triangle of the triangular lattice, 
with the  result:
\eqn\valq{q= \prod_{n=1}^{\infty} { (3n-1) \over \sqrt{ 3n(3n-2)} }=
{\sqrt{\Gamma(1/3)} \over \Gamma(2/3) }= {\sqrt{3} \over 2 \pi}
\Gamma(1/3)^{3/2}}
Its numerical value $q=1.208717...$ ($s={\rm Log} q=0.189560...$) is 
in very good agreement with the estimate of Ref.\KJ.
\fig{Phase diagram for the planar folding in the presence
of a bending energy term $E_b=-K \cos(\theta)$ and a 
symmetry breaking field $F$.
A first order transition at $K_{\rm c}\sim 0.1$ separates the crumpled phase
where the average projected area $M\equiv \langle \sigma \rangle =0$ 
from a flat phase
where $M=1$ (or $-1$). This latter phase is frozen in the pure state
of the completely flat surface. Within the crumpled phase,
a continuous piling-up transition now occurs at $K_{\rm p}\sim -0.28$ 
[\xref\QBR,\xref\BGGM,\xref\CGP] between a disordered phase with
$M_{\rm st}=0$ and an ordered phase with $M_{\rm st}\neq 0$.}
{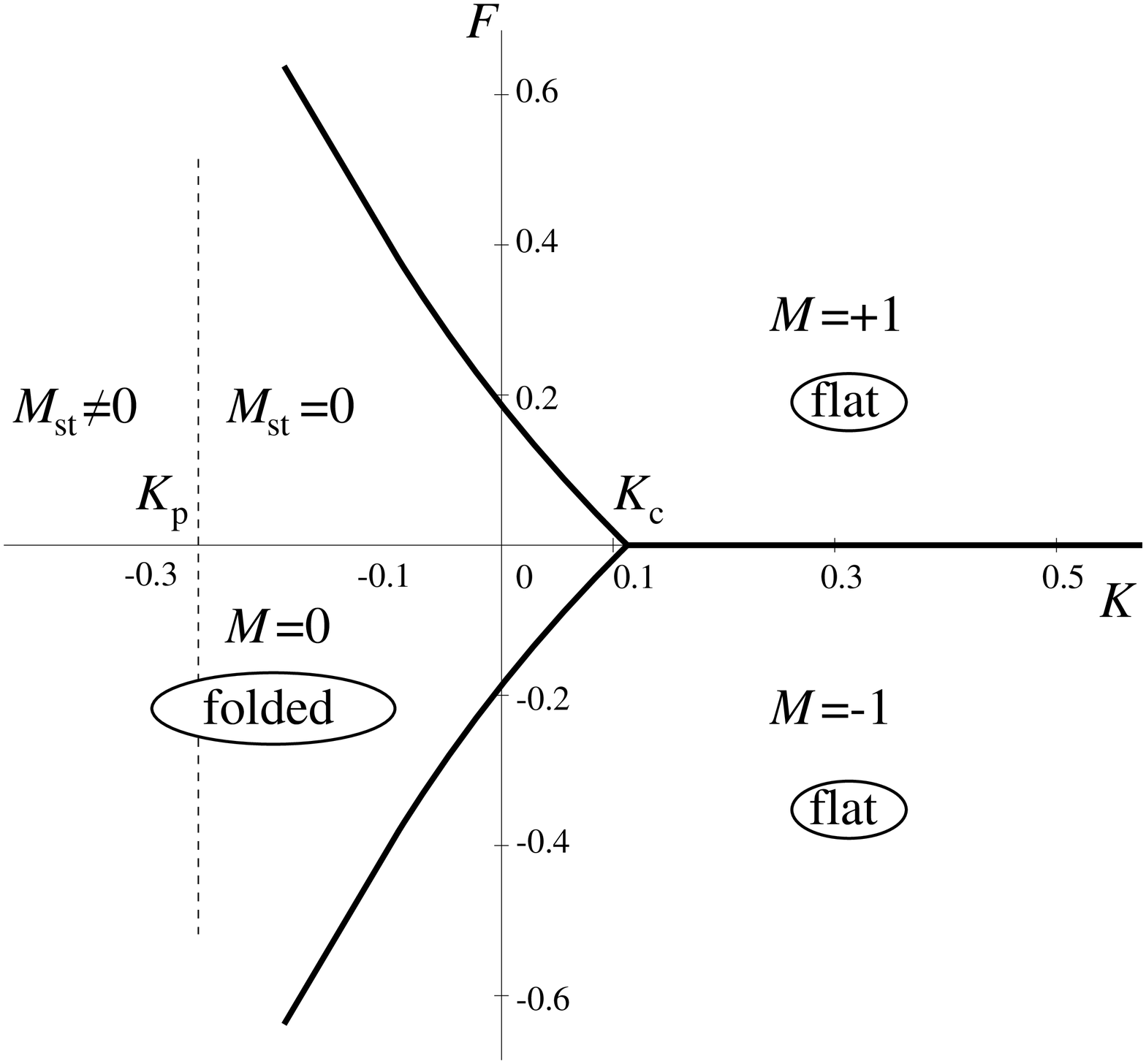}{10.cm}
\figlabel\phases
We may finally consider the phase diagram obtained in the presence 
of a bending energy $E_b=-K \cos(\theta)$ per link with folding
angle $\theta=0^\circ$ (no fold) or $180^\circ$ (complete fold).
Like in the square lattice case, we also introduce a symmetry breaking
field $F$ conjugate to the projected area ($E_p=-F\sigma$ per face)
and which plays the role of a lateral tension applied to the surface.
The resulting phase diagram, depicted in Fig.\phases, was 
obtained by a transfer matrix approach \Foltran\ and later corroborated by a 
variational study \CGP.

Comparing this result to that of the square lattice (Fig.\phasediagsq),
we note a qualitative resemblance, with three first order transition
lines separating regions with $M=0,\pm 1$. There are however two 
important differences. Firstly, the phase $M=0$ is no longer frozen in
a single state due to the existence of elementary excitations
of the fully folded state. More precisely, let us introduce the 
``piling-up" (staggered) order parameter $M_{\rm st}=\langle ({\cal M}_\vartriangle
-{\cal M}_\triangledown)/2 \rangle$ where ${\cal M}_\vartriangle$ 
(resp. ${\cal M}_\triangledown$) is the average projected area of the 
triangles pointing up (resp.  down) in the original triangular lattice.
A continuous piling-up transition now occurs at a negative value 
$K_{\rm p}\sim -0.28$ above which the
crumpled phase is disordered ($M_{\rm st}=0$) and below which the lattice tends
to be piled-up onto a single triangle ($M_{\rm st}\neq 0$). Note that 
the transition lines
between the crumpled and flat phases are no longer straight lines. 
Secondly, the triple point where 
the three transition lines meet is at a positive value $K_{\rm c}\sim 0.1$.
This shows the existence of a first order {\it flattening} or {\it crumpling transition}
between a crumpled phase for $K<K_{\rm c}$ and a {\it completely flat} phase
for $K>K_{\rm c}$ corresponding again to a {\it pure state} without any fold,
with $M=1$ (or $-1$). The existence of such a crumpling transition
was also predicted in continuous models of tethered membranes
[\xref\NP-\xref\DG]. The first order nature of the present transition 
seems however to be due to the discrete (spin-like) nature of
the (normal vector) degrees of freedom. 
For $K=0$, comparing the energy per triangle $-F$ 
in the flat state (no entropy) and the entropy ${\rm Log}\, q$
in the folded state (no energy since $M=0$), we deduce that the
transition lines cross the vertical axis at values $F=\pm {\rm Log}\, q
\sim \pm 0.189$. The existence of a strictly positive value of
the crumpling transition rigidity is therefore directly linked to the
existence of a non-zero entropy of folding. In some sense, the
critical point $K=0$ of the square lattice folding has been split
into two transition points $K_{\rm p}<0<K_{\rm c}$ (details may
be found in Refs. [\xref\QBR,\xref\BGGM,\xref\CGP]).

\subsec{Discrete folding of the triangular lattice in $d=3$: a 96-vertex model}

\fig{The FCC lattice viewed as a piling-up of octahedra completed
by tetrahedra.}{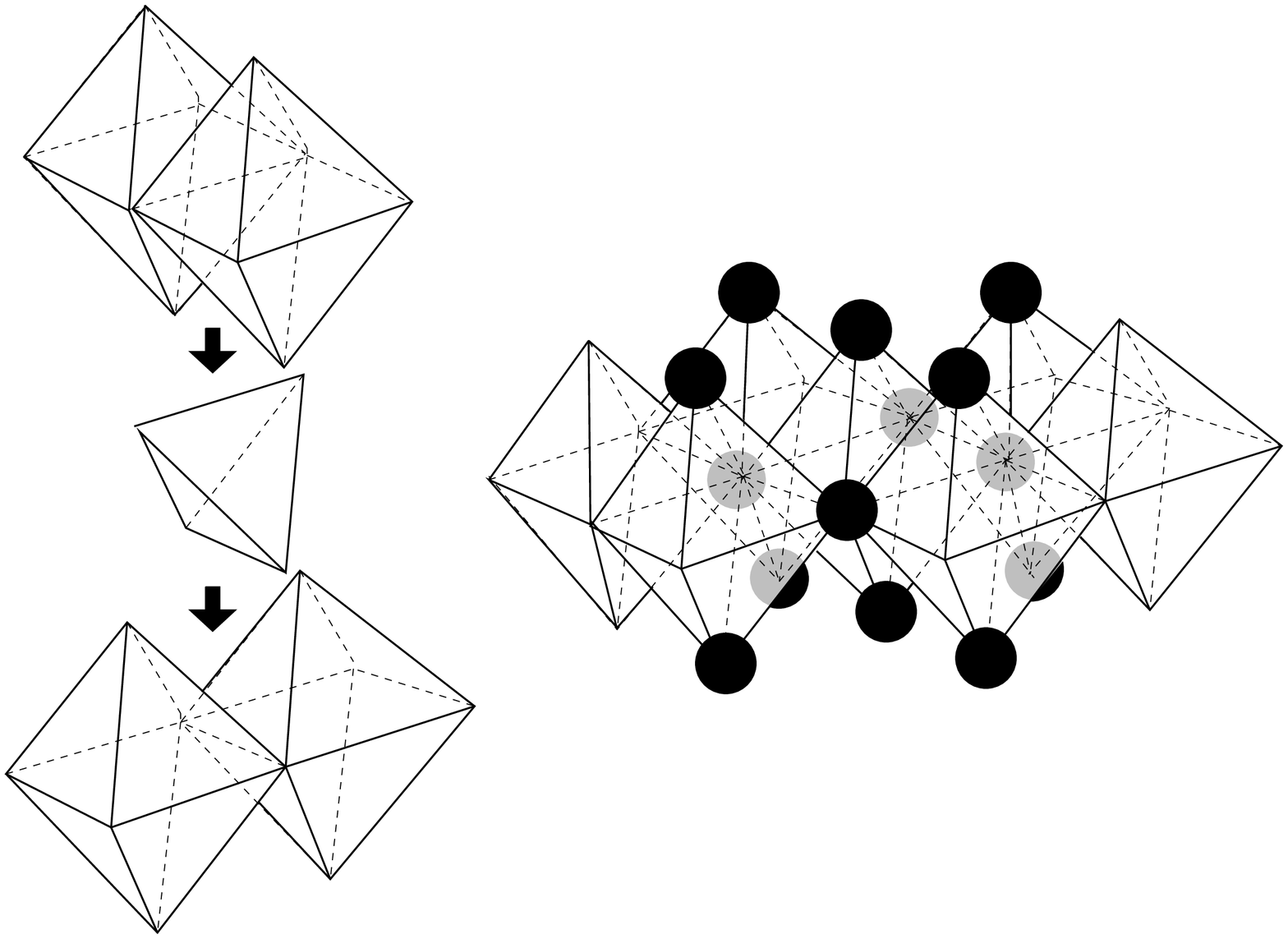}{10.cm}
\figlabel\FCC
A second simple example of folding of the triangular lattice consists
in picking the Face Centered Cubic (FCC) lattice for target space
\BFGG.
The latter may be seen as a discretization of $\IR^3$ for 
which the faces of all the elementary 2-cells are equilateral triangles.
More precisely, the FCC  lattice may be viewed as a regular piling-up of
elementary octahedra, completed by tetrahedra, as indicated
in Fig.\FCC.
\fig{The $12$ oriented edges of an elementary octahedron
of side $1$ provide a set of $12$ unit vectors for the allowed 
link variables of the FCC folding of the triangular lattice.
We attach colors A,B,C to these vectors as indicated. We
have three color planes, each containing four vectors. We 
further attach variables ${\cal A}$, ${\cal B}$ and ${\cal C}\in 
\{\pm 1\}$. Each vector is coded by a triplet of the form
$({\cal A},{\cal B},\cdot)$, $({\cal A},\cdot,{\cal C})$
or $(\cdot,{\cal B},{\cal C})$, where the position of the 
missing variable (coded by a dot) corresponds to the color
of the vector at hand. Note that the three vectors around each face
share the same ${\cal A},{\cal B},{\cal C}$ variables.}{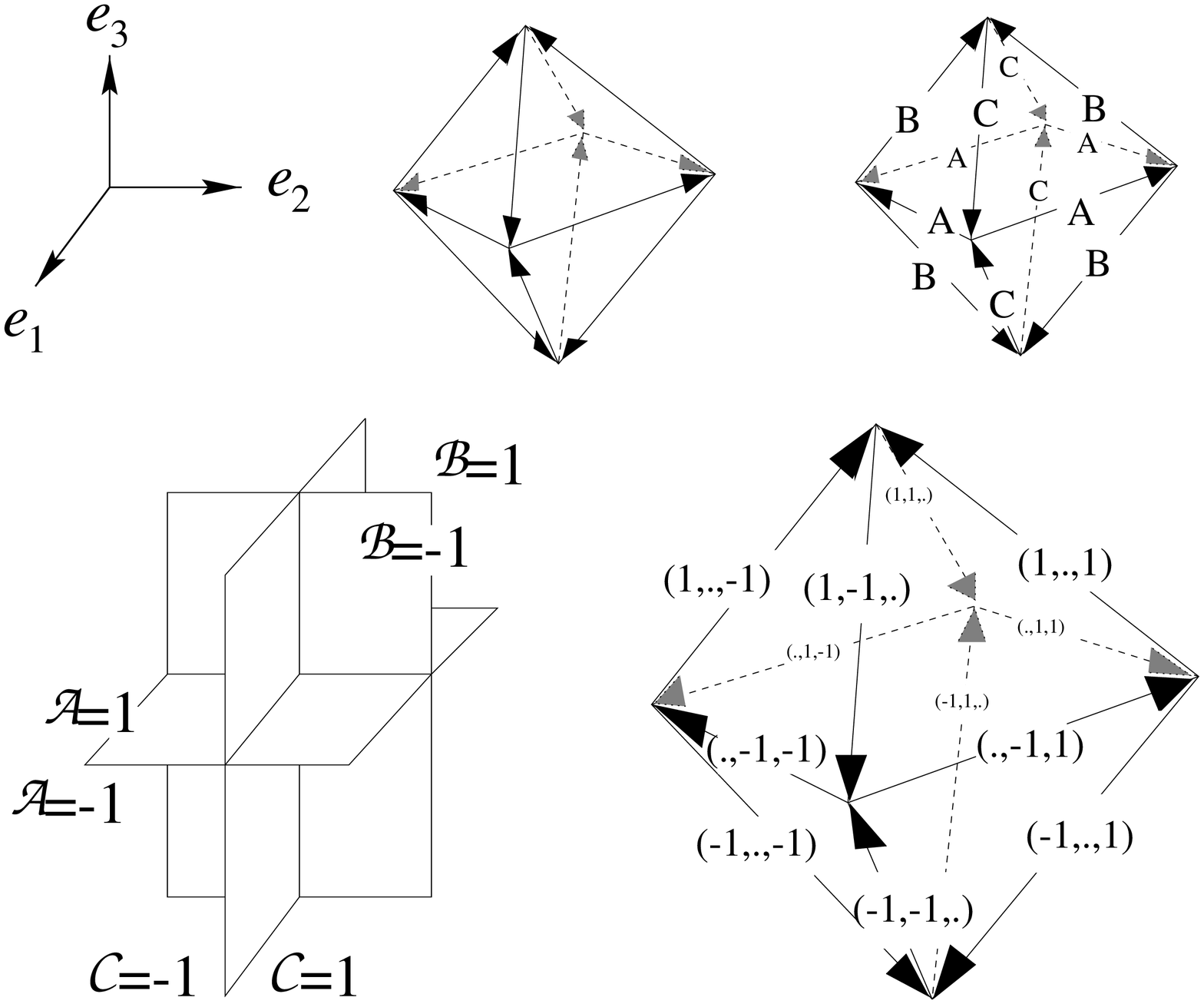}{
12.cm}
\figlabel\octa
The folding of the triangular lattice on the FCC lattice authorizes
four folding angles: $0^\circ$ (no fold) $180^\circ$ (complete fold)
$109^\circ 28'$ (acute fold, i.e.\ on the same tetrahedron) and $70^\circ 32'$
(obtuse fold, i.e on the same octahedron).
The link variables $t_\ell$ then take their values
in a set of $12$ allowed values $(\pm e_i \pm e_j)/\sqrt{2}$
with $1\leq i<j\leq 3$. These $12$ vectors are in one-to-one
correspondence with the oriented edges of an elementary octahedron 
of side $1$ (see Fig.\octa).  A simple way to label them 
is to first color them in three colors A,B,C such that all four 
vectors in a given plane of the octahedron of Fig.\octa\ 
be of the same color. We then complete the color by a complementary variable
defined as follows: for each vector of color A or B, we define
a variable ${\cal C}=\pm 1$ indicating on which side of the plane of
color C on the elementary  octahedron this vector lies
(see Fig.\octa). Similarly, we define a variable ${\cal B}=\pm 1$
for the vectors colored A or C and a variable ${\cal A}=\pm 1$ for
those colored B or C.  Each of the $12$ unit vectors is entirely
specified by its color and the value of the two complementary variables.
\fig{The $6$ link variables and the $6$ pairs of spins around
a node.}{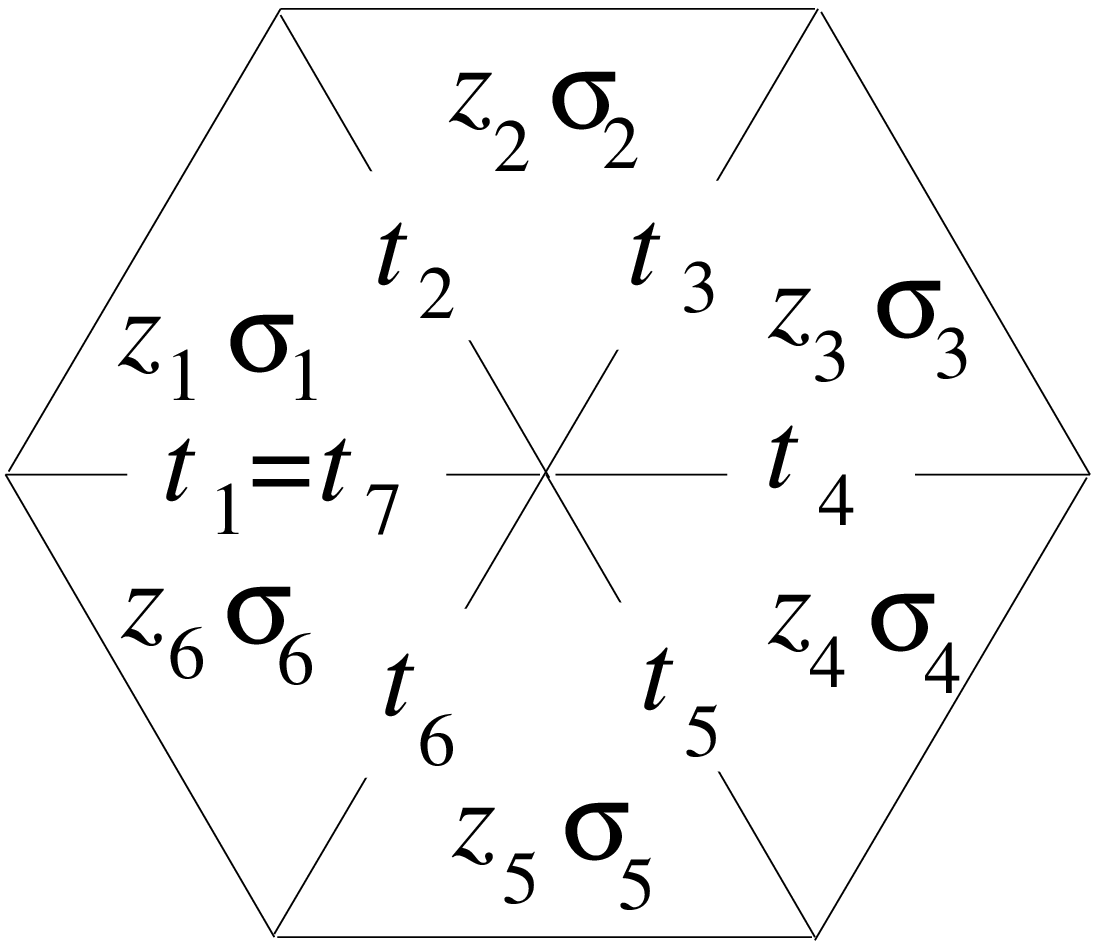}{4.cm}
\figlabel\spin
The three vectors $t_{\ell_i}$, $i=1,2,3$ around a given face must 
have a vanishing sum: they must be chosen among the $8$ triplets
of unit vectors with vanishing sum (in correspondence with 
the $8$ faces of the octahedron of Fig.\octa), which, with the 
$3!$ permutations of the three vectors, leads to $48$ possibilities.
For each triplet, the three vectors have different colors,
and have the same values of ${\cal A}$, ${\cal B}$ and ${\cal C}$,
i.e are of the form $({\cal A},{\cal B},\cdot),({\cal A},\cdot,{\cal C}),
(\cdot,{\cal B},{\cal C})$.

We may then follow the $6$ link variables $t_i$ around a node
in clockwise direction (see Fig.\spin).
Starting from $t_1$ which takes one of the $12$ allowed values,
the vector $t_2$ may take only $48/12=4$ possible values 
coded by {\it two} face spin variables $\sigma=\pm 1$ and $z=\pm 1$.
The variable $\sigma$ indicates the change of color from $t_1$
to $t_2$, with $\sigma=+1$ if the color ``increases" in cyclic
order ABC and $\sigma=-1$ otherwise. The variable $z$ is simply
the product ${\cal A}{\cal B}{\cal C}$ corresponding to the face
$\{t_1,t_2,-t_1-t_2\}$ on the octahedron of Fig.\octa. 

Going back to the original folding problem, we may now characterize
entirely the relative folding state of any two neighboring triangles 
by the relative values of $(\sigma_1,z_1)$ vs $(\sigma_2,z_2)$ 
on these two faces. We have the following correspondence
$$\vbox{\offinterlineskip
\halign{\tv\quad # & \quad\tv \quad
# & \quad \tv \quad  # & \quad \tv #\cr
\noalign{\hrule}
\tvi $z_2/z_1$ & $\sigma_2/\sigma_1$ & type of fold &\cr
\noalign{\hrule}
\tvi  $\phantom{-}1$ &  $\phantom{-}1$  &    no fold  &\cr
\tvi  $\phantom{-}1$ &  $-1$  &    complete fold  &\cr
\tvi  $-1$ &  $\phantom{-}1$  &    acute fold  &\cr
\tvi  $-1$ &  $-1$  &    obtuse fold &\cr
\noalign{\hrule} }} $$
After a complete turn around a node, the constraint that
we recover the same tangent vector translates into
the following two constraints on the spins. The
first constraint
\eqn\spinrul{\sum_{i=1}^6\sigma_i=0 \quad{\rm mod}\quad 3}
is the same as the constraint \spinrule\ for
planar folding, and ensures that we recover the same color
of the tangent vector. 
The second constraint deals with the $z$ variables\foot{Note 
that the relation for $x=3$ is a consequence of those at
$x=1,2$.} and reads
\eqn\zrul{\prod_{i\ {\rm such\ that:}\atop \sum_{j=1}^i\sigma_j=x \
 {\rm mod}\ 3}\kern -20pt z_iz_{i+1}
 =1, \quad x=1,2,3}
The left hand side simply counts the number of sign changes 
of the $z$ variable across the links of color A, 
B or C (one color for each value of $x$), which are nothing but 
the sign changes 
of the complementary variable ${\cal A}$, ${\cal B}$ and
${\cal C}$ respectively. Eq. $\zrul$ ensures that the
we recover the same value of the complementary variables
after a complete turn.
\fig{The 96 folding configurations of an elementary hexagon.
The complete folds are indicated by thick lines, the
obtuse folds by thin lines and the acute folds by dashed lines.
For each environment, we have indicated its multiplicity corresponding
to rotations of the hexagon. We also indicated the vertices which have
to be retained for the three sub-models corresponding to the planar 
folding (11 vertex),
the folding on a single tetrahedron (11 vertex) and that on a single octahedron
(16 vertex).}{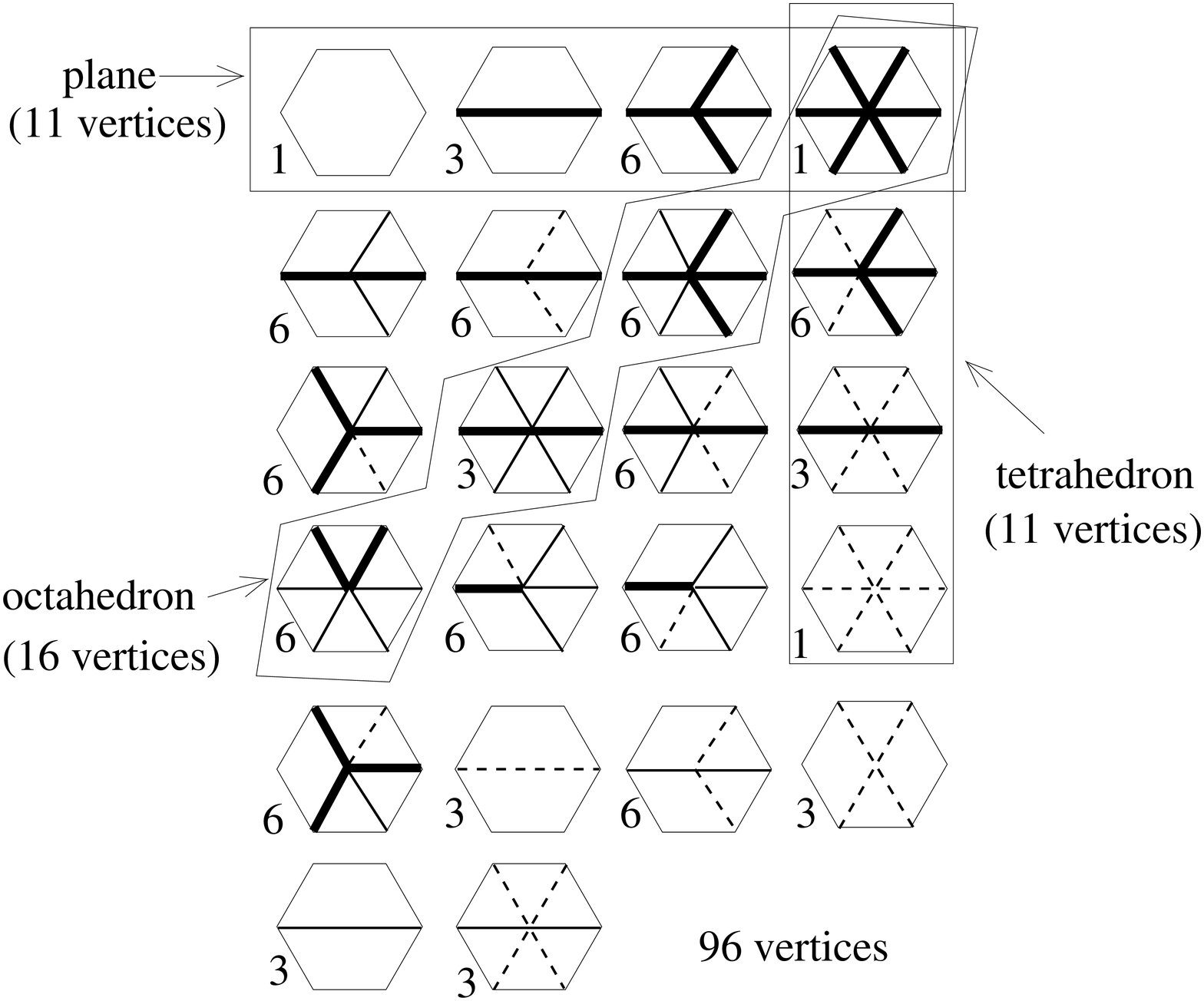}{12.cm}
\figlabel\ninetysix
In terms of constrained spins, it is easy to count all the
environments allowed by the constraints \spinrul\ and \zrul\
around a vertex. There are $96$ such environments, which are
represented in Fig.\ninetysix. As such, the FCC folding of
the triangular lattice is a 96-vertex model.

It is interesting to note that this model
has the following three sub-models (see Fig.\ninetysix):
\item{1.} An $11$-vertex model corresponding to the  {\it planar folding}
obtained by keeping only either unfolded or
completely folded links. This amount to requiring that
$z=1$ for all the triangles, in which case \zrul\ is automatically
satisfied and one recovers the constraint \spinrule.
\item{2.} An $11$-vertex corresponding to the
{\it folding on a single target tetrahedron}, obtained by keeping only
acute or complete folds.
This amounts to requiring that $z=\epsilon \sigma$ where $\epsilon$
is $1$ on triangles pointing up and $-1$ on triangles
pointing down. Once again, the relation \zrul\ is automatically
satisfied.
\item{3.} A $16$-vertex  model corresponding to the
{\it folding on a single target octahedron}, obtained by keeping only
obtuse or complete folds.
This amounts to having a perfect anti-ferromagnetic order
$\sigma=\epsilon$, with $\epsilon$ as defined above,
in which case the relation \zrul\ becomes $\prod_{i=1}^6z_iz_{i+1}=1$.
\par
\noindent Note that the constraints \spinrul\ and \zrul\ are
invariant under the change $(\sigma,z)\to(\sigma,\epsilon z \sigma)$.
One thus gets a duality relation which consists in exchanging
globally all the unfolded edges with those carrying an acute fold,
as apparent in Fig.\ninetysix. The sub-model 3
above is self-dual while the two
sub-models 1 and 2 are dual of one another.
In particular, the entropy of folding on a single tetrahedron
is the same as that for the planar folding, and given by Eq.\valq.

The entropy of folding on the FCC lattice is not know so far.
Numerical estimates \BFGG\ show that $q\sim 1.43(1)$. We may easily
show that $q>\sqrt{2}=1.414\cdots$ by estimating the
entropy of the sub-model 3 of folding on a single octahedron.
Indeed, the constraint
$\prod_{i=1}^6z_iz_{i+1}=1$ amounts to requiring that $z$ is the
product $\eta_{v_1}\eta_{v_2}\eta_{v_3}$ on the three nodes
$v_1$, $v_2$ and $v_3$ adjacent to the face at hand of a
node variable $\eta_v$ equal to $\pm1$ and
{\it independent} on each node of the lattice. The model 3 has
therefore an entropy of $2$ per node, i.e.\ $\sqrt{2}$ per triangle.

\fig{The proportion of the different types of folds as a function
of the rigidity modulus $K$ (energy $E_b=-K \cos(\theta)$) in the range
$K<0$.}
{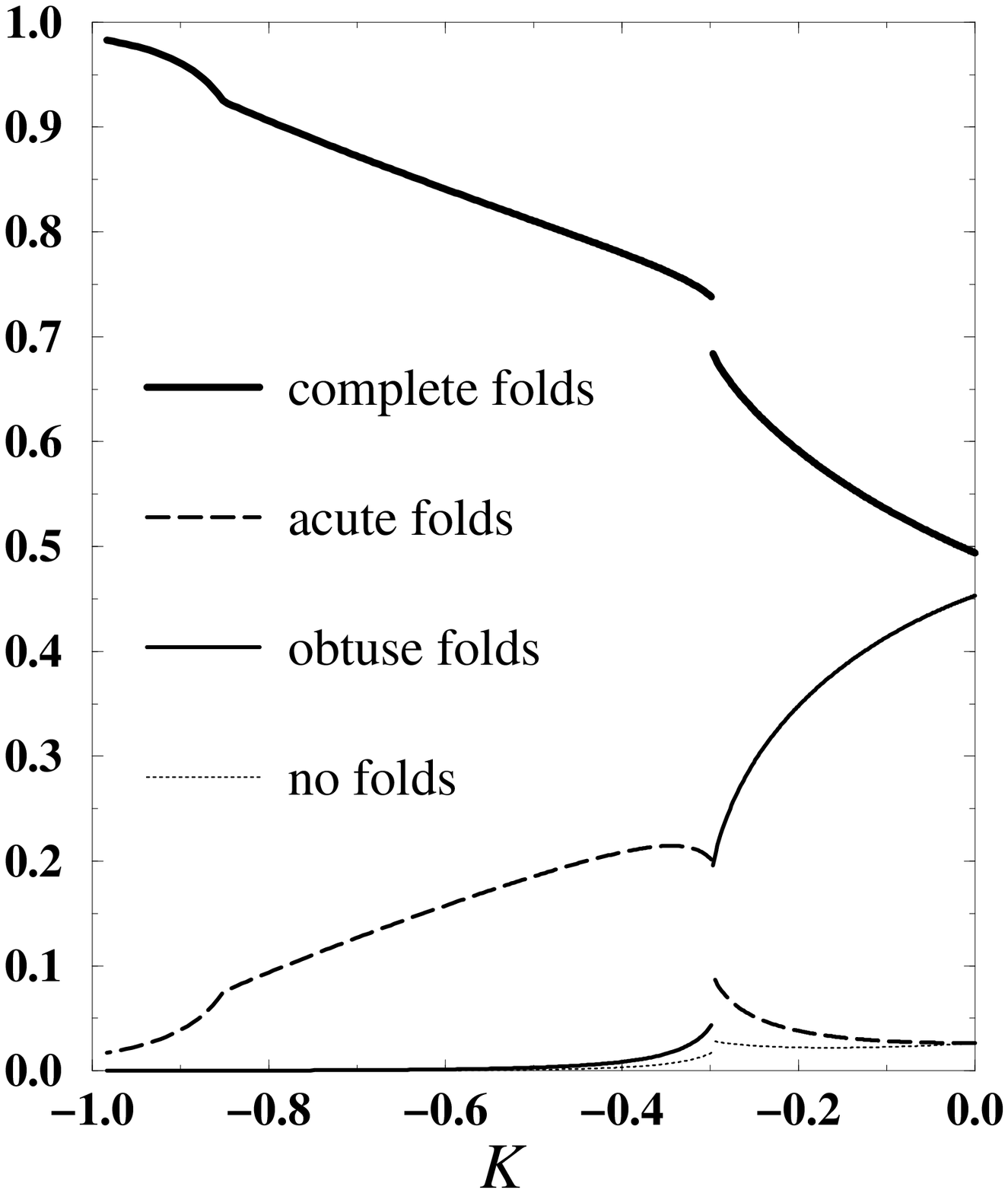}{8.cm}
\figlabel\propfr
We may finally consider the phase diagram obtained by
introducing a bending energy, again in the form
of a link energy $E_b=-K \cos(\theta)$ where $\theta$
is the folding angle (none, complete, acute or obtuse).
This phase diagram was obtained in Ref.\BGGM\ by use of a variational 
method (see also Ref.\CIGOPE). As for planar folding, 
one finds in the regime $K\geq 0$
a first order flattening/crumpling transition between a crumpled phase
for small values of $K$ and a {\it completely flat} phase
for large values of $K$. The situation for $K\leq 0$ (i.e.\ 
for which folds are favored) is richer:
the figure \propfr\ displays the proportion of the different types of folds
as a function of $K$.
One clearly distinguishes two successive folding transitions.
A first, discontinuous transition at $K\sim -0.3$ separates a regime 
where the folding 
occurs preferentially on octahedra from a regime where it occurs 
preferentially on tetrahedra. A second, continuous 
transition at $K\sim -.85$ separates this last regime
from a regime where the folding is essentially maximal with
dominance of complete folds.
\fig{The three order parameters $O$, $T$ and $P$ as defined in the text
as a function of the rigidity modulus $K$ (energy $E_b=-K \cos
(\theta)$) in the range $K<0$.}
{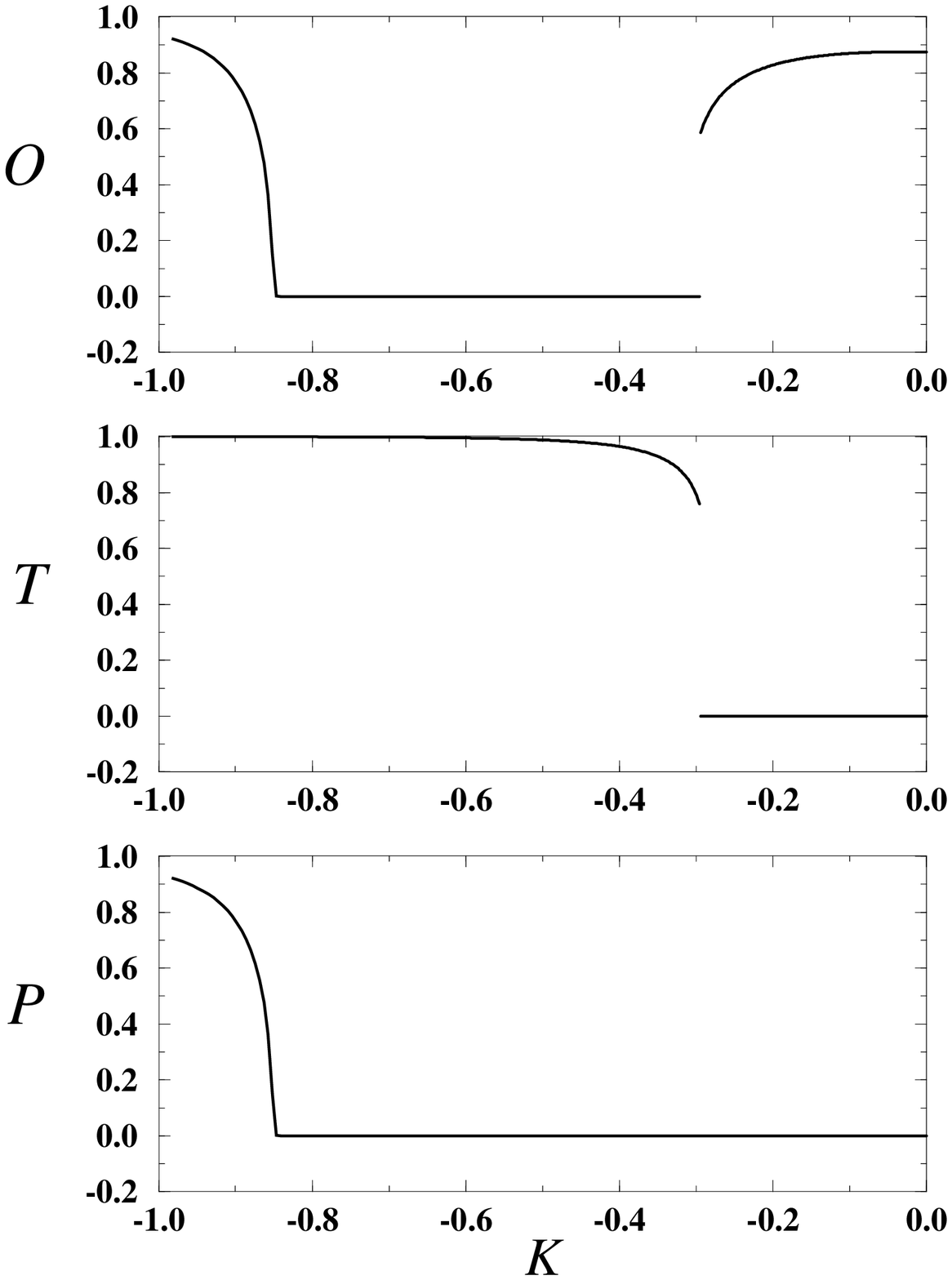}{8.cm}
\figlabel\OP
These transitions may also be viewed on the three order parameters
$O$, $T$ and $P$ defined as
\eqn\OPs{\eqalign{O&\equiv \langle \epsilon\, \sigma\rangle\cr T&\equiv
\langle z\, \epsilon\, \sigma\rangle\cr
P&\equiv \langle z \rangle\cr}}
and represented in Fig.\OP. The interpretation of these
order parameters is that a non-vanishing value of $O$ (respectively
of $T$) shows a tendency of the lattice to fold
preferentially around octahedra (respectively around
tetrahedra). Similarly, a non-vanishing value of $P$
shows a tendency of the lattice to preferentially
remain within planes of the FCC lattice.

This completes our study of phantom folding of 2D regular lattices.
Here we have concentrated on the folding of the triangular lattice
in $d=2$ and $3$ dimensions. We expect the same type of conformational
transitions to take place for higher target dimensions $d$, with
a first order crumpling transition to a completely flat state
at a positive value of the bending rigidity $K$, a continuous (Ising type)
piling-up transition at a negative value of $K$ toward a limiting 
completely folded state,
and a fine structure of intermediate transitions of wrapping around
intermediate-dimensional regular solids. Finally, we expect a similar
scenario to happen for all the compactly foldable lattices of 
Fig.\classi\ (including the rectangular lattice case as a
degenerate limit with zero entropy). This scenario is confirmed in the
case of the square-diagonal lattice in Ref.\otherfold.
\bigskip
\vfill\eject
\leftline{\bf PART B: LOOP MODELS ON REGULAR LATTICES}
\bigskip
In this part, we show that folding problems are particular 
instances of a larger class of problems, so-called fully packed
loop models \BN, with possible reformulations as height models for
which the loops play the role of contour plots. 
We focus here again on regular lattices 
and illustrate the techniques of solution in the case of
fully packed loops on the hexagonal lattice. These include
Bethe Ansatz calculations \Baxbook\ and effective Coulomb gas 
descriptions \Nien.

\newsec{Loop gas and height model reformulations of the triangular 
lattice folding}

Very generally, all lattice folding problems can be reformulated
as fully packed loop gases on the dual lattice. We will illustrate 
this property in the two cases of the $d=2$ (planar) and $d=3$ 
(FCC) foldings described in the previous Section.

\subsec{Fully packed loop gas formulations of the 2d folding}
\fig{Example of tricoloring of the links of the hexagonal lattice and
the associated configuration of oriented fully packed loops.}
{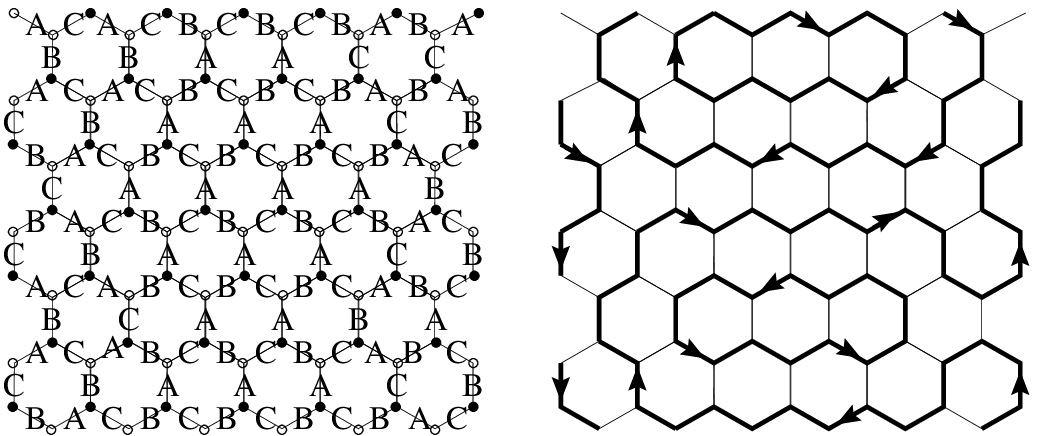}{12.cm}
\figlabel\troiscol

Let us first consider the planar folding of the triangular lattice. 
As seen in Section 2.2, it
is equivalent to the tricoloring of the links of the {\it hexagonal} 
lattice with three colors A, B and C, required to be distinct on
the three links adjacent to any node of the lattice. Note that the 
hexagonal lattice is bipartite, i.e.\ it may be decomposed into 
two sub-lattices of black, resp. white nodes, such that
the neighbors of a black node are white and conversely. The
black or white color of a node is equivalent to the ``pointing up" or
``pointing down" nature of the dual triangle. Ignoring the links of color A, 
the links of color B or C form loops on the hexagonal lattice along which 
the colors B and C alternate (see Fig.\troiscol).
By orienting all the links of color B from the adjacent
black node to the adjacent white node, and the links of color C
from the adjacent white node to the adjacent black node, we orient
each loop in a coherent way. Changing  the orientation of a loop
simply amounts to interchanging the colors B and C along this loop. 
The gas of oriented loops thus obtained is self-avoiding 
and {\it fully packed}, in the sense that each node of the lattice 
is visited by a loop.

Conversely, given any set of fully packed oriented loops, it is possible
reproduce the associated colors for all the links, and therefore rebuild
the associated folding configuration. This shows that the
planar folding of the triangular lattice is equivalent to a gas 
of fully packed oriented loops on the hexagonal lattice. 

The orientation of the loops may be rephrased into a weight $n=2$ per 
unoriented loop. More generally, we may consider the so-called
$FPL(n)$ model of a gas of fully packed, self-avoiding and  unoriented loops
on the hexagonal lattice with a weight $n$ per loop \BN. This 
is to be compared with the loop gas formulation of the so-called $O(n)$ 
model on the hexagonal lattice which also amounts to consider
self-avoiding loops with a weight $n$ per loop but without the 
requirement that every node of the lattice be visited by a loop \Nien.
\fig{Equivalence between (a) antiferromagnetic groundstates of the
Ising model on the triangular lattice, (b) $FPL(1)$ loop configurations
on the dual hexagonal lattice, and (c) Solid-On-Solid (SOS) interfaces
for piling-ups of cubes viewed in perspective. In (a), there is exactly one
frustrated link (with equal adjacent spins) per triangle (dashed lines). 
The frustrated links are dual to the unoccupied edges in (b) while
the links dual to the $+\vert-$ links form a gas of
fully packed loops on the hexagonal lattice. Erasing 
the frustrated links yields a rhombus tiling of the plane, also interpreted 
as a piling-up of cubes viewed in perspective (c).}{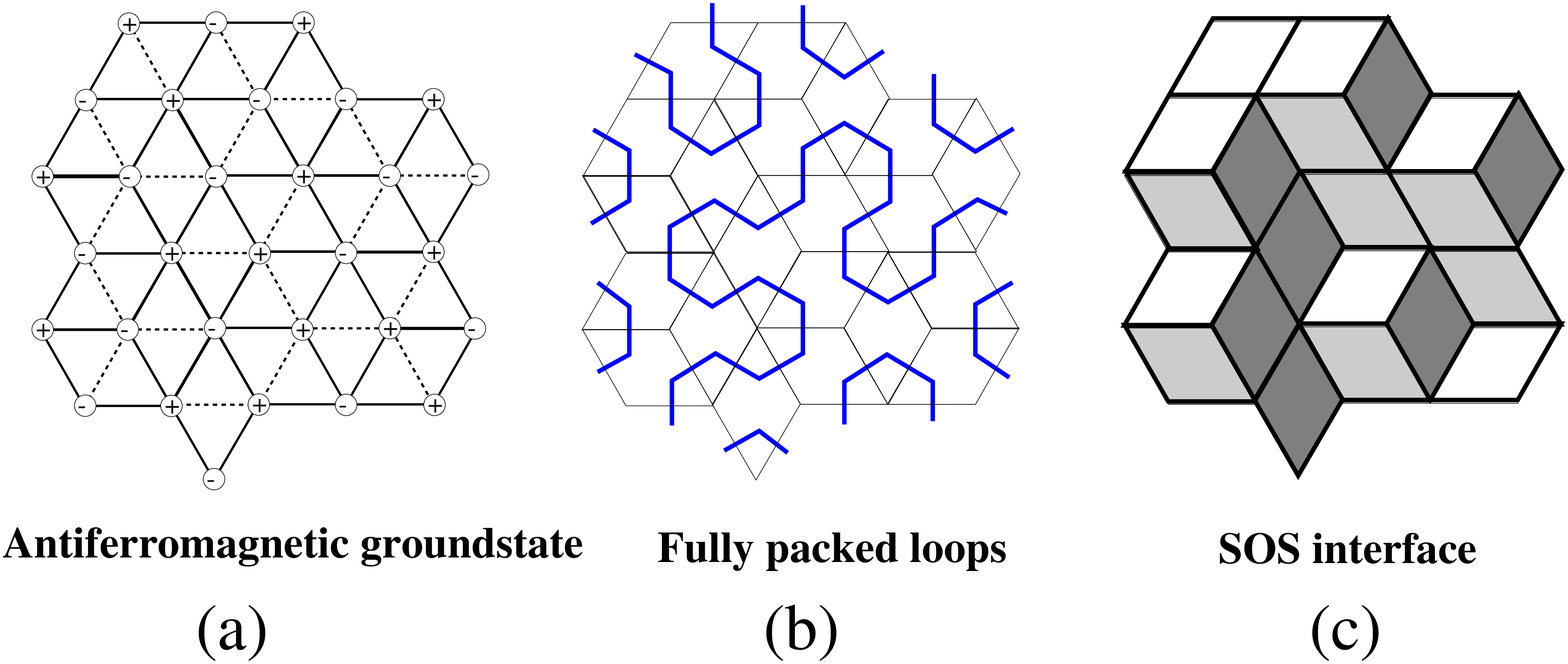}{15.cm}
\figlabel\af

Note that the $O(n)$ model is originally defined as a spin model
with $O(n)$ symmetry. Similarly, the $FPL(2)$ model alternatively 
describes the groundstates of the antiferromagnetic XY model (spins
with $O(2)$ symmetry) on the Kagome lattice \origami. Indeed, the nodes of 
the Kagome lattice are in one-to-one correspondence with the edges
of the triangular lattice. The tangent vectors of the folding configuration
yield spins on the nodes of the Kagome lattice. The constraint \constlien\ of
vanishing sum around a triangle is nothing but the condition for 
the minimization of the antiferromagnetic energy 
$t_{\ell_1}\cdot t_{\ell_2}+t_{\ell_2}\cdot t_{\ell_3}+t_{\ell_3}
\cdot t_{\ell_1}$ (together with $\vert t_{\ell_i}\vert =1$) 
around each elementary triangle of the Kagome lattice.
In the same spirit, the $FPL(1)$ model describes
the antiferromagnetic groundstates of the Ising model on the
triangular lattice, as illustrated in Fig.\af-(a,b). The loops on
the (dual) hexagonal lattice are easily identified with the links 
dual to the $+\vert-$ (i.e.\ energetically favored) links on the 
triangular lattice. The fully packed requirement amounts to maximizing
the number of these favored links (exactly 2 per triangle).

\subsec{Fully packed loop gas formulations of the 3d folding}
In the case of the folding of the triangular lattice on 
the FCC lattice, we have seen that the
link variables again induce a tricoloring of the links of the triangular
lattice, and thus, by duality, of the links of the hexagonal lattice.
This coloring by three colors A, B and C, as we just saw, is
equivalent to the gas of the fully packed BC loops.
We may consider as well the two other systems of fully packed loops
made of the links AC or the links AB. The three
systems of loops are strongly correlated as the knowledge of one of them
allows us to reconstruct the two others.
The colors are not sufficient however to characterize entirely the original
3d folding. To this end, we have introduced the additional link variables
${\cal A}, {\cal B},{\cal C}=\pm 1$ indicating, in the elementary octahedron,
the side of the plane of color A,B,C respectively on which the link variable 
at hand lies (see Fig.\octa). If we now consider any BC loop, 
it is easy to see that the value of ${\cal A}$ is {\it constant}
along the loop. Indeed, a change of sign of ${\cal A}$
requires to cross the plane of color A on the elementary octahedron
and therefore to pass through a link of color A, which in turn requires to
pass from one BC loop to another one. Moreover, the value of ${\cal A}$
is independent on
each of the BC loops. The same is true for the variable ${\cal B}$
on the AC loops and the variable ${\cal C}$ on the AB loops.
As a consequence, we may represent any folding on the FCC lattice
by a tricoloring of the hexagonal lattice, completed by
a $\IZ_2$ spin variable on each of the BC, AC and AB loops. 
The enumeration of 3d foldings is therefore performed by counting
tricolored configurations with a weight $2$ for each of the 
AB, BC and AC loops.

To conclude, we may write the partition functions for 2d and 3d
foldings as partition functions for edge-tricolorings or fully packed 
loop gases as
\eqn\partfu{\eqalign{Z_{\rm plane}&=\sum_{\rm Tricolorings} 1 =
\sum_{\rm Fully\ packed\ loops} 2^{\#\ \rm loops}\cr
Z_{\rm FCC}&=\sum_{\rm Tricolorings} 2^{\#\ \rm AB\ loops}\times
2^{\#\ \rm BC\ loops}\times 2^{\#\ \rm AC\ loops}}}

Other folding problems were considered \otherfold,
such as the planar folding of the square-diagonal lattice, or of the
triangular-diagonal lattice. All these problems are equivalent
to gases of fully packed loops on appropriate lattices.

\subsec{Height models}
\fig{Definition of the height variables $X$ for
fully packed loops on the hexagonal lattice. The rule explicitly uses
the bicoloring of the nodes of the hexagonal lattice. In the case
of fully packed loops, the only possible environment is (a) (up
to obvious symmetries), and the consistency of the definition of the heights
imposes $A+B+C=0$. In the case of loops not necessarily fully packed, the
presence of unvisited nodes with the environment (b) imposes moreover that $A=0$,
in which case $B=-C$ ($=1$ for instance) and the rules become insensitive
to the bicoloring of the nodes of the hexagonal lattice.}{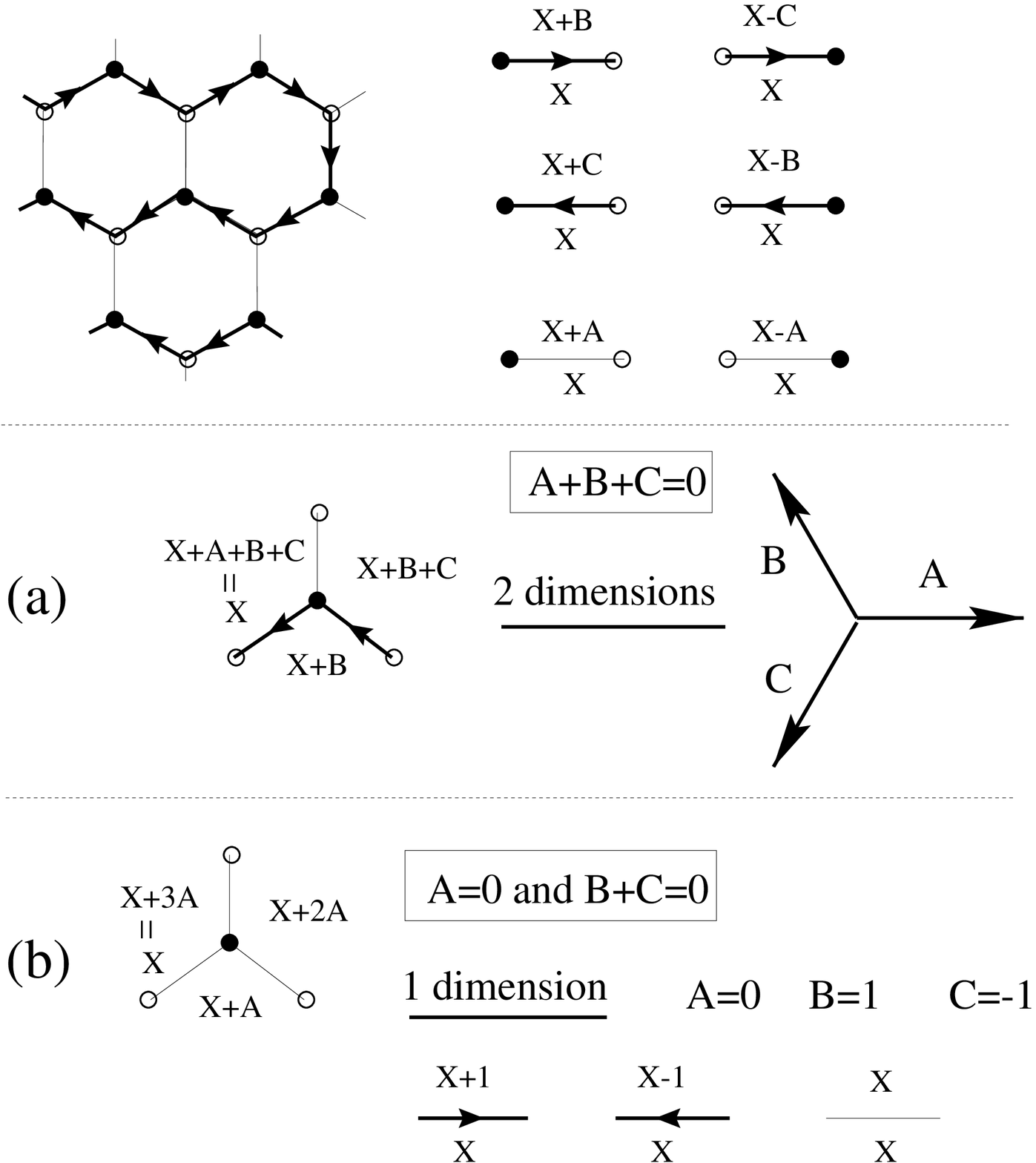}{11.cm}
\figlabel\Onrules

We shall now transform the $FPL(2)$ configurations into those of
a {\it height model} defined as follows \KGN. We start with a gas of
oriented fully packed self-avoiding loops. To each face of the
hexagonal lattice, we associate a ``height" $X$ in $\IR^d$ whose
variation from one face to another depends on whether the
separating edge is occupied or not by a loop. The precise rules
for the variation of heights are indicated in Fig.\Onrules\ and use explicitly
the bipartite nature of the hexagonal lattice,
with different conventions according to whether the link is oriented 
from a white or from a black node. These rules use three elementary height differences 
$A$, $B$ and $C$. If we now replace the oriented loops
by the corresponding tricoloring of the links as in Fig.\troiscol,
the height difference is simply given by the color A,B or C
of the crossed link.

Inspecting the allowed configurations around a node, we see that 
we must impose the constraint $A+B+C=0$ in order to have 
a well-defined height after one turn around a node (see Fig.\Onrules-(a)).
This naturally leads to a {\it two-dimensional} height $X$.
As we shall see below, the components of $X$ become two scalar
fields in a continuous effective description, and the $FPL(2)$ model 
is indeed a critical model described in the continuum by a conformal
field theory (CFT) of {\it central charge $2$} corresponding
to these two height components \CFT. A symmetric solution consists 
in choosing $A$ to be a unit vector, $B={\cal R} A$ and $C={\cal R}^2A$ 
where ${\cal R}$ is the rotation of angle $120^\circ$ in the plane. 
With this choice, for any given oriented loop configuration, we may identify 
the height $X$ of a face with the position of its dual node in the 
associated folded configuration of the triangular lattice in the plane. 
The above symmetric solution will be used from now on.

As described above, the $O(2)$ model corresponds to releasing 
the full-packing constraint by allowing for nodes which are not 
visited by loops. In this case, we must also include the configuration 
of Fig.\Onrules-(b) corresponding to an unvisited node. 
The consistency of heights after one turn around the node now imposes 
the extra constraint $A=0$, and thus $B+C=0$. This now leads to a
{\it one-dimensional} height. The corresponding loop gas is critical
for a suitable choice of edge weight. At this critical point \Nien,
the model is described in the continuum by a CFT of central charge 
$1$. Notice that the rules defining the heights in this case 
are now insensitive to the bipartite nature of the lattice
(see Fig.\Onrules-(b)).

Surprisingly, the full-packing constraint influences the
value of the central charge, and thus the universality class of the model.
More precisely, as first recognized in Ref.\BN, the net effect of the 
full-packing constraint has been to increase the central charge by $1$. 
We shall recover a similar property below in the case of random lattices.

For $n=1$, the $FPL(1)$ model may be described by a {\it one-dimensional}
height. This is readily seen on Fig.\af-(c). Indeed, any
configuration of the model may be interpreted as a piling of unit cubes 
in $\IR^3$ viewed in perspective from the direction $(1,1,1)$ and
whose free surface is connected and has no overhangs. This surface
may be entirely described by a single height $h=X\cdot A$ with $X$ 
obtained from the rules of Fig.\Onrules\ for some
arbitrary orientation of the loops (indeed this height is insensitive
to the choice of orientation as $B\cdot A=C \cdot A$).
As we shall see in the next Section, the model is critical with central 
charge $1$, as opposed the case of the $O(1)$ model which has central
charge $0$ (in its dense phase).

\newsec{Exact solutions via Bethe Ansatz: the example of the $FPL(1)$ model}

Exact solutions for the general $FPL(n)$ models with arbitrary $n$
may be obtained by transfer matrix methods together with
Bethe Ansatz type assumptions on the eigenvectors. The existence 
of such solutions is granted by the integrability of the models.
The Bethe Ansatz needed in the general solution is of ``nested"
type, which makes the explicit solutions quite technical \BSY.
In the case $n=1$ however, the problem simplifies drastically
and reduces to a free fermion model. The latter may be treated
via a simple coordinate Bethe Ansatz. For pedagogical purposes,
we choose to describe only this case in detail in the following
Sections. In particular, we will compute the value $c=1$ of
the central charge for the conformal field theory that describes
the large distance behavior of the model. We will return 
to the general case in Section 5 where we present an effective
field theoretical description of $FPL(n)$ via Coulomb gas.
 
\subsec{$FPL(1)$ and rhombus tiling of the plane}
In this Section, we address the $FPL(1)$ model, which,
as illustrated in Fig.\af, may be rephrased as the 
problem of tiling a domain of the plane by means of any number of the
three following rhombic tiles: 
\eqn\troislosanges{\alpha=\vcenter{\hbox{\epsfbox{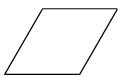}}} 
\qquad  \beta=\vcenter{\hbox {\epsfbox{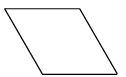}}} \qquad  
\gamma=\vcenter{\hbox{\epsfbox{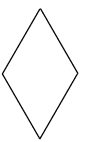}}}}
with edges of unit length and angles of $60^\circ$ and $120^\circ$,
all weighted by $1$.
As it turns out, the specific shape of the domain to be tiled is extremely
important. One can show for instance that a large rhombus of size
$N\times M$ may be tiled in exactly $N+M \choose N$ distinct manners,
resulting in a vanishing thermodynamic tiling entropy
$s=\lim_{N,M\to \infty}{1\over NM}{\rm Log} {N+M\choose N}=0$. 
On the other hand, a large hexagon with size $N\times M\times P$ may be
tiled in many more ways, namely \MacM
\eqn\macma{\prod_{i=1}^N\prod_{j=1}^M\prod_{k=1}^P {i+j+k-1 \over i+j+k-2}}
which results in a non-vanishing thermodynamic entropy of tiling.
It is not our purpose here to study the influence of boundary conditions 
on the entropy of tiling, but just to compute it in some generic situation, 
namely that of
a large cylinder represented by a rhombus of width $L$ and length 
$M\to \infty$, the two longitudinal sides of the rhombus being glued.

In the next Sections, we use the following strategy, quite standard
for solving integrable lattice models (the particular example at hand
is borrowed from a course delivered by B. Nienhuis at the 1997 Beg-Rohu
school ``\'Ecole de physique de la mati\`ere condens\'ee").
We first define the row-to-row transfer matrix $T_L$ of the model,
with periodic
boundary conditions along a row of $L$ tiles. The partition function of the model on a
rhombus of size $L\times M$, say with doubly periodic boundary conditions (torus)
reads then
\eqn\parZf{Z_{L,M}={\rm Tr}(T_L^M)}
To access the thermodynamic properties of the model,
we must take a large $L,M$ limit. Concentrating on a strip of fixed width $L$,
the thermodynamics ($M\to \infty$) is simply governed by the largest 
eigenvalue of $T_L$,
with a thermodynamic tiling entropy per tile of
\eqn\freter{s=\lim_{M\to \infty} {1\over LM} {\rm Log}\, Z_{L,M}}
We then use the expected conformal limit of the model to identify the central
charge by studying the finite size effects on a cylinder 
of finite but large width $L$. 

\subsec{Transfer matrix and eigenvalue equations}

Consider a horizontal row, tiled with a number of tiles 
$\alpha,\beta,\gamma$, and with total
length $L$. For simplicity, we deform slightly the three tiles into the following
\eqn\newlos{\alpha=\vcenter{\hbox{\epsfbox{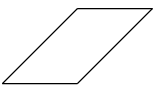}}} 
\qquad  \beta=\vcenter{\hbox {\epsfbox{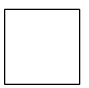}}} \qquad  
\gamma=\vcenter{\hbox{\epsfbox{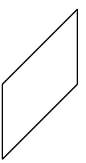}}}}
This amounts to consider the partition function of the rhombus tiling model on a
rectangle of size $L\times M$ with its vertical edges identified so as to form a cylinder.
The transfer between a tiled horizontal  row to the next (say from bottom to top)
is then described by assigning an occupation number $n_i=0,1$ (resp. $n_i'$)
to each lower (resp. upper) horizontal edge, resulting in the transfer matrix element
$(T_L)_{\{n_1',n_2',...,n_L'\},\{n_1,n_2,...,n_L\}}$ from
a configuration $\vec{n}$ of occupation numbers of a horizontal row of edges
to the next $\vec{n}'$, that typically reads
pictorially
\eqn\beloo{\eqalign{
&(T_L)_{\{1,0,1,0,...,0,1,1\},\{1,1,0,0,...,1,0,1\}}\cr
&=\figbox{8.cm}{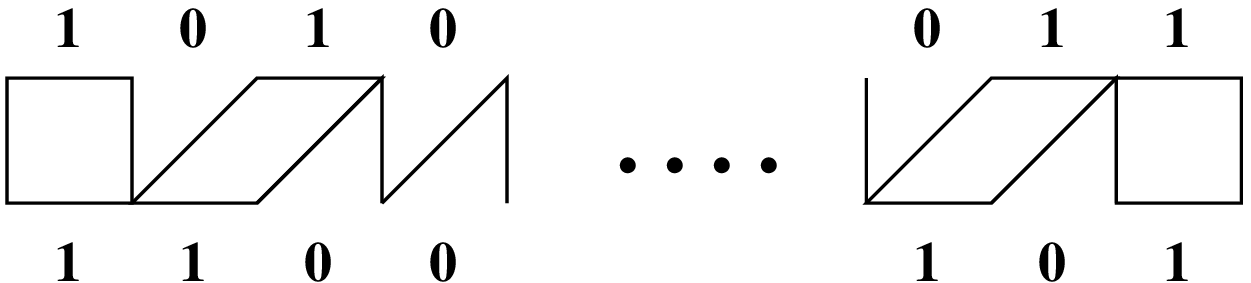}\cr} }
Note that the number $k$ of $1$'s is conserved from a row to the next,
and counts the total number of $\alpha$ or $\beta$ tiles in the row. 

Starting from a row with only $\gamma$ tiles, the next row can only 
consist of $\gamma$ tiles as well. This
forms the ``fundamental" invariant sector, with $k=0$. 
This unique possibility is summarized by the transfer matrix element
\eqn\fundmat{(T_L)_{\{0,0...,0\},\{0,0...,0\}} = \figbox{6.cm}{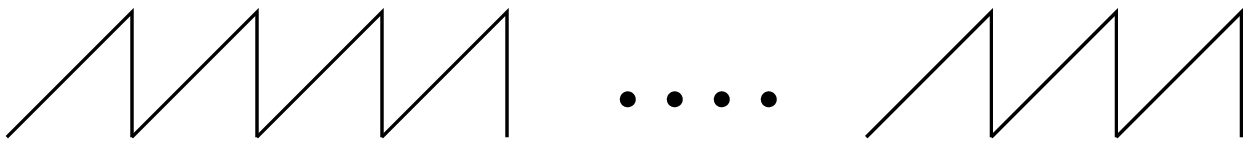} =1}
Note that all the upper and lower horizontal edges are unoccupied.
We conclude that the "no-occupied" edge state is an eigenstate of $T_L$ with eigenvalue
$\Lambda_0=1$.

An excitation of this groundstate is obtained by replacing one pair of adjacent
halves of $\gamma$ tiles along the row by either an $\alpha$ or a $\beta$ tile.
Assume the occupied lower edge of the tile is in position $x\in
Z_L$ (with periodic boundary conditions $L\equiv 0$) along the row, 
then the upper one is necessarily in position $x$ ($\alpha$ tile) 
or $x+1$ ($\beta$ tile).
Let us denote by $v(x)$ the lower edge configuration
vector $\{0,0,...,0,1,0,...,0\}$ with a $1$ in position $x$.
We then have
\eqn\tmvec{\eqalign{
T_L \, {v}(x) &=\qquad\qquad {v}(x)\qquad\qquad +\qquad\qquad {v}(x+1)  \cr &
\ \vcenter{\hbox{\epsfbox{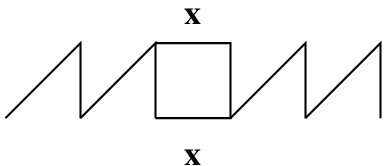}}}\qquad  \vcenter{\hbox{\epsfbox{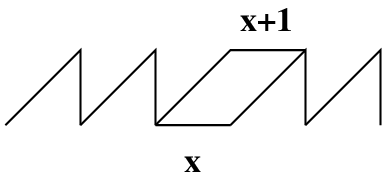}}}\cr}}

We must consider multiple excitations of the groundstate as well.
In the $k$-excitation sector, $k$ of the pairs of adjacent $\gamma$ 
half-tiles must be replaced by $\alpha$ or $\beta$ tiles. 
This results in the appearance of $k$ occupied lower
and upper edges. This conservation of the number of occupied edges from one row to the next
is an important feature of the model. It allows to break the action of the
transfer matrix $T_L$ into $L+1$ sectors, according to the number of excitations
$k\in\{0,1,2...,L\}$. The invariant space corresponding to $k$ excitations has
dimension ${L \choose k}$.
For instance,
in the case of $k=2$ excitations,  we must replace two of the pairs of adjacent
$\gamma$ half-tiles by $\alpha$'s or $\beta $'s. 
The corresponding lower occupied edges
are in positions say $x$ and $y$, $0\leq x<y \leq L-1$.
Let us denote by $v(x,y)$ the edge configuration vector
with only $O$'s except for two $1$'s in positions $x$ and $y$.
Just like in the 1-excitation case, the occupied upper edges must be in positions
$x$ or $x+1$ and $y$ or $y+1$, with moreover the constraint that they be distinct.
Forgetting about this constraint, we would simply have
\eqn\twomat{
T_L \, v(x,y)=  v(x,y) + v(x+1,y)+v(x,y+1)+ v(x+1,y+1) }
Imposing the constraint on the right hand side requires that, whenever
$y=x+1$, the second term should be removed, as well as the third one
whenever $y=L-1$ and $x=0$.
There is a much simpler way of automatically ensuring this: we simply enhance 
the domain of definition to $x \in [0,L]$ and demand that
$v(x,x)=0$ for all $x$, and that $v(x,L)=v(0,x)$ for all $x$.

The $k$-excitation sector is easily constructed. Let $v(x_1,x_2,...,x_k)$ be the
edge-configuration vector with $1$'s in positions $x_1\leq x_2\leq ... \leq x_k$,
with the convention that it vanishes whenever any two consecutive positions are equal,
and that $v(x_1,x_2,...,x_{k-1},L)=v(0,x_1,...,x_{k-1})$.
Let us also introduce the translation operator $\tau_j$ that adds $1$ to the
$j$-th argument of such a vector, namely
\eqn\translatau{\tau_j v(x_1,x_2,...,x_{j-1},x_j,x_{j+1},...,x_k)
=v(x_1,x_2,...,x_{j-1},x_j+1,x_{j+1},...,x_k)}
Note that the result is the zero vector when $x_{j+1}=x_j$ or $x_j+1$.
The transfer matrix then acts in the $k$-excitation sector as
\eqn\ktrans{
T_L \, v(x_1,...,x_k) = \prod_{j=1}^k (I+ \tau_j)\  v(x_1,...,x_k)}
where $I$ is the identity and the various translation operators commute 
with one-another. The vanishing constraint on the vectors takes care of all the steric
constraints for the $\alpha$ and $\beta$ tiles.

We now look for eigenvectors of $T_L$ in the form
\eqn\eigform{V^{(k)}=\sum_{0\leq x_1<x_2<\cdots x_k\leq L-1}f(x_1,\cdots,x_k)
v(x_1,\cdots,x_k)}
in the $k$-excitation sector.
The corresponding eigenvalue equation reads 
\eqn\evaleq{\Lambda_k V^{(k)} = \prod_{j=1}^k (I+ \tau_j) \ V^{(k)}}
This equation must be solved within the invariant ${L\choose k}$-dimensional 
vector space of the $k$-th sector, with the boundary conditions
\eqn\condiv{
\eqalign{
f(...,x,x,...)&=0 \cr
f(x_1,x_2,...,x_{k-1},L)&=f(0,x_1,...,x_{k-1})\cr}}

\subsec{Bethe Ansatz}

In view of the equations \evaleq, we may naturally think of trying some
simple Ansatz for the candidate eigenvectors, based on the eigenvectors of the
translation operator $\tau$. Indeed, for $k=1$, $f(x)=z^x$ produces 
an eigenvector for $\tau$, for
any non-zero $z$. The $k$-variable version is
$f(x_1,...,x_k)=z_1^{x_1}z_2^{x_2}....z_k^{x_k}$, for some non-zero complex
numbers $z_1,z_2,...,z_k$.
This Ansatz would work perfectly except that it violates the boundary
condition, that $f$ vanishes whenever two consecutive arguments are equal.

The next fundamental remark is that if we still ignore the boundary condition, any
permutation $\sigma\in S_k$ of the $x$'s would yield an equally acceptable
candidate $f_\sigma(x_1,x_2,...,x_k)=z_1^{x_{\sigma(1)}}
z_2^{x_{\sigma(2)}}....z_k^{x_{\sigma(k)}}$, with the same eigenvalue.

The idea of the Bethe Ansatz is then to combine these two ideas and look for
a solution of the eigenvalue equation \evaleq\ that is a linear
combination of the $f_\sigma$, $\sigma\in S_k$, and  that incorporates the
boundary conditions \condiv.
In the present problem, the answer is unique and reads
\eqn\solbeth{ f(x_1,x_2,...,x_k)=\det( z_i^{x_j} )_{1\leq i,j \leq k}}
Moreover, the periodic boundary conditions (second line of \condiv) 
along the strip are satisfied iff
\eqn\betaneq{\eqalign{z_i^L&=(-1)^{k-1} \qquad {\rm for} \ \ {\rm all}\ \ i=1,2,...,k\cr
z_r &\neq z_s \qquad {\rm for} \ \ {\rm all}\ \ r,s=1,2,...,k, \ \ r\neq s\cr}}
where the extra condition
that the $z_i$ must be distinct ensures that the eigenvectors are non-zero.
These are the celebrated Bethe Ansatz equations, in our particularly simple case.
Note that we indeed get a basis of the ${L\choose k}$-dimensional invariant space
by taking $f$ as in Eq.\solbeth\ with
\eqn\solzi{
\eqalign{ z_j &=e^{2i\pi (n_j-{k-1\over 2})/L} \ \ j=1,2,...,k\cr
0&\leq n_1<n_2<...<n_k\leq L-1\cr}}
The corresponding eigenvalues read
\eqn\evalues{
\Lambda_k(n_1,...,n_k)= \prod_{j=1}^k (1+e^{2i\pi (n_j-{k-1\over 2})/L}) }

\subsec{Continuum limit: largest eigenvalue}

The continuum thermodynamic limit of the model corresponds to $L,M \to \infty$.
For $L$ finite and $M\to \infty$, the
thermodynamic limit of the free energy of the model
is governed by the largest eigenvalues of $T_L$. More precisely,
we may compute the entropy of tiling per row of a strip of infinite length
and finite width $L$ by taking the large $M$ limit of \parZf\ 
$ S_L=\lim_{M\to \infty} {1\over M} {\rm Log}\, Z_{L,M}$.
Using the explicit form \evalues, the largest eigenvalues
are obtained by having the maximum number of terms $(1+z_j)$ that are
larger than one in module. 

\fig{The roots $z_j$ of the Bethe Ansatz equation \betaneq\ lie 
in the complex plane on the unit circle (dashed circle). We have 
also represented the location of their shifted values $(1+z_j)$ 
(solid circle). The maximal eigenvalue among the values \evalues\ 
is obtained by retaining only those $z_j$'s whose shifted value lies
outside of the unit circle (black dots) and having as many of these
as possible. These correspond
to $z_j=e^{i \theta_j}$ with $\theta_j\in [-2\pi/3,2\pi/3]$.}{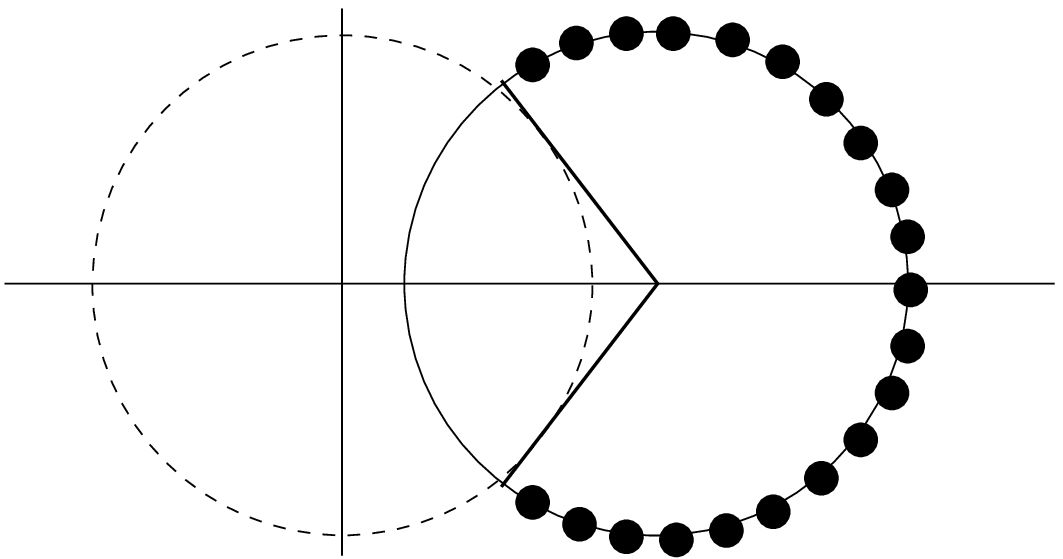}{7.cm}
\figlabel\ucircles

We have represented in Fig.\ucircles\ the unit circle and its shifted
image under $z\to 1+z$. Only a portion of this image lies outside of the unit
circle. Let us denote by $z_\pm =e^{\pm 2i\pi /3}$
the two intersections between the two circles of Fig.\ucircles.
We will have a maximum number of terms $(1+ z_j)$ with modulus larger than
one for the value of $k_0$ such that
\eqn\vakgener{ {k_0-1 \over 2L} < {1\over 3} \leq {k_0\over 2L}}
For simplicity, let us assume that $L=3 n$, in which case \vakgener\ gives
$k_0=2 n$. 
The largest eigenvalue reads then 
\eqn\larevgen{
\Lambda_{2n}(0,1,...,2n-1)=
\prod_{j=0}^{2n-1} (1+e^{2i\pi (j-{2n-1\over 2})/(3n)}) }
The corresponding tiling entropy reads
\eqn\entrogen{
S_L=S_{3n}=\sum_{j=0}^{2n-1}
{\rm Log}(1+e^{2i\pi (j-{2n-1\over 2})/(3n)})}

\subsec{Thermodynamic entropy and central charge}

To get the thermodynamic tiling entropy per tile \freter, we must
also take the large $L$ limit of $S_L$: $s=\lim_{L\to \infty} S_L/L$.
We immediately get 
\eqn\entrop{s=\int_{-1/3}^{1/3} dx\,{\rm Log}(1+e^{2i\pi x})=\sum_{p\geq 1}
{(-1)^{p-1}\over \pi p^2} \sin {2\pi p\over 3}}

The central charge may be obtained from the large $L$ corrections to the
thermodynamic entropy. From general principles, the large $L$ limit is 
indeed expected to be described by a conformal field theory
if the model is critical and translationally and rotationally invariant
as it is presently. 

Moreover, the central charge $c$ is extracted by computing the anomaly of free energy
of the model on a cylinder of width $L$, that behaves like
\eqn\freanom{ f_L= -S_L= -L s -{\pi c \over 6 L}{\sqrt{3}\over 2} +o({1\over L})}
for large $L$, with a bulk free energy per tile $f=-s$ as above and a central charge $c$.
The non-conventional factor ${\sqrt{3}\over 2}$ is simply due to our deformation
of the original rhombus of size $L\times M$ into a
rectangle with size $L\times M\sqrt{3}/2$ in the same units. 

The large $L=3n$ expansion of $S_L$ is easily derived using the following expansion,
valid for any sufficiently differentiable function $f$ 
\eqn\utilint{\eqalign{
\int_{-1/3}^{1/3} &f(x) dx -{1\over 3n} \sum_{j=-{2n-1\over 2}}^{2n-1\over 2}
f({j\over 3n}) = \sum_{j=-{2n-1\over 2}}^{2n-1\over 2} \int_{2j-1 \over 6n}^{2j+1\over 6n}
(f(x)-f({j\over 3n})) \cr
&=\sum_{j=-{2n-1\over 2}}^{2n-1\over 2} \int_{2j-1 \over 6n}^{2j+1\over 6n}
\big( (x-{j\over 3n}) f'({j\over 3n}) +{1\over 2}(x-{j\over 3n})^2 f''({j\over 3n})\big) dx+
O({1\over n^3})\cr
&={1\over 3 (6n)^3} \sum_{j=-{2n-1\over 2}}^{2n-1\over 2} f''({j\over 3n}) +O({1\over n^3})\cr
&={1\over 6 (6n)^2} \int_{-1/3}^{1/3} f''(x) dx +O({1\over n^3}) \cr
&={f'(1/3)-f'(-1/3) \over 6 (6n)^2}\cr}}
Applying this to the function
\eqn\funcf{f(x)={\rm Log}(1+e^{2i\pi x})}
we finally get the expansion of $S_L$:
\eqn\expansl{
{1\over L} S_L= s + {\pi \sqrt{3} c \over 12 L^2}+O({1\over L^3})}
with
\eqn\cenchar{ c= -{1\over 2 \pi \sqrt{3}} (f'(1/3)-f'(-1/3))=1}
The above computation therefore confirms the result
announced in Section 4.1 that the $FPL(1)$ model is described in the
continuum by a CFT with central charge $c=1$.

\fig{The triangular lattice ${\cal T}$ of mesh size $a$, dual
to the hexagonal lattice on which loops are drawn, is the locus of the
heights $X$ allowed by the rules of Fig.\Onrules-(a).
The ideal lattice ${\cal I}$, corresponding to the values of
$h=\langle X\rangle$ where the free energy is minimal, is an
hexagonal lattice of mesh size $a\sqrt{3}$.
The triangular sub-lattice ${\cal R}$ of ${\cal I}$
corresponds to ideal states associated with the {\it same} coloring. The
mesh size of ${\cal R}$ is $\sqrt{3}a$. Passing from triangle M to triangle N
by successive reflections along the edges brings the triangle back in the
same orientation. The reciprocal lattice ${\cal R}^*$ of the lattice
${\cal R}$ is a triangular lattice of mesh size $2/(3a)$ (here represented
for $a=1/\sqrt{3}$). The smallest non-zero vector of the reciprocal lattice
${\cal I}^*$ of ${\cal I}$ is the second smallest vector of ${\cal R}^*$,
for instance that joining M to P in the figure. Its length
is $2/(\sqrt{3}a)$.}{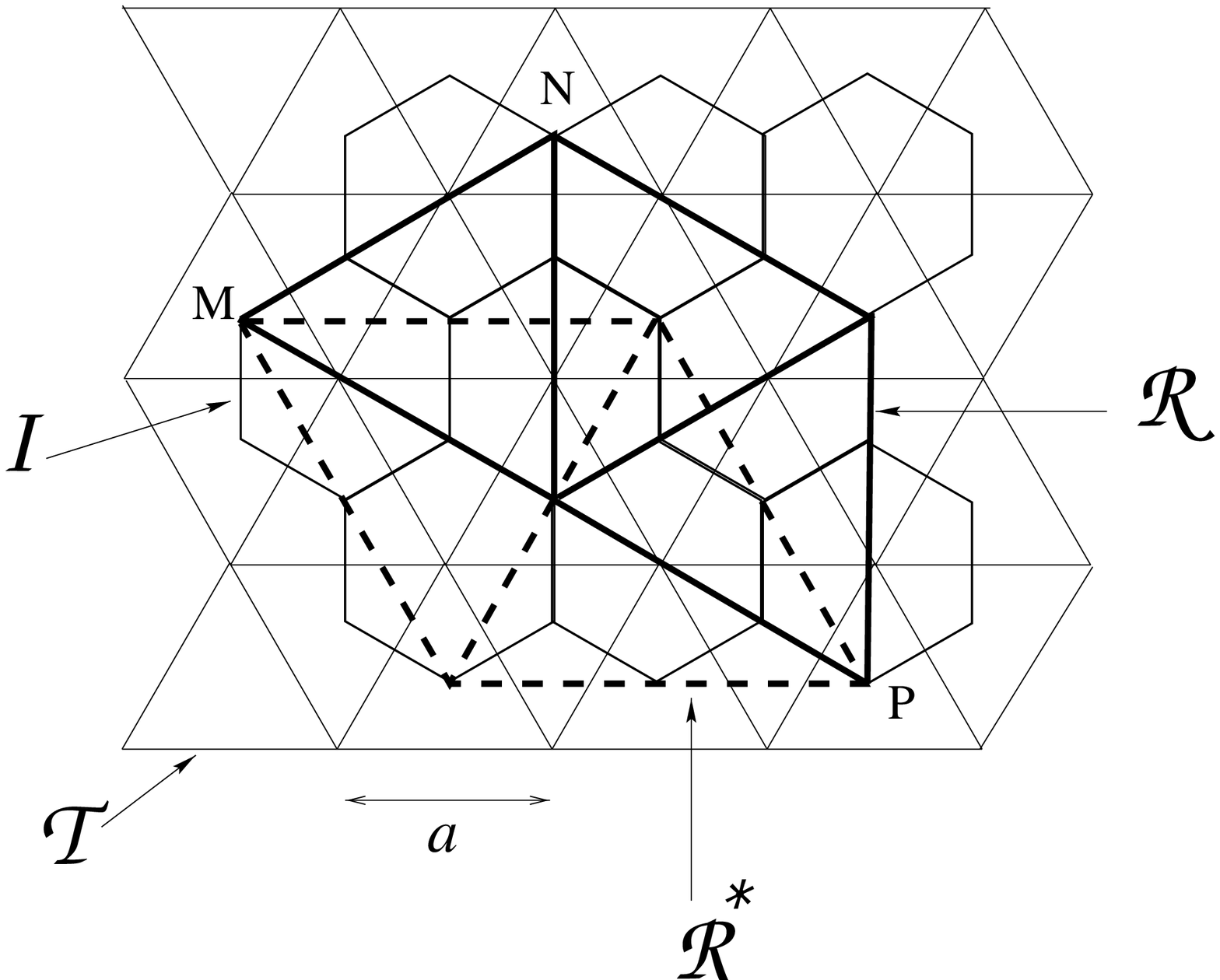}{8.cm}
\figlabel\ideal

\newsec{Effective field theory description via Coulomb gas}

\subsec{Two-component Coulomb gas for $FPL(2)$}
Let us now see how to construct an effective continuum field theory 
for the $FPL(n)$ model at large distances.
We introduce a continuous description of the heights in the form of
a two-dimensional two-component field $h(r)=(h_1(r),h_2(r))$  where
$r=(r_1,r_2)$ is the continuous version of the two coordinates of a node
on the hexagonal lattice while $h(r)$ is a locally averaged value of 
the two-dimensional height $X$
around the point $r$. A precise definition of $h$ is not necessary
as only the symmetries of the problem will indeed matter.
The approach presented here follows Ref.\KGN\
and is based on two-dimensional Coulomb gas techniques \Nien.

A first step consists in identifying the so-called ``ideal" configurations,
i.e.\ those with a maximal entropy of excitations. We shall indeed assume that
the large scale statistics of the problem is dominated by 
the fluctuations around these ideal states.
In the dual language of the planar folding of the triangular lattice, 
these {\it ideal configurations} correspond to {\it maximal foldings},
i.e.\ configurations where the lattice is completely folded onto a 
single triangle. These compactly folded states are indeed those with 
the largest number of elementary excitations obtained by unfolding
the six links bordering an elementary hexagon. To each ideal configuration, 
we may associate the corresponding averaged value of $h$,
equal to the position  of the center of mass of the triangle
on which the lattice is folded. The corresponding allowed values of $h$
form themselves a hexagonal lattice ${\cal I}$ of mesh size $a/\sqrt{3}$ 
if $a$ is  the mesh size of the folded triangular lattice
(see Fig.\ideal). We finally write the effective free energy as
\eqn\free{f=\int d^2r \big[\pi g \left((\nabla h_1)^2+(\nabla h_2)^2\right)
+V(h)\big]}
with a Gaussian part describing the height fluctuations
and an effective ``locking" potential $V(h)$ having the periodicity of 
the ideal lattice ${\cal I}$ and enhancing the weight of the ideal states. 
The ``roughness" $g$ will be determined below in a self-consistent way.

In terms of loops, each ideal state corresponds to a regular covering of
the hexagonal lattice by means of elementary hexagonal loops of length
six, all oriented in the same direction and centered on one of the
three sub-lattices of the dual (tripartite) triangular lattice. 
There are six such tilings corresponding to the six permutations of 
the three colors ABC. The vertices of the ideal lattice ${\cal I}$
that correspond to the {\it same} covering, i.e.\ the {\it same coloring}, 
form a sub-lattice ${\cal R}$ of the ideal lattice ${\cal I}$. 
This so-called ``repeat" lattice ${\cal R}$ is a triangular lattice of
mesh size $\sqrt{3} a$ (see Fig.\ideal).
Any local observable of the loop gas must therefore have the periodicity
of the lattice ${\cal R}$ and may thus be written as
\eqn\opelec{\Phi(r)=\sum_{G\in {\cal R}^*} \Phi_G {\rm e}^{2{\rm i} \pi G.h(r)}}
where $G$ runs over the vectors of the reciprocal lattice ${\cal R}^*$ of
${\cal R}$ (defined by $G.b=$ integer for any $b$ in ${\cal R}$).
The lattice ${\cal R}^*$ is a triangular lattice of mesh size
$2/(3a)$ (see Fig.\ideal).
On the other hand, the locking potential 
$V(h)=\sum_{G\in {\cal R}^*} v_G {\rm e}^{2{\rm i} \pi G.h(r)}$ 
has non-vanishing Fourier components $v_G$ only for $G\in {\cal I}^*$,
the sub-lattice of ${\cal R}^*$ equal to the reciprocal lattice of
the ideal lattice ${\cal I}$.
The value of $g$ may now be obtained by assuming that $V(h)$ is
a {\it marginal} perturbation of the Gaussian free field,
i.e.\ with dimension $x=2$. 

Very generally, the dimension of the (electric) ``vertex operator"
$V_G(r)\equiv {\rm e}^{2{\rm i} \pi G.h(r)}$, which describes the 
large distance behavior of the correlation function
$\langle V_G(r) V_{-G}(0)\rangle \sim r^{-2x(G)}$ is equal to \Nien\
\eqn\xg{x(G)={G^2\over 2g}}
The dimension $x$ of $V(h)$ corresponds to that of the most
relevant electric operator (with the smallest  $x(G)$) occurring in the
Fourier decomposition of $V(h)$, i.e.\ to the vector $G$ of ${\cal I}^*$ 
with {\it minimal} norm  (vector $MP$ in Fig.\ideal). 
The norm of this vector being equal to $2/(\sqrt{3}a)$, we finally obtain
\eqn\valg{g={1\over  3 a^2}}

Once $g$ is fixed, we can obtain exponents describing the large distance
behavior of correlations between {\it coloring defects}. These take the 
form of ``magnetic operators" defined as follows.
A coloring defect around a node of the hexagonal lattice
corresponds to a dislocation-type defect in the height $X$ after
a complete turn around this node. This height defect reads 
$\Delta X= M=m_1(A-B)+m_2(A-C)$ for two relative integers $m_1$ and $m_2$ 
(the vector $M$ is thus a vector of the lattice ${\cal R}$). 
In order to have a configuration
without defect at infinity, we must ensure {\it magnetic neutrality},
for instance by introducing another defect with magnetic charge $-M$ 
at some other node. In this case, the variable $X$ has a jump discontinuity 
equal to $M$ along a line joining these two defects. 
The defect/anti-defect correlation then behaves at large separation
$r$ as $r^{-2 x(M)}$ where the magnetic dimension reads
\eqn\xm{x(M)={g M^2\over 2}={1\over 6}\left({M\over a}\right)^2}
A first example of defect consists in having two unvisited sites
at distance $r$ from one another. This corresponds to a coloring defect
characterized by $M=3A$ with norm $3a$ and with therefore $x=3/2$.
The fact that $x<2$ shows that this is indeed a relevant perturbation
and that the fully packed loop gas corresponds to an {\it unstable} phase
with respect to such defects.
The introduction of coloring defects is also useful to describe open 
lines in the loop gas.
For instance, we may describe an {\it open} BC line linking two
points at (an odd) distance $r$ by introducing coloring defects
at these two points, each with its three incident edges colored A,A,B. 
The corresponding value of $M$
is $M=2A+B$, with norm $\sqrt{3}a$, which gives an exponent $x_1=1/2$.
Very generally, one finds an exponent $x_\ell$ associated
with the large distance behavior of correlations of the
form $r^{-2 x_\ell}$ for a set of $\ell$
BC lines starting from the vicinity of a given point and 
ending at the vicinity of another at distance $r$, with the value \KGN
\eqn\xl{\eqalign{x_{2k}&={1\over 2} k^2\cr x_{2k-1}&={1\over 2}(k^2-k+1)\cr}}
according to the parity of $\ell$.

\subsec{General $FPL(n)$ and $O(n)$ cases}
Beyond the case $n=2$, the $FPL(n)$ model for $-2\leq n\leq 2$ 
may be described by configurations of oriented self-avoiding
fully packed loops, with weights ${\rm e}^{+{\rm i}\pi e/6}$,
resp. ${\rm e}^{-{\rm i}\pi e/6}$ per right, resp. left turn
of the loop. Summing over the two possible orientations 
of each loop results in a weight $n=2\cos(\pi e)$ per loop. 
This holds only for contractible loops while loops escaping at 
infinity receive a weight $2$ instead. To understand how to cure
this problem, it is more convenient to define the model on an infinite 
cylinder, 
in which case the non-contractible loops wind once around the cylinder 
and receive a wrong weight $2$ instead of $n$. This is corrected by 
introducing two electric operators with (two-dimensional) charges $E$ 
and $-E$ at both ends of the cylinder, which implies an additional
energy term of the form $2{\rm i}\pi E.(h(+\infty)-h(-\infty))$.
Each non-contractible loop induces a discontinuity $h(+\infty)-h(-\infty)$ 
of $\pm B$ (or $\mp C$) and it is thus sufficient to choose $E$
such that $E\cdot B=-E\cdot C=e/2$, i.e.\ $E=e (B-C)/(3 a^2)$.
Summing over the two orientations of the non-contractible loops
now yield the correct weight $n$ while contractible loops are
unaffected. The introduction of the ``background" charge $E$ modifies
the dimensions of the electric and magnetic operators into
\eqn\dimelma{x(G,M)={1\over 2g}G\cdot(G-2E)+{g\over 2}M^2.}
The value of $g$ is again determined by the requirement that $V$ be
marginal. Choosing the determination $0\leq e\leq 1$ of $e$, one finally gets
\eqn\valgn{g={(1-e)\over 3 a^2}.}
One then finds an exponent $x_\ell(n)$ associated
to the correlation of  $\ell$ BC lines given by
\eqn\xln{\matrix{&\displaystyle{x_{2k}(n)={1-e\over 2} k^2-{e^2\over 2(1-e)}
(1-\delta_{k,0})}\cr
&\displaystyle{x_{2k-1}(n)={1-e\over 2}(k^2-k+1)-{e^2\over 2(1-e)}}\cr}
}
Finally, the central charge of the model now reads $c=2+12 x(E,0)$,
i.e.:
\eqn\ccc{c_{\rm fully\ packed}(n)=2-6 {e^2\over 1-e},\qquad n=2 \cos(\pi e)}

To conclude this study, let us now see how the above formulas 
are modified in the case of the dense phase of the $O(n)$ model, 
which as we already saw for $n=2$, consists in allowing  for the 
presence of unvisited nodes,
in which case the heights become one-dimensional. The above analysis
transcribed to this much simpler case leads finally to
a central charge
\eqn\ccon{c_{\rm dense}(n)=1-6 {e^2\over 1-e},\qquad n=2 \cos(\pi e)}
and to exponents
\eqn\xlon{x_{\ell}(n)={1-e\over 8} \ell^2-{e^2\over 2(1-e)}(1-\delta_{\ell,0})}
for any $\ell$.
\fig{Renormalization flow in the $(n,u^{-1})$ plane. We have indicated
an example of flow at constant $n$. The fully packed loop gas corresponds
to an unstable fixed point at $u^{-1}=0$. The dense loop gas
corresponds to a stable fixed point. The central charge decreases
by $1$ between these two fixed points 
$c_{\rm dense}(n)=c_{\rm fully\ packed}(n)-1$.
}{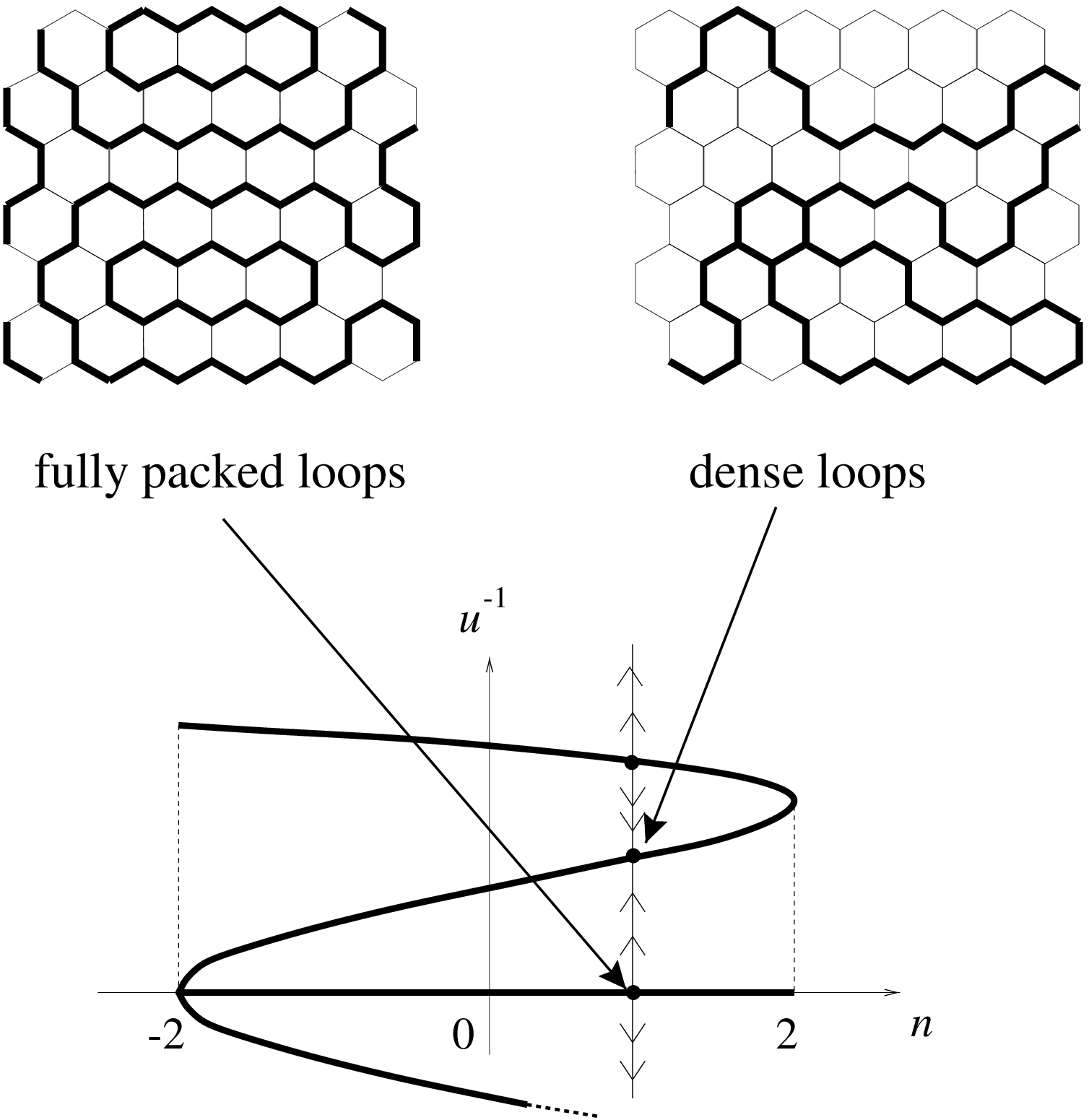}{10.cm}
\figlabel\diagflot
The lesson to draw from the above is that, at fixed weight $n$ per loop,
the fully packed loop gas has a central charge $1$ more
than the dense loop gas. This result is illustrated in
Fig.\diagflot\ where we display a diagram of renormalization group flows
in the $(n,u^{-1})$ plane, where $u$ is the weight per visited site
and $-2\leq n\leq 2$ \BN. All the flows follow lines of constant $n$ and
the problem is symmetric under $u\to -u$ (since the number of
sites visited by loops is even). The $u^{-1}=0$ line,
which corresponds to the fully packed loop gas, is a line of
unstable fixed points (see Fig.\diagflot). The presence of defects
(which, as we already saw, are relevant) drives the model away from this line
towards the line of stable fixed points describing the dense phase
of the $O(n)$ model. One also finds another line of unstable
fixed points at finite values of $u^{-1}=\sqrt{2+\sqrt{2-n}}$ 
corresponding to the transition points of the $O(n)$ model \BN.

\bigskip
\vfill\eject
\leftline{\bf PART C: LOOP MODELS ON RANDOM LATTICES}
\bigskip
We would like to generalize the models of 2D folding described in part A
and those of fully packed loops described in part B to the case of random 
2D lattices. By random 2D lattices, we mean tessellations
made of 2D polygonal rigid tiles glued together so as to form discrete surfaces 
with possible curvature defects concentrated at the nodes. 
We are interested in the phantom folding configurations of statistical 
ensembles of 2D lattices. These form discrete models for ``fluid membranes",
i.e. membranes without internal elastic skeleton and whose internal
metric fluctuates \SMMS. Beyond folding, we will also study fully packed 
loop models on random lattices. All these models are examples of
a larger class of statistical models describing the coupling of ``matter"
degrees of freedom to the fluctuations of ``space", used in the 
quantization of general relativity. More precisely, the present 2D models
correspond to discrete realizations of the so-called 
two-dimensional quantum gravity (2DQG) \QGRA.

In this part, we concentrate on the case of random triangulations or, 
dually, random trivalent graphs. Various results are presented, relying 
either on matrix integral 
techniques [\xref\MSRI,\xref\BIPZ-\xref\EY] or more general
effective 2DQG descriptions \KPZ. Exact predictions are tested
against numerical enumerations \GKN.
The lesson of this study is a clear distinction between
ordinary 2DQG and so-called Eulerian 2DQG corresponding
to a restriction of the set of dynamical lattices to Eulerian
tessellations.

\newsec{Folding of random lattices}

\subsec{Foldability of triangulations}

\fig{Example of triangulation with spherical topology.}{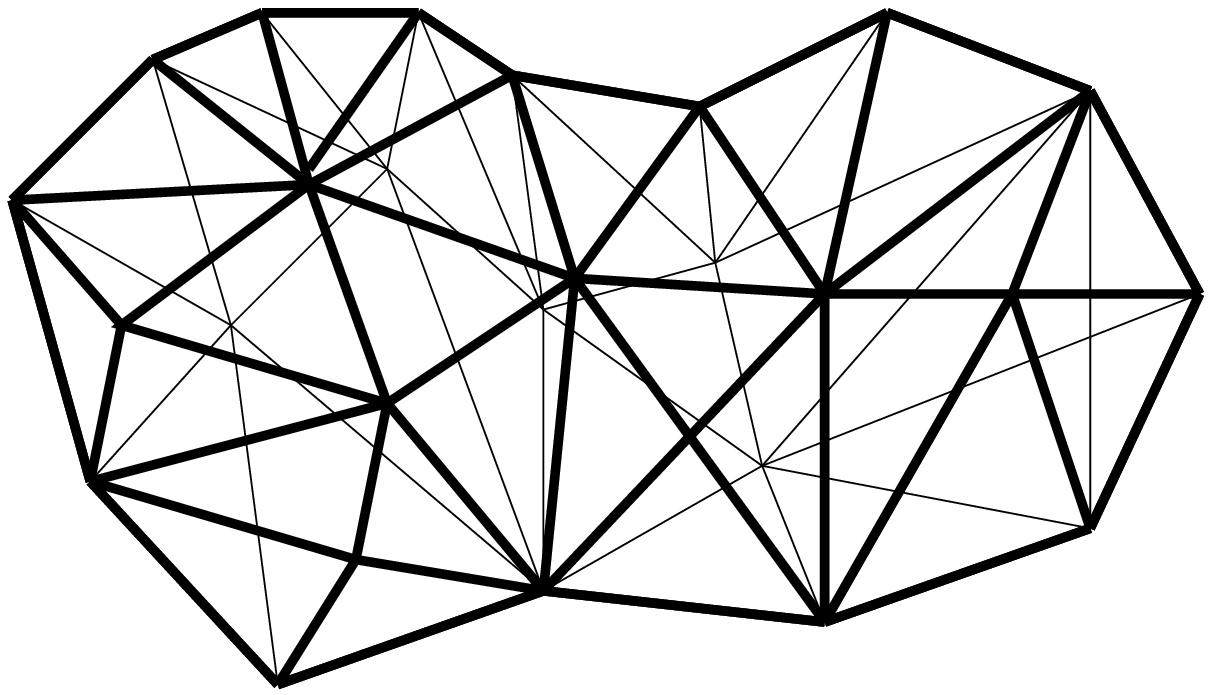}{8.cm}
\figlabel\triangu

In this part C, we shall consider only the
simplest case of triangulations, for which all the tiles are
equilateral triangles with edges of unit length. 
We will moreover restrict ourselves
to {\it planar} triangulations, i.e.\ tessellations of a surface with the
topology of the sphere (genus $0$, no handles). An example of such a
triangulation is given in Fig.\triangu. 

Given such a triangulation, we define as before a folding as a map
from the triangulation into $\IR^d$ such that its restriction
to any elementary triangle is an isometry. 
In particular, each triangle is again mapped onto an equilateral 
triangle in $\IR^d$.
\fig{Two examples of planar triangulations: the planar representation
(left) keeps track of the connectivity of the triangles only at the
expense of deforming them. Note that the external face is also a triangle. 
The triangulation (a) is foldable into $\IR^3$ on a tetrahedron, but
it is {\it not} foldable in the plane because its nodes are not tricolorable. 
The triangulation (b) is foldable into $\IR^3$ on an octahedron and 
it is foldable in the plane as it is node-tricolorable 
(here with three colors e,f,g). An example of folding is indicated on 
the right in the planar representation by thickening the folded links.
}{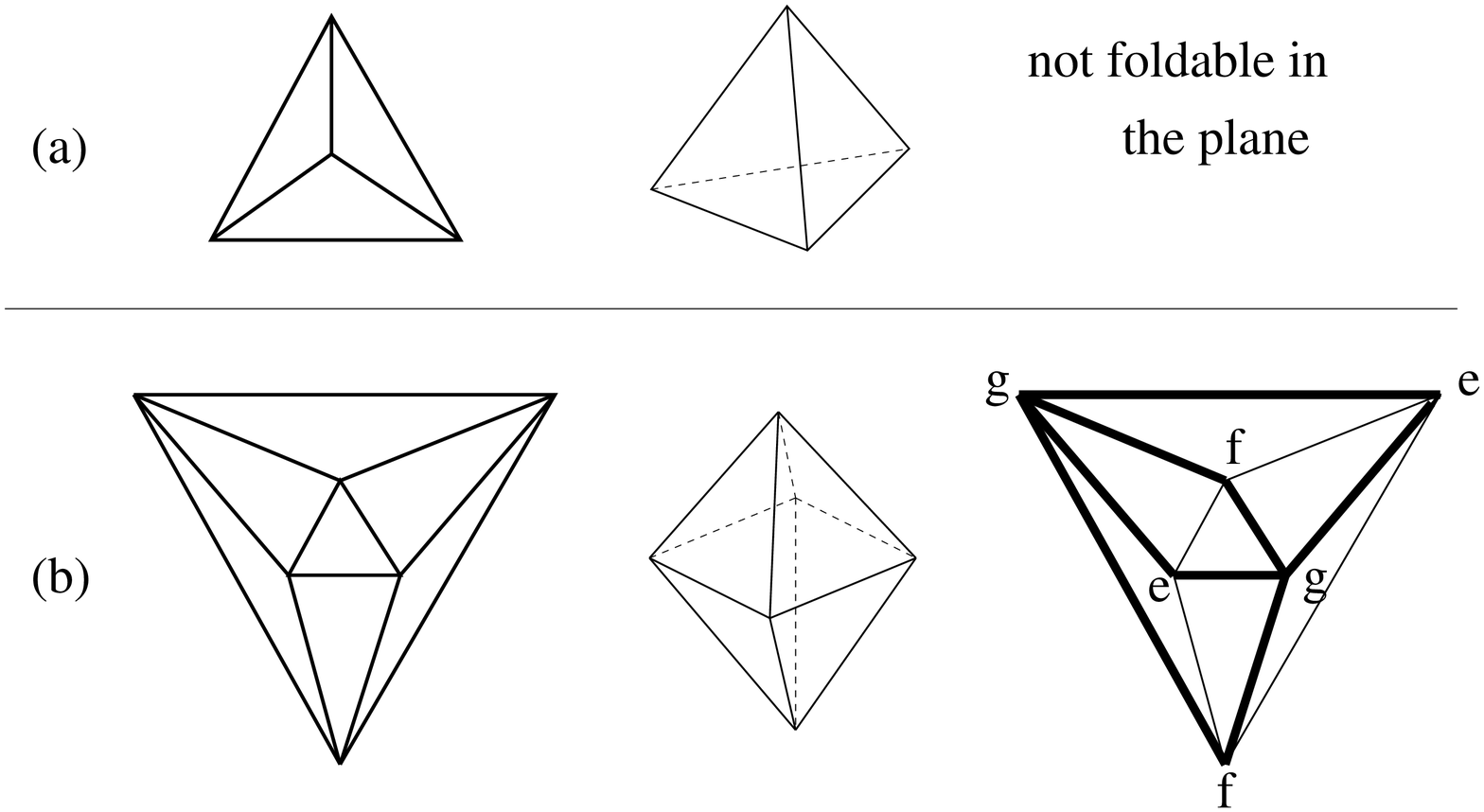}{11.cm}
\figlabel\pliables
In the simplest case of planar folding ($d=2$), a preliminary 
question concerns the very existence of such foldings.
Indeed, not all triangulations may be folded into the plane. 
For instance, the triangulation on the top of Fig.\pliables\ 
(which may be embed in $\IR^3$  so as to
form a tetrahedron) cannot be folded into the plane.
On the other hand, the triangulation at the bottom of Fig.\pliables\ 
(which may be embedded in $\IR^3$ so as to form an octahedron) 
can be folded into the plane.

To answer this question, we note that, whenever a folded state exists, 
the positions in the plane of the nodes of the triangulation necessarily 
belong to a regular triangular lattice of mesh size $1$ in the plane $\IR^2$.
The triangular lattice is tripartite, i.e.\ its nodes may be colored
with three colors (e,f,g) so that no two adjacent nodes be of the
same color. This tricoloring is unique (up to a global permutation
of the three colors). If a folded state exists, it therefore induces
a tricoloring of the nodes of the triangulation we started from. 
This yields a necessary condition for the triangulation to be foldable 
in the plane, namely that it be {\it node-tricolorable}.
Conversely, starting from a node-tricolorable triangulation, its
tricoloring is unique (up to a global permutation
of the colors). Such a triangulation may be folded on a single equilateral 
triangle by sending each node of a given color onto one of the three vertices 
of this triangle. Such folding corresponds to the {\it complete} folding 
of the triangulation from which other folded configurations may
be obtained by partial unfolding.

We obtain finally the following equivalent characterizations:
\item{1.} The triangulation is foldable into the plane;
\item{2.} The triangulation is node-tricolorable.
\par
\noindent In the case of a triangulation with spherical topology,
we also have the following equivalent alternative characterizations \Intel:
\item{3.} Its faces are {\it bicolorable} (with distinct colors 
on adjacent triangles);
\item{4.} Its edges may be oriented so that the boundary of each triangle
receives a well-defined (clockwise or counterclockwise) orientation;
\item{5.} The number of triangles around each node is even;
\item{6.} The number of edges adjacent to each node is even.

This last property justifies the denomination ``Eulerian" for
such triangulations as it ensures the existence on
the triangulation of a closed Eulerian path (i.e.\ a path visiting
all edges exactly once). In other words, it ensures the possibility of drawing 
the triangulation by a single (closed) path without lifting the pen.

To summarize, {\it the planar foldable triangulations are
the Eulerian triangulations}, characterized by any of
the above properties 1-6.

\subsec{Enumeration of foldable triangulations}

The enumeration of Eulerian triangulations was first carried out by
W. Tutte \Tutt\ in its dual version under the denomination of ``bicubic maps".
This enumeration may be generalized so as to keep track
of the three colors, e.g. with different weights
$p$, $q$ and $z$ for the nodes of color e, f or g respectively \CRT.
In other words, we enumerate random triangulations which are completely
folded on a single triangle, keeping track of the numbers of
nodes sent onto each of the three vertices of the triangle.

\fig{Equivalence between (a) a bicolored graph with arbitrary valences
and (b) a tricolored triangulation. From (b) to (a), we simply erase 
all the nodes of the triangulation of a given color, say, g
as well as all the links connected to them.
Conversely, from (a) to (b), we note that the nodes around each
face have alternating colors e and f. We then re-introduce
at the center of each face a node of color g and connect
it to all nodes around the face, thus creating triangles.
}{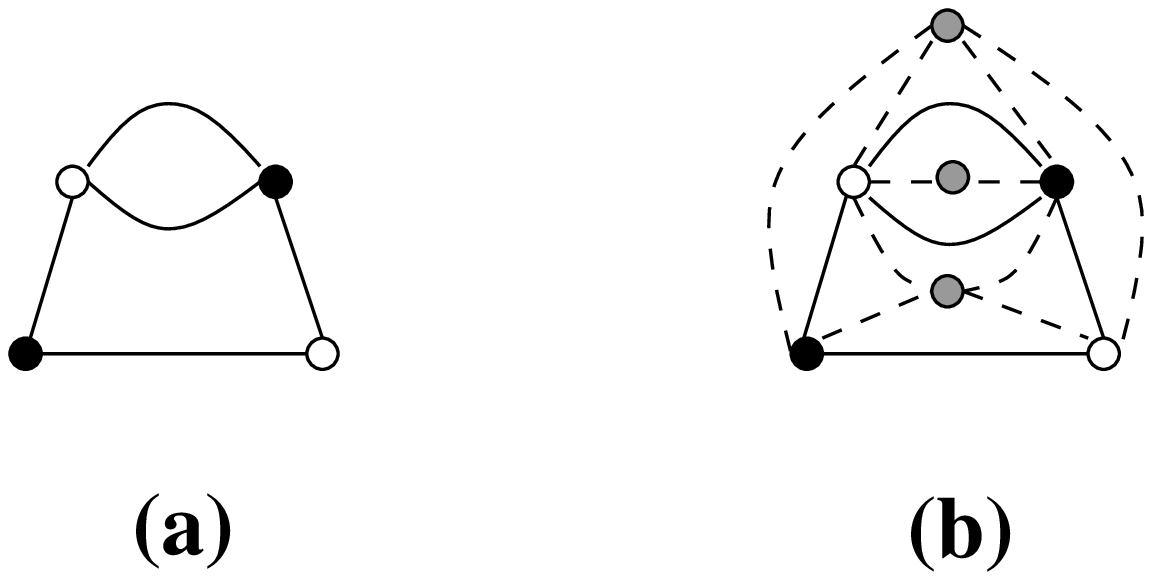}{10.cm}
\figlabel\sypqz
This may be performed by use of matrix integral techniques. 
More precisely, we may compute the generating function 
$Z(p,q,z;t;N)$ for possibly disconnected vertex-tricolored 
triangulations of arbitrary genus, with weights $p$, $q$,
$z$ for the three colors and a weight $t$ per edge of color e-f. 
The parameter $N$ governs the genus via a factor $N^{\chi}$ where
$\chi$ is the Euler characteristic of the triangulation at hand.
The partition function is defined as
\eqn\gentrico{Z(p,q,z;t;N) = \sum_{{\rm node}-{\rm tricolored}\atop
{\rm triangulations}\ T} p^{n_e(T)} q^{n_f(T)} z^{n_g(T)}
{t^{A(T)\over 2}  N^{\chi(T)}\over |{\rm Aut}(T)|} }
where $n_{e,f,g}(T)$ denote the total numbers of vertices of color 
e,f,g, where $A(T)$ denotes the total number of faces (twice
the number of e-f edges), and where $\chi(T)$ denotes the Euler
characteristic of $T$. The factor $|{\rm Aut}(T)|$ is the
order of the symmetry group of the tricolored triangulation $T$.

The construction of a matrix model to represent $Z(p,q,z;t;N)$ is
based on the following simple remark: in a given tricolored triangulation $T$
if we remove say all the vertices of color g and all edges connected to them,
we end up with a bicolored graph, with unconstrained vertex valencies
(see Fig.\sypqz).
The problem is therefore reduced to the enumeration
of bicolored graphs with a weight $p$ (resp. $q$) per node of color
e (resp. f), a weight $z$ per face and a weight $t$ per link.
Such bicolored graphs are easily built out of the Feynman graphs of a two
Hermitian matrix model, say $M_e$ and $M_f$, the index standing for
the color. The colored $m$-valent vertices of the Feynman diagrams read 
\eqn\verticolo{\eqalign{
{\rm Tr}(M_e^m)\ = \ \figbox{1.5cm}{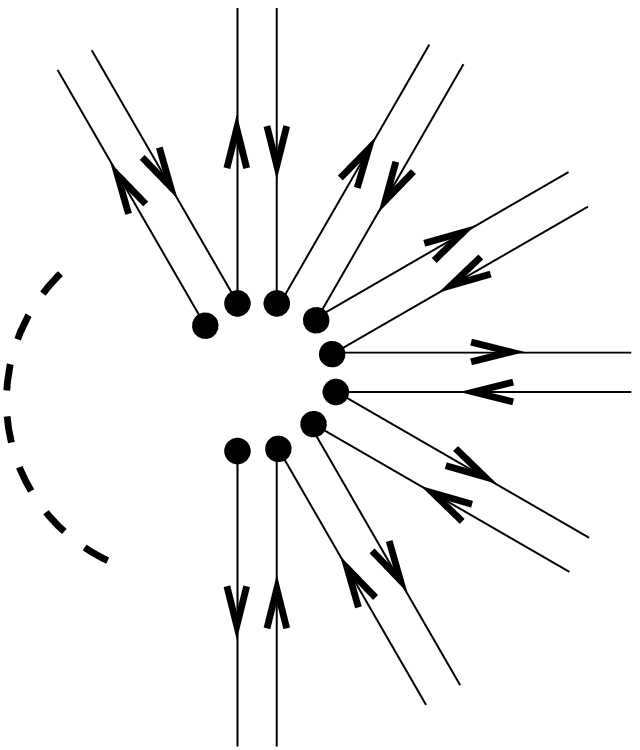}\ \  &\leftrightarrow N\, p \cr
{\rm Tr}(M_f^m)\ = \ \figbox{1.5cm}{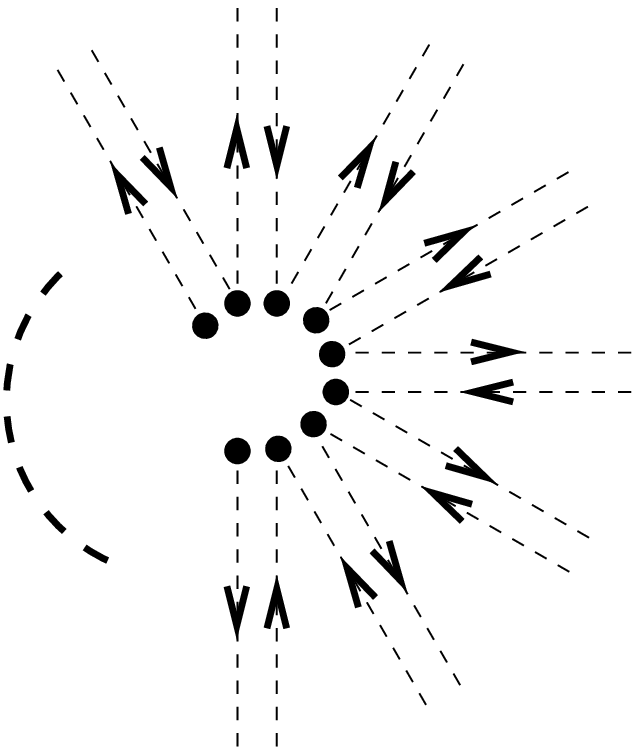}\ \  &\leftrightarrow N\, q 
\cr}}
for all $m$, and are to be connected via propagator edges with
weight $t/N$ while ensuring the alternation of colors e-f 
\eqn\propacol{ \langle (M_a)_{ij} (M_b)_{kl} \rangle\
=\ (1-\delta_{ab})\delta_{jk}\delta_{il} {t \over N}
= \ \figbox{1.5cm}{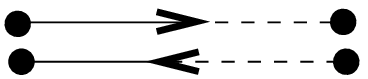}\ }
where $a,b=$ $e$ or $f$. Let us introduce the corresponding matrix integral, but
keep $N$ fixed while the matrices are taken of size $n\times n$, $n$ possibly
different from $N$. This gives the partition function
\eqn\matrep{\eqalign{
Z_n(p,q;t;N)~&=~{1\over \varphi_n(t,N)}
\int dM_e dM_f e^{-N\, {\rm Tr}\, V(M_e,M_f;p,q,t)}\cr
V(M_e,M_f;p,q,t)~&=~p{\rm Log}(1-M_e)+q{\rm Log}(1-M_f) +{1\over t} M_eM
_f \cr}}
where the normalization factor $\varphi_n(t,N)$ ensures that $Z_n(0,0;t;N)=1$.

The Feynman graph expansion of the free energy reads
\eqn\fretrico{\eqalign{
F_n(p,q;t;N)~&=~{\rm Log}\, Z_n(p,q;t;N)\cr
&=\sum_{{\rm bicolored}\ {connected}\atop
{\rm graphs}\ \Gamma} {1\over |{\rm Aut}(\Gamma)|} p^{n_e(\Gamma)}
q^{n_f(\Gamma)} t^{E(\Gamma)} N^{(V(\Gamma)-E(\Gamma))} n^{F(\Gamma)}\cr
}}
where we have denoted by $n_a(\Gamma)$ the number of vertices of color $a$,
$V(\Gamma)=n_e(\Gamma)+n_f(\Gamma)$, $E(\Gamma)$ the number of edges,
$F(\Gamma)$ the number of faces, and $|{\rm Aut}(\Gamma)|$ the
order of the symmetry group of the bicolored graph $\Gamma$.
Adding a central vertex of color $g$ in the middle of each face of $\Gamma$,
and connecting it to all the vertices around the face with edges will result
in a vertex-tricolored triangulation $T$. The number of such added vertices
is nothing but $n_g(T)=F(\Gamma)$. Introducing
\eqn\link{ z={n\over N} }
we may rewrite
\eqn\rewfnco{ F_n(p,q;t;N) = F(p,q,z;t;N) }
by use of the Euler relation 
$2-2h(\Gamma)=2-2h(T)=V(\Gamma)-E(\Gamma)+F(\Gamma)$
and the fact that $A(T)=2 E(\Gamma)$, as each edge of $\Gamma$ gives 
rise to two triangles of $T$, one in each of the two faces adjacent 
to the edge. It is also
a simple exercise to show that $|{\rm Aut}(T)|=|{\rm Aut}(\Gamma)|$.
Hence computing $F_n(p,q;t;N)$ through
the integral formulation \matrep\ will yield the generating function for
compactly foldable triangulations. In particular,
the generating function $f(p,q,z;t)$ for {\it planar} tricolored triangulations
with a weight $p$, $q$, $z$ for the nodes of the three colors and $t^{1/3}$
per edge is obtained by taking the $N\to \infty$  limit so as to
select planar graphs only, and with $n/N=z$ fixed, i.e.\ 
\eqn\fF{f(p,q,z;t)\equiv \lim_{N\to \infty} {1\over N^2} {\rm Log}Z(p,q,t;zN,N)}
An explicit calculation gives \CRT  
\eqn\finF{t\partial_t f(p,q,z;t)~=~
{U_e U_f U_g\over t^2} (1-U_e-U_f-U_g) }
where the functions $U_e$, $U_f$ and $U_g$ are determined as formal 
power series of $t$ by the equations
\eqn\generp{\eqalign{
U_e(1-U_f-U_g)~&=~pt\cr
U_f(1-U_g-U_e)~&=~qt\cr
U_g(1-U_e-U_f)~&=~zt\cr}}
with the condition $U_a={\cal O}(t)$, $a=e,f,g$.
This solution was alternatively recovered in Ref.\RECT\ by solving
a rectangular matrix model.
Remarkably, the functions $U_a/t$ are nothing but generating functions
for rooted planar trees with tricolored vertices and a root vertex
colored $a$. For instance, the first line of \generp\ reads
$U_e/t=p/(1-t(U_f/t+U_g/t))$ which is easily seen to generate
all trees with root colored e weighted by $p$ and arbitrary 
many descending subtree of color f or g attached to the root 
by inner edges weighted by $t$. The reason for this apparently
mysterious coincidence between generating functions for node-tricolored
triangulation and node-tricolored trees may be explained in a purely
combinatorial way. Indeed, a series of works [\xref\SCH,\xref\CENSUS]
has established bijections between various classes of planar graphs
and possibly decorated trees. In the present case, such a
bijection exists which consist in cutting the triangulations so as to form 
tricolored trees [\xref\COL,\xref\CONST].

By eliminating $U_f$ and $U_g$, we get the following 5th order equation 
for $U_e$
\eqn\sixo{
U_e^2(1-U_e)^2(1-2U_e+2(p-q-z)t)~
=~t^2\big((1-U_e)^2 p^2-U_e^2(z-q)^2\big)}
of which we should retain the unique solution with behavior $U_e\sim pt$
at small $t$. The values of $U_f$ and $U_g$ are obtained
from $U_e$ by permuting the weights $p,q,z$.
\fig{The value of $t_\star(p,q,z)$ in the domain $z\leq q\leq p$.}
{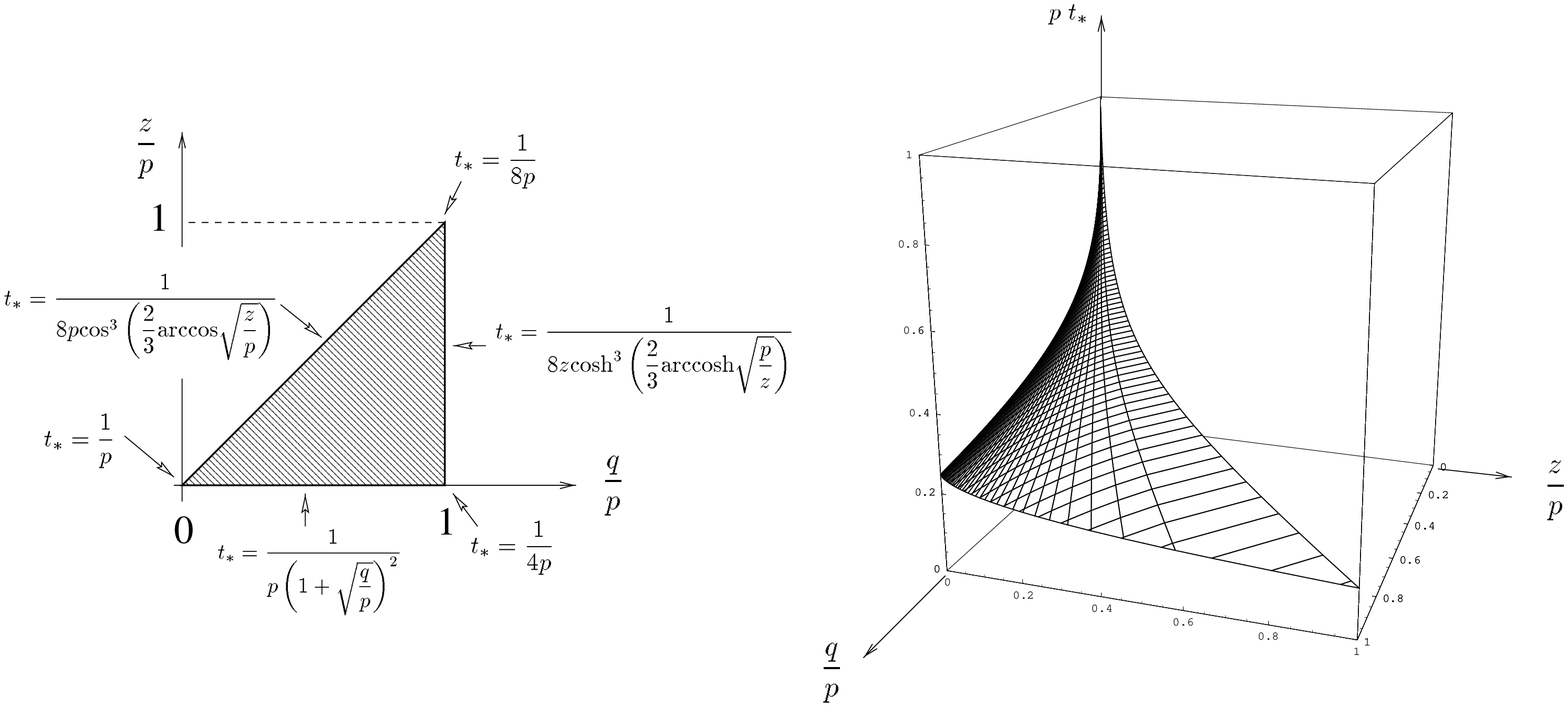}{13.cm}
\figlabel\resdeux
The case $p=q=z$ is much simpler as we may write $U_e=U_f=U_g\equiv U$, where
$U$ is the solution of the quadratic equation $U(1-2U)=zt$, namely
\eqn\Ueq{U~=~{1\over 4}(1-\sqrt{1-8zt})}
The generating function $f(z,z,z;t)$ therefore reads
\eqn\frepqeg{t \partial_t f(z,z,z;t)~=~{U^3\over t^2}(1-3U)~=~
{z\over 8t}\left(\left(2-{1\over 4zt}\right)(1-\sqrt{1-8zt})+1-6zt\right)}
from which we deduce the number $N_s$ of Eulerian triangulations
with $3s$ edges (i.e.\ $2s$ faces) and a marked oriented edge.
\eqn\reeseule{N_s={3\over 2} 2^{s}{(2s)!\over s!(s+2)!} \sim
{3\over 2\sqrt{\pi}}{8^s\over s^{5/2}}}
in agreement with Tutte's result \Tutt.
This result is to be compared to the case of arbitrary 
rooted planar triangulations
with $2s$ faces, which behaves as $(12\sqrt{3})^s/s^{5/2}$
[\xref\BIPZ,\xref\Tutt].

We deduce from the solution \sixo\ that with arbitrary
positive weights $p$, $q$, $z$, the free energy
$f_s(p,q,z)$ of tricolored  triangulations with $3s$ edges behaves
generically for large $s$ as
\eqn\partfs{f_s(p,q,z)\sim C(p,q,z) {\big(t_\star(p,q,z)\big)^{-s}
\over s^{7\over 2}}}
where  $t_\star(p,q,z)$ is the value of $t$ at which the function 
$U_e$  becomes singular.
By homogeneity, we have $t_\star(p,q,z)=(1/p)T(z/p,q/p)$
and, by symmetry, we can restrict ourselves to the domain where, for instance,
$z\leq q\leq p$.
Figure \resdeux\ represents $t_\star(p,q,z)$ in this domain
and gives some explicit formulas on the boundary of the domain.
The value $7/2$ of the exponent in formula \partfs\ is 
a general feature of so-called ``pure gravity" models that
describe the universality class of fluctuating 2D-space without matter. 
It is a universal result, common to all planar graph enumeration problems 
at generic values of their parameters, i.e.\ without fine-tuning. 
This is a particular application of the so-called KPZ relations
\KPZ\ described in detail in Section 7 below, for the coupling to gravity
of a trivial conformal matter theory with central charge $c=0$. 

The above enumeration of Eulerian triangulations solves the problem 
of compact folding, i.e.\ enumerating configurations of random 
triangulations {\it completely} folded onto a single triangle.
The Eulerian nature of a triangulation allows to
orient all its links in a coherent way, which in turn allows
for a proper definition of the link variables in the form of tangent 
unit vectors with a vanishing sum around each face. 
In order to describe the most general foldings of the triangulation, 
we simply have to assign colors to the links representing the orientation
of the corresponding tangent vector in the folded configuration.
To obtain all folded states, we have to consider all the
tricolorings of these links with three colors A, B or C,  
distinct around each triangle.
In other words, the folding of random planar triangulations
is equivalent to the simultaneous tricoloring of the nodes and of the links.
As opposed to the tricoloring of the nodes which is essentially unique,
the tricoloring of the links is a source of entropy.
While the problem of tricoloring of the {\it links only }
was solved [\xref\EK,\xref\IK], the simultaneous tricoloring
of both the links and the nodes is still an open question.

\newsec{Statistical models coupled to 2D Quantum Gravity}

The problem of folding of random triangulations studied in Section 6 above is
a particular case of statistical model coupled to 2D Quantum Gravity (2DQG).
This simply means that we replace the underlying regular lattice of
some ordinary 2D statistical model by a somewhat arbitrary
tessellation of the sphere or of some  higher genus surface.
The tessellation becomes therefore part of the configuration to be summed over.
Such models were introduced as discrete descriptions of 2D quantum gravity,
where ``matter" models are coupled to the quantum fluctuations of the underlying 
2D ``space", represented in the discrete by the tessellations 
(see e.g. Ref.\DGZ\ for a review and more references).
To each tessellation $\Theta$, we associate a statistical weight directly borrowed
from the Einstein action in 2D, namely a weight $N^{\chi(\Theta)} g^{A(\Theta)}$
where $\chi$ and $A$ are respectively the Euler characteristic and area of the
tessellation (respectively measured via $\chi=\#$faces$-\#$edges$+\#$vertices and
$A=\#$faces or vertices), and where $N$ and $g$ are the discrete counterparts of the 
Newton constant and cosmological constant.
For each tessellation $\Theta$ together with a matter configuration, 
realized for instance via a set of spins $\{\sigma\}$ on $\Theta$,
we also have a weight $e^{-E(\{\sigma\})}$ for some energy functional 
$E$. We also make the standard choice of dividing the resulting weight 
by the order $|{\rm Aut}(\Theta,\{\sigma\})|$ of the automorphism
group of the tesselation together with its spin configuration.
This leads for instance to the discretized
partition function of any 2D statistical lattice model coupled to 2D quantum gravity
\eqn\twodqgpf{ Z=\sum_{{\rm tessel.}\ \Theta}
N^{\chi(\Theta)} g^{A(\Theta)} \sum_{\{\sigma\}}
e^{-\beta E(\{\sigma\})} /|{\rm Aut}(\Theta,\{\sigma\})| }
Note that the parameter $N$ allows to isolate the contributions of tessellations
of fixed genus. In the following we will be mainly interested in the genus zero
contributions, obtained by letting $N\to \infty$.
Note also that the ``free energy" $F={\rm Log}\, Z$ selects only the connected tessellations
in the sum \twodqgpf.

Like in the fixed lattice case, we are interested in the thermodynamic limit of the system,
in which say the average area or some related quantity diverges, ensuring that the
dominant contributions to $Z$ come from large tessellations. This is guaranteed
in general by the existence of a critical value $g_c$ of the cosmological constant $g$
at which such divergences take place. This value is a priori a function of the type
of random lattices we sum over as well as of the various matter parameters. We may now
attain interesting critical points by also letting the matter parameters approach critical
values, a priori distinct from those on fixed lattices. The result is well described
by the coupling of the corresponding CFT's with quantum fluctuations of space, namely
by letting the metric of the underlying 2D space fluctuate. Such fluctuations may be
represented in the conformal gauge by yet another field theory, the Liouville field theory,
which is coupled to the matter CFT. This field theoretical setting allows for a complete
understanding of the various critical exponents occurring at these critical points
\KPZ.
For instance, one defines the (genus zero) string susceptibility exponent $\gamma_{str}$
as the exponent associated to the cosmological constant singularity, namely
by writing the singularity of the free energy as
\eqn\frensing{ F\vert_{\rm sing} \propto (g_c-g)^{2-\gamma_{str}} }
In the case of coupling of a matter theory with central charge $c$ to 2DQG, one has the
exact relation \KPZ\
\eqn\kpz{ \gamma_{str}\equiv 
\gamma_{str}(c)={c-1-\sqrt{(1-c)(25-c)} \over 12} }
In the case of ``pure gravity", namely when the matter is trivial and has $c=0$,
we get $\gamma_{str}=-1/2$, while for the critical Ising model with $c=1/2$ we
have $\gamma_{str}=-1/3$.

Upon coupling to gravity the spinless operators of the CFT ($\Phi_h(z,{\bar z})$)
get ``dressed" by gravity ($\Phi_h \to {\tilde \Phi_h}\equiv \Psi_\Delta$) and acquire
dressed dimensions $\Delta$, given similarly by \KPZ\
\eqn\kpzdim{\Delta={\sqrt{1-c+24 h}-\sqrt{1-c} \over \sqrt{25-c}-\sqrt{1-c}} }
As opposed to the fixed lattice case, where conformal dimensions govern
the fall-off of correlation functions of operators with distance, the dressed operators
of quantum gravity do not involve distances, as their position is integrated over
the surfaces, but rather only involve changes of area at fixed genus. More precisely
the general genus zero correlators behave in the vicinity of $g_c$ \KPZ\ as
\eqn\corrgra{ \langle \Psi_{\Delta_1} \Psi_{\Delta_2} ... \Psi_{\Delta_k} \rangle
\sim (g_c-g)^{2-\gamma_{str}+\sum_{1\leq i\leq k} (\Delta_i-1)} }

These results may be easily translated into the large (but fixed) area $A$ behavior
of the various thermodynamic quantities, upon performing a Laplace transform,
which selects the coefficient of $g^A$ in the various expansions. Let $F_A$
denote the partition function for connected tessellations of genus zero and area $A$,
we have
\eqn\largA{ F_A \sim {g_c^{-A} \over A^{3 -\gamma_{str}}} }
while if $\langle ...\rangle_A$ denotes  any genus zero correlator at fixed area, we have
\eqn\corrA{  \langle \Psi_{\Delta_1} \Psi_{\Delta_2} ... \Psi_{\Delta_k} \rangle_A
\sim {g_c^{-A} \over A^{3-\gamma_{str}+\sum_{1\leq i\leq k} (\Delta_i-1)} } }
In Ref.\KPZ, all these formulas were also generalized to higher genus as well.
Note finally that Eqs.\kpz\ and \kpzdim\ are valid only as long as $c\leq 1$.
This corresponds to the famous ``$c=1$ barrier" beyond which the gravitational system
degenerates into infinitely branched structures (branched polymer phase of 2DQG).

As an illustration of Eq.\largA, recall that the number of Eulerian triangulations of the
sphere with $A$ triangular tiles and with a marked edge reads from
Eq.\reeseule
\eqn\quadA{ 
N_A={3\over 2} 2^{A/2}{(A)!\over (A/2)!(A/2+2)!} \sim
6\, \sqrt{2\over \pi} {8^{A/2}\over A^{5/2}}}
with $A=2s$ the number of triangles.  
Noting that the rooting simply amounts to $N_A \propto A F_A$, hence
the asymptotics \quadA\ correspond to $g_c=1/\sqrt{8}$ and $\gamma_{str}=-1/2$, 
hence $c=0$ according to \kpz.
This is one of the various examples where we attain the universality class of pure gravity, 
namely by summing over some bare tessellations without matter on them.

The asymptotic enumeration for most combinatorial problems involving planar 
(or more generally fixed genus) graphs is encoded in
Eqs.\largA\ and \corrA\ or their higher genus generalizations. This allows
to predict the corresponding configurational exponents provided
one is able to identify the central charge $c$ of the underlying CFT. This latter
step however may prove to be quite involved.
In fact, a lesson of the forthcoming Section is that the naive
application of these formulas may lead to wrong results for statistical models
whose definition strongly relies on the structure of the underlying (whether fixed or
random) lattice. This is precisely the case for the problems of folding
or of fully packed loops that we are interested in.

\newsec{One-flavor fully packed loops}

\subsec{Fully packed loops on random trivalent graphs}

\fig{An example of configuration of fully packed loops
on a random planar trivalent graph.}{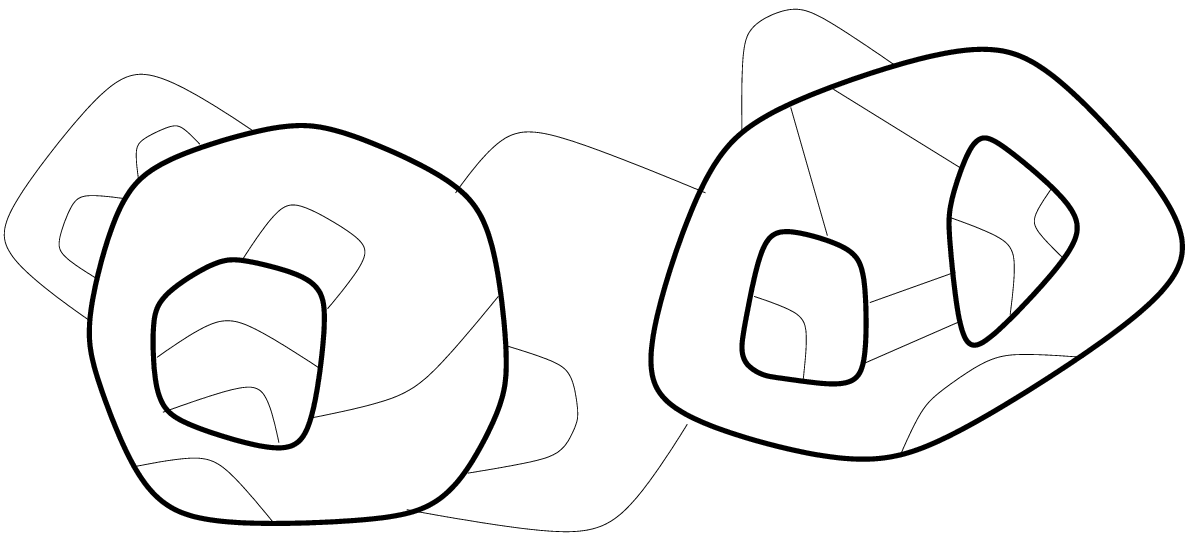}{10.cm}
\figlabel\Ongravi
We now come to the random version of the $FPL(n)$ model, i.e.\ 
to a gas of fully packed loops on random planar trivalent graphs
(dual to random triangulations). The most natural version consists 
in summing over all planar trivalent graphs. A typical, connected
configuration of this model is represented in Fig.\Ongravi. 
It is made of
a set of disjoint or nested loops, each with a weight $n$, linked 
together by non-crossing {\it arches}, corresponding
to the links not visited by loops.

The partition function $Z_s$ of the model, which counts the number
of configurations with $2s$ nodes behaves as
\eqn\zs{Z_{s}\sim C\,{R^s\over s^{3-\gamma}}}
where $C$, $R$ and $\gamma$ depend on $n$.
The value of $\gamma$ was obtained exactly by matrix integral techniques
(see Ref.\ON) with the result
\eqn\valgamma{\gamma=-{e\over 1-e}\qquad n=2\cos(\pi e),\qquad
0\leq e\leq 1}
We may also consider the partition sum $\langle \Psi_l \Psi_{-l} \rangle$
counting $FPL(n)$ configurations in the presence of 
two marked points connected by $l$ open lines. The latter is characterized
by a configurational exponent $\alpha_l(n)$ through 
$\langle \Psi_l \Psi_{-l} \rangle_s \sim R^s/ s^{\alpha_l(n)}$, 
with the result \DK
\eqn\alphal{\alpha_l(n)=1+{l\over 2}}
independently of $n$.
\fig{An example of oriented Hamiltonian cycle on a planar random trivalent
graph (left). By cutting the link and stretching the loop into
an oriented infinite straight line, one obtains a system of arches on top
and below the line (right).}{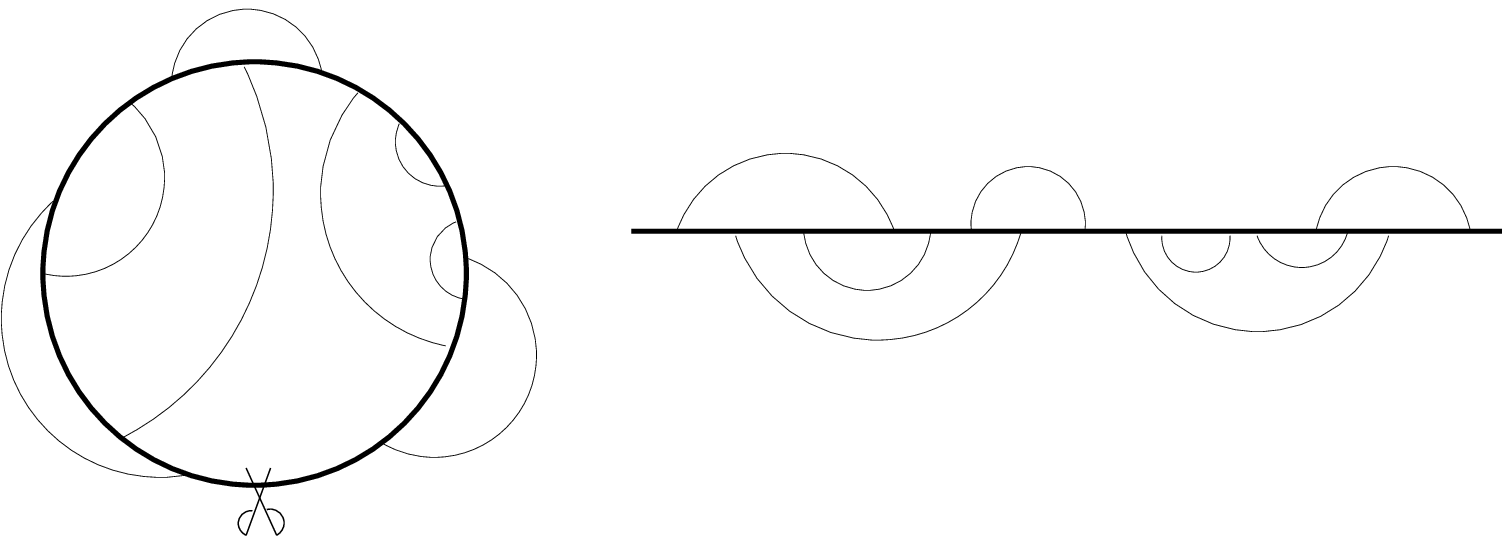}{11.cm}
\figlabel\Onarches
As an illustration, a particularly interesting case is
that of $n\to 0$ for which configurations have only one loop.
This case therefore describes the problem of enumeration of 
{\it Hamiltonian} cycles, i.e.\ closed loops passing through all 
the nodes of the random trivalent graph \EGK.
If we open the loop at any of the $2s$ visited links and
stretch it into an infinite oriented line as in Fig.\Onarches, 
we simply obtain a configuration with $2s$ nodes on the line 
connected by pairs via a set of, say $k$ non-intersecting arches 
in the upper half plane and $s-k$ in the lower half plane. We deduce
that \CDV\
\eqn\zsnzero{2s Z_s=\sum_{k=0}^s{2s \choose 2k} c_k c_{n-k}=c_s c_{s+1}
\sim {4\over \pi} {16^s\over s^3}}
where the Catalan number $c_s=(2s)!/(s!(s+1)!)$ counts the number
of possible configurations of $s$ non-intersecting arches.
One thus finds an exponent $\gamma=-1$, which is compatible with Eq.\valgamma\ for
$n=0$ ($e=1/2$).
\fig{Examples of configurations with two marked points
connected by $l$ open lines, for $l=1$ (top) and
$l=2$ (bottom)}{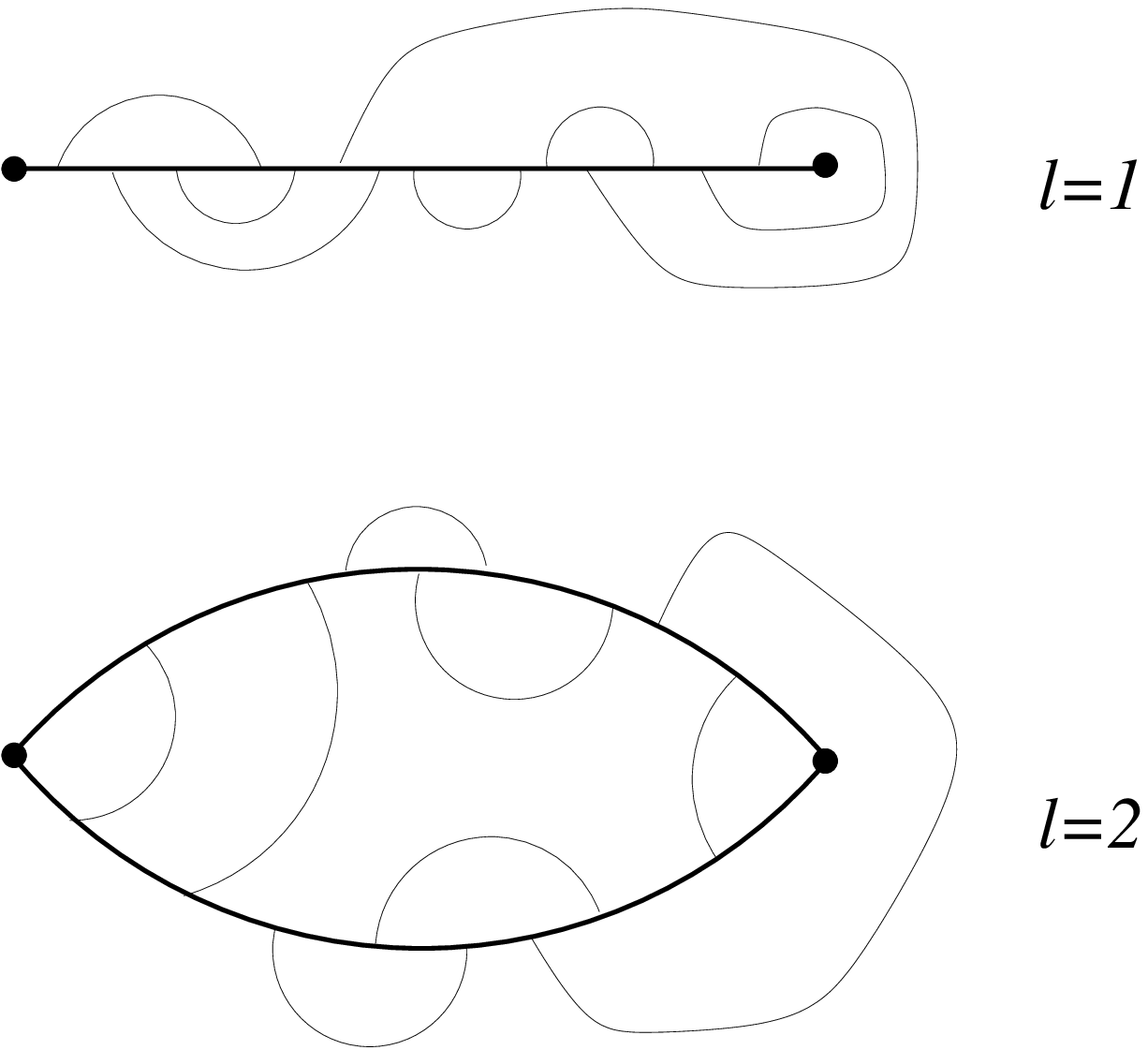}{7.cm}
\figlabel\Onl
We may similarly count configurations with two marked points 
connected by $l$ open lines. For instance, 
the case $l=1$ corresponds to having a segment dressed by arches
(see Fig.\Onl). The number
of such configurations with $2s$ trivalent nodes is clearly
given by $2^{2s} c_s\sim 16^s/s^{3/2}$, in agreement with Eq.\alphal\ for $l=1$.
Similarly, for $l=2$, one describes configurations of a Hamiltonian cycle
with two marked links (see Fig.\Onl), in number
$4s^2 Z_s\sim 16^s/s^2$, in agreement with Eq.\alphal\ for $l=2$.

By inverting the relation \kpz, we see that the value
\valgamma\ of $\gamma$ is compatible with a central
charge $c=c_{\rm dense}(n)=1-6 e^2/(1-e)$.
Similarly, by using Eq.\corrA\ with $k=2$, i.e.\ 
$\alpha_l(n)=1-\gamma+2 \Delta_l$, and by inverting the relation \kpzdim,
one finds that the result \alphal\ for $\alpha_l(n)$ is
compatible with a conformal dimension $2h=x_l(n)$ as given by Eq.$\xlon$.
To conclude, the model $FPL(n)$ defined on arbitrary random trivalent 
graphs corresponds to the coupling to gravity of a conformal theory of central charge
$c_{\rm dense}(n)$ as if the loops were not fully packed, a result further
confirmed by the identification of the spectrum of dimensions $x_l(n)$. 
Naively, we would have expected a central charge equal to that,  $c_{\rm fully\ packed}(n)$,
of the $FPL(n)$ model on a regular lattice. 
In other words, we do not observe here the phenomenon $c\to c+1$ which
we found for the regular lattice when passing from dense loops
to fully packed loops.

The explanation for this clearly comes from the strong coupling of the
matter model to the symmetries of its underlying space. For instance,
fully packed loops on the regular lattice all have even length, a
necessary condition for viewing them as the BC loops of some ABC tricoloring.
We can therefore consider a slightly constrained model in which 
we impose that loops be of {\it even} length on the random lattice.
It was shown however \EK\ that the $O(n)$ model
with loops of even length on random graphs is in the universality class
of the $O(n/2)$ model. This does not lead to the desired $c\to c+1$ phenomenon. 
As we shall see just below, we need to apply a much more
drastic constraint by reducing the {\it class} of random graphs itself 
in order to recover the desired central charge $c_{\rm fully\ packed}(n)$
of the regular case.

\subsec{Fully packed loops on random trivalent bipartite graphs}

On the regular lattice, the central charge $c=2$ of the $FPL(2)$ model 
as opposed to $c=1$ for the $O(2)$ model had a simple geometrical 
interpretation, as describing the $2$ degrees of freedom of the
2d positions of nodes for the equivalent 2d folding problem of the
triangular lattice. On the other hand, we also observed that
this 2d folding picture could be extended to the case of random
triangulations provided we restricted their class to the {\it Eulerian}
ones. Indeed the latter are precisely the triangulations foldable
onto the plane. In the dual language of trivalent graphs,
this condition amounts to requiring that the graph be bipartite, i.e.\  
may be node bicolored. Beside the folding interpretation,
we also noticed in the Coulomb gas approach of Section 5.1 that 
the rules for defining a two-dimensional height 
use explicitly the bipartite nature of the hexagonal lattice
(see Fig.\Onrules). We therefore expect this bipartite nature 
to be also crucial in the random case.

In this Section, we concentrate on the two cases $n=0$ and $n=1$
for which we show that the $FPL(n)$ model, when defined on trivalent
node-bicolored graphs, does indeed have the ``increased" central charge
$c_{\rm fully\ packed}(n)=c_{\rm dense}(n)+1$ as given by Eq.\ccc. 
In the case $n=0$, this translates into a remarkable
{\it irrational} critical exponent for a very simple apparently harmless
combinatorial problem.

For $n=0$, the $FPL(0)$ model describes the configurations
of a Hamiltonian cycle, i.e.\ a single loop visiting all the nodes 
of the random bicolored trivalent graph (or equivalently visiting 
all faces of the Eulerian triangulation). We would like to compute 
the number $Z_s^E$ of such configurations with $2s$ trivalent nodes.
\fig{An example of configuration of arches around a line obtained
by cutting the Hamiltonian cycle. The white and black nodes
alternate along the line. Each arch connects a white node to
a black one.}{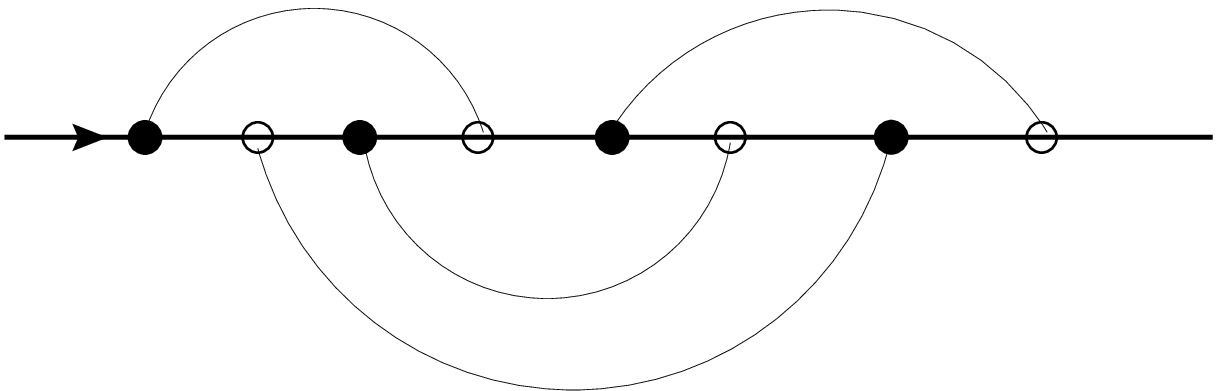}{7.cm}
\figlabel\Oneuler
Once again we may open the Hamiltonian cycle at any of its
$2s$ links so as to build an oriented line containing all
the original nodes, together with a system of arches connected them
by pairs either in the upper- or lower-half plane delimited by the line
(see Fig.\Oneuler). Comparing this situation with that of
previous Section, we now have two additional constraints:
\item{(i)} the colors of the nodes along the line
alternate between black and white;
\item{(ii)} each arch connects nodes of opposite colors.
\par
This combinatorial problem, although extremely simply stated,
is still open. We can however estimate the configuration
exponent of the problem with reasonable accuracy by direct 
enumeration for small enough sizes $s$. 
A first approach, explained in Ref.\GKN,
consists in first building the two systems
of arches on top and below the line and to count the number of ways
to interlock them so that the black and white nodes alternate.
A second approach consists in using a transfer matrix to generate 
the systems of arches from left to right. A similar technique will be 
discussed in detail in part D devoted to meanders. 
The first values of $s Z_s^E$ for $s=1,2,\cdots ,22$ are given
in the table below:

\centerline{\vbox{\epsfxsize=6.cm\centerline{\epsfbox{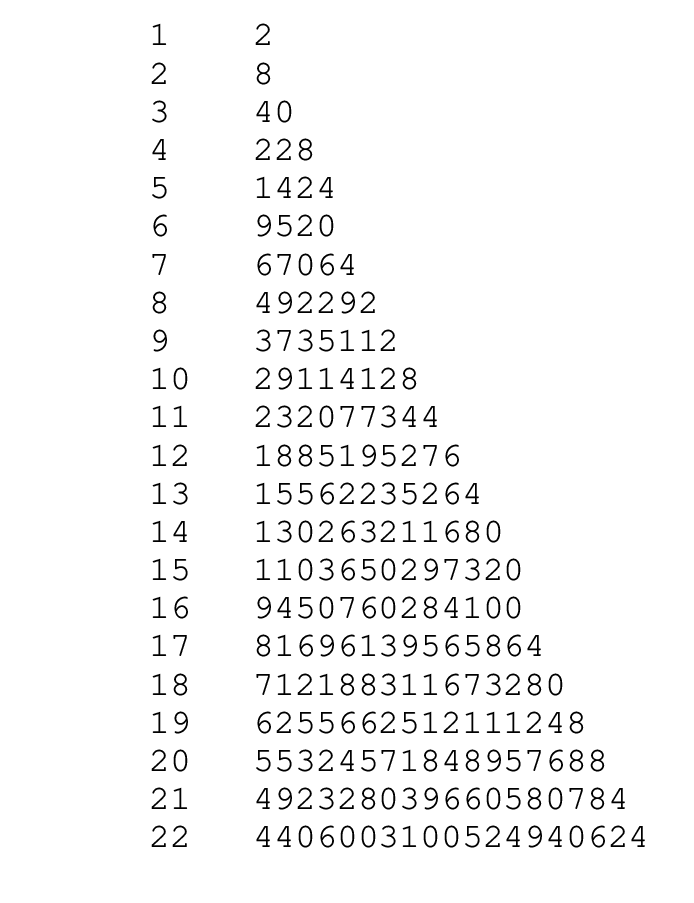}}}}
\fig{Estimates of $\gamma$ extracted from the exact enumeration for
$s Z_s^E$ (see table) in the Eulerian ($\circ$) case and from
$2s Z_s=c_s c_{s+1}$ in the non Eulerian case  ($\times$). The x-axis
coordinate corresponds to the minimal value of $s$ used in the determination.
The  different curves correspond to successive iterations
of the convergence algorithm used (see Ref.\GKN) . 
The value of $\gamma$, equal to $-1$ in the non Eulerian  case, 
is equal to  $\sim -0.77(1)$ in the Eulerian case.}{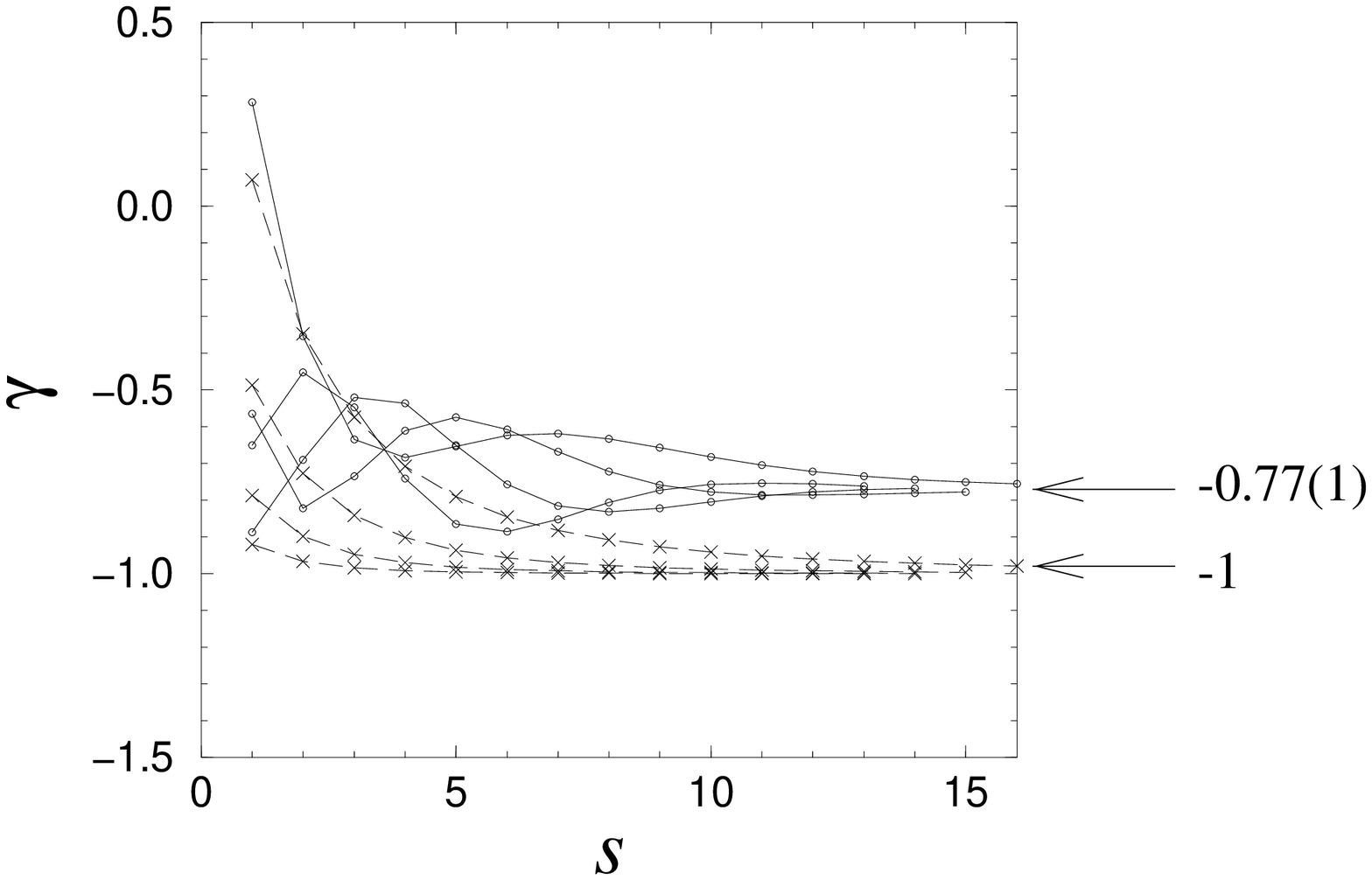}{13.cm}
\figlabel\gammaeuler

We may easily extract from these data an estimate for
the exponent $\gamma$ (see Fig.\gammaeuler), namely
$\gamma=-0.77(1)$. This value differs clearly from the value 
$\gamma=-1$ obtained in previous Section for non-Eulerian triangulations. 
Moreover, this new value is now compatible with a central charge 
$c=c_{\rm fully\ packed}(0)=-1$, for which, according to \kpz, $\gamma$ reads
\eqn\gammacmone{\gamma=-{1+\sqrt{13}\over 6}\sim -0.76759\cdots}
More precisely, by inverting the relation \kpz, our estimate
for $\gamma$ yields a central charge $c=-1\pm 0.05$. This leaves 
not much doubt on the fact that $c=c_{\rm fully\ packed}(0)=-1$ as expected.
As a consequence, we predict that our simple
combinatorial problem displays a quite remarkable
irrational configuration exponent, given by Eq.\gammacmone.

\fig{Mapping of the $FPL(1)$ model on node-bicolored trivalent
graphs onto a particular point of the gravitational six-vertex model
on random tetravalent graphs. Top: orienting all links toward black nodes 
and contracting unvisited links, we obtain tetravalent nodes with two
consecutive ingoing and two consecutive outgoing arrows. 
Bottom: the six local environments of the ``six-vertex model" 
and their respective weights on the regular square lattice.
Once defined on a random tetravalent graph, the first two pairs
of environments are indistinguishable, hence we must set $w_1=w_2$.
The constraint of having two consecutive ingoing arrows imposes
$w_3=0$.}{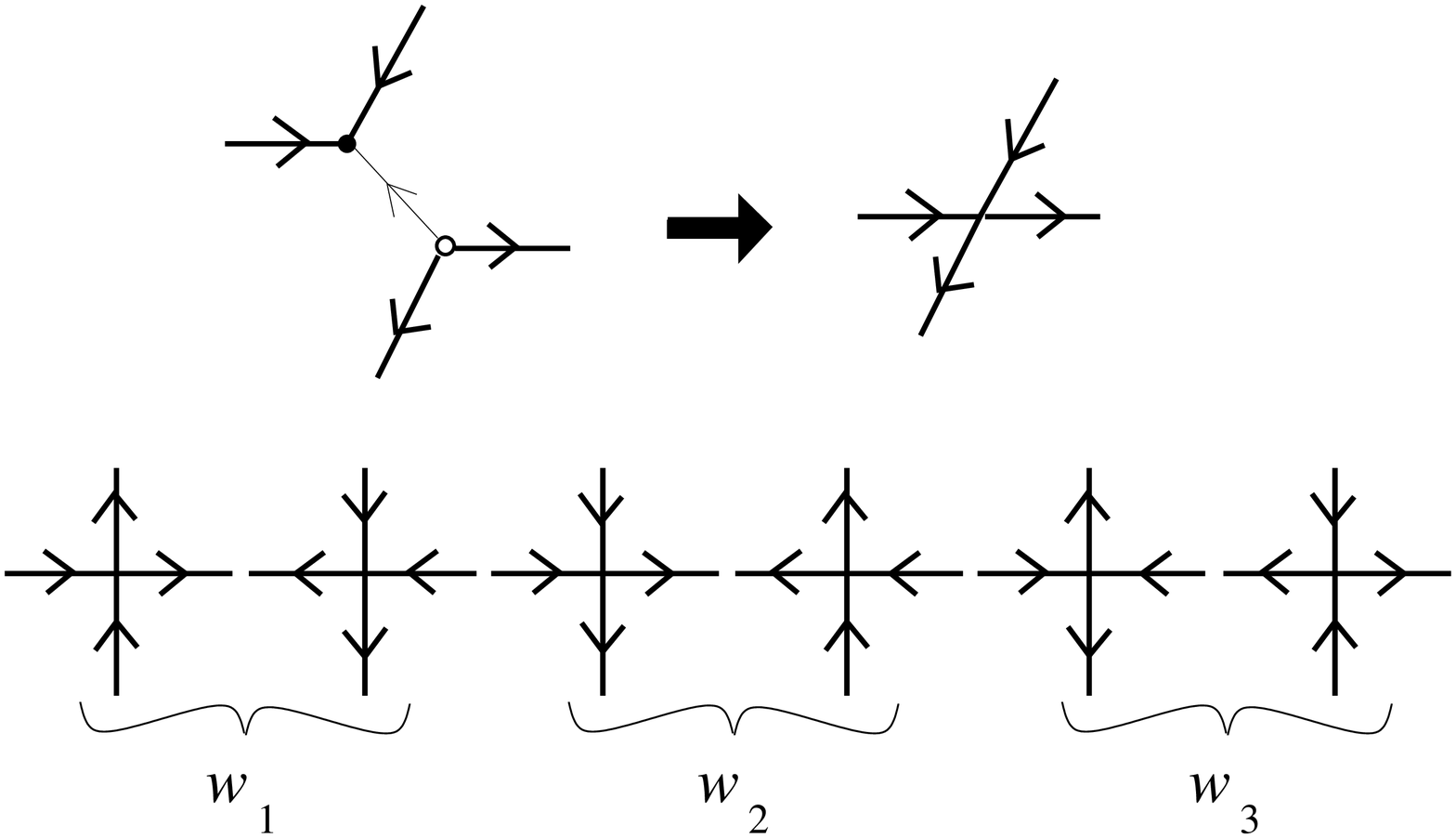}{10.cm}
\figlabel\sixvertex
We now turn to the case of the $FPL(1)$ model on node-bicolored 
trivalent random graphs, discussed in Ref.\DGK. For $n=1$, 
and only at this value, 
the connectivity of the loops plays no role and allows for local
transformations. Let us orient all the links toward their adjacent
black node. By contracting into a single tetravalent node each pair
of trivalent nodes separated by an unvisited link, we end up with a
tetravalent graph with oriented links. Moreover the orientations 
obey the so-called ice-rule of the six-vertex model, namely that 
each node have exactly two ingoing and two outgoing arrows 
(see Fig.\sixvertex). More  precisely, the two ingoing arrows must be
{\it consecutive} around the node, which corresponds to the particular 
point of the six-vertex model where one of the weights ($w_3$) is zero
(see Fig.\sixvertex). The six-vertex model was solved by use 
of matrix integrals in Ref.\KZJ, where it was shown on one hand that the
$w_3=0$ point corresponds to a critical point, and on the other hand that 
the latter is described by the coupling to gravity of a particular
CFT with central charge $c=1$. This corresponds precisely 
to the expected result $c=c_{\rm fully\ packed}(n=1)=1$.

To conclude this part, the above analysis of the cases $n=0$ and $n=1$
suggests that the specific universality class of fully packed loops 
observed on the regular lattice is preserved provided we impose
the bicolorability of the underlying trivalent graph, i.e.\ the 
Eulerian  nature of the dual random triangulation. Summing over such 
triangulations will be called Eulerian gravity for obvious reasons.
Now we may define two possible ``gravitational" models of fully packed 
loops on random trivalent graphs.

$\bullet$ We may sum over {\bf arbitrary} trivalent graphs. 
As the dual triangulation will in general {\it not} be Eulerian, 
it will not be foldable, and the extra degree of freedom
will be lost. The string susceptibility of the corresponding gravitational 
model will be computed using Eq.\kpz\ with the dense central charge
\eqn\orgra{ {\rm ordinary}\ {\rm gravity}: 
c=1-6 {e^2\over 1-e} , \quad n=2\cos\pi e}

$\bullet$ We may sum over {\bf bipartite} trivalent graphs, whose dual 
triangulation is automatically foldable, thus preserving the height variable 
in $\IR^2$. The string susceptibility must be computed using Eq.\kpz\ with 
the fully packed central charge
\eqn\charfuly{ {\rm Eulerian} \ {\rm gravity}: 
c=2 -6 {e^2\over 1-e} , \quad n=2\cos\pi e}

In the case $n=2$ describing the 2d folding of (foldable) random 
triangulations, we expect a CFT with central charge $c_{\rm fully\ packed}(2)=2$
whose coupling to gravity would lead us beyond the above-mentioned $c=1$ barrier
above which the relation \kpz\ no longer applies. On heuristics grounds,
we expect that the statistics of folding should be dominated by configurations
where the folded surface degenerates into a highly branched
one-dimensional structure. A confirmation of this image would require
solving the simultaneous tricoloring of both nodes and links
of random triangulations, an open question.

\bigskip
\vfill\eject
\leftline{\bf PART D: MEANDERS}
\bigskip
This part is devoted to the study of meanders,
also equivalent to self-avoiding folding configurations of a
one-dimensional lattice. Beyond exact results for
meander-related problems, we present a random-lattice 
fully packed loop description of the problem leading to precise
predictions for the asymptotics of various meandric numbers. These 
predictions are tested numerically with remarkable agreement.
A lesson of this study is the equivalence between the 1D self-avoiding
folding and the 2D phantom folding of random foldable but 
non-unfoldable quadrangulations.

\newsec{1D self-avoiding folding: Meanders}

The study of previous Sections was limited to {\it phantom} folding
problems, with lattices made of interpenetrable cells.
In particular, we concentrated on statistical models of
folded states without any reference to an actual folding process
in a possibly higher dimension. The question of self-avoidance,
where we now prevent cells from interpenetrating one-another is
extremely difficult. As we shall see now, even in the 
simplest 1D case, the problem is highly non-trivial and turns out to
belong to the same class as the so-called {\it meander problem},
a notoriously difficult subject. 

\subsec{The meander problem} 

\fig{The $M_3=8$ configurations of meanders for $2n=6$
bridges.}{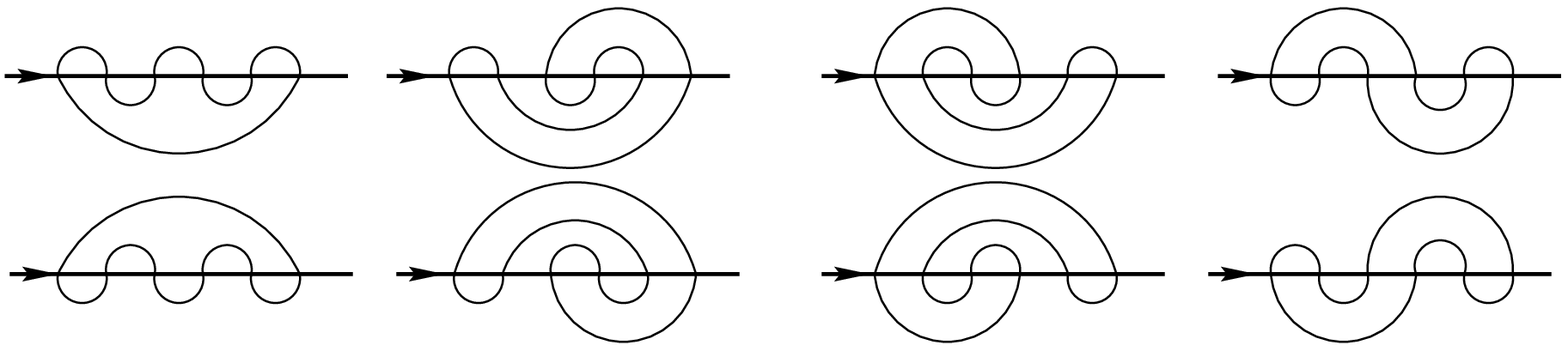}{12.cm}
\figlabel\meandre
The problem of meanders is one of the combinatorial problems
which, although very simply stated, still resist any attempt 
to an exact enumeration, would it be only for the asymptotics.
The meander problem may be stated as follows: {\it find the number 
$M_n$ of topologically inequivalent configurations of a closed 
non-intersecting circuit
crossing a river through $2n$ bridges}. In this formulation,
the river is assumed to be an infinite oriented line. Note that,  
by closing the river into a loop and by opening and deforming the 
circuit into a line, the circuit and the river play symmetric roles. 
The denomination ``meanders" refers rather to the dual 
picture where the river meanders around the circuit but, for historical
reasons, we shall work within the first formulation.
For illustration, the figure \meandre\ represents the $M_3=8$ 
configurations of meanders with $6$ bridges. In the following,
we shall be mainly interested in the asymptotics of $M_n$ for
large numbers of bridges, as well as various generalizations.

\subsec{Meanders as a 1D self-avoiding folding problem}

\fig{Correspondence between (left) the folding of a {\it closed} strip of
$2n$ stamps and (right) a meander with $2n$ bridges. For clarity,
the stamps are represented with different lengths and the folding
is slightly undone. }{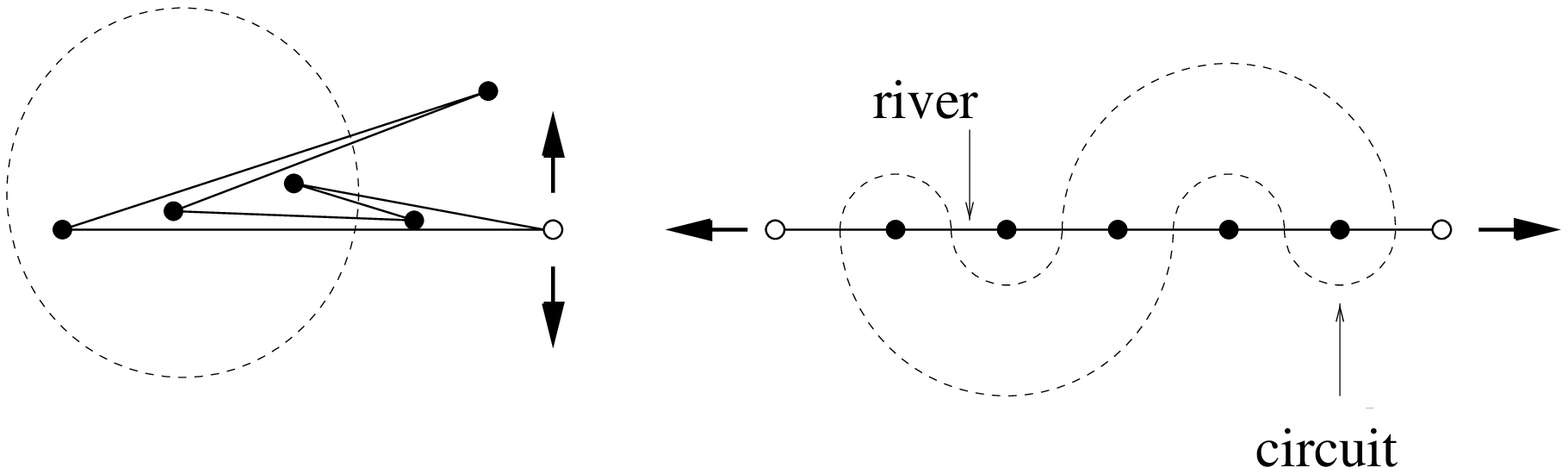}{12.cm}
\figlabel\foldmea

The problem of meanders is related to that of the self-avoiding folding of
a closed chain of segments of unit length. Imagine for instance a strip 
of $2n$ post-stamps attached so as to form a loop. We want to count
the distinct ways of maximally folding this strip on top of a single
distinguished stamp. The main difficulty of the problem comes from the 
avoidance of the stamps which cannot interpenetrate one-another. 
Such a folding is represented on the left in Fig.\foldmea.
We may now imagine piercing the stamps with a needle and a piece of 
thread and then knotting the thread into a loop (see Fig.\foldmea).
Let us then open the strip at the level of the marked
stamp and unfold it into a straight line (river), while the thread 
(circuit) meanders around it (see Fig.\foldmea). 
This transformation is clearly a one-to-one mapping between 
the 1D self-avoiding foldings of a closed chain of length $2n$ on
top of one of its segments and meanders with $2n$ bridges.

\fig{The four ways (a) of folding a strip of $3$ stamps (here viewed
from the side) on a single stamp. The transformation (b) of a compact
folding of a strip of $n-1$ stamps (onto a single stamp) into a semi-meander
with $n$ bridges. The semi-meanders correspond to the case of a
semi-infinite river with a source ($\bullet$) around which the circuit
may wind.}{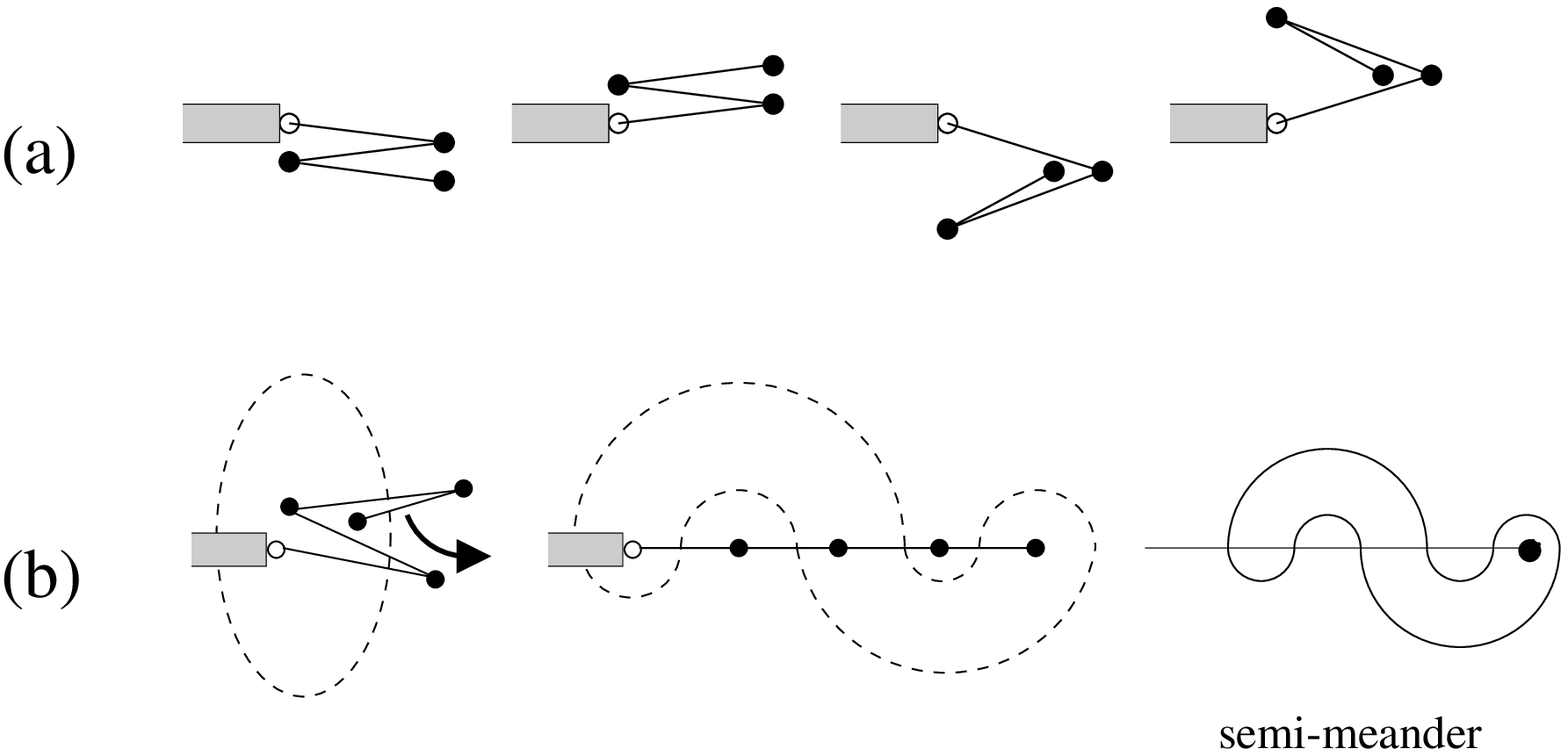}{12.cm}
\figlabel\foldstamp

The same correspondence applies to the folding of an {\it open}
chain of, say $n-1$ segments attached to a fixed support 
(see Fig.\foldstamp-(a)). In this case, the transformation
leads to what is called a {\it semi-meander} with $n$ bridges
(in correspondence with the $n-1$ segments and the support),
i.e.\ a configuration of a closed non-intersecting circuit crossing
a semi-infinite river (with a source) through $n$ bridges. 
Note that in this case, the circuit may wind freely around the source.
Both meanders and semi-meanders will be studied below in
a unified framework.

\subsec{A brief history of meanders}

The meander problem is quite old. It is mentioned
as early as 1891 under the name of ``probl\`eme des timbres-poste"
(post-stamp problem) by E. Lucas \Lucas. Later on, Sainte-Lag\"ue 
devotes a chapter of his book ``Avec des nombres et des lignes: 
r\'ecr\'eations math\'ematiques" (with numbers and lines: 
mathematical entertainments, 1937) \Saintelag. Several combinatorial 
approaches of this problem are discussed in Refs.[\xref\Touc-\xref\Lunn]. 
More recently, the problem re-appeared with 
the work of Arnol'd \Arno\ in connection with the 16th Hilbert problem
(the enumeration of ovals of algebraic planar curves). The modern
formulation of the problem, as just presented above, is due to Lando 
and Zvonkin \LZ, who also introduced the name ``meander".
Let us finally mention that the meander problem has many facets
in relation with: 
mathematical problems such as the classification of 3-surfaces \KOSMO,
computer science problems such as the study of planar permutations 
\HMRT, or even artistic questions such as the description of mazes
in Roman mosaics \Phil.

Despite a number of attempts, the problem is still unsolved to this day.
The most important result of this part D is the prediction of a 
number of configurational exponents, all irrational, for meanders and 
such.

\newsec{Solvable cases}

A number of exact results are known for variants of the meander
problem. These results rely on various descriptions using either
direct combinatorics via arch statistics \MFAS, explicit evaluations via
matrix models \CK\ or algebraic formulations via the Temperley-Lieb algebra
\Detmean.

\subsec{Generalization: multi-circuit meanders}
\fig{A typical meander (a) with $2n=16$ bridges and $k=4$ circuits.
A typical semi-meander (b) with $n=11$ bridges,
$k=3$ circuits and winding $w=3$.}{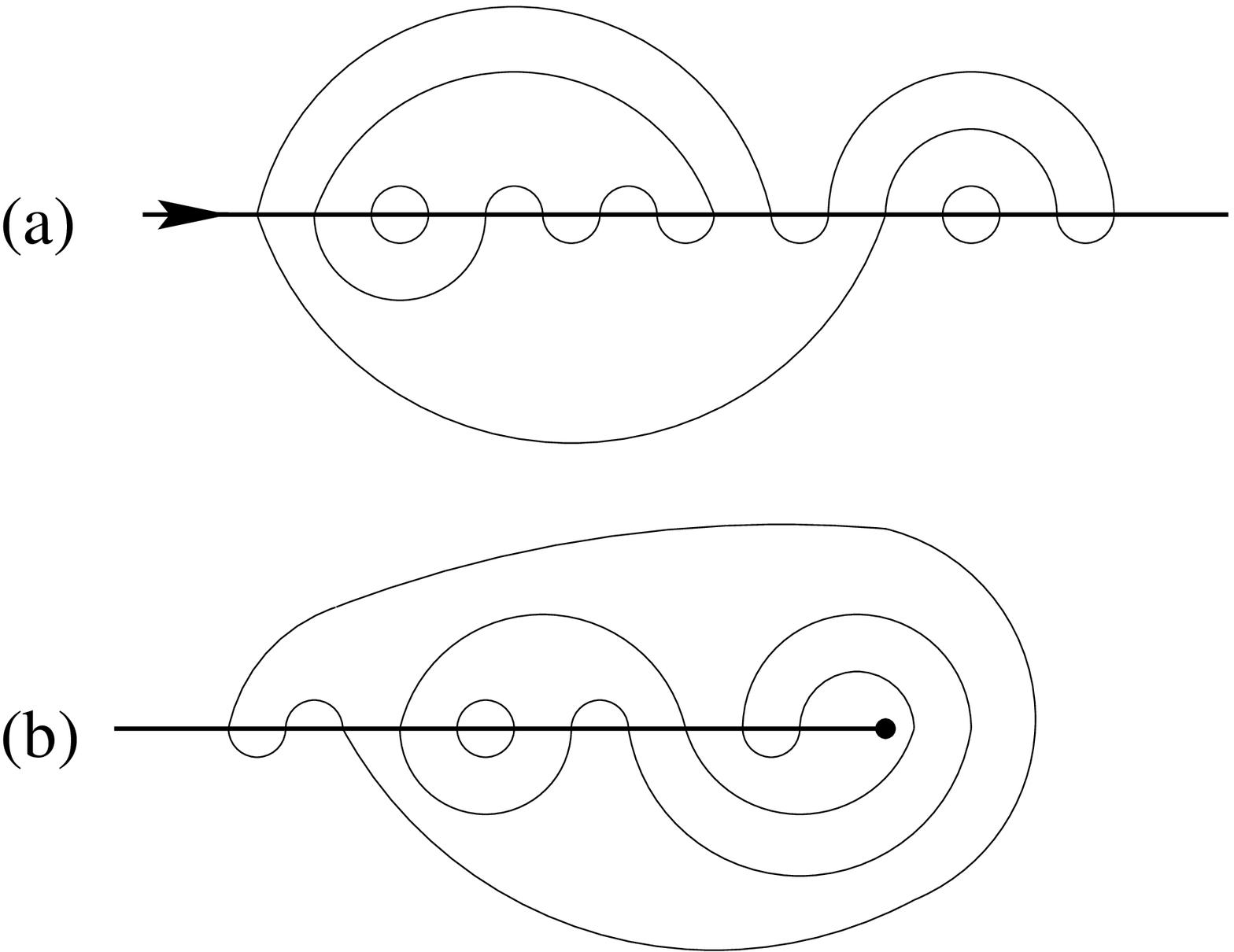}{9.cm}
\figlabel\measemimea
In the following, we consider the generalized case of {\it meanders
with $k$ circuits} still made  of {\it a single river}
but with a number $k$ of non-intersecting, possibly interlocking, 
circuits. We shall denote by $M_n^{(k)}$ the number of topologically 
inequivalent configurations of $k$ non-intersecting circuits which cross 
a river through a total number of $2n$ bridges, with
clearly $k\leq n$ (we require that each circuit crosses the river).
More simply, we shall refer to these configurations as meanders 
with $2n$ bridges and $k$ circuits (see Fig.\measemimea). 
Similarly, we shall denote by 
$SM_{n}^{(k)}$ the number of {\it semi-meanders} with $n$ bridges
and $k$ circuits corresponding to the case where the river is
a semi-infinite line with a source around which the $k$ circuits
may wind (see Fig.\measemimea). Here again, one necessarily has
$k\leq n$. With these definitions, the
original problems of meanders and semi-meanders correspond
to $k=1$.

Instead of working at constant $k$, it is easier
to let $k$ vary and to introduce a weight $q$ per circuit.
One then defines the partition functions
\eqn\partmean{\eqalign{M_n(q)&\equiv \sum_{k=0}^n q^k\, M_n^{(k)}\cr
SM_n(q)&\equiv \sum_{k=0}^n q^k\, SM_n^{(k)}\cr}}
The original cases of meanders and semi-meanders can be recovered
by considering the limit $q\to 0$ of $M_n(q)/q$ (resp. $SM_n(q)/q$).

For large $n$, we expect asymptotic behaviors of the form:
\eqn\asymmean{\eqalign{M_n(q)&\sim C(q)\, {R(q)^{2n}\over n^{\alpha(q)}}\cr
SM_n(q)&\sim {\bar C}(q)\, {{\bar R}(q)^n\over n^{{\bar \alpha}(q)}}\cr}}
where $\alpha(q)$ and ${\bar \alpha}(q)$ are the configurational
exponents. In the case of semi-meanders, we can also define
the {\it winding} $w$ as the depth of the source, namely
the minimal number of circuit crossings in a path from the
source to infinity (see Fig.\measemimea).
We may then consider the average value of $w$ for
configurations with $n$ bridges and a weight $q$ per circuit.
We expect the following behavior at large $n$:
\eqn\wasy{\langle w\rangle_n(q)\sim n^{\nu(q)}}
with a winding exponent $0\leq \nu(q)\leq 1$.
It is clear that $n$ and $w$ have the same parity and that
meanders correspond to semi-meanders with $2n$ bridges
and with a winding $w=0$.

\fig{By opening the river, we transform any semi-meander with 
$n$ bridges into a meander with $2n$ bridges. The opened circuits
are completed by connecting diametrically opposite bridges via
nested arches forming a rainbow, thus keeping track of their 
connectivity. The winding (here $w=3$) corresponds to the number of 
arches above the middle point ($\bullet$).}{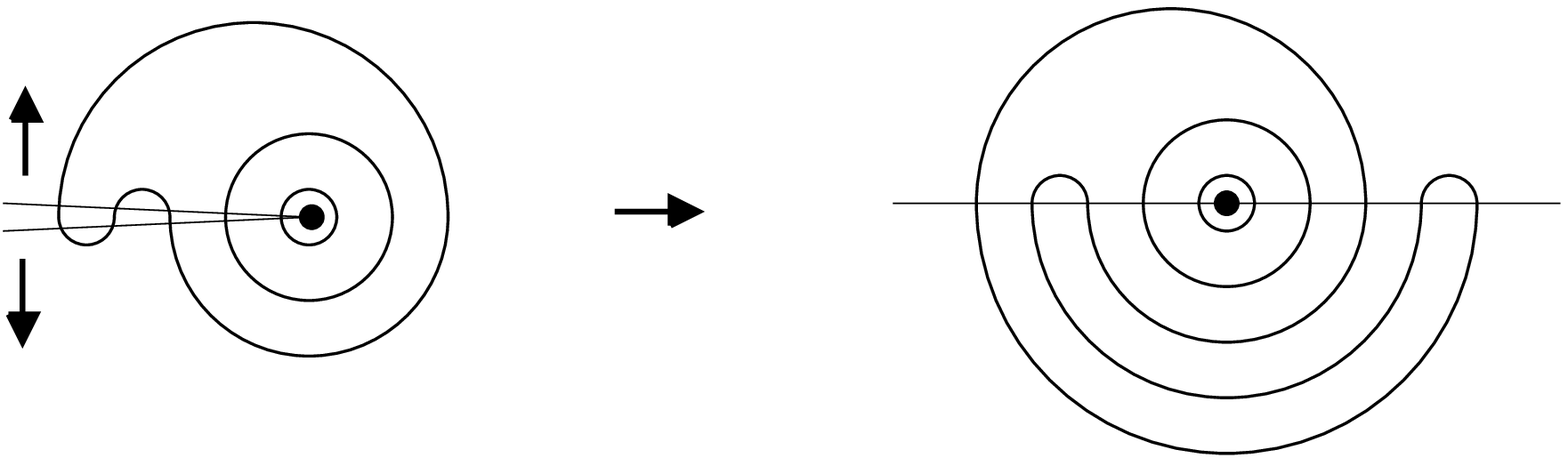}{10.cm}
\figlabel\opensemi
Note conversely that a semi-meander with $n$ bridges may be viewed as
a particular meander with $2n$ bridges by ``opening" the river 
(see Fig.\opensemi), hence splitting each bridge in two.
The connectivity of the loop is preserved by connecting 
the newly formed pairs of bridges via a ``rainbow" of nested arches.

Two situations are {\it a priori} possible:
\item{(i)} $\nu(q)<1$ and $R(q)={\bar R}(q)$.
This case corresponds to a situation where the winding
becomes negligible for large $n$ and the entropy per bridge
for meanders and semi-meanders is thus the same;
\par
\noindent or:
\item{(ii)} $\nu(q)=1$ and $R(q)<{\bar R}(q)$.
This case corresponds to a situation where the possibility of winding
generates an extra configurational thermodynamic entropy per bridge.
\par
\noindent We shall see in next Section that both situations
(i) and (ii) actually occur, depending on the value of $q$.

\subsec{Combinatorial solutions at $q=\infty$, $1$ and $-1$: exact enumeration 
via arch statistics}
The list of exactly known enumerations of multi-circuit meanders
reduces to the three cases $q\to \infty$, $q=1$ and $q=-1$.
In all cases, the problem is reduced to that of {\it arch statistics}.
A multi-circuit meander with $2n$ bridges may indeed be seen 
as the juxtaposition
of two arbitrary systems of $n$ arches, one on each side
of the river, and connected at the bridges. Similarly,
by opening the river (see Fig.\opensemi), any multi-circuit semi-meander
is the juxtaposition of a system of $n$ arches
on one side and the particular system made of a rainbow of $n$ nested arches
on the other side.

The number of possible configurations of $n$ non-intersecting arches is
given by the celebrated Catalan numbers:
\eqn\nbcata{c_n\equiv {(2n)!\over n!\, (n+1)!}}

In the case $q=1$ where we ignore the number of circuits,
we deduce that:
\eqn\meanqone{\eqalign{M_n(q=1)&=(c_n)^2\sim {1\over \pi}{4^{2n}\over n^3}\cr
SM_n(q=1)&=c_n\sim {1\over \sqrt{\pi}}{4^n\over n^{3\over 2}}\cr}}
from which we get $R(1)={\bar R}(1)=4$, $\alpha(1)=3$  and
${\bar\alpha}(1)=3/2$. The winding number $w$ corresponds to the number
of arches passing above the source of the river 
(middle-point in Fig.\opensemi). Its average value is easily evaluated,
with the asymptotic behavior:
\eqn\wone{\langle w\rangle_n(q=1)\sim {2\over \sqrt{\pi}} n^{1\over 2}}
i.e.\ $\nu(q=1)=1/2$. This is a particular instance of the situation (i)
of previous Section, in which the winding is negligible.

In the case $q\to \infty$, we have to maximize the number $k$
of circuits for fixed $n$, i.e.\ take $k=n$. In the case
of meanders, this amounts to requiring that the systems of arches 
above and below the river be identical up to reflection. Similarly,
in the case of semi-meanders, the system of arches above
the (opened) river must be made of $n$ nested arches.
By re-closing the river, this selects a unique configuration
made of $n$ nested circles winding around the source.
This gives:
\eqn\meanqinf{\eqalign{M_n(q)&\buildrel {q\to \infty} \over
\sim c_n q^n\sim {1\over \sqrt{\pi}} {(2\sqrt{q})^{2n}\over n^{3/2}}\cr
SM_n(q)&\buildrel {q\to \infty} \over
\sim q^n\cr}}
and thus $R(q)\to 2\sqrt{q}$, ${\bar R}(q)\to q$, $\alpha(q)\to 3/2$,
$\bar{\alpha}(q)\to 0$. One has clearly $\langle w\rangle_n(q)\to n$
as the unique semi-meander with $n$ circuits has
winding $n$, and thus $\nu(q)=1$.
This now corresponds to the situation (ii) described in previous Section.
We shall see later how this $q\to \infty$ result may be taken as the starting
point of a systematic $1/q$ expansion.

Finally, a last solvable situation concerns the case $q=-1$ for which
one shows that
\eqn\meanminusone{\eqalign{M_n(q=-1)&=\left\{\matrix{0&\quad n \quad
{\rm even}\cr
-(c_p)^2&\quad n=2p+1\cr}\right.\cr
SM_n(q=-1)&=\left\{\matrix{0&\quad n \quad {\rm even}\cr
-c_p&\quad n=2p+1\cr}\right.\cr}}
This result may be proved by use of a simple involution for arches \MFAS,
or by a more technical supersymmetric matrix model approach \MAK.

\subsec{Multi-river, multi-circuit meanders: asymptotic enumeration 
via matrix models}

We now address a generalization of the meander problem in which 
we allow for arbitrarily many rivers forming a set of nested
or disjoint loops, crossed by arbitrarily many non-intersecting
circuits, each with a weight $q$. We shall denote by $M_n(1,q)$ 
the generating function of the corresponding {\it connected}
configurations with one of the rivers opened into a line.

Following the work of Ref.\CK, we may use a ``black and white" matrix model 
to evaluate the exact asymptotics of $M_n(1,q)$.
We proceed by computing the generating function $Z_{1,q}(N;x)$ 
for possibly disconnected configurations of arbitrary topology, 
with a weight $x$ per bridge and the standard gravitational weight $N^\chi$,
$\chi$ the Euler characteristics of the corresponding graph.
For integer $q$, this function is given by the multi-matrix integral
\eqn\multimat{Z_{1,q}(N;x)=\int dW dB_1\cdots dB_q\,
e^{-{\rm Tr}\left( {W^2\over 2}+\sum\limits_{a=1}^{q} {B_a^2\over 2} -x\,
\sum\limits_{a=1}^{q} {(B_aW)^2\over 2}\right)}}
where the integral is over $q+1$, $N\times N$ Hermitian matrices
and normalized so that $Z_{1,q}(N;0)=1$. The quantity
$M_n(1,q)$ is recovered in the planar limit $N\to \infty$ as
the coefficient of $x^{2n}$ of the quantity $\partial_x {\rm Log}(Z_{1,q}
(N;x))/N^2$. Indeed, the diagrammatic expansion of \multimat\ involves
connecting black or white half-edges representing matrix 
elements of $B_a$ or $W$ into a closed graph via the propagators
\eqn\bwprop{\eqalign{
{\rm white} \ {\rm edges}&:
\ \ \ \langle W_{ij} W_{kl} \rangle~=
{1 \over N} \delta_{il}\delta_{jk}=\figbox{1.3cm}{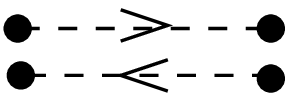} \cr
{\rm black}\ {\rm edges}&:
\ \ \ \langle (B_a)_{ij} (B_b)_{kl} \rangle ~=~
{1 \over N} \delta_{a,b}\delta_{il}\delta_{jk} =\figbox{1.3cm}{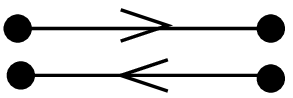}\cr}}
and with simple intersection vertices
\eqn\intervert{{\rm Tr}(W B_a W B_a) =\figbox{2.cm}{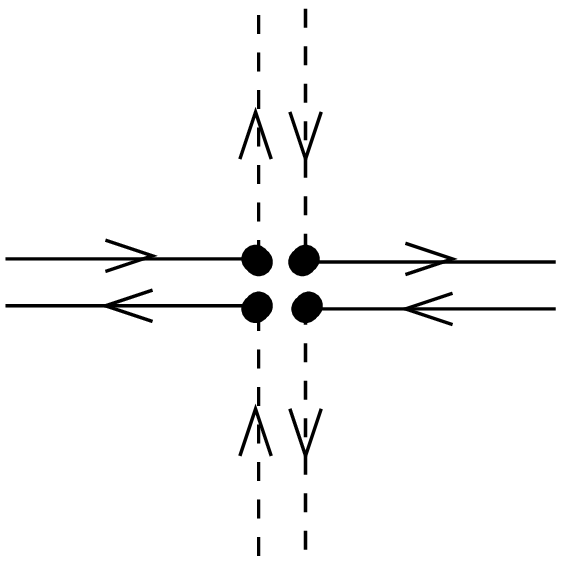} }
This results in graphs made of black loops intersecting white ones,
and the sum over the index $a$ produces a weight $q$ per black loop.
This allows to identify white loop with rivers and black ones with circuits.

Performing the Gaussian integration over all $B_a$ matrices,
we are then left with
\eqn\onew{
Z_{1,q}(N;x)= {\int dW e^{-N{\rm Tr}(W^2/2)} \det(1\otimes 1-x W\otimes W^t)^{-q
/2} \over
\int dW e^{-N{\rm Tr}(W^2/2)} } }
This is just a Gaussian average over one Hermitian matrix $W$.

The critical singularity of the genus zero free energy $f=\lim_{N\to \infty}
{\rm Log}(Z_{1,q}(N;x))/N^2\sim (x(q)-x)^{2-\gamma}$ is found 
\CK\ to lie at a critical value $x(q)$ given by
\eqn\critix{ x(q)={e^2 \over 2 \sin^2(\pi {e \over 2}) } }
where we have set $q=2 \cos(\pi e)$ ($0\leq e < 1$), 
while the corresponding critical 
exponent $\gamma$ reads
\eqn\gavalon{ \gamma=-{e \over 1 - e} }
This exponent is precisely that expected from the KPZ formula \kpz\
for the coupling to 2DQG of the dense $O(q)$ model with
central charge $c_{\rm dense}(q)=1-6 e^2/(1-e)$  as given
by Eq.\ccon. This shows that meander problems may be viewed as
the coupling to gravity of particular critical loop models,
and that the universality class of the present case coincides
with that of the dense $O(q)$ loop gas.

The above results translate into the multi-river meander asymptotics
\eqn\asymenla{M_{n}(1,q)\sim {R(1,q)^{2n}\over n^{\alpha(1,q)} }}
with
\eqn\randalph{\eqalign{
R(1,q)&={1\over x(q)}=2 {\sin^2(\pi {e\over 2}) \over e^2} \cr
\alpha(1,q)&={2-e \over 1-e} \cr}}
In particular, we recover from these values the case of meanders with 
arbitrarily many rivers and one single circuit by taking $q=0$, 
$e=1/2$, and $R(1,0)=4$, $\alpha(1,0)=3$. These values coincide
with those of previous Section for one river and arbitrarily many circuits,
as it should by river-circuit duality.
We list a few of the values $R(1,q)$ and $\alpha(1,q)$ for various 
fractions $e$ in the table below.

$$\vbox{\font\bidon=cmr8 \bidon
\offinterlineskip
\halign{\tv \quad \hfill # \hfill &\tv \hfill  \ # \hfill &\tv
\hfill # \hfill & \tv  \hfill #
\hfill \tv \cr
\noalign{\hrule}
$q$ & $e$ & $\ \ \ R(1,q)\ \ \ $ & $\ \ \ \alpha(1,q)\ \ \ $  \cr
\noalign{\hrule}
0  &  ${1\over 2}$  & 4  &  3  \cr
1  & ${1\over 3}$   & ${9\over 2}$  & ${5\over 2}$  \cr
$\sqrt{2}$  & ${1\over 4}$   & $16-8\sqrt{2}=4.68...$ & ${7\over 3}$ 
  \cr
$\sqrt{3}$  & $ {1\over 6}$  & $36-18\sqrt{3}=4.82...$  & ${11\over 5}$  \cr
2  &  0  & ${\pi^2\over 2}=4.93...$  & 2    \cr
\noalign{\hrule}
}}$$

\subsec{Meander determinant}
\fig{The ${\cal G}_n(q)$ matrix (here for $n=3$, $c_3=5$) obtained by
juxtaposing all pairs $({\cal A},{\cal B})$ of systems of $n$
arches. The matrix element
${\cal G}_n(q)_{{\cal A}{\cal B}}=q^{c({\cal A},{\cal B})}$ encodes the
number of circuits $c({\cal A},{\cal B})$ of the obtained
meander.}{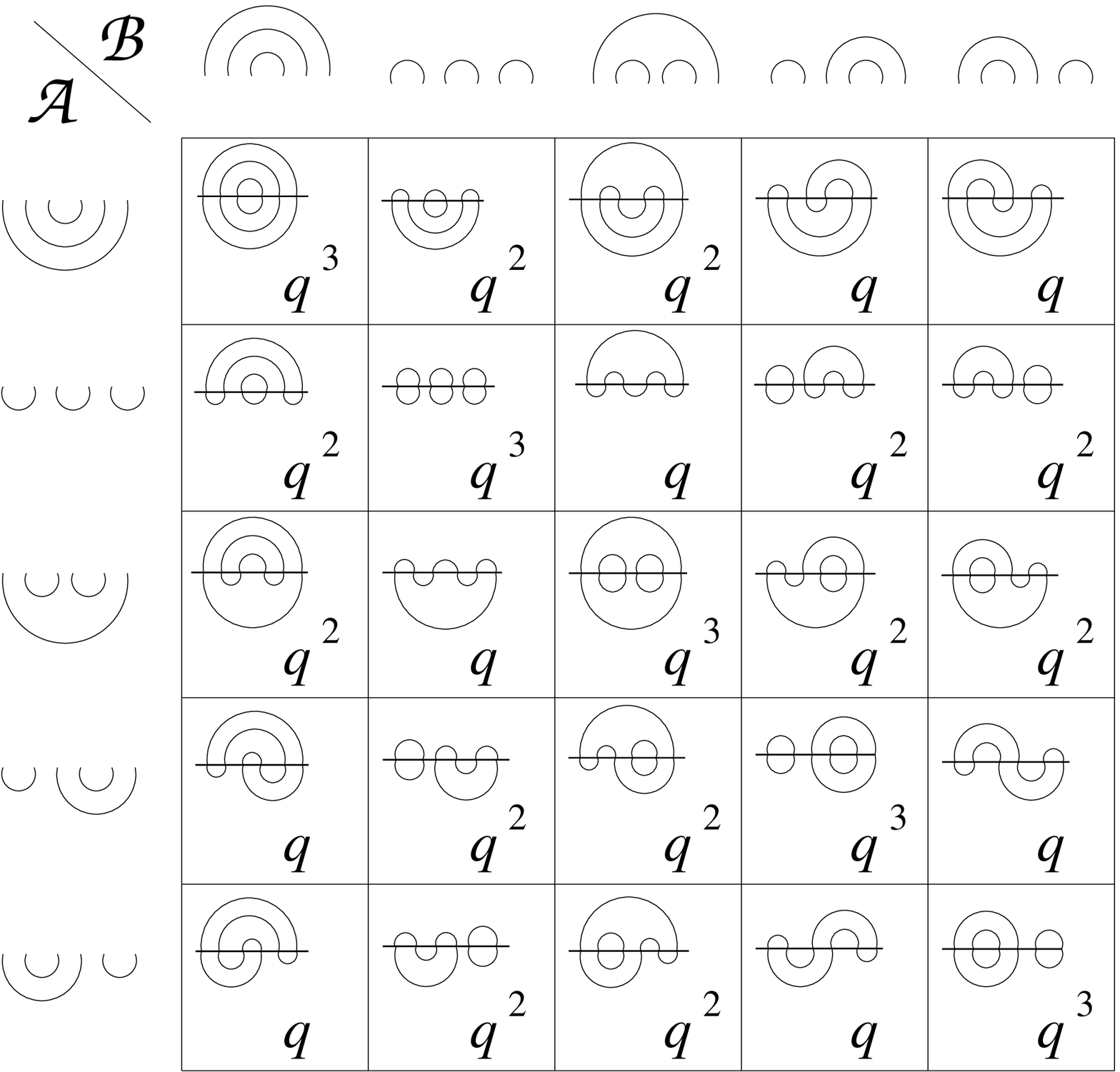}{10.cm}
\figlabel\detmea
Beyond the particular enumerations above, another
exact result concerns the computation {\it for any $q$} of
the {\it meander determinant} defined as follows.
We have seen that any meander with $2n$ bridges (and with an
arbitrary number of circuits) may be viewed as
the juxtaposition of two systems of $n$ arches, say
${\cal A}$ and ${\cal B}$. By taking all pairs of
systems of $n$ arches, we define a symmetric square matrix
${\cal G}_n(q)$ of size $c_n\times c_n$ whose element
${\cal G}_n(q)_{{\cal A}{\cal B}}$, indexed by the pair
$({\cal A},{\cal B})$, is equal to
\eqn\gab{{\cal G}_n(q)_{{\cal A}{\cal B}}=q^{c({\cal A},{\cal B})}}
where $c({\cal A},{\cal B})$ denotes the number of circuits
of the meander obtained by juxtaposing the arch systems
${\cal A}$ and ${\cal B}$ (see Fig.\detmea). 
We shall call ``meander determinant" the determinant of the
matrix ${\cal G}_n(q)$. It is remarkable that this determinant
may be computed exactly, with the result \Detmean:
\eqn\detmanres{\eqalign{
{\rm det}({\cal G}_n(q))&=\prod_{i=1}^n U_i(q)^{a_{n,i}}\cr
a_{n,i}&={2n \choose n-i}-2{2n \choose n-i-1}+{2n \choose n-i-2}\cr}}
where the $U_i$'s are Chebyshev polynomials defined
recursively as $U_{j+1}(q)=q U_j(q)-U_{j-1}(q)$,
$U_0(q)=1$, $U_1(q)=q$. Equivalently, we have
$U_j(2\cos(\theta))=\sin((j+1)\theta)/\sin(\theta)$, which leads to
the equivalent formula for the determinant
\eqn\detmeanbis{{\rm det}({\cal G}_n(q))=\prod_{1\leq l\leq i\leq n}
\left(q-2\cos\left(\pi {l\over i+1}\right)\right)^{a_{n,i}}}
For instance, for $n=3$, the determinant of the matrix
${\cal G}_3(q)$ represented in Fig.\detmea\ is equal to
$q^5(q^2-1)^4(q^2-2)$. The proof of the above formulas in Ref.\Detmean\ 
makes use of the intimate link between the meander problem and
the Temperley-Lieb algebra \TL.
The equivalence between arch configurations and reduced
elements of the Temperley-Lieb algebra $TL_n(q)$ goes as follows.
The Temperley-Lieb algebra is expressed in its pictorial form, 
as acting on a ``comb" of $n$ strings, with the $n$
generators $1$, $e_1$, $e_2$, ..., $e_{n-1}$ defined as
\eqn\braid{ 1~=~\figbox{2.cm}{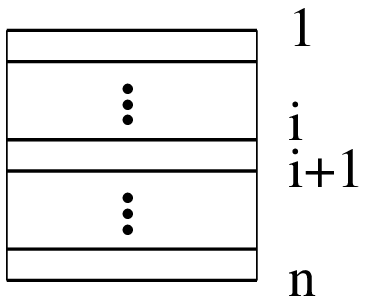}
\qquad e_i~=~\figbox{2.cm}{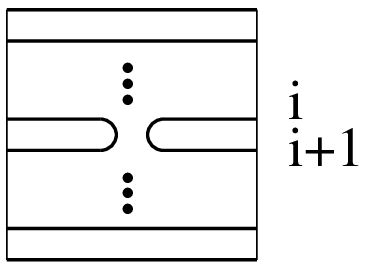} }
The most general element $e$ of $TL_n(q)$ is obtained by
composing the generators \braid\ like dominoes.
The algebra is defined through the following
relations between the generators
\eqn\tla{\eqalign{(i)\ \ \ \ \ \ \ \ \  e_i^2 ~&=~ q \, e_i
\quad i=1,2,...,n-1\cr
(ii)\ \ \ \ [e_i,e_j]~&=~0 \quad {\rm if}\ |i-j|>1 \cr
(iii)\ e_i\, e_{i \pm 1}\, e_i~&=~ e_i  \quad i=1,2,...,n-1\cr}}
The relation (ii) expresses the locality of the $e$'s,
namely that the $e$'s commute whenever they involve distant strings.
The relations (i) and (iii) read respectively
\eqn\unbraid{\eqalign{(i)\ \ \ \ \ \ \ \ \ e_i^2~&=~
\figbox{2cm}{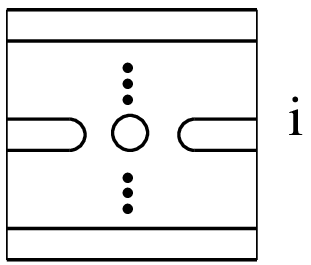}~=~q~\figbox{2.4cm}{ei.eps}~=~q
\, e_i\cr (iii)\ e_i\, e_{i+1}\, e_i~&=~
\figbox{2.4cm}{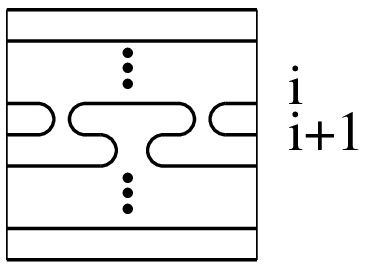}~=~\figbox{2.4cm}{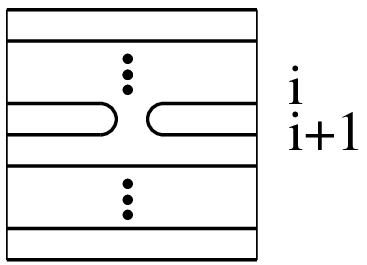}~=~ e_i\cr}}
In (i), we have replaced a closed loop by a factor $q$.  Therefore
we can think of $q$ as being a weight per circuit
of string. In (iii), we have simply ``pulled the string" number $i+2$.

An element $e\in TL_n(q)$ is said to be reduced if all its
strings have been pulled and all its loops removed, and if
it is further normalized so as to read $\prod_{i\in I} e_i$
for some minimal
finite set of indices $I$. A reduced element is formed of
exactly $n$ strings.
\fig{The transformation of a reduced element of $TL_9(q)$
into an arch configuration ${\cal A}$ of $9$ arches. The reduced
element reads $e_{\cal A}\equiv e_3 e_4 e_2 e_5 e_3 e_1 e_6 e_4
e_2$.}{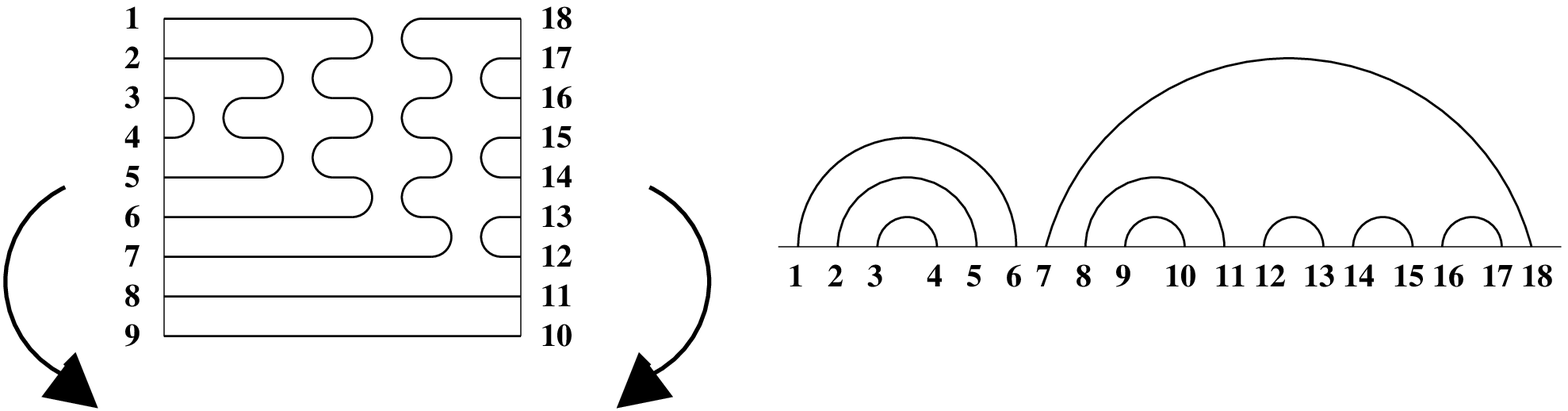}{12.cm} \figlabel\artla
There is a bijection between the reduced elements of
the Temperley-Lieb algebra
$TL_n(q)$ and the arch configurations with $n$ arches.
Starting from a reduced element of $TL_n(q)$, we index the
left ends of the $n$ strings by $1,2,...,n$, and the right
ends of the strings
$2n,2n-1,...,n+1$ from top to bottom (see Fig.\artla\ for an
illustration). Interpreting these ends as bridges, and
placing them on a line, we obtain a planar pairing of
bridges by means of non-intersecting strings,
equivalent to a configuration of $n$ arches.
Conversely, we can deform the arches of any arch configuration
so as to form a reduced element of $TL_n(q)$.
As a consequence, we have dim$(TL_n(q))=c_n$, as vector space
with a basis formed by all the reduced elements.

In the language of the Temperley-Lieb algebra, the meander matrix 
${\cal G}_n(q)$ is precisely interpreted as the Gram matrix of the 
basis of reduced elements of $TL_n(q)$ (labeled by the corresponding
arch configuration) with respect to the
bilinear form $(e_{\cal A},e_{\cal B})={\cal G}_n(q)_{{\cal A}{\cal B}}$
of Eq.\gab. 
The determinant of ${\cal G}_{n}(q)$ is obtained as a
by-product of the Gram-Schmidt orthogonalization of this matrix,
readily performed by use of the representation theory of the
Temperley-Lieb algebra \Detmean.

The formula \detmanres\ may be extended to the more general case of 
meanders with open arches \MEDET\ or meanders based on the Hecke algebra 
\SUM.

\newsec{Two-flavor fully packed loops: exponents of the meander problem}

\subsec{Generalized meanders as random lattice loop models}

\fig{An example of generalized meander with three loops 
of river (solid lines) and two circuits (dashed lines). The obtained graph
is automatically bipartite.}{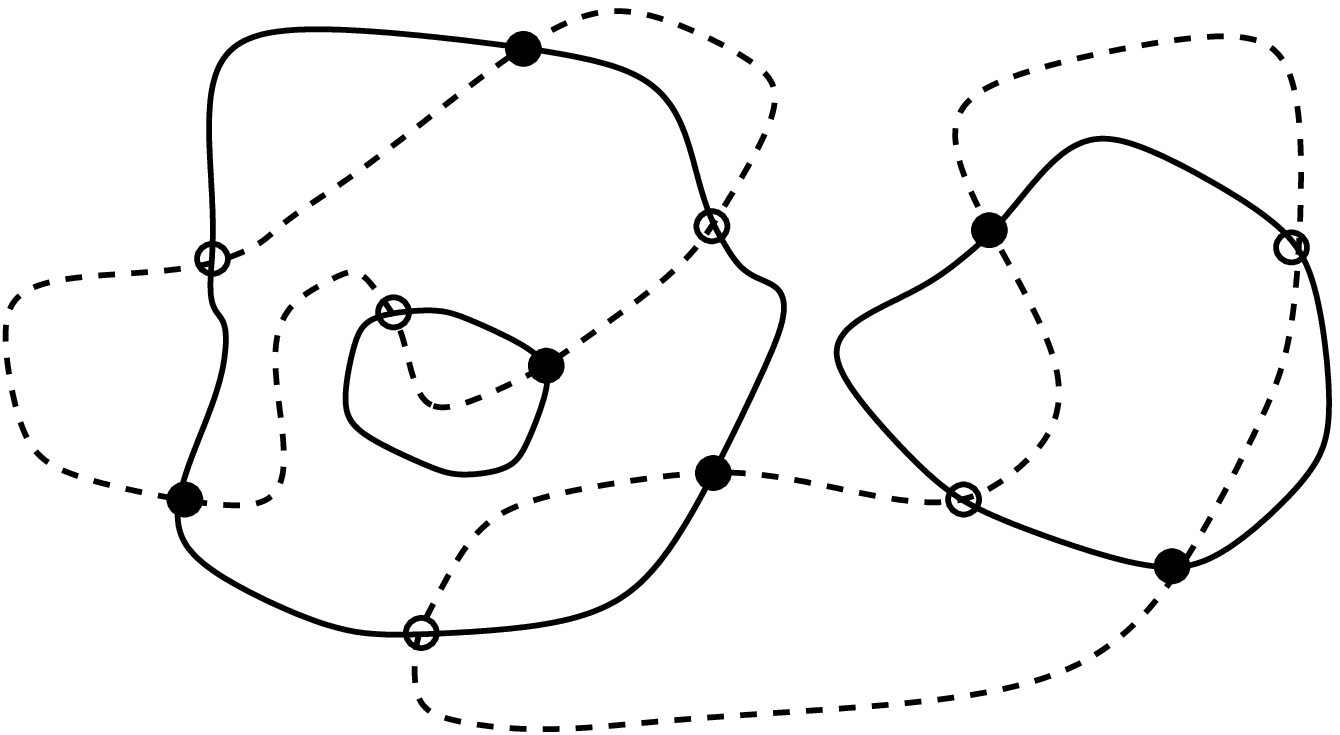}{9.cm}
\figlabel\gmean
The most elaborate description of the meander problem consists
in viewing it as the ``gravitational" version (i.e.\ 
defined on random lattices) of a fully packed loop model.
As in Section 10.3, we consider the general case of multi-river
multi-circuit meanders (see Fig.\gmean). We moreover attach a weight $n_1$ per
loop of river and $n_2$ per circuit. The case of multi-circuit
meanders with a single river of Section 10.1 corresponds to taking 
$n_1\to 0$, $n_2=q$, while that of multi-river multi-circuit
meanders of Section 10.3 amounts to taking $n_1=1$ and $n_2=q$.

We may therefore rephrase the problem as that of enumerating 
the configurations of planar graphs made of a set of river loops
and circuits with the constraints that:
\item{(a)} The rivers are self- and mutually-avoiding;
\item{(b)} The circuits are self- and mutually-avoiding;
\item{(c)} A river and a circuit may cross at a "bridge" node;
\item{(d)} The resulting graph is connected.
\par
To avoid problems of symmetry factors, it is as usual
convenient to consider configurations with a marked and oriented edge
(say corresponding to a portion of river).
The above model is therefore a model of fully packed loops
on random {\it tetravalent} graphs whose nodes correspond to the bridges, 
with two types of loops (the rivers and the circuits) weighted
by $n_1$ and $n_2$ respectively. We will  denote
this model by $GFPL^2(n_1,n_2)$ (for gravitational fully packed loops).

It is interesting to note that, as all nodes correspond to crossings 
of river loops and circuits, the corresponding tetravalent graphs
are automatically bipartite, i.e.\ may have their nodes bicolored 
(see Fig.\gmean).
\fig{Examples of configurations of tangent meanders with two
possible vertices, either the crossing of two loops, or a tangency point
(contact without crossing). The underlying tetravalent graph
may be arbitrary (a) or constrained to be bipartite
(b).}{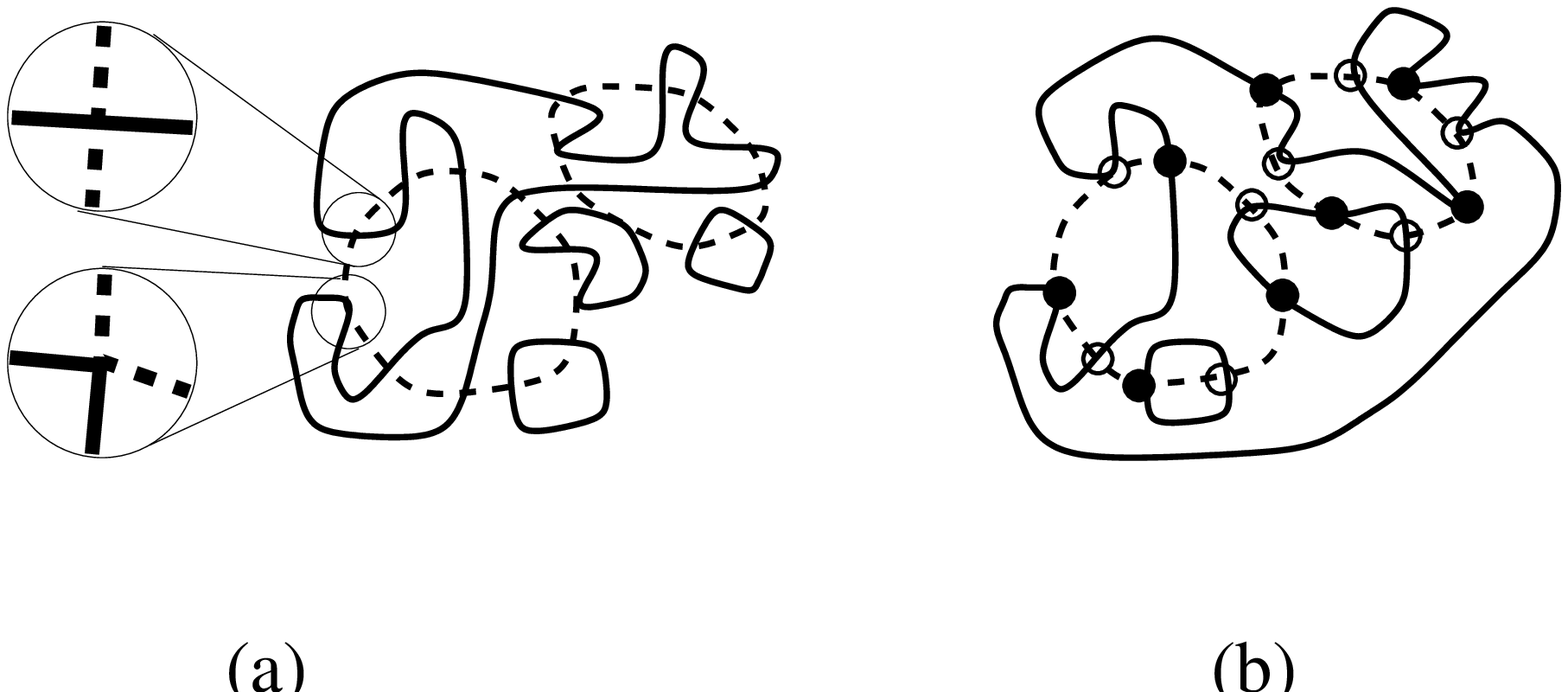}{14.cm}
\figlabel\tgmean

By anticipating the coming discussion, we may introduce
two other versions of the problem, which consist in allowing for
the presence, in addition to the crossings, of {\it tangency points} 
at which a circuit and a river come in contact but do not cross (see
Fig.\tgmean). We shall refer to this case as {\it tangent meanders}. 
The bicolorability of the underlying graph is no longer ensured and,
by analogy with the one-flavor loop case of Section 8, we may consider 
two different versions of the problem: the ordinary gravity version where
we sum over all tetravalent graphs or the Eulerian gravity version where
we restrict the summation to node-bicolored tetravalent graphs only.

\subsec{The $FPL^2$ model on the square lattice}
\fig{Example of configuration of the $FPL^2$ model. The
black and white (here represented as dashed) loops
may either cross or avoid each other. The two allowed vertices
(up to rotations and symmetries) are represented
on the right.}{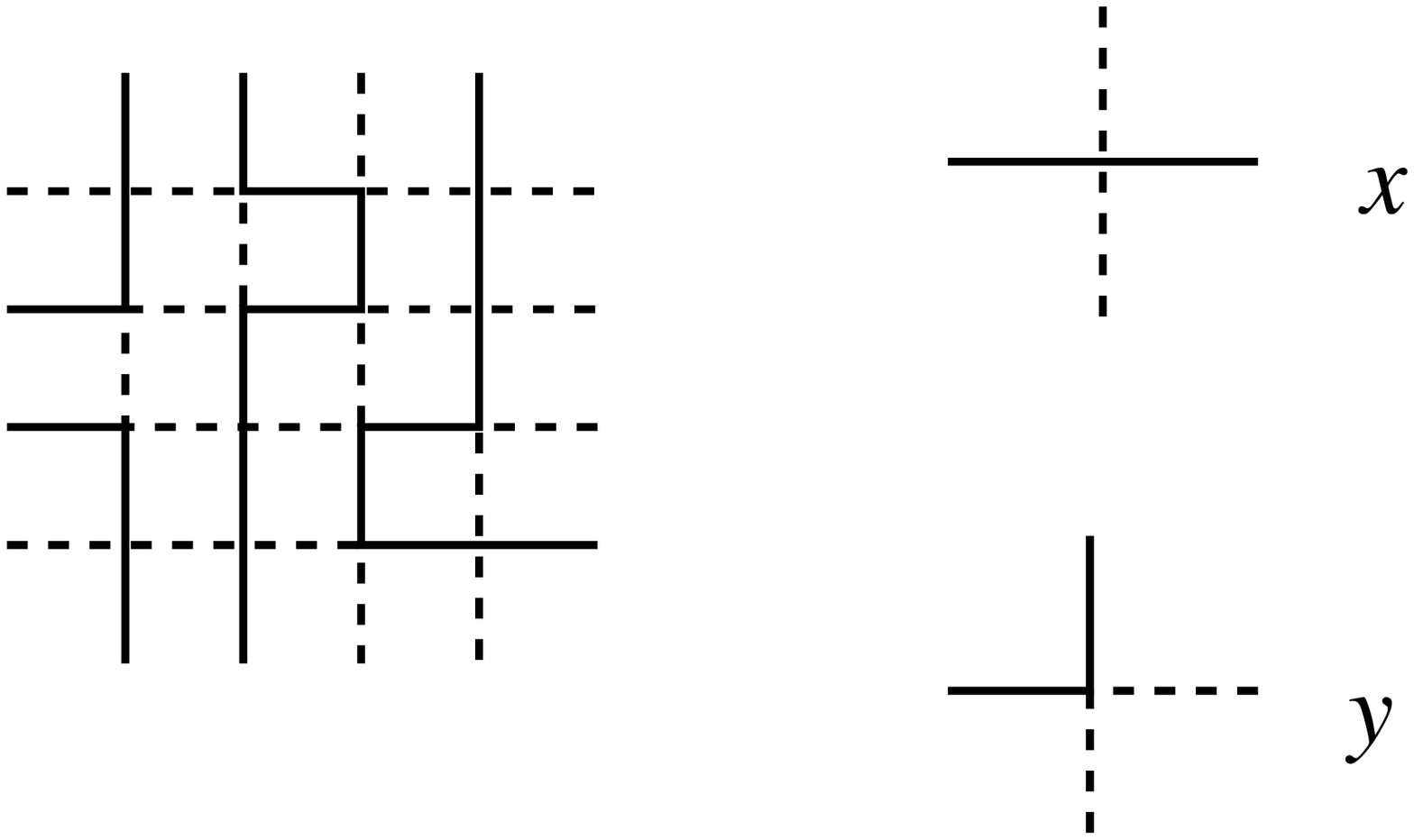}{10.cm}
\figlabel\fpldeux
We now would like to interpret the $GFPL^2(n_1,n_2)$ model
or its generalizations with tangency points as gravitational versions
of models defined on the regular lattice. In the case of
tetravalent graphs, the regular lattice to be considered is the
square lattice. A natural candidate is the $FPL^2(n_1,n_2)$ model,
introduced in Ref.\JACO, which consists of a gas involving 
two types of loops, say black and white, with respective weights 
$n_1$ and $n_2$ and the constraints that:
\item{1.} The loops are fully packed, i.e.\ each of the two systems 
of loops is self-avoiding and visits all the nodes of the lattice;
\item{2.} Each link is occupied by one type of loop only, the loops
being in contact only at the nodes.
\par
The $FPL^2(n_1,n_2)$ model allows for two types of vertices
(up to rotations and symmetries), represented
in Fig.\fpldeux\ and corresponding respectively to a
crossing of the two types of loops (crossing vertex) or to a contact 
with mutual avoidance (tangent vertex).
Very generally, we may introduce different weights,
say $x$ and $y$ respectively for the two types of vertices.
The $FPL^2(n_1,n_2)$ model of Ref.\JACO\ concerns
the particular case where $x=y$.

Before we study this model, let us note that the suppression
of the tangent vertex ($y=0$) leads to a quite trivial model 
on the square lattice since the only allowed loop configuration 
is that where the black ``loops" occupy the horizontal links of 
the lattice and the white loops the vertical links (or conversely).
Still, by taking $n_1=n_2=2$, which amounts to orienting
independently upwards or downwards each vertical
(white) line and to the left or to the right each horizontal
(black) line, the configurations that we obtain are in one-to-one 
correspondence with the foldings of the (dual) square lattice
in $d=2$, as studied in Section 1.1.
The correspondence is simply that a vertical (resp. horizontal) 
line of the (dual) square lattice is folded if and only if the 
orientations of the two vertical (resp. horizontal) lines on each side
are opposite. This remark, although elementary, prefigures the 
link between the $GFPL^2(2,2)$ model and the folding of random 
quadrangulations.  We will discuss this link more precisely in 
Section 13 below.

\fig{Definition of the heights $X$ in the $FPL^2$ model. The
definition uses the bipartite nature of the lattice. The
consistency after one turn imposes the constraint $A+B+C+D=0$. The
height is therefore three-dimensional.}{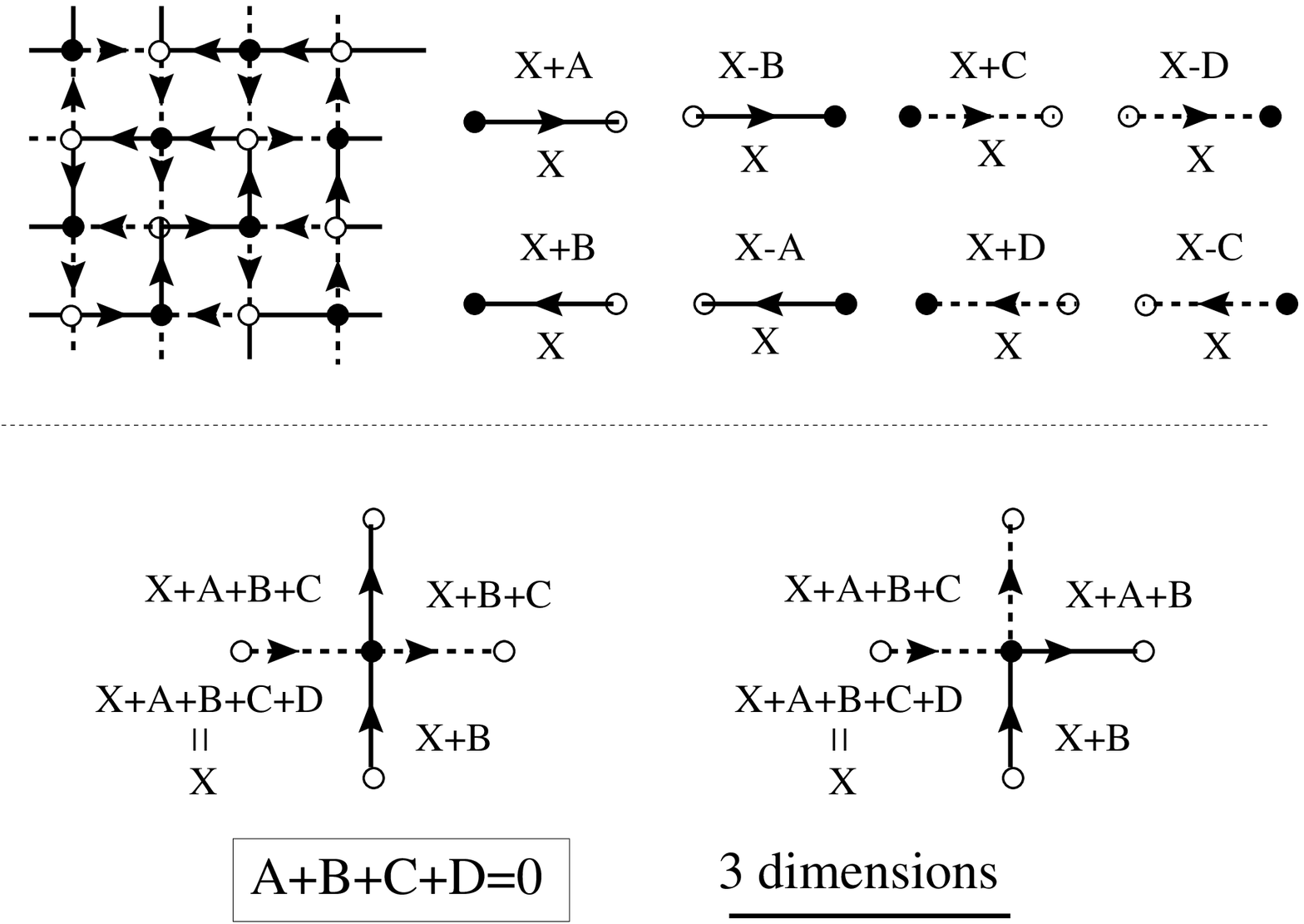}{14.cm}
\figlabel\fpldeuxrules
Let us now consider the $FPL^2(n_1,n_2)$ model with $x=y=1$.
This model was studied in Ref.\JACO\ by Coulomb gas
techniques similar to those presented in Section 5. Some properties 
could also be derived exactly by Bethe Ansatz [\xref\DCN,\xref\JZ], 
in the same spirit as in Section 4.

We start from the case $n_1=n_2=2$, which amounts to considering 
oriented loops with no extra weight.
The model is then equivalent to a four-color model with colors, say A, B, C 
and D on the links of the square lattice and the constraint
that the four colors are present around each node.
The links of color A or B (resp. C or D) form the black
(resp. white) loops and alternate along these loops. The two
possible choices for the alternation of colors correspond to the two 
orientations of the loop.

As in Section 3.3, we may transform the model into a model of heights $X$
on the faces of the square lattice with the transition rules represented 
in Fig.\fpldeuxrules. We need {\it a priori} four height differences 
$A$, $B$ (black loops) and $C$, $D$ (white loops), with the constraint 
that $A+B+C+D=0$ to ensure a well defined height on each face. 
The obtained height variable $X$ is therefore three-dimensional
and corresponds to a three-component scalar field theory with central 
charge $3$ \JACO. 

As in Section 5.2, the introduction of the weights $n_1$ and $n_2$ is 
performed by introducing vertex weights $\exp(\pm \pi e_1/4)$ (resp. 
$\exp(\pm \pi e_2/4)$) per left or right turn of the black 
(resp. white) loops, with $n_1=2 \cos(\pi e_1)$ and $n_2=2 \cos(\pi e_2)$.
For $0\leq n_1,n_2 \leq 2$ ($0\leq e_1, e_2 \leq 1/2$), 
the theory is conformal with central charge \JACO:
\eqn\ccfpl{c_{\rm FPL}(n_1,n_2)=3-6 {e_1^2\over 1-e_1} -6
{e_2^2\over 1-e_2}=1+c_{\rm dense}(n_1)+c_{\rm dense}(n_2)}
with $c_{\rm dense}$ given by Eq.\ccon.
One can also compute the exponents $x_{l,m}$
associated with the long distance behavior $r^{-2 x_{l,m}}$ for 
the correlation of a set of $l$ black lines and $m$ white lines connecting
two points at distance $r$, with the value \JACO
\eqn\xlonrap{\eqalign{x_{l,m}(n_1,n_2)
& ={1-e_1\over 8} l^2-{e_1^2\over 2(1-e_1)}
(1-\delta_{l,0}) \cr &
+{1-e_2\over 8} m^2-{e_2^2\over 2(1-e_2)}(1-\delta_{m,0})\cr
& +{1\over 16} \delta_{l+m,{\rm odd}}+\delta_{l,{\rm odd}}
\delta_{m,{\rm odd}} {(1-e_1)(1-e_2)\over (1-e_1)+(1-e_2)}\cr}}

We may finally consider a ``dense" version of the problem,
denoted by $DPL^2(n_1,n_2)$ (for densely packed loops) by allowing for
the presence of sites not visited by black loops
or white loops \JKdense. In terms of height, this requires
to impose the additional constraint $A+B=0=C+D$, which
leads to  two-dimensional heights and reduces the central charge
to a value
\eqn\ccdpl{c_{\rm DPL}(n_1,n_2)=2-6 {e_1^2\over 1-e_1} -6
{e_2^2\over 1-e_2}=c_{\rm dense}(n_1)+c_{\rm dense}(n_2)}
In this case, the exponents $x_{l,m}$ reduce to
\eqn\xlonden{\eqalign{x_{l,m}(n_1,n_2)&={1-e_1\over 8} l^2-{e_1^2\over 2(1-e_1)}
(1-\delta_{l,0})\cr & +{1-e_2\over 8} m^2-{e_2^2\over 2(1-e_2)}(1-\delta_{m,0})
\cr & = x_l(n_1)+x_m(n_2)\cr}}
with $x_l$ given by Eq.\xlon.

\subsec{Coupling $FPL^2(n_1,n_2)$ to gravity: Effective field theory 
of meanders}

The coupling of the $FPL^2(n_1,n_2)$ model to gravity corresponds 
{\it stricto sensu} to the case of multi-river multi-circuit tangent 
meanders with the two types of vertices, namely crossings and tangencies. 
For Eulerian gravity, realized by summing only over node-bicolored
tetravalent graphs, the three-dimensional degrees of freedom of the
regular lattice model are preserved and the critical behavior of the
model corresponds to the coupling to gravity of a CFT of central 
charge $c_{\rm FPL}(n_1,n_2)$, as given by Eq.\ccfpl.
We may in particular use Eq.\kpz\ to derive the configuration exponent
governing the large $n$ behavior $\mu_n \sim \rho^{2n}/n^\alpha$ of 
the number $\mu_n$ of tangent meanders with one river line and one 
circuit meandering around it with $2n$ contact (crossing or tangency) points 
of alternating color along both circuit and river. 
This corresponds to taking $n_1,n_2\to 0$,
in which case $c_{\rm FPL}(0,0)=-3$, leading to 
\eqn\alphatgmea{\alpha={7+\sqrt{7}\over 3}} 
\fig{Typical height configurations around (a) a crossing vertex and
(b) a tangency vertex. When going diagonally say from the NW to the SE face (shaded), 
the height increases by an amount restricted to $\pm (A+C)$ or $\pm (A+D)$
in the case (a) and only to $\pm (A+B)$ in the case (b). Forbidding (b)
will therefore amount to a reduction of the height variable range from 3 to 2
dimensions.}{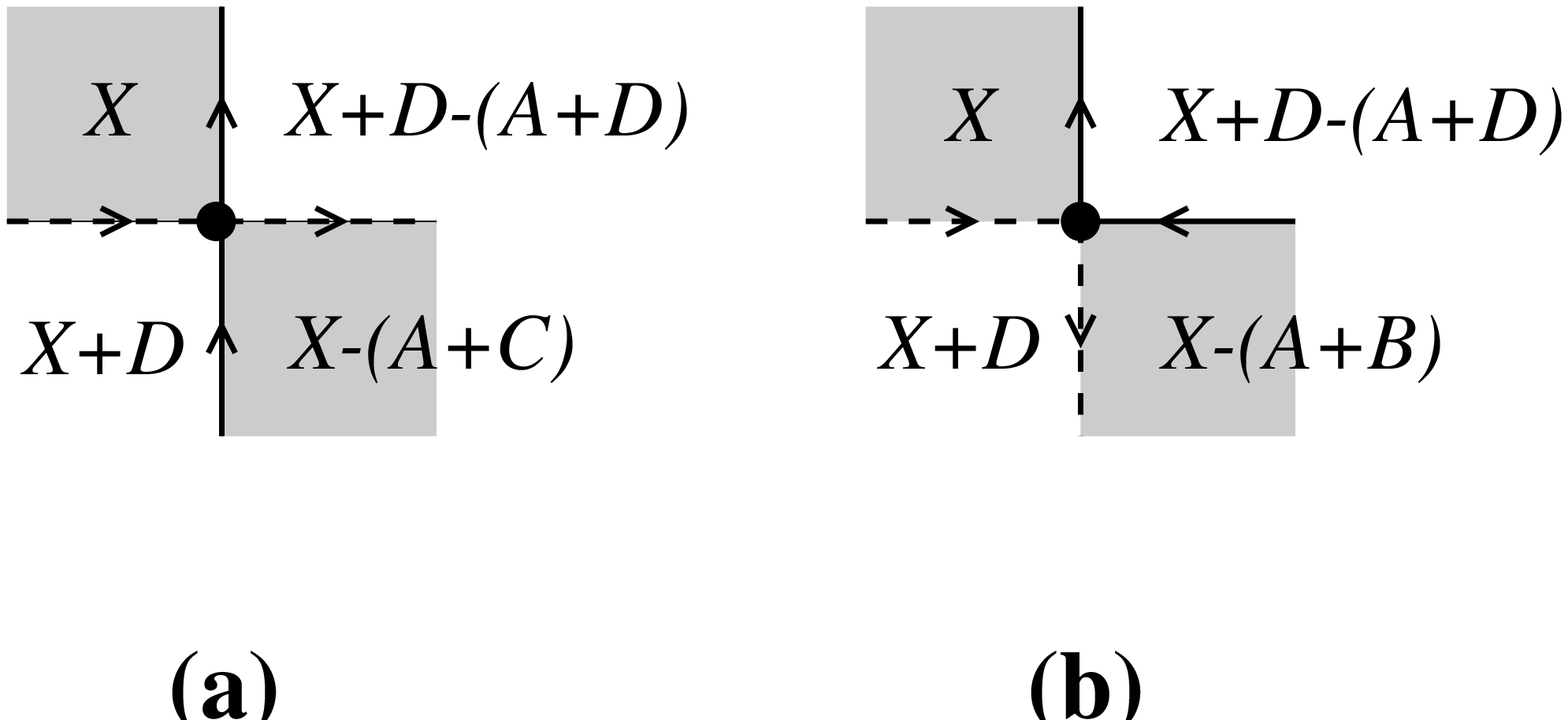}{10.cm}
\figlabel\forbid
This is not however the original meander problem we started with.
To get the $GFPL^2(n_1,n_2)$ model, we have indeed to forbid 
tangency points. We shall now argue that tangency points are relevant
and that their suppression reduces the value of the central 
charge. Assuming indeed that only crossing vertices are allowed, in
which case the bicolorability of the nodes of the graph 
is automatic, we can still build a three-dimensional height as before.
Note that the graph is also automatically face bicolorable 
(as is any tetravalent graph) and we may thus define a sub-lattice 
${\cal W}$ of white faces and a sub-lattice ${\cal B}$ of black faces. 
It is then easy to see that the heights on two neighboring faces on the 
sub-lattice ${\cal B}$, i.e.\ faces diametrically opposite around a node,
may differ only by $\pm (A+C)$ or $\pm (A+D)$ (see Fig.\forbid-(a)) 
but that the third direction $\pm (A+B)$ never occurs (as opposed to what
happens with the tangency vertex - see Fig.\forbid-(b)). The heights
of the sub-lattice ${\cal B}$ are {\it de facto} two-dimensional
in the $(A+C,A+D)$ plane and the same is true for
those of the sub-lattice ${\cal W}$ . It is therefore harmless to
set $A+B=C+D=0$, which takes us back to the $DPL^2$ model.
This leads to the prediction \ASY\ that the multi-river 
multi-circuit meander problem lies in the universality class
of the $DPL^2(n_1,n_2)$ model, therefore a CFT with central charge
\eqn\cmeandres{c_{\rm meander}(n_1,n_2)=c_{\rm DPL}(n_1,n_2)
=c_{\rm dense}(n_1)+c_{\rm dense}(n_2)}

\fig{In the case of tangent meanders on arbitrary
tetravalent graphs, the tangent vertices
may be untied without affecting the heights.}{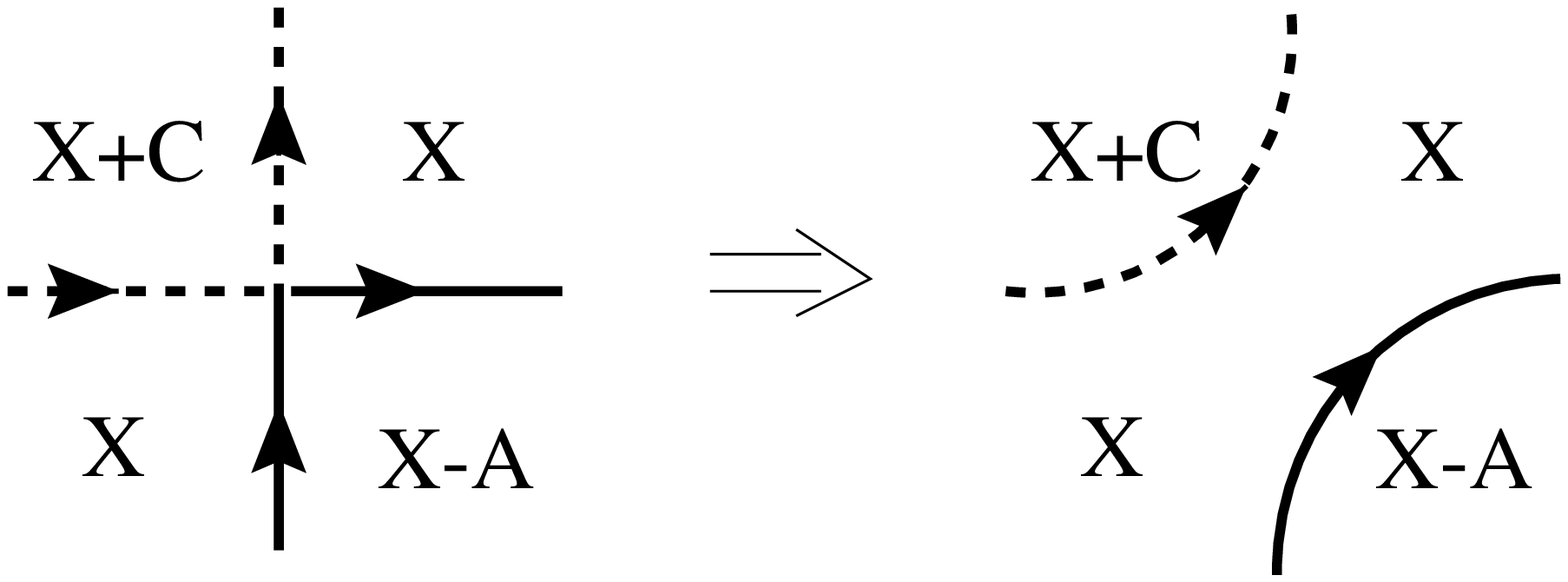}{9.cm}
\figlabel\irrtg
The same conclusion may be reached by a different reasoning.
Considering now tangent meanders on arbitrary tetravalent
graphs (ordinary gravity version), we may still use the
height transition rules of Fig.\fpldeuxrules\ but we 
can no longer distinguish between $A$ and $-B$ (resp. $C$ and $-D$), 
which equivalently amounts to imposing $A+B=0=C+D$ too. 
This also leads to a two-dimensional height and we again
have to use the reduced central charge $c_{\rm DPL}$ of Eq.\cmeandres.
To interpret this central charge as that for meanders, 
we still have to show that tangency points are {\it irrelevant}
for ordinary gravity as opposed to the case of Eulerian gravity.
We may understand this phenomenon heuristically by noticing that,
in the effective height language, any tangent vertex may be
safely ``untied" as shown in Fig.\irrtg\ {\it without modifying the 
heights}.

To conclude, we have predicted in two different ways that
meanders correspond to the coupling to gravity of a CFT with
central charge given by Eq.\cmeandres. For the original problem of meanders
with one river line and one circuit, this gives $c=-4$ and
leads to a configuration exponent \ASY\
\eqn\alphamean{\alpha\equiv \alpha(0)=2-\gamma_{str} (-4)=
{29+\sqrt{145}\over 12}}
by use of Eq.\kpz.

\subsec{More meander exponents}

We have now identified the CFT underlying meanders, as the dense two-flavor
loop model with $n_1=n_2=0$ coupled to ordinary gravity. 
The complete knowledge of the conformal operator content of this CFT 
via the Coulomb gas picture gives access to a host of meandric numbers 
which we describe now.

The important operators for our present purpose are those 
identified as creating oriented river vertices, namely the operators 
$\phi_{k}$ (resp. $\phi_{-k}$), $k=1,2,...$ which
correspond to the insertion of a $k$-valent source (resp. sink)
vertex at which $k$ oriented river edges originate (resp. terminate). 
In the Coulomb gas picture, these operators create ``magnetic" defect lines
along which the height variable has discontinuities.
The operator $\phi_k$ has conformal dimension \JACO
\eqn\dimpsi{ h_k = {x_{k,0}(0,0)\over 2}= {k^2-4 \over 32} }
$k=\pm1,\pm2,...$, with $x$ as in Eq.\xlonden.
When coupled to gravity, these operators get dressed (into $\Psi_k$)
and acquire, according to Eq.\kpzdim, the dimension:
\eqn\acqdim{ \Delta_k={{1\over 2}\sqrt{8+3 k^2}- \sqrt{5}\over 
\sqrt{29}-\sqrt{5}} }
As a preliminary remark, we note that $h_{\pm 2}=\Delta_{\pm 2}=0$.
The operators $\Psi_2$, $\Psi_{-2}$ indeed correspond to the marking of 
an edge of the river in meanders, and moreover such operators must 
go by source/sink pairs 
for the orientations of the pieces of river connecting them to be compatible.
Applying Eq.\corrA\ to the two-point correlator 
$\langle \Psi_2 \Psi_{-2}\rangle_A$ at fixed large area $A=2n$, we find
\eqn\markpoi{ \langle \Psi_2 \Psi_{-2}\rangle_A \sim 
{g_c^{-A} \over A^{1-\gamma_{str}} }}
while the meander counterpart (with a closed river) behaves as
$M_A/(2A)\sim g_c^{-A}/(A^{3-\gamma_{str}})$. We see that the net effect of the insertion
of the operators $\Psi_{\pm 2}$ is an overall factor
proportional to $A^2\propto n^2$, which confirms their interpretation
as marking operators. 

We may now turn to the case of semi-meanders, for which the river is a
semi-infinite line around the origin of which the road may freely wind.
Considering the point at infinity on the river as just another point, 
the semi-meanders
may equivalently be viewed as meanders whose river is made of a segment. 
Sending one of the ends of the segment to infinity just resolves the ambiguity of
winding around either end. Using the above river insertion operators, we immediately identify
the generating function for semi-meanders as
\eqn\serexP{ \langle \Psi_1 \Psi_{-1}\rangle = \sum_{n\geq 1} {SM}_n g^n}
Using again Eq.\corrA\ and the explicit values of $\Delta_{\pm 1}$ via \acqdim,
we arrive at the large $n$ asymptotics \ASY\
\eqn\larnsem{ {SM}_n\sim  {{g_c}^{-n}\over n^{\bar \alpha}}
\qquad {\bar \alpha}=1+2 \Delta_1-\gamma_{str}= 1+{\sqrt{11}\over 24}(\sqrt{5}+\sqrt{29}) }
Note that we expect the value of $g_c$ to be the {\it same} for
meanders and semi-meanders, as both objects occur as thermodynamic quantities in the {\it
same} effective field theory. In the language of Section 10.1, this
means that we are in regime (i).

\fig{Three types of meandric configurations in which the river has the geometry of (a)
a $k$-valent star (b) an ``eight" (c) a ``cherry". The vertices corresponding to
river sinks or sources are represented by filled circles ($\bullet$). The edges
of river in-between them are oriented accordingly. The road (dashed line)
may freely wind around univalent vertices.}{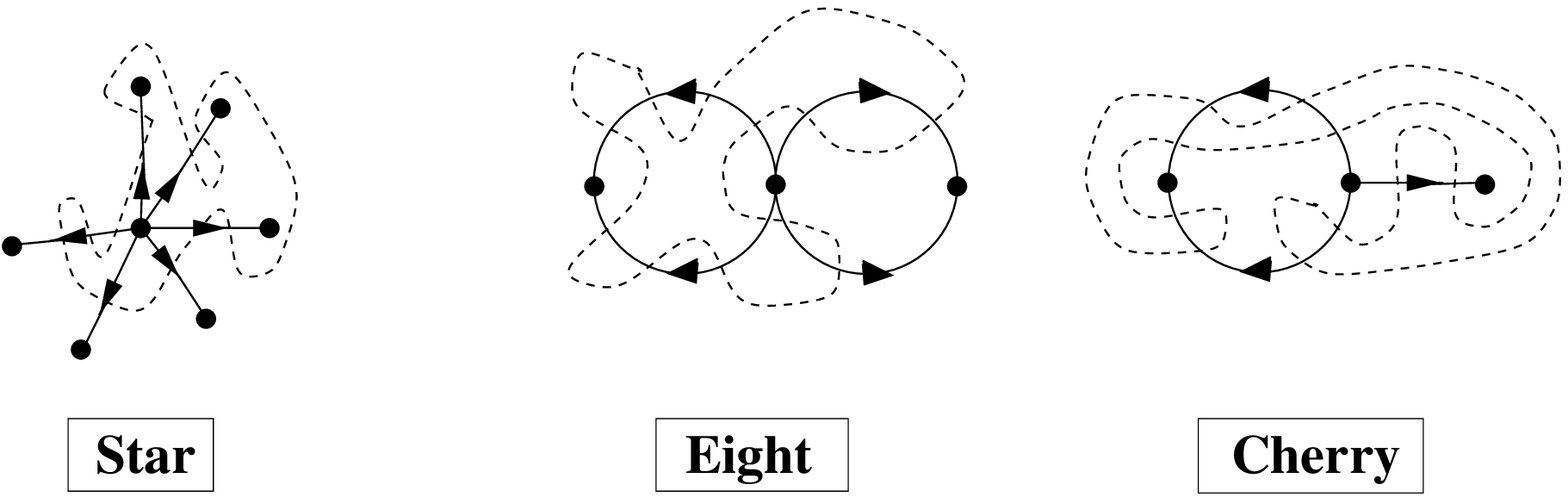}{13.cm}
\figlabel\meandric

We may now generate many more meandric numbers by considering more general correlators.
To name a few (all depicted in Fig.\meandric),
we may generate rivers with the geometry of a star with one
$k$-valent source vertex and $k$ univalent sink vertices generated
by $\langle \Psi_k (\Psi_{-1})^k\rangle$, rivers with the geometry of an ``eight"
with one tetravalent source vertex and two loops, each containing a bivalent sink vertex
generated by $\langle \Psi_4 (\Psi_{-2})^2\rangle$, or rivers with the geometry of
a ``cherry" with one trivalent source vertex, one univalent sink, and one loop,
marked by a bivalent sink vertex generated by
$\langle \Psi_3 \Psi_{-1}\Psi_{-2}\rangle$, etc ... For each of these situations,
we get the corresponding configuration exponent
$\alpha=3-\gamma+\sum_i (\Delta_i-1)$ by applying Eq.\corrA\ with the dimensions
of Eq.\acqdim.
We get respectively \ASY\ 
\eqn\respget{\eqalign{
\alpha_{\rm k-star}&={1\over 48}(\sqrt{5}+\sqrt{29})(\sqrt{3k^2+8} +k(\sqrt{11}-2 \sqrt{29})
+4 \sqrt{29}-2\sqrt{5}) \cr
\alpha_{\rm eight}&={1\over 24}(\sqrt{5}+\sqrt{29})(\sqrt{14}+\sqrt{5})\cr
\alpha_{\rm cherry}&=
{1\over 48}(\sqrt{5}+\sqrt{29})(\sqrt{11}+\sqrt{35}) \cr }}

\subsec{Multi-circuit meander exponents}
More generally, we may consider the one-river multi-circuit case of 
Section 10.1
by taking $n_1\to 0$ and $n_2=q$. Our predictions require moreover
$0\leq q\leq 2$ for the field theoretical description to hold.
Using similar argument as in previous Section, we predict
a critical exponent $\alpha(q)=2-\gamma_{str}(c)$ with
the central charge 
\eqn\cmq{c=c_{\rm meander}(0,q)=-1+6 {e^2\over 1-e}, \quad
q=2\cos(\pi e), \quad 0\leq e\leq {1\over 2}}
We arrive at
\eqn\predalpha{\alpha(q)= 2+{1-e+3e^2+
\sqrt{(1-e+3e^2)(13-13e+3e^2)}\over 6(1-e)}\qquad q=2\cos(\pi e) }

Similarly, we can compute the exponent ${\bar \alpha}(q)$
by inserting two defects corresponding to the two extremities
of an open segment. This leads to the prediction
${\bar \alpha}(q)=\alpha(q)-1+2\Delta_1(q)$ with $\Delta_1(q)$
related to $h_1(q)\equiv x_{1,0}(0,q)/2$ of \xlonden\ through \kpzdim.
We finally get \ASY\ 
\eqn\alsm{ {\bar \alpha}(q)=1+{\sqrt{2(24e^2+e-1)}
(\sqrt{1-e+3e^2}+\sqrt{13-13e+3e^2})\over
24(1-e)} \qquad q=2\cos(\pi e)}
The prediction for ${\bar \alpha}(q)$ requires that
$q$ be less than a critical value given by $q_c=2\cos(\pi e_c)$
where $e_c$ is the positive root of $24 e_c^2+e_c-1=0$, namely \ASY\
\eqn\valqc{q_{\rm c}=2\cos\left(\pi {\sqrt{97}-1\over 48}\right)
=1.6738\cdots}

At this value,
we have ${\bar \alpha}\to 1$. A heuristic argument shows that
the entropically favored semi-meanders are then those
with a ``Russian doll" structure, i.e.\ those made of a first finite
semi-meander using only the, say $s_1$ first bridges
{\it closest to the source}, of a second semi-meander
using only the $s_2$ next bridges (and possibly winding around
the set source/first semi-meander), etc... with
the $s_i$'s all finite. We thus expect at $q_c$ a transition to a regime
where the winding becomes extensive. The critical value $q_c$ 
is thus a good candidate for a transition point between the regimes 
(i) and (ii) of Section 10.1. 

\newsec{Numerical checks}

The above predictions may be tested numerically by performing a direct 
enumeration of meanders, semi-meanders and other related configurations. 
Several algorithms have been used, the best ones producing results
up to about 50 bridges. We shall present here two particular algorithms, 
one based on arch growth \NOUS, well adapted to the case of semi-meanders, and 
the other based on a transfer matrix formalism 
[\xref\JEN,\xref\POLEM,\xref\DGJ]. 
Beside exact enumerations, 
other statistical approaches such as Monte Carlo algorithms were also used
in Ref.\Goli\ to investigate large meander statistics.
\subsec{Arch growth algorithm}

\fig{The arch growth algorithm: the two transformations
(I) and (II) are described in the text. X, Y et Z denote arbitrary
systems of arches and R$_{\rm n}$ a rainbow made
of $n$ nested arches. Repeated application of (I) and (II)
on the empty configuration generates all semi-meanders, 
represented here up to $n=4$.}{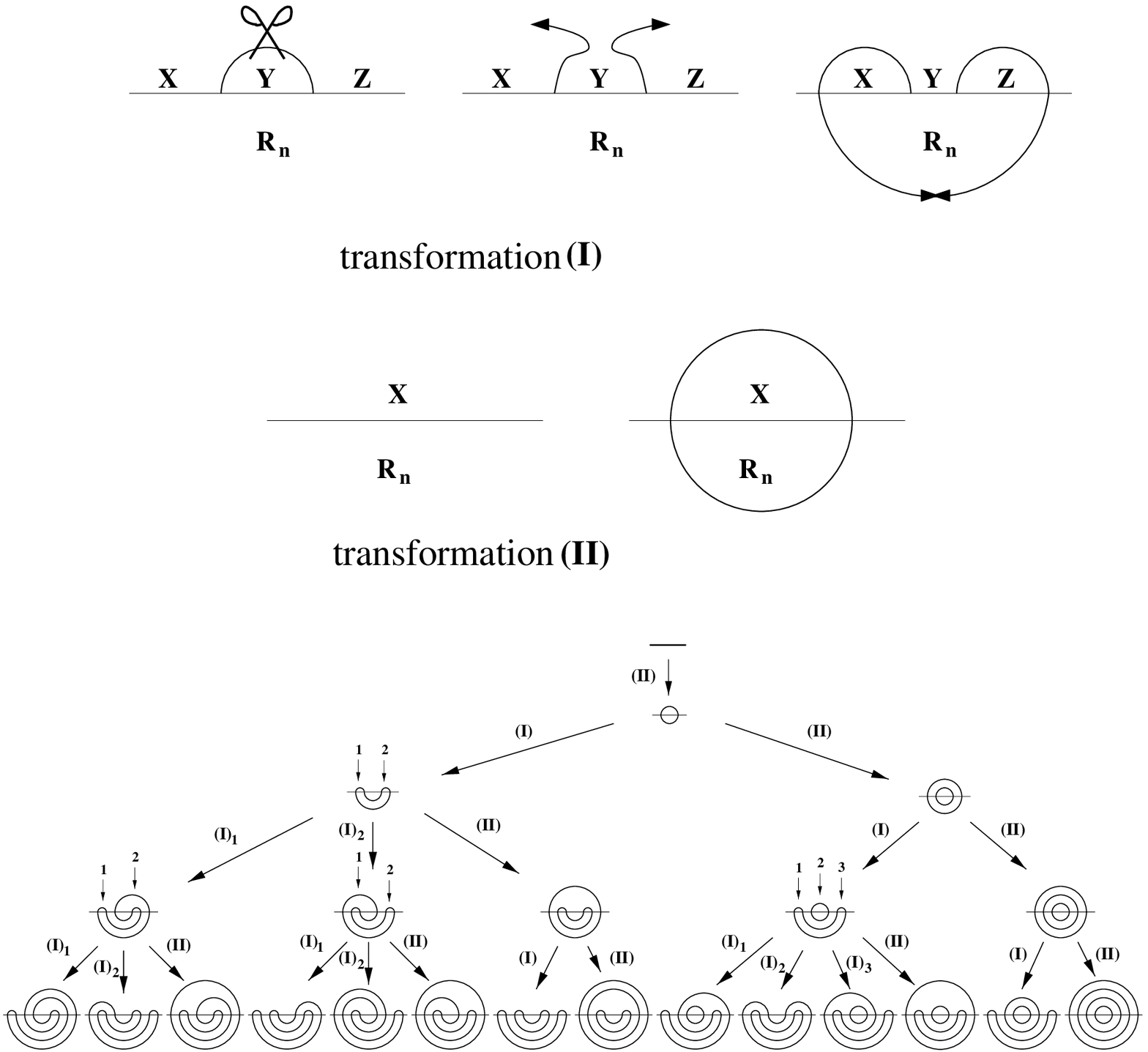}{13.cm}
\figlabel\croisarch
This algorithm generates configurations of {\it semi-meanders}. The
meanders are then recovered as a particular case of semi-meanders
with vanishing winding.
We use the ``open" representation of semi-meanders obtained by
opening the river as in Fig.\opensemi.
A semi-meander with $n$ bridges is therefore represented by a system
of $n$ arches above the (opened) river closed by a rainbow R$_{\rm n}$ 
made of $n$ nested arches below the (opened) river. 
The algorithm generates all semi-meanders with $n+1$ bridges from those 
with $n$ bridges by applying to the arch system one of the following two 
transformations, as illustrated in Fig.\croisarch:
\item{(I)} we select one of the {\it outer-most} upper arches, cut it 
and re-close it below the river by encircling the original semi-meander,
thus creating two arches above the river, two new crossings with
the river and one new arch of rainbow below the river.
In the original semi-meander picture, this transformation corresponds 
to the creation of one new bridge and {\it does not modify the number of
circuits}.
\item{(II)} we surround the existing semi-meander by a big circle,
thus creating an outer arch above the river, two new crossings
with the river and one new arch of rainbow below the river.
In the original semi-meander picture, this transformation corresponds 
to the creation of one new bridge and {\it increases by $1$ the number 
of circuits}.
\par
It is easy to see that the above algorithm generates all the
semi-meanders with $n+1$ bridges once and only once
from the set of semi-meanders with $n$ bridges. Indeed,
the transformation is easily inverted by cutting the external
arch of the rainbow below the river and gluing it above. 
Each semi-meander is thus obtained from the empty configuration 
{\it in a unique way} by the application of a succession of 
transformations (I) and (II) (see Fig.\croisarch).
An immediate consequence of this property is that  the average 
number $\langle {\rm out}\rangle_n(q)$ of outer-most arches 
above the river for semi-meanders with $n$ bridges weighted 
by a factor $q$ per connected component is equal to
\eqn\archext{\langle {\rm out}\rangle_n(q)={SM_{n+1}(q)\over SM_n(q)}-q
\buildrel {n\to \infty} \over \sim {\bar R}(q)-q}
The arch growth algorithm was used in Ref.\NOUS\ to enumerate 
semi-meanders up to $27$ bridges. 

Beyond mere enumeration, this algorithm has the advantage of 
allowing for a step by step tracking of the number of circuits,
thus giving access to the analytic structure of the $SM_n^{(k)}$.
More precisely, the number of circuits corresponds to the number 
of times the transformation (II) was used. For large $n$ and 
$k=n-l$ with $l$ {\it finite}, we deduce for instance that
\eqn\asymmnl{SM_n^{(n-l)}\buildrel {n\to \infty} \over \sim {n^l\over l!}}
corresponding to the ${n \choose l}\sim n^l/l!$ choices of the 
transformation (I). 
More precisely, we have \NOUS\
\eqn\premval{\eqalign{SM_n^{(n)}&= 1 \cr
SM_n^{(n-1)}&= n-1 \quad n\geq 1 \cr
SM_n^{(n-2)}&= {1\over 2} (n^2+n-8) \quad n\geq 3 \cr
SM_n^{(n-3)}&= {1\over 6} (n^3+6 n^2 -31 n -24) \quad n\geq 5 \cr 
SM_n^{(n-4)}&= {1\over 24} (n^4 +14 n^3 -49 n^2 -254 n) \quad n\geq 7 \cr}}
obtained by analyzing all possible shapes of semi-meanders with
only $1,2,3,4$ applications of (I). 

\subsec{Transfer matrix algorithm}
\fig{Transfer matrix algorithm. The operators
O, C, U and D are described in the text.}{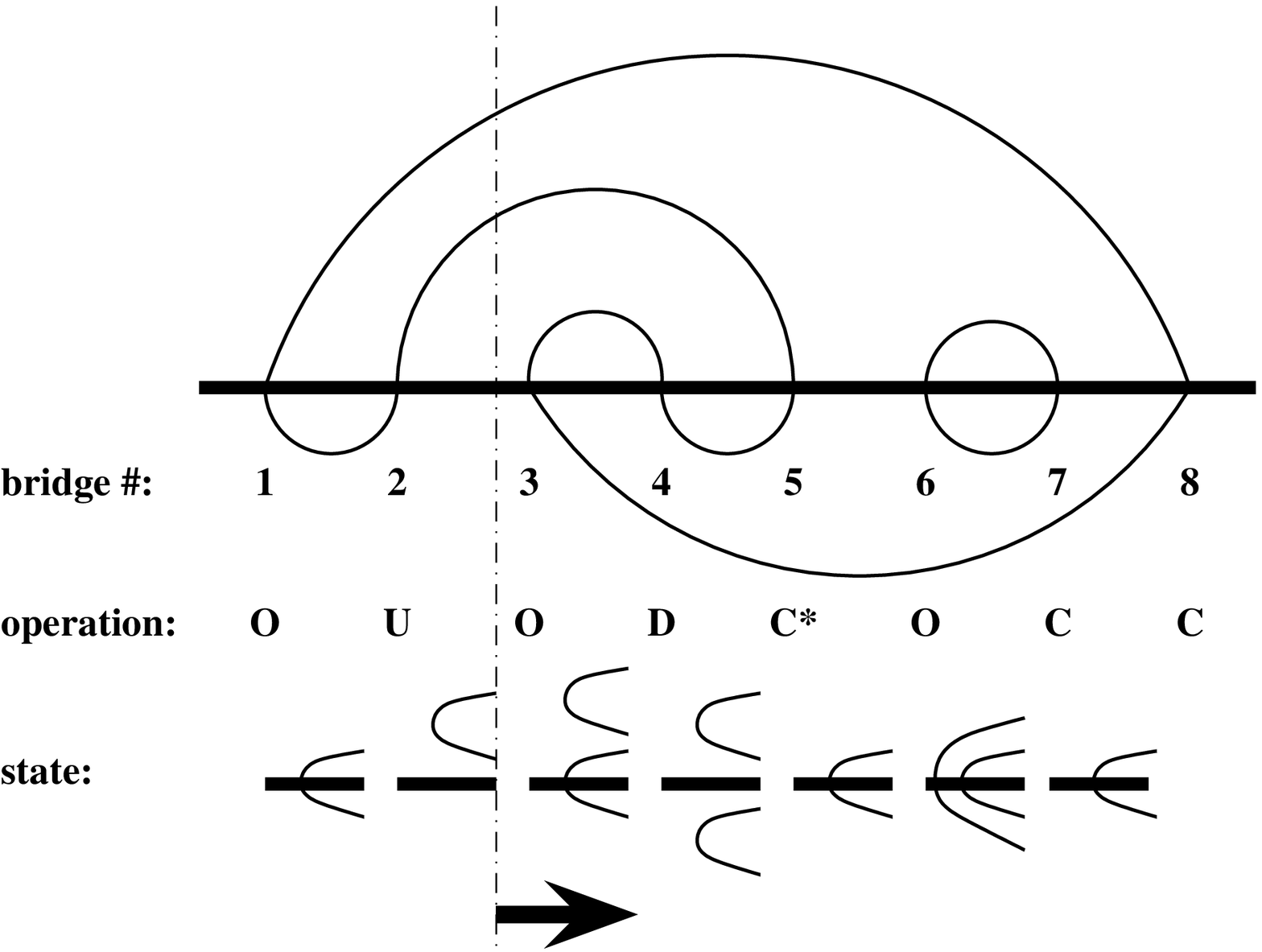}{10.cm}
\figlabel\transfer
This particularly efficient algorithm was first used
by Jensen in Ref.\JEN, where meanders were enumerated up 
to $2n=48$ bridges.
The idea consists in generating the meanders from left to right
one bridge after the other by means of a transfer matrix. The state
of the system between the $m$-th and the $(m+1)$-th bridge
is characterized by the connectivity of those arches above
and below the river which have been opened but not yet closed (see
Fig.\transfer). These arches are indeed connected by pairs
in the left part of the meander. These connections form themselves 
a {\it transverse system of arches}, which, together with
the relative position of the river, entirely characterize the state of the
system. The application of the transfer matrix corresponds to the
crossing of a bridge where one of the following four operations takes place
(see Fig.\transfer):
\item{O:} Opening of a new transverse arch with extremities
immediately on both sides of the river.
\item{C:} Connection of the two extremities of transverse arches
which are immediately on both sides of the river.
If these two extremities belong to the same transverse arch,
a new circuit of meander has been created on the left of that
bridge. Otherwise, the number of circuits remains unchanged 
(operation denoted by C$^*$ in Fig.\transfer).
\item{U:} Upward migration (with respect to the river) of the system
of transverse arches.
\item{D:} Downward migration (with respect to the river) of the system
of transverse arches.
\par
Starting from the empty state, the repeated application of the transfer 
matrix allows to generate meanders or semi-meanders
according to the final state reached after $2n$ or $n$ iterations.

\subsec{Numerical results}
The above algorithms lead to the following exact enumerations.
The table below gives the numbers $SM_n^{(k)}$ of semi-meanders 
with $n$ bridges and $k$ connected components for $n$ up to $27$:

\centerline{\vbox{\epsfxsize=16.cm\centerline{\epsfbox{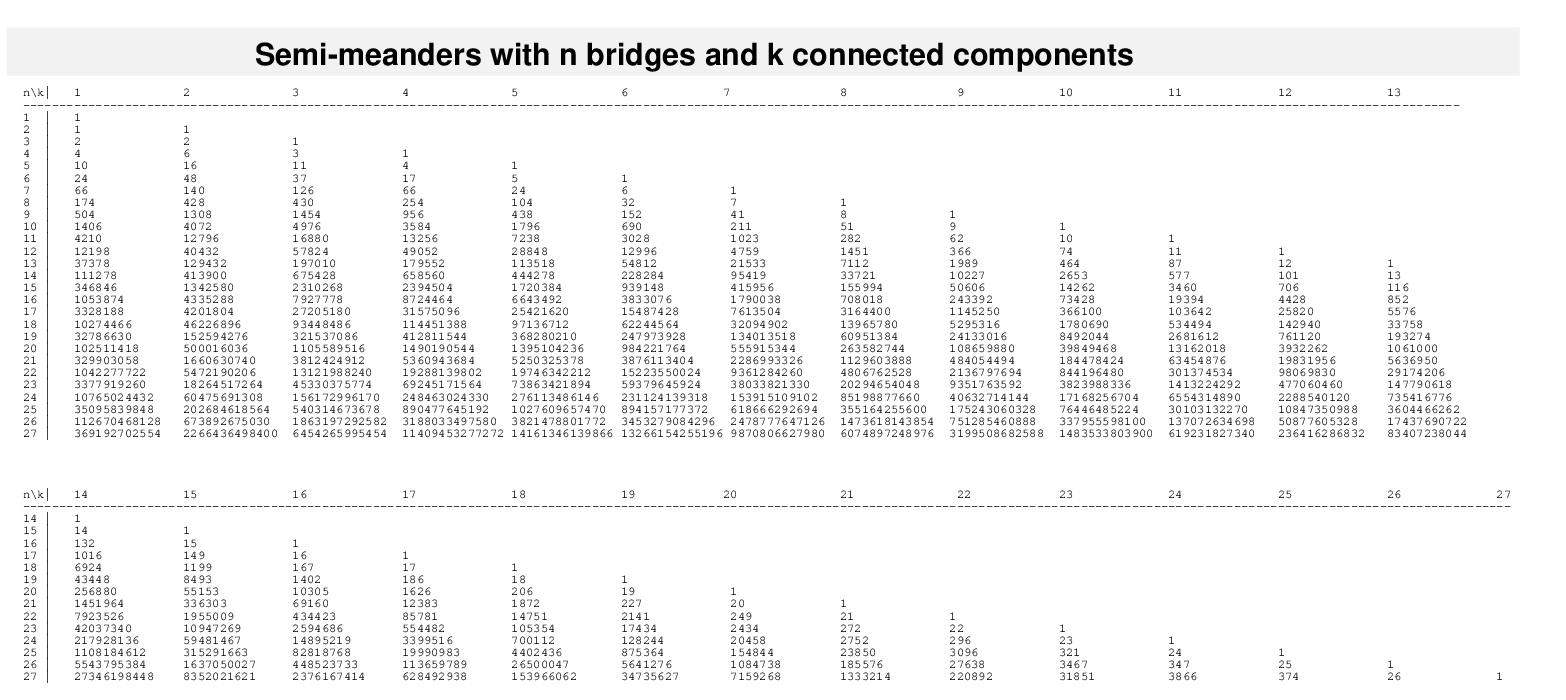}}}}

Similarly, the table  below gives the numbers $M_n^{(k)}$ of meanders
with $2n$ bridges and $k$ connected components for $n$ up to $20$:

\centerline{\vbox{\epsfxsize=16.cm\centerline{\epsfbox{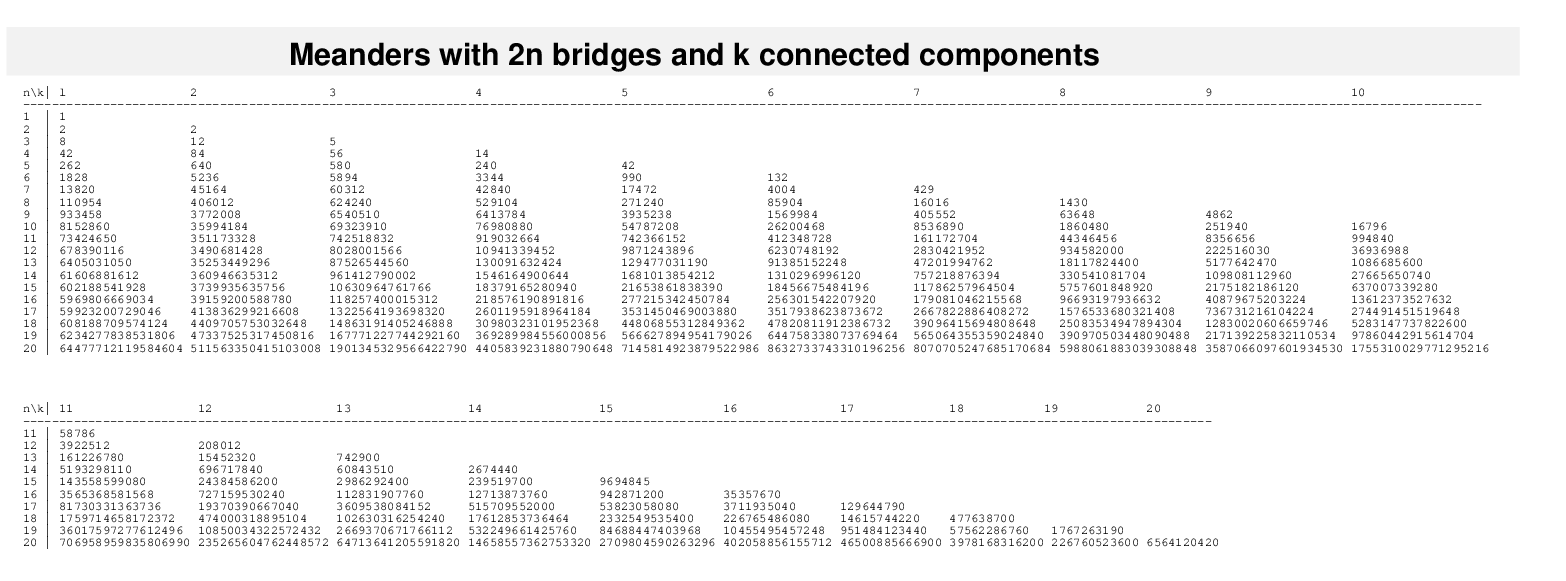}}}}

These data may be analyzed in two different ways, either numerical, 
or analytic. The first approach, purely numerical, consists in building
$M_n(q)$ and $SM_n(q)$ for the values of $n$ accessible from
the data and to directly estimate $R(q)$
(resp. ${\bar R}(q)$) by taking for instance the ratio of
two consecutive values of $M_n(q)$ (resp. $SM_n(q)$)
and by using appropriate convergence algorithms to extract
a limiting value at large $n$.
\fig{Estimates of $R(q)$ and ${\bar R}(q)$ for $q$ between $0$ and
$6$. The numerical errors are smaller than the thickness of the 
curves.}{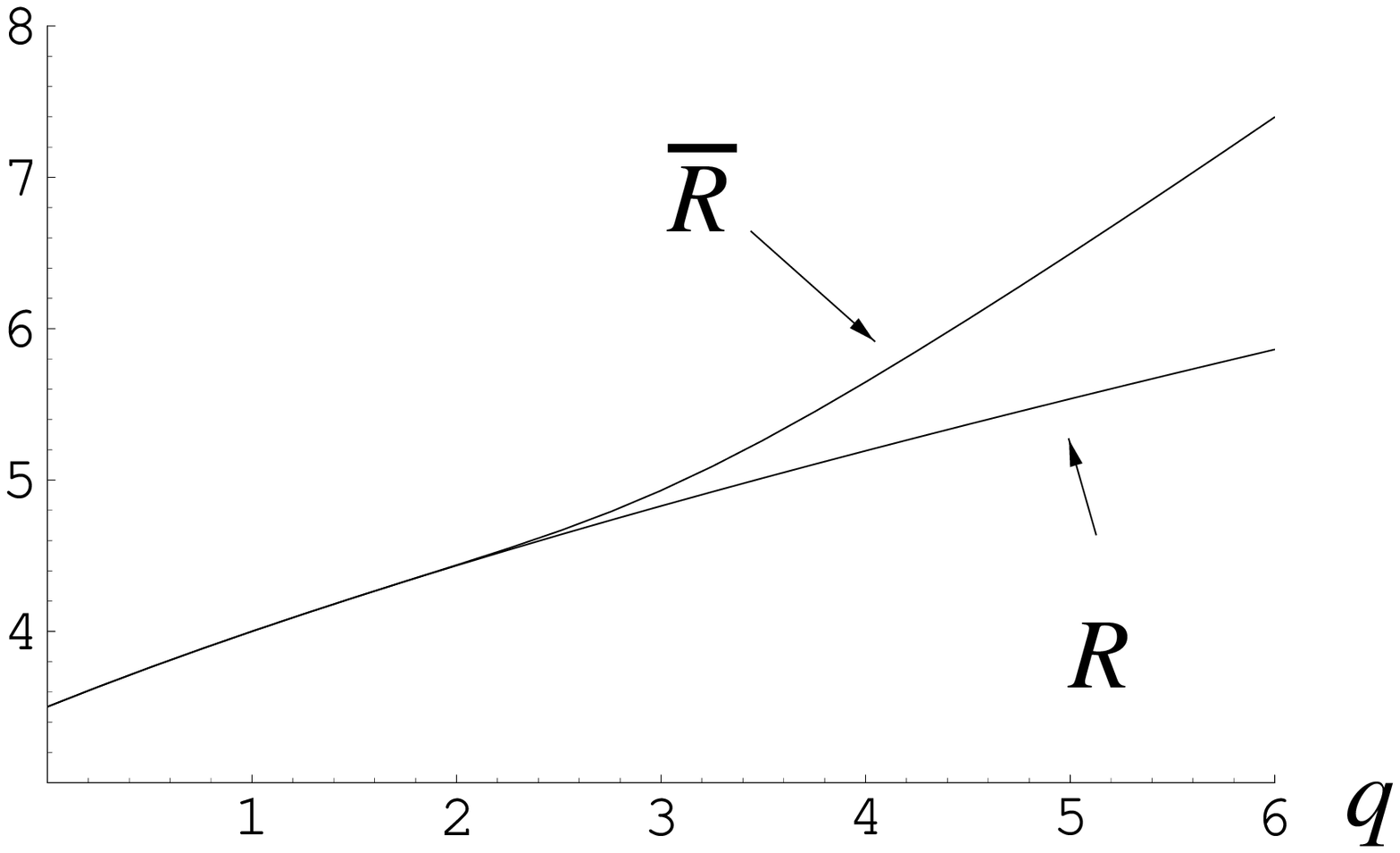}{9.cm}
\figlabel\RetR
The figure \RetR\ shows the obtained estimates of $R(q)$ and ${\bar R}(q)$
for values of $q$ between $0$ and $6$. One clearly distinguishes
two regimes, a large $q$ regime where ${\bar R}(q)>R(q)$
and a low $q$ regime where ${\bar R}(q)=R(q)$ up to numerical errors
(recall that at $q=1$, we already know that the two are indeed equal). 
This confirms the existence of a transition point $q_{\rm c}$ between 
a low $q$ phase corresponding to the situation (i) of Section 10.1
of negligible winding, and a large $q$ phase corresponding to the situation 
(ii) of extensive winding. The precise value of $q_{\rm c}$ 
is difficult to estimate as the approach of the two curves is tangential. 
Still, the graph of Fig.\RetR\ is compatible with the analytic prediction 
of Eq.\valqc. 
\fig{Estimates of the exponents $\alpha(q)$, ${\bar \alpha}(q)$ and $\nu(q)$
for $q$ between $0$ and $8$. The different curves correspond to different
numbers of iterations of the convergence algorithm.}{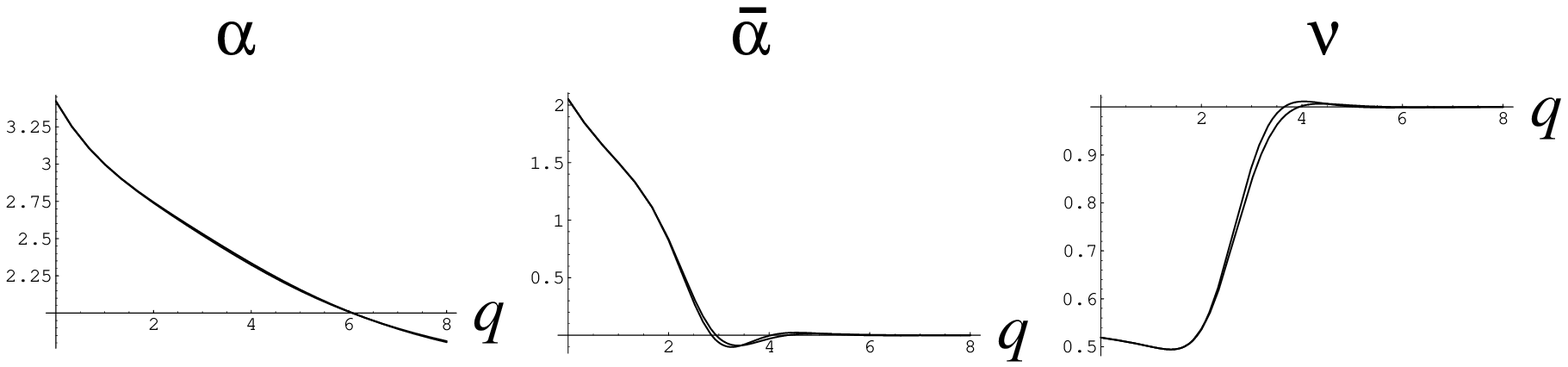}{14.cm}
\figlabel\expo
We may similarly estimate the exponents $\alpha(q)$, ${\bar \alpha}(q)$
and even $\nu(q)$ by taking appropriate combinations of (semi-)meander 
polynomials which converge to these exponents at large $n$. 
These estimates are represented
in Fig.\expo\ for $q$ between $0$ and $8$. The estimated
value of $\nu(q)$ has a large variation  around $q\sim q_{\rm c}$, 
which suggest a discontinuity of $\nu$ at the transition point.
We therefore expect that $\nu(q)$ varies continuously for 
$q<q_{\rm c}$ (regime (i)) and has a jump discontinuity beyond which
$\nu(q)=1$ (regime (ii)).
The exponent ${\bar \alpha}(q)$ follows the same scenario
with a continuous variation for $q<q_{\rm c}$ and a jump discontinuity
to a {\it constant} value ${\bar \alpha(q)}=0$ (we already know that
${\bar \alpha}(q)\buildrel {q\to \infty} \over \sim 0$).
Finally, the exponent $\alpha(q)$ seems to decrease without discontinuity
toward its limiting value $\alpha(q)\buildrel {q\to \infty} \over \sim
3/2$.

As far as the original problem of meanders and semi-meanders is concerned
($q\to 0$), a refined analysis of the numerical data is presented in
Refs.[\xref\JEN,\xref\POLEM] and leads to the precise estimates
\eqn\valjensR{R(q=0)={\bar R}(q=0)=3.501837(3)}
\eqn\valjens{\alpha(q=0)=3.4208(6) \quad {\bar\alpha}(q=0)=2.0537(2)}

These exponents agree remarkably with the predicted values
$\alpha(0)=3.420132\ldots$ of Eq.\alphamean\ and 
${\bar \alpha}=2.053198\ldots$ of Eq.\larnsem.

\fig{Comparison between the theoretical value $c_{\rm meander}(0,q)$
of the central charge and its estimated value from the numerical data
of meanders (a) and semi-meanders (b).}{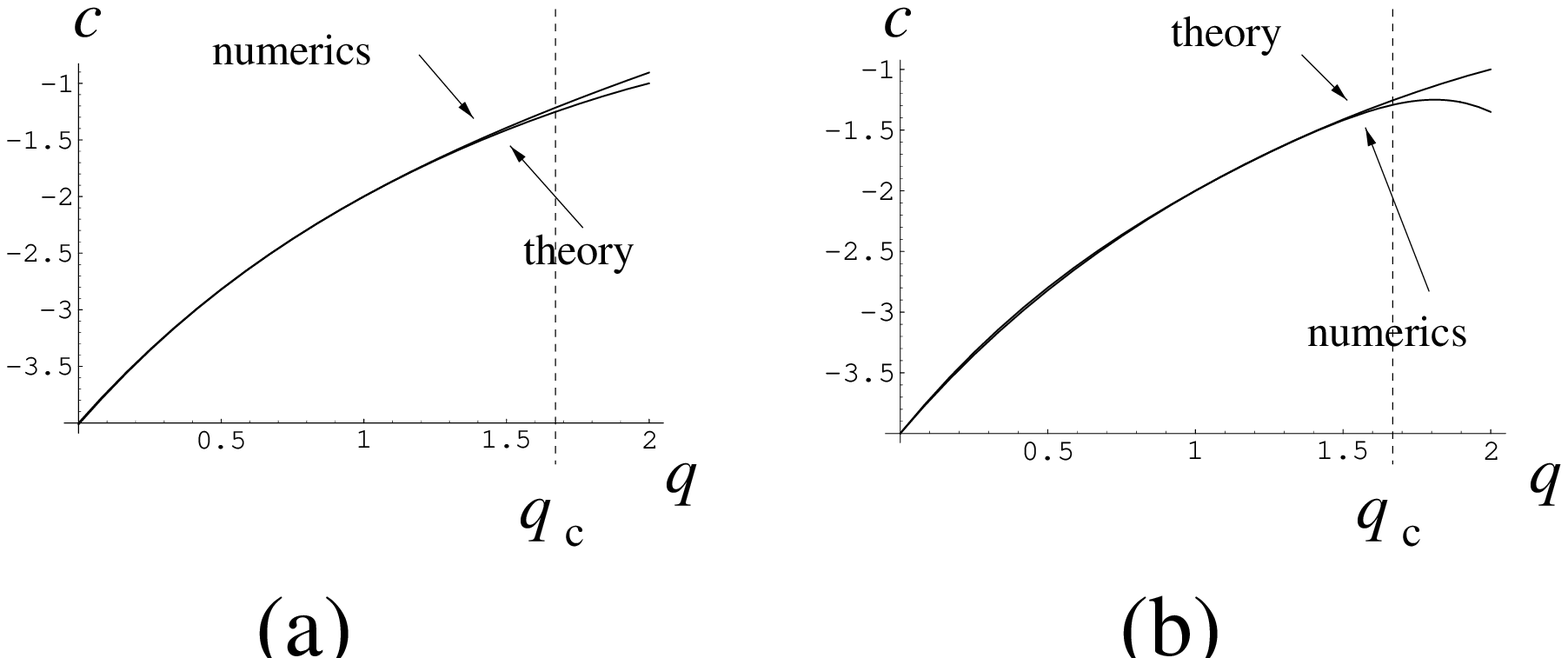}{12.cm}

\figlabel\cvsth
For generic $q$, we have represented in Fig.\cvsth\ the comparison between
the theoretical central charge $c_{\rm meander}(0,q)$ of Eq.\cmq\ 
and its two numerical estimates: (a) that obtained
by inverting the relation \predalpha\ from the measured value of
$\alpha(q)$ and (b) that obtained by inverting
the relation \alsm\ from the measured value of ${\bar \alpha}(q)$.
For $q=0$, the estimates \valjens\ lead
respectively in the cases (a) and (b) to $c=-4.003(3)$ and
$c=-4.002(1)$. 

Note finally that the above transfer matrix method was also adapted
to estimate configurational exponents for other meandric numbers
such as those with ``eight" or ``cherry" river configurations
(see Fig.\meandric). The agreement with the predictions \respget\ 
is excellent \DGJ. Using the same method, it was also checked that
the presence of the tangency vertex does not affect the exponents
for the ordinary gravity version of meanders, hence it corresponds 
to an irrelevant perturbation, as we argued.

\bigskip
A second way of exploiting the data consists in using them to obtain
a large $q$ expansion of various quantities. At large $q$, the expansion
of ${\bar R}(q)$ up to order $p$ in $1/q$ is governed by $SM_n^{(n-l)}$ 
for $l\leq p+1$. We have seen that for $l$ finite and $n$ large enough, 
$SM_n^{(n-l)}$ is a polynomial of $n$ of degree $l$. This property is a direct
consequence of the arch growth algorithm. The existence
of a thermodynamic limit for $SM_n(q)^{1/n}$ then guarantees that
${\bar \alpha}(q) =0$ as long as the $1/q$ expansion remains valid.
Moreover, imposing that the above polynomials actually reproduce the 
numerical data leads to their complete determination up to $l=19$.
The net result is finally a large $q$ expansion of ${\bar R(q)}$, 
given by \NOUS\
\eqn\devrbar{
\eqalign{
{\bar R}(q)~&=~ q+1+{2 \over q}+{2 \over q^2}
+{2 \over q^3}-{4 \over q^5}
-{8 \over q^6}-{12 \over q^7}-{10 \over q^8}-{4 \over q^9}
+{12 \over q^{10}}
+{46 \over q^{11}}\cr
&+{98 \over q^{12}}+{154 \over q^{13}}+{124 \over q^{14}}+{10\over q^{15}}
-{102 \over q^{16}}+{20 \over q^{17}}-{64 \over q^{18}}
+O({1 \over q^{19}})\cr}}
A similar analysis for meanders allows to determine
the analytic structure of the $M_n^{(n-l)}$ for large $n$ and finite $l$.
Instead of polynomials, one obtains rational fractions in this case.
This results in \NOUS\
\eqn\devr{\eqalign{R(q)~&=~2 \sqrt{q} \big(1+{1\over q}
+{3\over 2 q^2}-{3\over 2 q^3} -{29\over 8 q^4}
-{81\over 8 q^5}-{89\over 16 q^6} +O({1\over q^7})\big) \cr}}
\fig{Comparison between the numerical estimates and the
large $q$ expansions for $R(q)$ (a) and ${\bar R}(q)$ (b).
For $R(q)$, we have drawn the expansions truncated at order
$q^{-3}$ and $q^{-6}$. For ${\bar R}(q)$, we have drawn
the expansions truncated at order $q^{-6}$, $q^{-12}$ and $q^{-18}$.}
{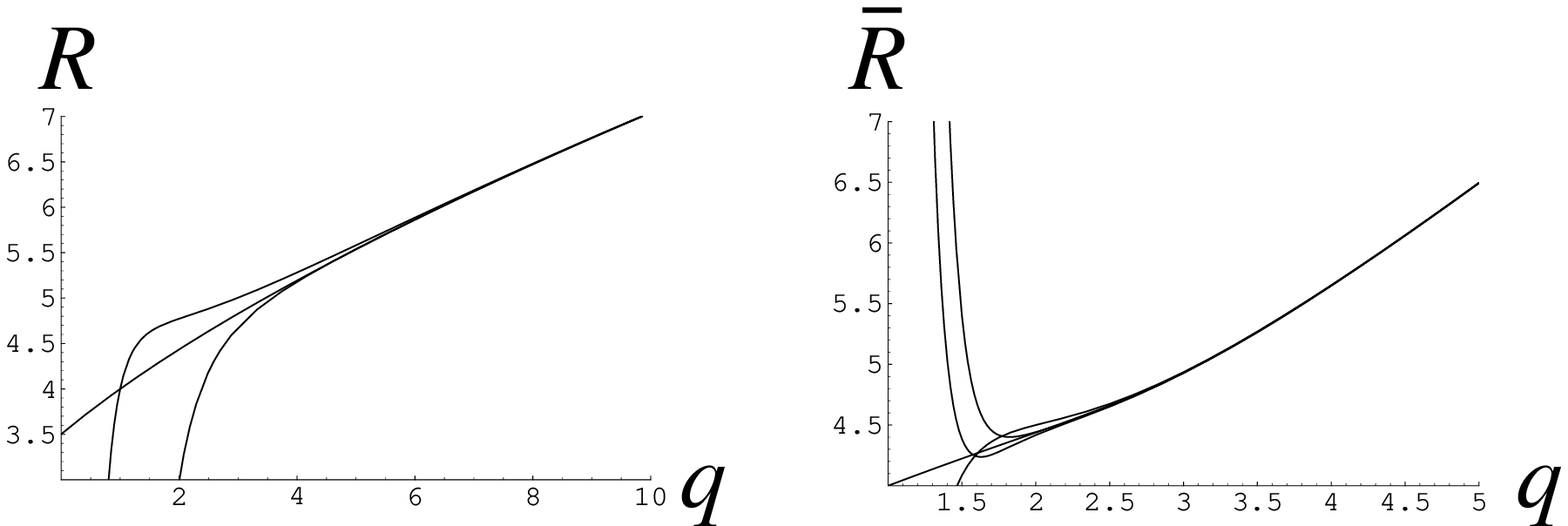}{12.cm}
\figlabel\devlpt
The figure \devlpt\ shows the comparison between the values of $R(q)$
and ${\bar R}(q)$ as estimated numerically and the expansions
\devr\ and \devrbar\ above.

The analysis of the corrections to the polynomial behavior for
the numbers $SM_n^{(n-l)}$ displays corrections
of the form $R(q)^n$ with $R(q)$ given by Eq.\devr.
It is thus natural to think that the expansions above are
valid as long as ${\bar R}(q)>R(q)$, i.e.\ in the whole
regime (ii), and in particular that $\bar{\alpha}(q)=0$, $\nu(q)=1$
for all $q>q_{\rm c}$. For $q<q_{\rm c}$, these expansions
are no longer valid and therefore do not allow to describe
the regime (i). Fortunately, this regime (i) is precisely that
for which we have obtained predictions from the KPZ formula.

\newsec{Meanders as folding of random quadrangulations}

In this paper, we started in part A by a study of the planar folding
of the triangular lattice and extended it in part C to the case of
random triangulations. We saw that this latter problem has a fully 
packed loop formulation on trivalent graphs. The above formulation
of the meander problem as a fully packed loop gas on random tetravalent 
graphs suggests a link between this problem and the folding of 
quadrangulations. 
In this Section we will ``close the loop" by showing that the meander 
problem is indeed the same problem as that of {\it planar folding
of random quadrangulations}. This may be seen as the non-trivial
gravitational version of the trivial correspondence between the 
$FPL^2(2,2)$ model with only the crossing vertex and the
planar folding of the regular square lattice.
This allows alternatively to view the folding of 
a {\it self-avoiding} one-dimensional chain as the {\it phantom} folding 
of a two-dimensional random quadrangulation, a somewhat surprising
result. 

Let us start by considering the set of planar random quadrangulations,
namely tessellations for which all the tiles are unit squares. 
By planar folding of such a quadrangulation, we again mean any map 
of the quadrangulation onto the plane whose restriction to each square
tile is an isometry. 
The nodes of the quadrangulation then necessarily have their image
on the nodes of a regular square lattice in the plane.
\bigskip
\noindent$\diamond$ \underbar{\sl Foldability of random quadrangulations}
\par\nobreak
\fig{Four-coloring of the nodes of the regular square lattice
}{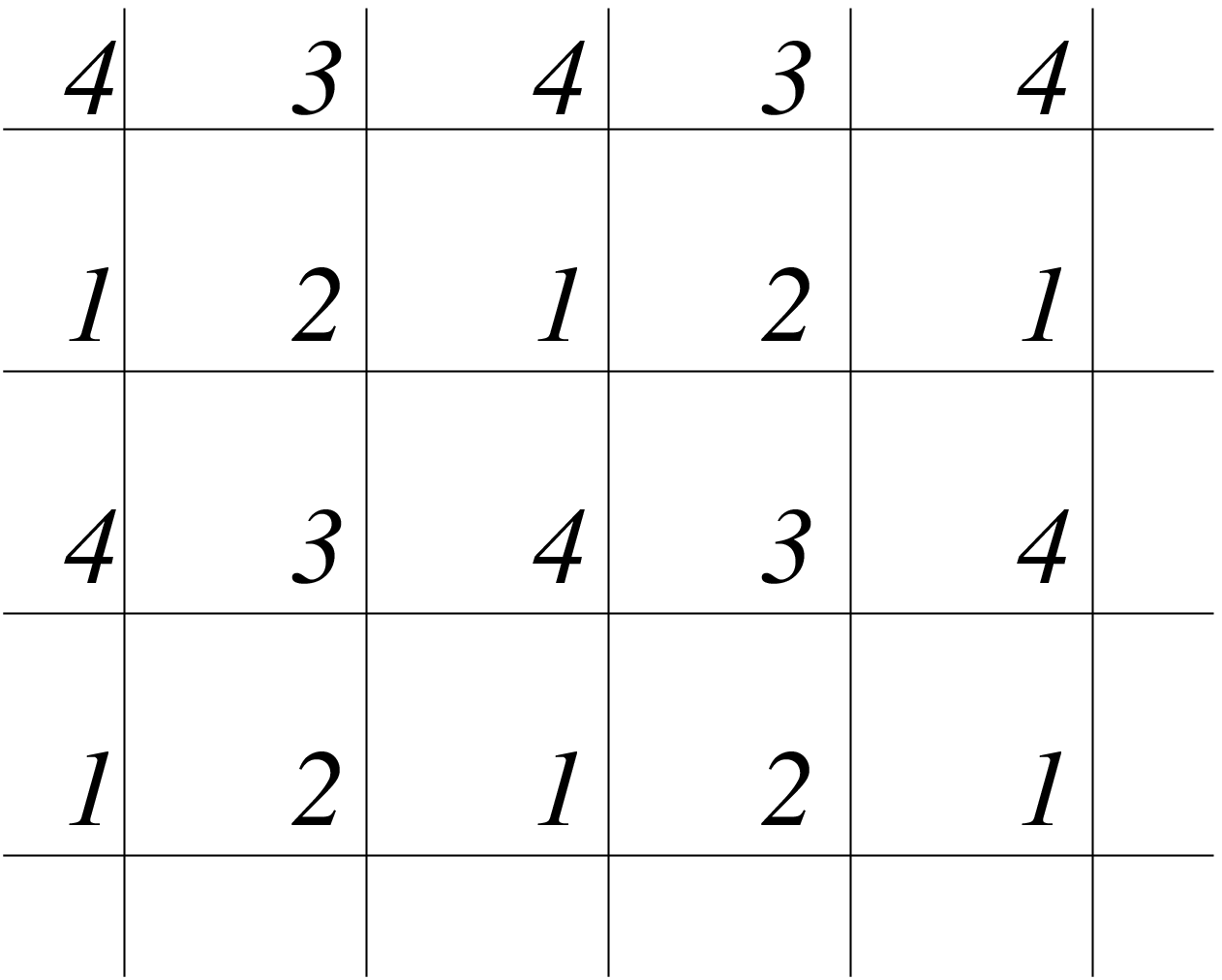}{4.cm}
\figlabel\quadrico
\fig{Example of a foldable quadrangulation (left),
i.e.\ a  quadrangulation whose nodes may be colored by four colors
$1$, $2$, $3$ and $4$  in such a way that the nodes around each
face are in cyclic order $1$,$2$,$3$,$4$ or $4$,$3$,$2$,$1$, also
corresponding to an orientation of the links. The squares have been deformed
in the planar representation. The equivalent dual tetravalent
graph (right) is node-bicolored. The links dual to the
$1$--$2$ or $3$--$4$ links form black lines which cross white
lines formed by the links dual to the
$2$--$3$ and $4$--$1$ ones.}{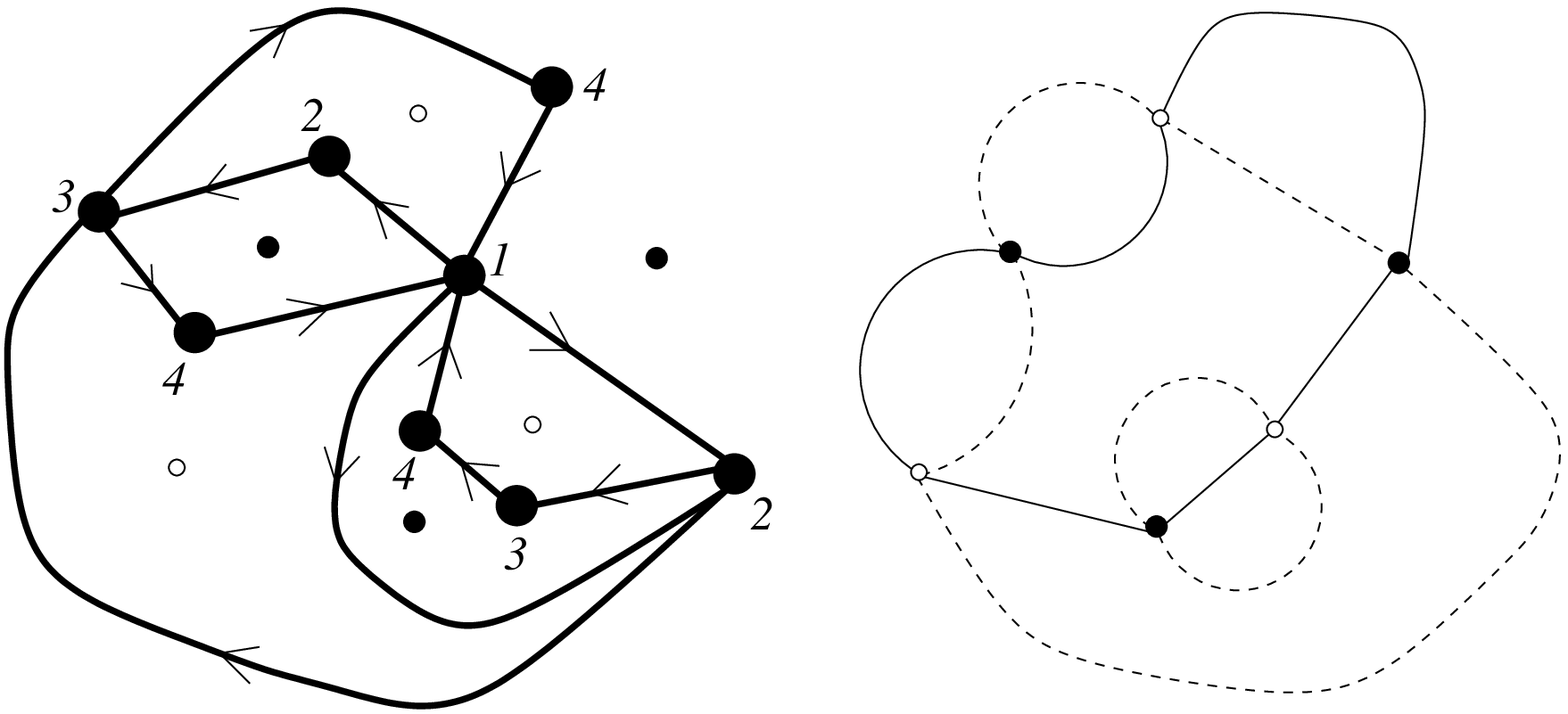}{12.cm}
\figlabel\quadpliable
As in the case of triangulations, we first address the question 
of foldability in the plane. By use of
the four-coloring of the nodes of the regular square lattice
(see figure \quadrico), we immediately see that,
if a folding exists, it naturally induces a four-coloring
of the nodes of the quadrangulation. In this coloring, the
four colors $1$, $2$, $3$, $4$ must appear in {\it cyclic} 
order $1$,$2$,$3$,$4$ or $4$,$3$,$2$,$1$ around each face. 
Conversely, a quadrangulation with such a four-coloring
(which is unique up to a cyclic permutation of the colors)
is clearly foldable onto a single square whose four vertices would have
been labeled $1$,$2$,$3$,$4$. We thus have the following
equivalent characterizations for genus zero quadrangulations:
\item{1.} The quadrangulation is foldable in the plane
\item{2.} Its nodes may be colored by four colors $1$, $2$,
$3$ and $4$ appearing in cyclic order $1$,$2$,$3$,$4$ or
$4$,$3$,$2$,$1$ around each face
\item{3.} Its faces are {\it bicolorable} by two colors
which must be distinct on two neighboring faces
\item{4.} Its edges may be oriented in such a way that each face
has a well-defined orientation
\item{5.} The number of faces around each node is even
\item{6.} The number of edges adjacent to each node is even
\par
This last property 6 justifies that we call such quadrangulations
Eulerian.

{\it The planar foldable quadrangulations are therefore
the Eulerian quadrangulations}, characterized
by any of the above equivalent properties.

The number of such quadrangulations with $4s$ edges (and
rooted, i.e.\ with a distinguished oriented link)
is given by
\PS
\eqn\biq{N_s=8\ {3^{s-1} (3s)!\over (2s+2)! s!}
\sim {1\over 6\sqrt{\pi}}{\left({9\over 2}\right)^{2s}\over s^{5\over 2}}}
to be compared with Eq.\reeseule.
\bigskip
\noindent$\diamond$ \underbar{\sl Planar folding of random quadrangulations}
\par\nobreak
The four-coloring of the nodes allows to classify all the
links of the  quadrangulation into the ``horizontal" links 
$1$--$2$ and $3$--$4$ and the ``vertical" links $2$--$3$
and $4$--$1$. These links are moreover naturally oriented in the
direction $1\to 2 \to 3 \to 4 \to 1$ (see figure \quadpliable).
The dual graph is a tetravalent node-bicolored graph, on which
the dual of horizontal links form a gas of, say black, fully packed loops, 
and the dual of vertical links a complementary gas of white fully packed 
loops. The two systems of loops are further constrained to {\it cross} 
each other at the nodes of the tetravalent graph.
\fig{Folding constraints for quadrangulations. In order for the folded
image of a face to be a square of side unity, we must have
$t_{1}=-t_{3}=\pm A$ and $t_{2}=-t_{4}=\pm C$
where $A$ and $C$ are two orthogonal unit vectors. These
constraints propagate to the neighboring squares so that
each square has its two horizontal links equal to
$A$ and $B=-A$, and its two vertical links equal to $C$ and $D=-C$.
}{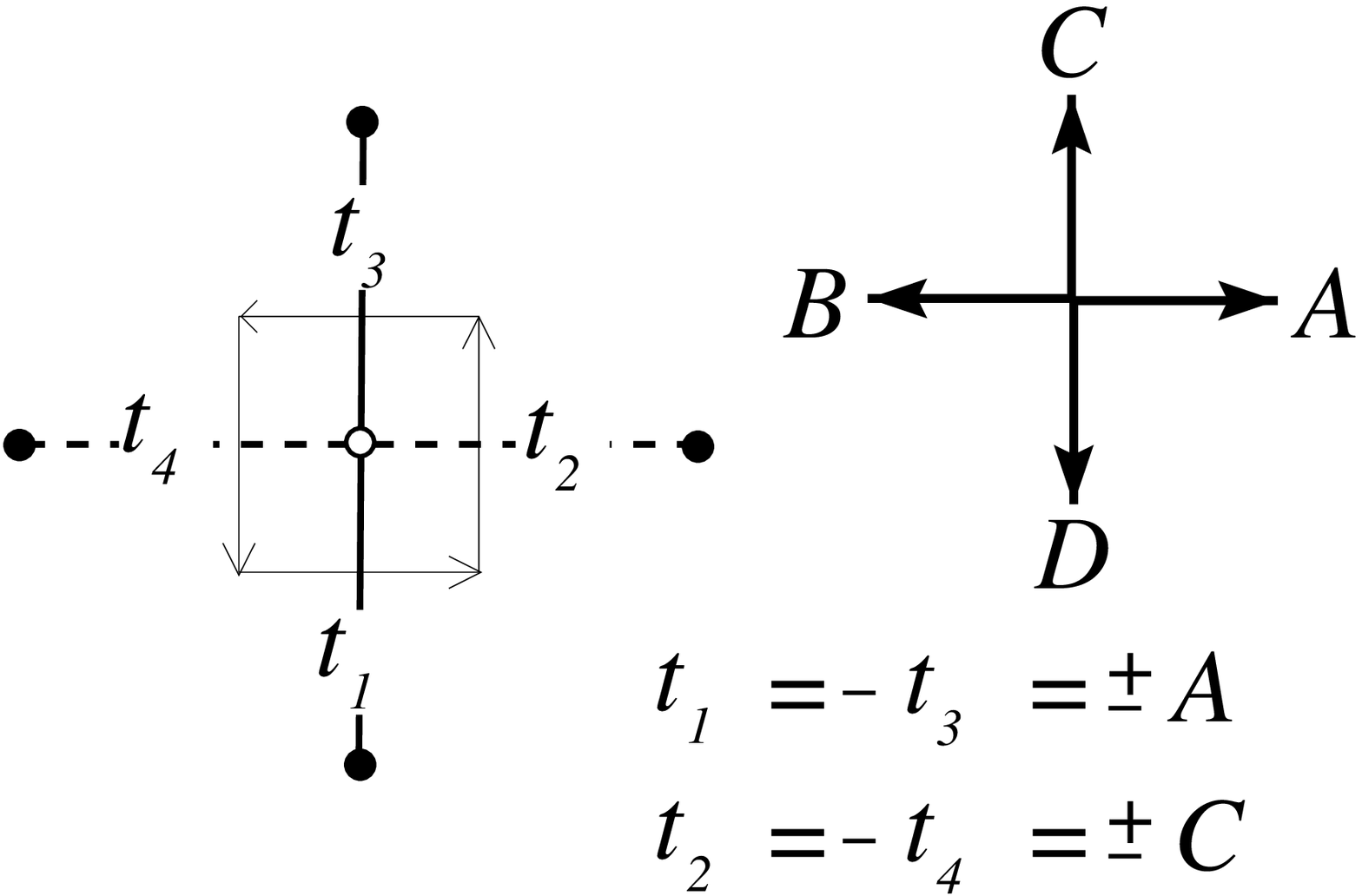}{8.cm}
\figlabel\pliquad
It is then clear that, in a folded configuration of the
quadrangulation, the faces are mapped on unit squares of
the square lattice, hence the link variables (defined on the oriented links) 
may take only two values, say $A$ and $B=-A$ on the horizontal links,
while those on the vertical links may take two values
$C$ and $D=-C$, where $A$ and $C$ are orthogonal unit vectors
(see figure \pliquad).
Moreover, the two horizontal (resp. vertical) links around a given
face have opposite values.
\fig{Example of planar foldable quadrangulation (a) in a spherical 
representation.
The corresponding black and white fully packed loops
(b) are entirely fixed. Only their orientation remains arbitrary.
The same loops in a planar representation (c) form a configuration
of the $GFPL^2(2,2)$ model. The orientation of the white loops determines
which horizontal links are folded, according to the rule 
displayed on the right.
A similar rule determines which vertical links which are folded
according to  the orientation of the black loops (not represented
here). For the particular
configuration of loops drawn here, the links of the ``perimeter"
joining the nodes labeled $1$ to $14$ are necessarily folded
{\it irrespectively of the orientation of the black and white loops}.
The other links may be folded or not. For instance,
the particular choice of orientation of the white loops represented here in
(c) corresponds to having a horizontal fold between the nodes $12$ and
$8$ in the representation (b), with thus four folded links.
}{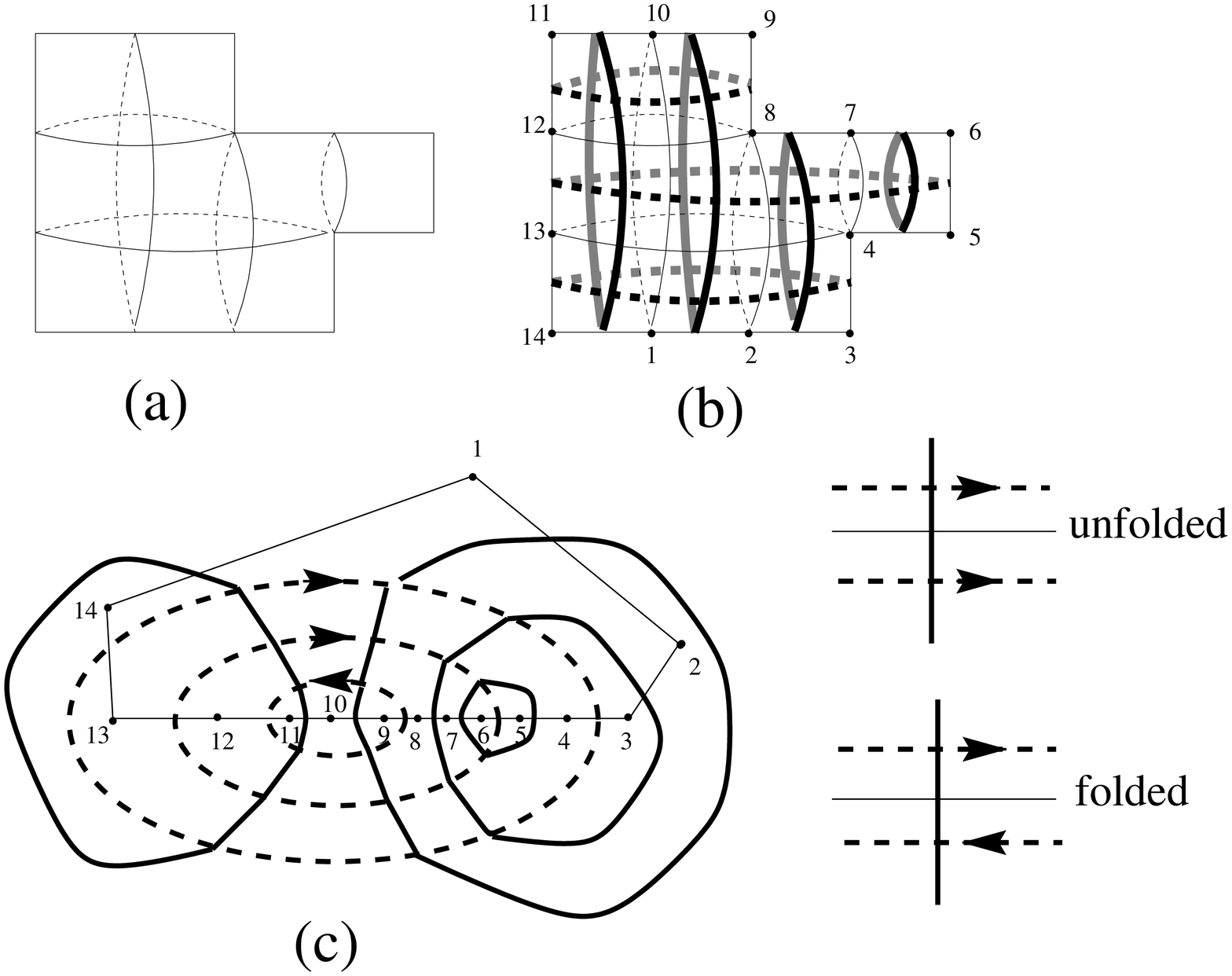}{12.cm}
\figlabel\pliagequadran
In a dual representation, we can transfer
the ``color" A, B, C or D of the link variable to the dual link, in which
case the colors alternate between A and B along the black loops,
and between C and D along the white ones. 
Orienting each link of color A (resp. C)
from the adjacent black node,
and each link of color B (resp. D) from the adjacent white 
node, assigns a well-defined orientation to each loop.
The orientations are independent on each loop, 
which precisely defines the $GFPL^2(2,2)$ model. 
The orientation of the loops corresponds
to the folding degree of freedom, with the correspondence that
a horizontal (resp.
vertical) link is folded if and only if the two white
(resp. black) loops immediately on each side of the dual link have opposite
orientations (see figure \pliagequadran).
Conversely, the height variable
associated to a configuration of the $GFPL^2(2,2)$ model
(which by definition allows only for crossing vertices)
with the rules of figure \fpldeuxrules\ and the choice $B=-A$ and $D=-C$
is nothing but the position of the nodes in the corresponding folded
configuration.
To summarize, the $GFPL^2(2,2)$ model describes the planar
folding of random quadrangulations.
\bigskip

\noindent$\diamond$ \underbar{\sl Complete folding, partial folding and
non-unfoldable quadrangulations}
\par\nobreak
The situation is simpler than that of triangulations
for the following reason. Given any foldable quadrangulation,
or rather its dual tetravalent node-bicolored graph, there
exists clearly a {\it unique configuration of fully packed loops}
(up to a global interchange of black and white loops) compatible 
with the rule that loops have to cross each other
at each vertex. For a given graph, there is no entropy associated with 
the positioning of the loops. The folding entropy is therefore entirely encoded
in the choice of orientations for the loops.
\fig{Orientation of the loops leading to the complete folding. The black loops
are oriented alternatively in the counterclockwise and clockwise direction
as we go deeper in the graph from the outermost loops. In this way,
black loops on both sides of any white link
always have opposite orientations, hence all vertical links are folded. 
The complete folding is obtained by orienting the white loops
accordingly so that all horizontal links are 
folded as well.}{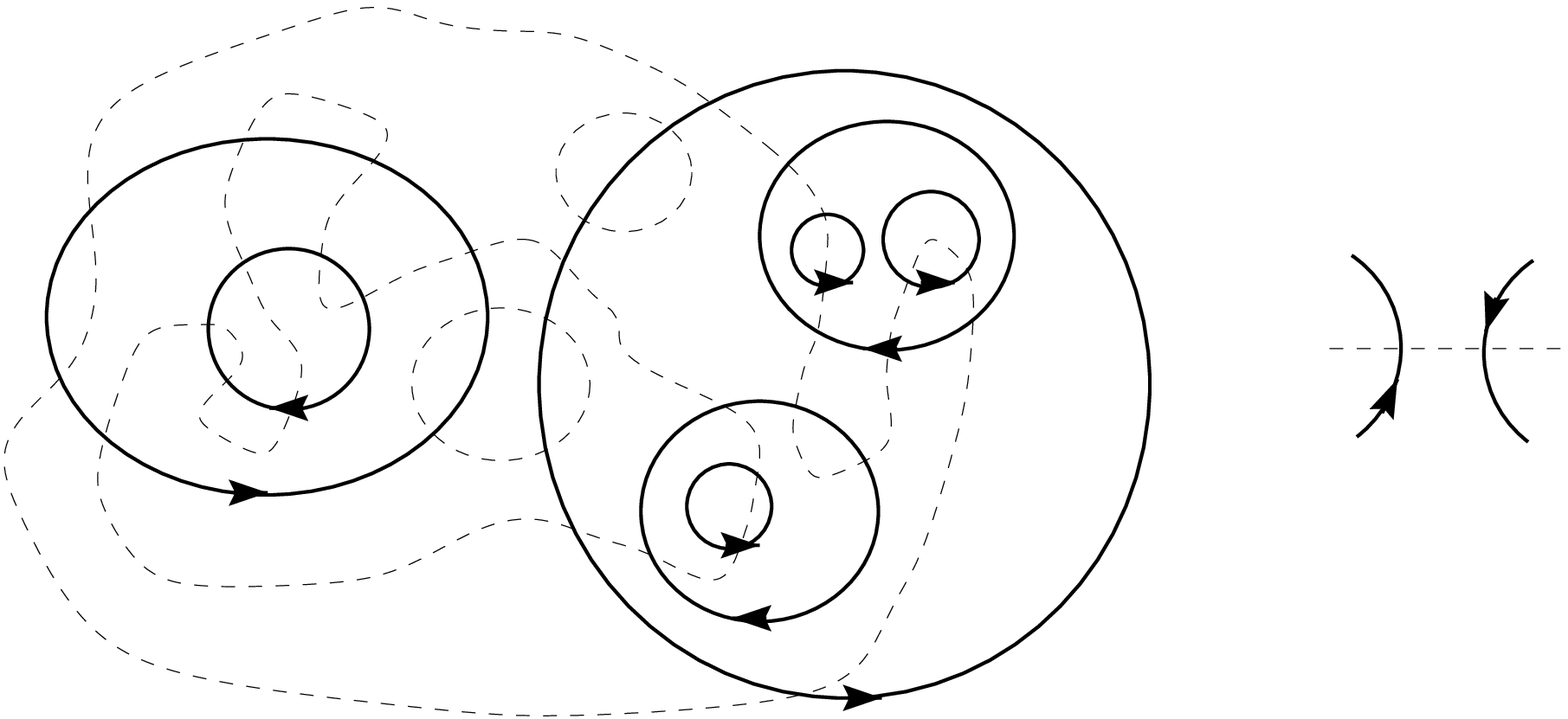}{12.cm}
\figlabel\completfold
A first consequence is that the {\it complete} folding of quadrangulations, 
i.e.\ their folding onto a single square is simply described by the 
$GFPL^2(1,1)$ model. Indeed, the folding onto a single square always exists
and is unique for each foldable quadrangulation. On the tetravalent graph,
it is obtained by the unique orientation of the loops defined 
as follows:
\item{1.} We orient the outermost black loops counterclockwise;
\item{2.} Inside each of these loops, we orient clockwise 
the outermost black loops;
\item{3.} We continue iteratively for deeper loops
by alternating the counterclockwise and clockwise orientations;
\item{4.} We repeat the procedure 1-3 for white loops.
\par
One clearly sees (see figure \completfold) that all the loops
of a given color on neighboring nodes have opposite orientations,
hence all links are folded.

The foldable quadrangulations being in one-to-one correspondence with
their complete foldings, we deduce that the $GFPL^2(1,1)$ model
also counts Eulerian quadrangulations, whose number is given by Eq.\biq.

Similarly, the $GFPL^2(1,2)$ model
describes the {\it partial} folding of quadrangulation, i.e.\ 
the {\it complete folding in one direction only}, with
all vertical links folded, while the horizontal ones may be folded or not.
The above interpretations are compatible with
the values of the central charge  $c_{\rm meander}(2,2)=2$
for the planar folding, $c_{\rm meander}(1,1)=0$ for
the complete folding, and $c_{\rm meander}(1,2)=1$ for
the partial folding in one direction.

The $GFPL^2(1,q)$ model was discussed in Section 10.3
with in particular the prediction \asymenla-\randalph\ 
for the asymptotics of the number of folded configurations.
In particular for $q=1$, one recovers $R(1,1)=9/2$ as
apparent on formula \biq. For $q=2$, the
number of configurations of partial folding in one
direction growths like $(\pi^2/2)^{2s}$ which,
divided by the $(9/2)^{2s}$ foldable quadrangulations
gives an average number of foldings varying as $(\pi^2/9)^{2s}$,
i.e.\ $(\pi^2/9)$ per horizontal link.

Finally, the original meander problem with
$n_1=n_2=0$ may be interpreted as describing the {\it foldable but
``non-unfoldable" quadrangulations},
i.e.\ the quadrangulations for which the complete folding is the
only possible one. Indeed, requiring that no unfolding
is possible amounts to requiring that all the relative orientations
of the loops are fixed, which is true if and only
if there is a unique loop of either species.

The mixed situation where $n_1=0$ only amounts to selecting
the quadrangulations which are non-unfoldable on vertical links,
but possibly unfoldable on the horizontal links. For
$n_2=q=1$, one imposes a complete folding of these horizontal links
(although some of them could a priori be unfolded) while for
$n_2=q=2$, one allows for unfolding the horizontal unfoldable links.

To conclude, the passage via the meander picture has allowed
to prove 
that the enumeration of foldings on a single stamp of a closed,
one-dimensional and self-avoiding strip of $2n$ stamps is
equivalent to the enumeration of planar, two-dimensional phantom
random quadrangulations with $2n$ faces, which are both foldable
and ``non-unfoldable", a quite amazing result.

\bigskip
\vfill\eject
\leftline{\bf CONCLUSION}
\bigskip
In this paper, we tried to give a unified presentation
of a number of problems arising from the statistical physics of folding. 
We explained the intimate links between foldings of lattices or graphs, 
colorings of their links and/or nodes, as well as fully packed loop gases. 
Beside phantom folding, the introduction of self-avoidance in one
dimension led us to the meander problem, itself related to phantom folding of
random quadrangulations.
We have introduced various formulations of these problems:
constrained spin variables, height models, Coulomb gas description,
all giving different angles of approach.

In the case of regular lattices, the foldings in the presence
of bending energy give rise to rich phase diagrams with conformational 
transitions intimately linked to the underlying geometry. 
In the case of random lattices, a similar sensitivity to the underlying 
graph was observed. It leads in particular to different universality 
classes according to the nature of the underlying lattice, distinguishing
between ordinary and Eulerian gravity. More precisely, a shift of
the central charge $c\to c+1$ of the associated field theory is 
observed when going from dense loops to fully packed loops provided
the latter are defined on bipartite lattices.

This phenomenon seems to contradict the more familiar notion of universality
in statistical physics. Note however that a similar breakdown of universality
occurs in other systems for which
the symmetry of the problem is linked to the ambient geometry.
A particular example is that of ``hard objects", say particles occupying
the nodes of a lattice so that no two adjacent nodes be simultaneously
occupied [\xref\GF,\xref\RUN]. On the regular lattice, these include the 
famous ``hard square" problem on the square lattice \BAXHS, and the 
``hard hexagon" problem on the triangular lattice \BAXHH\ (see also \BAX). 
These two problems 
display a crystallization transition between an ordered phase at high density 
of particles and a disordered phase of low density. 
The universality class of the transition 
point differs however for these two problems. This is easily understood
by observing that the symmetry of the high density groundstates 
(corresponding to having particles on one sub-lattice made 
of next-nearest neighbors on the original lattice) differs.
The square lattice is bipartite and has two such sub-lattices,
leading to an Ising-like transition point, while the triangular lattice 
is tripartite with three such sub-lattices, leading to a transition
point in the universality class of the critical 3-state Potts model.
When going to random lattices, a similar sensitivity was observed
in the following sense: it was shown \HARD\ that hard particles on 
random tetravalent graphs display no crystallization transition 
when defined on arbitrary graphs (ordinary gravity) while the
transition is restored for bipartite graphs (Eulerian gravity).
A simple heuristic explanation for this is that ordered groundstates
exist only when the graph is bipartite.
Remarkably, the ordinary or Eulerian nature of the graph is again 
what matters.

Another surprise of our study is the appearance of
irrational critical exponents for {\it a priori} simple combinatorial 
problems. Here again, other examples have been found. For instance, it
was recently conjectured \SZ\ that the enumeration
of closed planar curves with $n$ intersections (a
problem related to the enumeration of alternating knots)
corresponds to the coupling to gravity of a
conformal theory with central charge $c=-1$, and that
the asymptotic large $n$  behavior is characterized
by the same irrational exponent \gammacmone\ as in Section 8.

In these notes, we have limited ourselves to the simplest
examples of folding. As we mentioned already, our description
may be extended to the case of all compactly 2D foldable lattices
of Fig.\classi, or higher $D$-dimensional simplices, all with
their own (multicolored) fully packed loop gas formulation and 
with their own symmetry. 
Similarly, meanders may be generalized in various ways: 
\item{-} meanders drawn on surfaces with higher genus \MFAS, 
accessible via topological
expansion of the black and white matrix model of Section 10.3; 
\item{-} colored meanders, based on multicolor generalizations of the
Temperley-Lieb algebra [\xref\FOLCO,\xref\BJ];
\item{-} $SU(N)$ meanders, based on higher quotients of the Hecke algebra 
generalizing the Temperley-Lieb algebra ($N=2$) \SUM.
\par
Finally, a very natural question is that of a proper description of 
self-avoidance for lattices of dimension $D\geq 2$. At this time, 
only enumerations for rectangular domains of the square lattice with
very small sizes are known \LUN\ and no systematic approach has been 
devised yet.  

The most advanced results presented here all rely on field-theoretical
effective descriptions. It would be desirable to obtain the same results
via a more traditional combinatorial approach. A first step in
this direction was performed recently [\xref\SCH,\xref\CENSUS,\xref\MOB] 
by solving various
planar graph enumeration problems with combinatorics of trees.
This approach is based on a proper cutting of the graphs into trees
with local characterizations. Alternatively, it relates large graph statistics
to properties of the so-called Integrated SuperBrownian Excursion (ISE),
a probabilistic description of tree embeddings \CS. 
In view of these developments, we may expect similar technique to apply to 
loop gas systems on random graphs. 
Another probabilistic direction concerns stochastic Loewner
evolution (SLE) in the context of which various exponents, predicted
in Ref.\DUP\ by use of the equation \kpz, were derived rigorously 
with the help of Brownian motion \LSW. 

As a challenge for the reader, the simple combinatorial problem
displayed in Fig.\Oneuler\ has a simple transfer matrix formulation
and we may hope for an exact analytic solution to be within reach. 

\listrefs
\end